\documentclass[11pt]{article}
\usepackage{amsgen,amsmath,amstext,amsbsy,amsopn,amssymb,mathabx,mathrsfs,amsthm,bm,bbm,romannum,enumitem,natbib,xr,multirow,makecell,setspace}
\usepackage[labelfont=bf]{caption}
\usepackage[dvips]{graphicx}
\graphicspath{{figures/}}
\usepackage{subcaption}
\usepackage[ruled,vlined,linesnumbered,resetcount]{algorithm2e}

\usepackage{hyperref}
\hypersetup{
	colorlinks=true,
	linkcolor=blue,
	urlcolor=blue,
	citecolor=blue
}
\usepackage[depth=2]{bookmark}

\usepackage[margin=1in]{geometry}

\usepackage[utf8]{inputenc}

\usepackage{booktabs}
\usepackage{longtable}
\usepackage{array}
\usepackage{multirow}
\usepackage[table]{xcolor}
\usepackage{wrapfig}
\usepackage{float}
\usepackage{colortbl}
\usepackage{pdflscape}
\usepackage{tabu}
\usepackage{threeparttable}
\usepackage{threeparttablex}
\usepackage[normalem]{ulem}
\usepackage{makecell}

\newtheorem{thm}{Theorem}
\newtheorem{lem}{Lemma}
\newtheorem{prop}[thm]{Proposition}

\theoremstyle{definition}
\newtheorem{dfn}{Definition}
\newtheorem{asm}{Assumption}

\newtheorem*{rmk}{Remark}

\theoremstyle{remark}

\newcommand{\fb}{\mathbf{f}}

\newcommand{\ba}{\bm{a}}

\newcommand{\bd}{\bm{d}}

\newcommand{\bbf}{\bm{f}}

\newcommand{\bh}{\bm{h}}

\newcommand{\bu}{\bm{u}}

\newcommand{\bx}{\bm{x}}

\newcommand{\Ib}{\mathbf{I}}

\newcommand{\bH}{\bm{H}}

\newcommand{\bS}{\bm{S}}

\newcommand{\bX}{\bm{X}}

\newcommand{\bbeta}{\bm{\beta}}
\newcommand{\bmeta}{\bm{\eta}}

\newcommand{\bomega}{\bm{\omega}}

\newcommand{\btau}{\bm{\tau}}

\newcommand{\bphi}{\bm{\phi}}

\newcommand{\bbE}{\mathbb{E}}

\newcommand{\bbI}{\mathbb{I}}

\newcommand{\bbP}{\mathbb{P}}

\newcommand{\bbR}{\mathbb{R}}

\newcommand{\cA}{\mathcal{A}}
\newcommand{\cB}{\mathcal{B}}

\newcommand{\cD}{\mathcal{D}}

\newcommand{\cH}{\mathcal{H}}
\newcommand{\cI}{\mathcal{I}}

\newcommand{\cN}{\mathcal{N}}
\newcommand{\cO}{\mathcal{O}}

\newcommand{\cV}{\mathcal{V}}

\newcommand{\cX}{\mathcal{X}}

\newcommand{\sA}{\mathscr{A}}
\newcommand{\sB}{\mathscr{B}}

\newcommand{\sX}{\mathscr{X}}

\newcommand{\sfA}{\mathsf{A}}
\newcommand{\sfB}{\mathsf{B}}

\newcommand{\sfF}{\mathsf{F}}

\newcommand{\sfH}{\mathsf{H}}
\newcommand{\sfI}{\mathsf{I}}

\newcommand{\sfO}{\mathsf{O}}

\newcommand{\sfV}{\mathsf{V}}
\newcommand{\sfW}{\mathsf{W}}

\newcommand\smallcO{
	\mathchoice
	{{\scriptscriptstyle\mathcal{O}}}
	{{\scriptscriptstyle\mathcal{O}}}
	{{\scriptscriptstyle\mathcal{O}}}
	{\scalebox{.5}{$\scriptscriptstyle\mathcal{O}$}}
}

\newcommand{\argmin}{\mathop{\mathrm{argmin}}}
\newcommand{\argmax}{\mathop{\mathrm{argmax}}}

\newcommand{\sfvar}{\mathsf{Var}}

\newcommand{\tr}{\mathsf{trace}}	    
\newcommand{\sfvec}{\mathsf{Vec}}

\newcommand{\bzero}{\mathbf{0}}	
\newcommand{\bone}{\mathbf{1}}	
\newcommand{\bbone}{\bbI}	

\newcommand\indep{\protect\mathpalette{\protect\independenT}{\perp}}
\def\independenT#1#2{\mathrel{\rlap{$#1#2$}\mkern2mu{#1#2}}}	

\def\ubar#1{\underbar{$ #1 $}}

\newcommand{\bern}{\mathsf{Bernoulli}}

\newcommand{\unif}{\mathsf{Uniform}}

\newcommand{\rd}{\mathrm{d}}
\newcommand{\eff}{\mathrm{eff}}
\newcommand{\opt}{\mathrm{opt}}

\title{Efficient Learning of Optimal Individualized Treatment Rules for Heteroscedastic or Misspecified Treatment-Free Effect Models}
\author{
	Weibin Mo and Yufeng Liu\thanks{\textit{Address for correspondence}: Yufeng Liu, Department of Statistics and Operations Research, Department of Genetics, Department of Biostatistics, Carolina Center for Genome Sciences, Lineberger Comprehensive Cancer Center, University of North Carolina at Chapel Hill. 354 Hanes Hall, Chapel Hill, NC 27599, USA. E-mail: \href{mailto:yfliu@email.unc.edu}{yfliu@email.unc.edu}.
	} 
	\\ 
	University of North Carolina at Chapel Hill
}
\date{}

\begin{document}
	\pagenumbering{arabic}
	\maketitle
	\begin{abstract}
		Recent development in data-driven decision science has seen great advances in individualized decision making. Given data with individual covariates, treatment assignments and outcomes, researchers can search for the optimal individualized treatment rule (ITR) that maximizes the expected outcome. Existing methods typically require initial estimation of some nuisance models. The double robustness property that can protect from misspecification of either the treatment-free effect or the propensity score has been widely advocated. However, when model misspecification exists, a doubly robust estimate can be consistent but may suffer from downgraded efficiency. Other than potential misspecified nuisance models, most existing methods do not account for the potential problem when the variance of outcome is heterogeneous among covariates and treatment. We observe that such heteroscedasticity can greatly affect the estimation efficiency of the optimal ITR. In this paper, we demonstrate that the consequences of misspecified treatment-free effect and heteroscedasticity can be unified as a covariate-treatment dependent variance of residuals. To improve efficiency of the estimated ITR, we propose an \textit{Efficient Learning (E-Learning)} framework for finding an optimal ITR in the multi-armed treatment setting. We show that the proposed E-Learning is optimal among a regular class of semiparametric estimates that can allow treatment-free effect misspecification. In our simulation study, E-Learning demonstrates its effectiveness if one of or both misspecified treatment-free effect and heteroscedasticity exist. Our analysis of a \textit{Type 2 Diabetes Mellitus (T2DM)} observational study also suggests the improved efficiency of E-Learning.
		
		\noindent \textbf{Keywords and Phrases:} Double robustness; Heteroscedasticity; Individualized treatment rules; Model misspecification; Multi-armed treatments; Semiparametric efficiency
	\end{abstract}
	
	\newpage
	
\section{Introduction}

Individualized decision making is very essential in various scientific fields. One of the common goals is to find the optimal \textit{individualized treatment rule (ITR)} mapping from the individual characteristics or contextual information to the treatment assignment, that maximizes the expected outcome, known as the \textit{value function} \citep{manski2004statistical,qian2011performance}. Such a goal can be seen from applications in many different areas. In disease management, the physician needs to decide whether to introduce or switch a therapy based on patients' characteristics in order to optimize his/her clinical outcome \citep{bertsimas2017personalized}. In public management, decision makers may seek for a policy that allocates resources based on individual profiles and maximizes the overall efficiency \citep{kube2019allocating}. In a context-based recommender system, contextual information such as time, location and social connection can be incorporated to increase effectiveness of the recommendation \citep{aggarwal2016recommender}. 

There is a vast literature on estimating an optimal ITR. Among various existing methods, there are two main strategies. The first strategy is to estimate the outcome mean model given covariates and treatment, which is often referred as the \textit{model-based} approach. The optimal ITR can be induced by maximizing the mean outcome over treatment conditional on covariates. Existing methods including Q-Learning \citep{watkins1989learning,qian2011performance}, A-Learning \citep{murphy2003optimal,lu2013variable,shi2018high}, \textit{dynamic Weighted Ordinary Least Square (dWOLS)} \citep{wallace2015doubly} and \textit{Robust D-Learning (RD-Learning)} \citep{meng2020robust} all fall into this category. Some related approaches focus on a transformed outcome model, such as the Subgroup Identification approach \citep{tian2014simple,chen2017general} and D-Learning \citep{qi2018d,qi2020multi}. The second strategy, known as the \textit{direct-search} approach, is to estimate the value function nonparametrically, and maximize the value function estimate over a prespecified ITR class to obtain an optimal ITR. A well-known example using this strategy is the \textit{inverse-probability weighted estimate (IPWE)} \citep{zhao2012estimating,zhou2017residual,liu2018augmented,kitagawa2018should}. For these two strategies, the model-based approach relies on a correctly specified outcome mean model, while the direct-search approach based on the IPWE requires correctly estimating the propensity score function. In order to improve these two strategies, various papers proposed to combine the strength of both. In particular,  \citet{zhang2012robust,zhao2019efficient,athey2021policy} considered to combine the outcome model with the IPWE to obtain the \textit{augmented IPWE (AIPWE)} of the value function. Such an estimate can be more robust to the model specification for the outcome model or the propensity score model.

Among the aforementioned approaches, the \textit{double robustness} property has been studied and advocated to protect from potential model misspecifications. In the model-based approaches, the optimal ITR only depends on the interaction effect between covariates and treatment within the outcome mean model. Then the treatment-free effect that only depends on covariates can be a nuisance component. \citet{robins2004optimal} investigated the incorrectly specified parametric model for the treatment-free effect, and introduced the G-estimating equation that can incorporate additional information from the propensity score. The G-estimator can be doubly robust in the sense that the estimate remains consistent even if one of the treatment-free effect model and the propensity score model is misspecified. As special cases, \citet{lu2013variable,ertefaie2021robust} developed least-squares approaches that can equivalently solve the G-estimating equation and enjoy double robustness.  \citet{wallace2015doubly,meng2020robust} took a different approach to hedge the risk of treatment-free effect misspecification. Specifically, they proposed the weighted least-squares problem that utilizes the propensity score information to construct balancing weights, and the resulting estimates can also be doubly robust. In the direct-search approaches, the AIPWE of the value function is doubly robust in a slightly different way. Specifically, the AIPWE incorporates the outcome mean function and the propensity score function. When estimating the outcome mean and propensity score functions, even if one of their model specifications is incorrect, the corresponding AIPWE can still remain consistent.

The double robustness property has also been widely studied in the causal inference literature \citep{robins1994estimation,robins1995analysis,ding2018causal}. One problem of particular interest is to study the case when one of or both model misspecifications happen. \citet{kang2007demystifying} provided a comprehensive empirical study on how model misspecification can affect the resulting estimates. They concluded that the misspecified outcome mean model can be generally more harmful than the misspecified propensity score model. When both models are misspecified, the doubly robust estimate can perform even worse than the IPWE. Later studies further developed improved estimates and inference procedures to overcome such challenges \citep{tan2010bounded,rotnitzky2012improved,vermeulen2015bias,benkeser2017doubly}. These studies have also motivated some improvement of the AIPWE for the ITR problem. Specifically, when the outcome mean model is incorrectly specified, \citet{cao2009improving} proposed an optimal estimation strategy for the misspecified outcome mean model in the sense that the resulting AIPWE can have the smallest variance. \citet{pan2021improved} further extended this work to the ITR problem, and utilized augmented inverse-probability weighted estimating equations for the outcome mean model estimation.

Motivated from \citet{kang2007demystifying} that the misspecified treatment-free effect can have more severe consequence, we focus on addressing this challenge. In our study, we find that the misspecified treatment-free effect in the model-based approach can have a consequence similar to heteroscedasticity \citep{carroll1982adapting}. More specifically, both misspecified treatment-free effect and heteroscedasticity can cause the variance of residuals being dependent on covariates and treatment. Therefore, we take the approach of semiparametric efficient estimation under heteroscedasticity \citep{ma2006efficient} and propose an \textit{Efficient Learning (E-Learning)} framework for the optimal ITR in the multi-armed treatment setting. Our proposed E-Learning can enjoy the following properties: 
\begin{enumerate}[leftmargin=*]
	\item When nuisance models are correctly specified, E-Learning performs semiparametric efficient estimation. Our framework can allow the variance of outcome depends on covariates and treatment, and hence is more general than existing semiparametric efficient procedures such as G-Estimation and its equivalents; 
	
	\item E-Learning is doubly robust with respect to the treatment-free effect model and the propensity score model; 
	
	\item In presence of misspecified treatment-free effect, E-Learning is optimal with the minimal $ \sqrt{n} $-asymptotic variance among a regular class of semiparametric estimates based on the given working treatment-free effect function. Our optimality incorporates the standard semiparametric efficiency \citep{tsiatis2007semiparametric} as a special case for the ITR problem.
\end{enumerate}
This paper contributes to existing literature in terms of the followings: 
\begin{enumerate}[leftmargin=*]
	\item Parallel to the improved doubly robust procedure in \citet{pan2021improved} for direct-search approaches, E-Learning is an improved doubly robust method for model-based approaches. Specifically, E-Learning performs optimal efficiency improvement when one of or both misspecified treatment-free effect and heteroscedasticity exist; 
	
	\item E-Learning incorporates many existing approaches as special cases, including Q-Learning, G-Estimation, A-Learning, dWOLS, Subgroup Identification, D-Learning and RD-Learning. It provides a more general framework to study the double robustness and estimation efficiency for these methods;
	
	\item We develop E-Learning for the setting with multiple treatments. In particular, E-Learning utilizes a generalized equiangular coding of multiple treatment arms to develop the efficient estimating function. This can be the first work to incorporate equiangularity in the semiparametric framework among those utilizing the equiangular coding \citep{zhang2014multicategory,zhang2020multicategory,qi2020multi,meng2020near,xue2021multicategory};
	
	\item In our simulation study, our proposed E-Learning demonstrates superior performance over existing methods when one of or both misspecified treatment-free effect and heteroscedasticity exist, which confirms the superior performance of the proposed E-Learning. In the analysis of a \textit{Type 2 Diabetes Mellitus (T2DM)} observational study, E-Learning also demonstrates its improved efficiency compared to other methods.
\end{enumerate}

The rest of this paper is organized as follows. In Section \ref{sec:method}, we introduce the methodology of E-Learning. In particular, mathematical setups and notations are introduced in Section \ref{sec:setup}. A motivating example is discussed in Section \ref{sec:eg} to demonstrate the consequence of misspecified treatment-free effect and heteroscedasticity. Semiparametric efficient estimating equation is developed in Section \ref{sec:eff}. E-Learning and its implementation details are proposed in Sections \ref{sec:eLearn} and \ref{sec:implement}. In Section \ref{sec:connect}, we discuss the connection of E-Learning with the existing literature. In Section \ref{sec:theory}, we establish theoretical results for E-Learning. Simulation studies and the application to the T2DM dataset are provided in Sections \ref{sec:simulation} and \ref{sec:t2dm} respectively. Some discussions are given in Section \ref{sec:discuss}. Additional discussions, including nonlinear simulation studies, an analysis of the ACTG 175 dataset, technical proofs, additional tables and figures can be found in the Supplementary Material. The \texttt{R} code for the implementation of this paper is available at \url{https://github.com/harrymok/E-Learning.git}.

\section{Methodology}\label{sec:method}

In this section, we first introduce the ITR problem as a semiparametric estimation problem. Then we study the semiparametric efficient estimation procedure and propose E-Learning.

\subsection{Setup}\label{sec:setup}

Consider the data $ (\bX,A,Y) $, where $ \bX \in \cX \subseteq \bbR^{p} $ denotes the covariates, $ A \in \cA = \{ 1,2,\cdots,K \} $ is the treatment assignment with $ K $ treatment options, and $ Y \in \bbR $ is the observed outcome. For $ 1 \le k \le K $, let $ Y(k) $ be the potential outcome under the assigned treatment $ k $. An ITR is a mapping from covariates to treatment assignment $ d: \cX \to \cA $. The \textit{value function} of an ITR is defined as $ \cV(d) := \bbE[Y(d(\bX))] $. Assuming that a larger outcome is better, the goal is to find the optimal ITR that maximizes the value function $ d^{\star} \in \argmax_{d: \cX \to \cA}\cV(d) $. 

Assume the identifiability conditions \citep{rubin1974estimating}: (\textit{consistency}) $ Y = Y(A) $; (\textit{unconfoundedness}) $ A \indep \{ Y(k) \}_{k=1}^{K} | \bX $; (\textit{strict overlap}) for $ 1 \le k \le K $, $ \bbP(A = k|\bX)  \ge \ubar{p}_{\sA} $ for some $ \ubar{p}_{\sA} > 0 $. Then the value function can be written as $ \cV(d) = \bbE[Y|A=d(\bX)] = \bbE\left\{ \sum_{k=1}^{K}\bbE(Y|\bX,A=k)\bbone[d(\bX) = k]\right\} $. Consequently, the optimal ITR satisfies $ d^{\star}(\bx) \in \argmax_{1 \le k \le K}\bbE(Y|\bX=\bx,A=k) $ for any $ \bx \in \cX $. This motivates us to study the following semiparametric model:
\begin{align}
	\begin{array}{rl}
		&Y = \mu_{0}(\bX) + \gamma(\bX,A;\bbeta) +  \epsilon,\\
		\text{subject to} & \sum\limits_{k=1}^{K}\gamma(\bX,k;\bbeta) = 0; \quad \bbE(\epsilon|\bX,A) = 0; \quad \sigma^{2}(\bX,A) := \bbE(\epsilon^{2}|\bX,A) < +\infty; \\
		& (\bX,A,\epsilon) \sim p_{\sX}(\bx)p_{\sA}(a|\bx)p_{\epsilon}(\epsilon|\bx,a).
	\end{array} \label{eq:model}
\end{align}
Here, $ \mu_{0}(\bX) $ is the \textit{treatment-free effect}, and $ \gamma(\bX,A;\bbeta) $ is the \textit{interaction effect} between $ \bX $ and $ A $ that is parametrized by the $ p $-dimensional parameter vector $ \bbeta \in \sB \subseteq \bbR^{p} $. In particular, it requires that the parametrized interaction effect satisfies a sum-to-zero constraint for identifiability. The dependency on $ \bbeta $ may be suppressed for ease of notation in our later presentation. Moreover, $ \sigma^{2}(\bX,A) $ is the \textit{variance function} of $ \epsilon $ that can depend on $ (\bX,A) $. Finally, $ p_{\sX}(\bx) $, $ p_{\sA}(a|\bx) $ and $ p_{\epsilon}(\epsilon|\bx,a) $ are density functions. Then the nuisance component $ \eta := (p_{\sX},p_{\sA},p_{\epsilon},\mu_{0}) $ is left unspecified only with the moment restriction $ \int \epsilon p_{\epsilon}(\epsilon|\bx,a)\rd \epsilon = 0 $.  

Given the true parameter $ \bbeta $ in Model (\ref{eq:model}), the optimal ITR is $ d^{\star}(\bx) \in \argmax_{1 \le k \le K}\gamma(\bx,k;\bbeta) $. In Theorem \ref{thm:regret_est} below, we  show that maximizing the value function can be directly related to finding a good estimate of the interaction effect $ \gamma(\bX,A) $ in Model (\ref{eq:model}). 

\begin{thm}[Estimation and Regret Bound]\label{thm:regret_est}
	Consider Model (\ref{eq:model}). Let $ \widehat{\gamma}_{n}(\bX,A) $ be an estimate of $ \gamma(\bX,A) $, $ \widehat{d}_{n}(\bx) \in \argmax_{1 \le k \le K}\widehat{\gamma}_{n}(\bx,k) $, and $ d^{\star}(\bx) \in \argmax_{1 \le k \le K}\gamma(\bx,k) $. Then
	\[ \cV(d^{\star}) - \cV(\widehat{d}_{n}) \le 2\max_{1 \le k \le K}\bbE\left |\widehat{\gamma}_{n}(\bX,k) - \gamma(\bX,k)\right |. \]
	Here, $ \widehat{\gamma}_{n} $ is fixed and $ \bbE $ takes expectation over $ \bX $.
\end{thm}

The proof of Theorem \ref{thm:regret_est} is similar to \citet[Lemma 2]{murphy2005generalization} and is included in the Supplementary Material. It implies that minimizing the estimation error of the interaction effects $ \{\gamma(\bX,k)\}_{k=1}^{K} $ can also minimize the regret. In this paper, we focus on finding an efficient estimate of the parametric interaction effect $ \gamma(\bX,A;\bbeta) $.

\subsection{A Motivating Example}\label{sec:eg}

We introduce a motivating example to demonstrate that several existing approaches, including Q-Learning, G-Estimation, A-Learning, dWOLS, Subgroup Identification, D-Learning and RD-Learning, may not be optimal if either the treatment-free effect $ \mu_{0}(\bX) $ is misspecified, or the variance function $ \bbE(\epsilon^{2}|\bX,A) $ depends on $ (\bX,A) $. In contrast, the E-Learning estimate can be much more efficient. All these methods are compared in Section \ref{sec:connect}.

Consider the covariate $ X $ with a symmetric distribution on $ \bbR $, the treatment $ A \sim \bern(1/2) $, and the error term $ \epsilon \sim \cN(0,1) $, where $ X,A,\epsilon $ are mutually independent. Suppose the outcome $ Y $ is generated by
\[ Y = \underbrace{c_{1}|X|}_{\text{treatment-free effect}} + \underbrace{(A-1/2)\beta_{0}}_{\text{interaction effect}} + \sqrt{\underbrace{1+2c_{2}^{2}AX^{2}}_{\text{variance function}}}\epsilon, \]
for some $ \beta_{0} \ge 0 $. When estimating from the training data, suppose that we specify $ X\eta $ for the treatment-free effect with $ \eta $ to be estimated, and $ (A-1/2)\beta $ for the interaction effect with $ \beta $ to be estimated. If $ c_{1} = 0 $, then the treatment-free effect is correctly specified, with the true parameter $ \eta = 0 $; otherwise, the treatment-free effect is misspecified. If $ c_{2} = 0 $, then the variance function is $ 1 $, and homogeneous with respect to $ (X,A) $; otherwise, we have a heteroscedastic model with the variance of error depending on $ (X,A) $.

Denote $ \bbE_{n} $ as the empirical average over the training dataset of size $ n $. Then for this particular example,  Q-Learning \citep{watkins1989learning}, G-Estimation \citep{robins2004optimal}, A-Learning \citep{murphy2003optimal}, dWOLS \citep{wallace2015doubly}, Subgroup Identification \citep{tian2014simple}, D-Learning \citep{qi2018d} and RD-Learning \citep{meng2020robust} are equivalent to the following \textit{Ordinary Least-Squares (OLS)} problem:
\begin{align}
	(\widehat{\eta}_{n},\widehat{\beta}_{n}) \in \argmin_{\eta,\beta \in \bbR}\bbE_{n}[Y - X\eta - (A-1/2)\beta]^{2}.\label{eq:eg_ols}
\end{align}
Note that if $ c_{1} = c_{2} = 0 $ with correctly specified treatment-free effect and homoscedasticity, then $ \widehat{\beta}_{n} $ is semiparametric efficient. For the general $ c_{1} $ and $ c_{2} $, the OLS estimates $ \widehat{\beta}_{n} $ and $ \widehat{\eta}_{n} $ are asymptotically independent, with $ \sqrt{n}\widehat{\eta}_{n} \stackrel{\cD}{\to} \cN(0,\nu^{2}) $ for some $ \nu^{2} > 0 $ and
\[ \sqrt{n}(\widehat{\beta}_{n} - \beta_{0}) = \sqrt{n}[\bbE(A-1/2)^{2} +\smallcO_{\bbP}(1)]^{-1}\bbE_{n}\left[(A-1/2)\left(c_{1}|X| + \sqrt{1+2c_{2}^{2}AX^{2}}\epsilon\right)\right] \stackrel{\cD}{\to} \cN(0,v^{2}), \]
where the $ \sqrt{n} $-asymptotic variance of $ \widehat{\beta}_{n} $ is given by $ v^{2} = 4\bbE[1+(c_{1}^{2} + c_{2}^{2})X^{2}] = 4\bbE(1+c^{2}X^{2}) $ with $ c^{2} := c_{1}^{2} + c_{2}^{2} $. 
Notice that the residual is $ \widehat{e} = Y - X\widehat{\eta}_{n} - (2A-1)\widehat{\beta}_{n} = c_{1}|X| + \sqrt{1+2c_{2}^{2}AX^{2}}\epsilon + \cO_{\bbP}(n^{-1/2}) $. Then we have $ \bbE(\widehat{e}^{2}|X) = 1+c^{2}X^{2} + \cO_{\bbP}(n^{-1}) $, which clearly depends on $ X $.

Motivated from the heteroscedastic residual, we define $ \widecheck{v}_{\epsilon}(x) := 4(1 + c^{2}x^{2}) $. Consider the solutions to the generalized least-squares problem
\begin{align}
	(\widehat{\eta}_{\eff,n},\widehat{\beta}_{\eff,n}) \in \argmin_{\eta,\beta \in \bbR}\bbE_{n}\Big\{ \widecheck{v}_{\epsilon}^{-1}(X)[Y - X\eta - (A-1/2)\beta]^{2}\Big\}.\label{eq:eg_gls}
\end{align}
Then $ \widehat{\beta}_{\eff,n} $ and $ \widehat{\eta}_{\eff,n} $ are asymptotically independent, with $ \sqrt{n}\widehat{\eta}_{\eff,n} \stackrel{\cD}{\to} \cN(0,\widetilde{\nu}^{2}) $ for some $ \widetilde{\nu}^{2} > 0 $,
\begin{small}
	\[ \sqrt{n}(\widehat{\beta}_{\eff,n} - \beta_{0}) = \sqrt{n}\left\{ \bbE\left[ {(A-1/2)^{2} \over 4(1 + c^{2}X^{2})} \right] + \smallcO_{\bbP}(1) \right\}^{-1}\bbE_{n}\left[ {(A-1/2)\left( c_{1}|X| + \sqrt{1+2c_{2}^{2}AX^{2}}\epsilon \right) \over 4(1 + c^{2}X^{2})} \right] \stackrel{\cD}{\to} \cN(0,v_{\eff}^{2}), \]
\end{small}%
where the $ \sqrt{n} $-asymptotic variance of $ \widehat{\beta}_{\eff,n} $ is given by $ v_{\eff}^{2} = 4\left[\bbE(1 + c^{2}X^{2})^{-1}\right]^{-1} $.
The asymptotic relative efficiency of $ \widehat{\beta}_{\eff,n} $ with respect to $ \widehat{\beta}_{n} $ is $ {v^{2}/v_{\eff}^{2}} = \bbE(1+c^{2}X^{2})\bbE\left(  {1 \over 1 + c^{2}X^{2}}  \right) \ge 1 $.
That is, $ \widehat{\beta}_{\eff,n} $ has a smaller $ \sqrt{n} $-asymptotic variance than $ \widehat{\beta}_{n} $. The strict inequality generally holds if $ c \ne 0 $ and $ X $ is non-degenerate. 

Next we consider an extreme case to illustrate that $ \widehat{\beta}_{\eff,n} $ can be much more efficient than $ \widehat{\beta}_{n} $. Suppose $ X \sim qf^{(M)}(x) + (1-q)f^{(\infty)}(x) $, where $ f^{(M)}(x) $ is a symmetric \textit{probability density function (PDF)} with compact support on $ [-M,M] $, $ f^{(\infty)}(x) $ is a symmetric PDF on $ \bbR $ with $ \int_{\bbR} x^{2}f^{(\infty)}(x)\rd x = +\infty $, and $ q \in (0,1] $ is the mixture probability. Then for $ c \ne 0 $, $ v^{2} \ge 4[1+c^{2}\bbE_{X \sim f^{(\infty)}}(X^{2})] = +\infty $, while $ v_{\eff}^{2} \le \left[ q\bbE_{X \sim f^{(M)}}(1+c^{2}X^{2})^{-1} \right]^{-1} \le 4(1 + c^{2}M^{2})/q $. Here, $ v^{2} = +\infty $ implies that $ \widehat{\beta}_{n} $ cannot even be $ \cO_{\bbP}(n^{-1/2}) $, while in contrast, $ \widehat{\beta}_{\eff,n} $ has a bounded $ \sqrt{n} $-asymptotic variance $ v_{\eff}^{2} $. Therefore, if either the treatment-free effect is misspecified $ (c_{1} \ne 0) $, or the variance function is not homogeneous $ (c_{2} \ne 0) $, then $ \widehat{\beta}_{n} $ can have much worse perforamance than the more efficient estimate $ \widehat{\beta}_{\eff,n} $.

From the motivating example above, we can conclude that the efficiency of many existing approaches can be improved when either misspecified treatment-free effect or heteroscedasticity happens. In fact, our example shows that misspecified treatment-free effect or heteroscedasticity can cause the dependency of $ \bbE(\widehat{e}^{2}|X) = 1 + c^{2}X^{2} $ on $ X $. Motivated from efficient estimation under heteroscedasticity \citep{ma2006efficient} and our motivating example, we introduce the working variance function $ v_{\opt}(X) = 1 + c^{2}X^{2} $, and consider the generalized least-squares estimate as in (\ref{eq:eg_gls}). The estimation efficiency can be greatly improved in this case. 

\citet[Theorem 6]{xiao2019robust} pointed out a phenomenon similar to our finding in Section \ref{sec:eg}, while their methodology and theoretical properties differ from ours. To be specific, \citet{xiao2019robust} replaced the squared loss by general robust loss functions. Under the assumption $ \epsilon \indep A|\bX $, their estimate based on the quantile loss function can be shown consistent and $ \sqrt{n} $-asymptotic normal. However, it remains unclear whether the $\sqrt{n}$-asymptotic normality still holds, and if so, how large the corresponding $ \sqrt{n} $-asymptotic variance is, when treatment-free effect misspecification and heteroscedasticity exist. In contrast, we show in Theorem \ref{thm:eff_opt} that, under a more general setting, our proposed estimation strategy using the working variance function $ \widecheck{v}_{\epsilon}(\bx) $ is optimal, with the smallest $ \sqrt{n} $-asymptotic variance, for heteroscedastic and misspecified treatment-free effect models. This implies that E-Learning is more general with better optimality guarantee than \citet{xiao2019robust}.

The methodology introduced in this section is special in the sense that the treatment assignment is binary, \textit{i.e.} $ A \in \{0,1\} $. For multiple treatment options $ A \in \{ 1,2,\cdots,K \} $ with $ K > 2 $, the estimation problem is no longer an inverse-variance weighted least-squares problem. We will motivate our general methodology from the semiparametric efficient estimate of Model (\ref{eq:model}).

\subsection{Semiparametric Efficient Estimate}\label{sec:eff}
In this section, we derive the semiparametric efficient estimate of $ \bbeta $ for Model (\ref{eq:model}). The efficient estimating function can be related to some existing methods in the literature. The connections are discussed in Sections \ref{sec:binary} and \ref{sec:multiple}.

\subsubsection{Efficient Score}\label{sec:score}
In order to obtain the corresponding estimating equation, we first show the procedures to calculate the semiparametric efficient score following \citet{tsiatis2007semiparametric}. To that end, we take the following steps to derive: 1) the nuisance tangent space; 2) the efficient score; 3) the efficient estimating function.

We first derive the nuisance tangent space with respect to $ \eta $ following \citet[Chapter 7]{tsiatis2007semiparametric}. The same result was also used in \citet{ma2006efficient,liang2020semiparametric}. 

\begin{lem}[Nuisance Tangent Space]\label{lem:tangent}
	Consider Model (\ref{eq:model}). Define $ \cH:=\{ \bh(\bX,A,\epsilon) \mid \bh: \cX \times \cA\times \bbR \to \bbR^{p}, ~ \bbE \bh(\bX,A,\epsilon) = \bzero, ~ \bbE\|\bh(\bX,A,\epsilon)\|_{2}^{2} < +\infty \} $, which is equipped with the norm $ \|\cdot\| := (\bbE\|\cdot\|_{2}^{2})^{1/2} $. Then the nuisance tangent space is
	\[ \Lambda = \Big\{ \bH \in \cH: \bbE(\bH\epsilon|\bX,A) = \bbE(\bH\epsilon|\bX)\Big\}. \]
\end{lem}

\noindent The proof of Lemma \ref{lem:tangent} is included in the Supplementary Material. 

Next we discuss how to obtain the efficient score of Model (\ref{eq:model}). The efficient score is defined as the projection of the score vector onto the orthogonal complement $ \Lambda^{\perp} $ of the nuisance tangent space. 
Notice that the moment restriction in Lemma \ref{lem:tangent} is equivalent to
\[ \bbE(\bH\epsilon|\bX,A=1) = \bbE(\bH\epsilon|\bX,A=2) = \cdots = \bbE(\bH\epsilon|\bX,A=K). \]
Then we can introduce a set of coding vectors $ \{ \bomega_{k} \}_{k=1}^{K} \subseteq \bbR^{K-1} $, such that $ \sum_{k=1}^{K}c_{k}\bomega_{k} = \bzero $ if and only if $ c_{1} = c_{2} = \cdots = c_{K} $. Equivalently, we can let $ \Omega := \sqrt{1-1/K}[\bomega_{1},\bomega_{2},\cdots,\bomega_{K}]^{\intercal} \in \bbR^{K \times (K-1)} $, and require that $ (1/\sqrt{K})\bone_{K\times 1} $ is the only left singular vector corresponding to the singular value $ 0 $ of $ \Omega $. In the following Lemma \ref{lem:equiang}, we show that any coding vectors satisfying such a requirement are equiangular up to normalization.

\begin{lem}[Equiangularity]\label{lem:equiang}
	Let $ \Omega := \sqrt{1-1/K}[\bomega_{1},\bomega_{2},\cdots,\bomega_{K}]^{\intercal} \in \bbR^{K \times (K-1)} $ such that $ (1/\sqrt{K})\bone_{K\times 1} $ is the only left singular vector corresponding to the singular value $ 0 $. Then $ \{ (\Omega^{\intercal}\Omega)^{-1/2}\bomega_{k} \}_{k=1}^{K} $ are equiangular. 
\end{lem}

The equiangular coding representation in \citet{zhang2014multicategory,zhang2020multicategory,qi2020multi} is an example that satisfies Lemma \ref{lem:equiang}. The equiangular coding vectors $ \{ \bomega_{k} \}_{K=1}^{K} $ can be useful to define the following $ \bbR^{K-1} $-valued decision function associated with the interaction effect.

\begin{lem}[Angle-Based Decision Function]\label{lem:ang}
	Consider Model (\ref{eq:model}). For the coding vectors $ \{\bomega_{k}\}_{k=1}^{K} \subseteq \bbR^{K-1} $ as in Lemma \ref{lem:equiang}, define an $ \bbR^{K-1} $-valued decision function $ \vec{\bbf}(\bx;\bbeta) := (\Omega^{\intercal}\Omega)^{-1}\sum_{k=1}^{K}\gamma(\bx,k;\bbeta)\bomega_{k} $. Then we have
	\[ \gamma(\bx,k;\bbeta) = \left( 1 - {1 \over K} \right)\langle \bomega_{k},\vec{\bbf}(\bx;\bbeta) \rangle; \quad 1 \le k \le K. \]
	Moreover, the optimal ITR is given by
	\begin{align}
		d^{\star}(\bx) \in \argmax_{1 \le k \le K}\langle \bomega_{k},\vec{\bbf}(\bx;\bbeta) \rangle.\label{eq:ang_opt}
	\end{align}
\end{lem}

Without loss of generality, assume that $ \| \bomega_{k} \|_{2} = 1 $ for $ 1 \le k \le K $. For ease of notation, we denote $ \vec{\bbf} = \vec{\bbf}(\bx;\bbeta) $. Then the angle between $ \bomega_{k} $ and $ \vec{\bbf} $ satisfies $ \cos\angle(\bomega_{k},\vec{\bbf}) = \langle \bomega_{k},\vec{\bbf} \rangle/\| \vec{\bbf} \|_{2} $. The decision rule (\ref{eq:ang_opt}) is equivalent to $ \argmin_{1 \le k \le K}\angle(\bomega_{k},\vec{\bbf}) $. That is, among $ K $ coding vectors $ \{\bomega_{k}\}_{k=1}^{K} $, the decision function $ \vec{\bbf} $ seeks for the arm that the corresponding coding vector has the least angle with respect to $ \vec{\bbf} $.

Based on the coding vectors, the tangent space in Lemma \ref{lem:tangent} can be rewritten as
\[ \Lambda = \left\{ \bH \in \cH: \sfO_{p\times(K-1)} = \sum_{k=1}^{K}\bbE(\bH\epsilon|\bX,A=k)\bomega_{k}^{\intercal} = \bbE\left( {\bH\bomega_{A}^{\intercal}\epsilon \over p_{\sA}(A|\bX)}\middle|\bX \right) \right\}. \]
Then we can obtain $ \Lambda^{\perp} $ and the projection operator onto it as in the following Lemma \ref{lem:tangent_proj}. For a vector $ \ba $, we denote $ \ba^{\otimes 2} := \ba\ba^{\intercal} $. 

\begin{lem}[Projection onto $ \Lambda^{\perp} $]\label{lem:tangent_proj}
	Let $ \Lambda $ be the tangent space in Lemma \ref{lem:tangent}, $ \{ \bomega_{k} \}_{k=1}^{K} \subseteq \bbR^{K-1} $ be the coding vectors satisfying $ \sum_{k=1}^{K}c_{k}\bomega_{k} = \bzero $ if and only if $ c_{1} = c_{2} = \cdots = c_{K} $. Then
	\[ \Lambda^{\perp} = \left \{{\sfH(\bX)\bomega_{A}\epsilon \over p_{\sA}(A|\bX)}\middle|\sfH:\cX\to \bbR^{p\times(K-1)} \right \}. \]
	Furthermore, the projection operator onto $ \Lambda^{\perp} $ is
	\[ \bbE(\bH|\Lambda^{\perp}) = \bbE\left\{ {\bH\bomega_{A}^{\intercal}\epsilon \over p_{\sA}(A|\bX)}\middle|\bX \right\}\sfV_{\epsilon}(\bX)^{-1}{\bomega_{A}\epsilon \over p_{\sA}(A|\bX)}, \]
	where $ \sfV_{\epsilon}(\bX) := \sum_{k=1}^{K}{\sigma^{2}(\bX,k)\bomega_{k}^{\otimes 2} \over p_{\sA}(k|\bX)} \in \bbR^{(K-1)\times(K-1)} $. Here, if $ \sfV_{\epsilon}(\bX) $ is degenerate, then $ \sfV_{\epsilon}(\bX)^{-1} $ represents its measurable generalized inverse.
\end{lem}

The efficient score of the semiparametric model \eqref{eq:model} is defined as the projection of the score vector, the gradient of the log-likelihood with respect to $ \bbeta $, onto $ \Lambda^{\perp} $ \citep{tsiatis2007semiparametric}. Proposition \ref{prop:eff_ang} provides the explicit form of the efficient score.

\begin{prop}[Efficient Score] \label{prop:eff_ang}
	Consider Model (\ref{eq:model}), the coding vectors $ \{ \bomega_{k} \}_{k=1}^{K} \subseteq \bbR^{K-1} $ as in Lemma \ref{lem:equiang}, and the angle-based representation in Lemma \ref{lem:ang}. The semiparametric efficient score is
	\[ \bS_{\eff}(\bbeta) = \dot{\sfF}(\bX;\bbeta)^{\intercal}\Omega^{\intercal}\Omega\sfV_{\epsilon}(\bX)^{-1} \times {\bomega_{A} \over p_{\sA}(A|\bX)} \times \epsilon,
	 \]
	where $ \dot{\sfF}(\bX;\bbeta) := (\partial/\partial \bbeta^{\intercal})\vec{\bbf}(\bX;\bbeta) \in \bbR^{(K-1)\times p} $, and $ \sfV_{\epsilon}(\bX)^{-1} $ is the same as in Lemma \ref{lem:tangent_proj}.
\end{prop}

As a consequence of Proposition \ref{prop:eff_ang}, we can finally define the efficient estimating function:
\begin{small}
	\begin{align}
		& \bphi_{\eff}(\bbeta;\widecheck{\mu}_{0},\widecheck{p}_{\sA},\widecheck{\sigma}^{2}) \nonumber \\
		:=& \underbrace{\left [Y - \widecheck{\mu}_{0}(\bX) - \left( 1 - {1 \over K} \right)\langle \bomega_{A},\vec{\bbf}(\bX;\bbeta) \rangle \right ]}_{\text{residual}} \times \underbrace{\dot{\sfF}(\bX;\bbeta)^{\intercal}\Omega^{\intercal}\Omega\left[ \sum_{k=1}^{K}{\widecheck{\sigma}^{2}(\bX,k)\bomega_{k}^{\otimes 2} \over \widecheck{p}_{\sA}(k|\bX)} \right]^{-1}{\bomega_{A} \over \widecheck{p}_{\sA}(A|\bX)}}_{\text{efficient instrument}},\label{eq:estim_fun}
	\end{align}
\end{small}%
which depends on the nuisance functions $ \widecheck{\mu}_{0}(\bX) $, $\widecheck{p}_{\sA}(A|\bX) $ and $ \widecheck{\sigma}^{2}(\bX,A) $. In particular, $ \bS_{\eff}(\bbeta) = \bphi_{\eff}(\bbeta;\mu_{0},p_{\sA},\sigma^{2}) $. That is, if the parameters $ \bbeta $ of interest and all nuisance functions $ (\mu_{0},p_{\sA},\sigma^{2}) $ match with the truth in Model (\ref{eq:model}), then the estimating function becomes the efficient score. 

\subsection{E-Learning}\label{sec:eLearn}
In Section \ref{sec:eff}, we have obtained the efficient estimating function $ \bphi_{\eff}(\bbeta;\widecheck{\mu}_{0},\widecheck{p}_{\sA},\widecheck{\sigma}^{2}) $ from (\ref{eq:estim_fun}). 
An E-Learning estimate of $ \bbeta $ solves 
\begin{align}
	\bbE_{n}[\bphi_{\eff}(\bbeta;\widehat{\mu}_{0,n},\widehat{p}_{\sA,n},\widehat{\sigma}_{n}^{2})] = \bzero,\label{eq:eLearn}
\end{align}
where $ (\widehat{\mu}_{0,n},\widehat{p}_{\sA,n}) $ are the finite-sample estimates of treatment-free effect and treatment assignment probability in Model \eqref{eq:model}. Furthermore, $ \widehat{\sigma}_{n}^{2}(\bX,A) $ is an estimate of the optimal variance function
\begin{align}
	\sigma_{\opt}^{2}(\bX,A;\widehat{\mu}_{0,n}) := [\widehat{\mu}_{0,n}(\bX) - \mu_{0}(\bX)]^{2} + \sigma^{2}(\bX,A).\label{eq:var_opt}
\end{align}
The optimality of $ \sigma_{\opt}^{2}(\bX,A;\widehat{\mu}_{0,n}) $ is justified in Theorem \ref{thm:eff_opt} in Section \ref{sec:asymptotic_misspecified}. However, \eqref{eq:var_opt} can depend on the true treatment-free effect function $ \mu_{0}(\bX) $ and variance function $ \sigma^{2}(\bX,A) $, which are unknown. Motivated from the example in Section \ref{sec:eg}, we can consider the working residual $ \widehat{e} := Y - \widehat{\mu}_{0,n}(\bX) - \gamma(\bX,A;\bbeta) $, such that $ \bbE(\widehat{e}^{2}|\bX,A,\widehat{\mu}_{0,n}) = [\widehat{\mu}_{0,n}(\bX) - \mu_{0}(\bX)]^{2} + \sigma^{2}(\bX,A) = \sigma_{\opt}^{2}(\bX,A;\widehat{\mu}_{0,n}) $. Therefore, $ \widehat{\sigma}_{n}^{2}(\bX,A) $ can be obtained by regressing $ \widehat{e}^{2} $ on $ (\bX,A) $.

Similar to the general methodology in \citet{davidian1987variance}, the E-Learning estimate of $ \bbeta $ can be solved by the following three steps:
\begin{enumerate}[label=\textbf{Step \arabic*.}]
	\item Obtain a consistent estimate $ \widehat{\bbeta}_{n}^{(0)} $ of $ \bbeta $. This can be done by solving $ \eqref{eq:eLearn} $ with $ \widehat{\sigma}_{n}^{(0)2} = 1 $ that results in a consistent estimate of $ \bbeta $. The consistency is guaranteed by Proposition \ref{prop:dr}; 
	
	\item \label{enum:var} Obtain $ \widehat{\sigma}_{n}^{2}(\bX,A) $. Specifically, we first compute the working residual $ \widehat{e} = Y - \widehat{\mu}_{0,n}(\bX) - \gamma(\bX,A;\widehat{\bbeta}_{n}^{(0)}) $, and then
	perform a nonparametric regression using $ \widehat{e}^{2} $ as the response and $(\bX,A)$ as the covariates to estimate the optimal working variance function;
	
	\item Solve \eqref{eq:eLearn} again using $ \widehat{\sigma}_{n}^{2}(\bX,A) $ from \textbf{Step 2} to obtain the E-Learning estimate $ \widehat{\bbeta}_{n} $.  
\end{enumerate}
More implementation details are discussed in Section \ref{sec:implement}. 

Note that the estimation procedure in this section relies on the parametric model for the interaction effect $ \gamma(\bX,A;\bbeta) $. A typical parametric assumption is $ \gamma(\bX,A;\sfB) := (1-1/K)\langle \bomega_{A},\sfB^{\intercal}\bX \rangle $, where the angle-based decision function in Lemma \ref{lem:ang} is modeled linearly as $ \vec{\bbf}(\bX;\sfB) = \sfB^{\intercal}\bX $ with $ \sfB \in \bbR^{p\times(K-1)} $. However, this should not restrict the applicability of E-Learning. When the true interaction effect is nonlinear, we can still consider the basis expansion of $ \bX $. For example, we can use $ \bX^{(3)} := (X_{1},X_{1}^{2},X_{1}^{3};\cdots,X_{p},X_{p}^{2},X_{p}^{3})^{\intercal} $ as the covariate vector instead. In the Supplementary Material Section \ref{sec:simulation_nonlinear}, we demonstrate the effectiveness of E-Learning with the linear and cubic polynomial basis. Although the true interaction effect may not be correctly specified by the linear or cubic polynomial, our results can still show the strong advantage of cubic E-Learning when compared to other methods based on the same corresponding function basis.

\subsection{Implementation}\label{sec:implement}
For the implementation of E-Learning, we first need to estimate the treatment assignment probabilities $ \{ p_{\sA}(k|\bX) \}_{k=1}^{K} $ and the treatment-free effect $ \mu_{0}(\bX) $. Then we follow the three-step procedures in Section \ref{sec:eLearn} for E-Learning estimation.

\subsubsection{Estimating the Propensity Score Function}\label{sec:prop}
Suppose the treatment assignment probability $ p_{\sA} $ is unknown. The first approach of estimating $ p_{\sA} $ is to consider the penalized multinomial logistic regression \citep{friedman2010regularization}. Specifically, consider the multinomial logistic working model $ \widecheck{p}_{\sA}(k|\bX;\btau_{1},\btau_{2},\cdots,\btau_{K}) := {\exp(\btau_{k}^{\intercal}\bX) \over \sum_{k'=1}^{K}\exp(\btau_{k'}^{\intercal}\bX)} $. The propensity score parameters $\btau_{1},\btau_{2},\cdots,\btau_{K} \in \bbR^{p} $ can be estimated by the following penalized log-likelihood maximization:
\[ \max_{\btau_{1},\cdots,\btau_{K}\in\bbR^{p}}\Bigg\{ \bbE_{n}\left[  \sum_{k=1}^{K}\btau_{k}^{\intercal}\bX\bbone(A=k) - \log\left(\sum_{k'=1}^{K}e^{\btau_{k'}^{\intercal}\bX}\right)\right] - \lambda_{\sA}\sum_{j=1}^{p}\left( \sum_{k=1}^{K}\tau_{jk}^{2} \right)^{1/2}\Bigg\}, \]
where the group-LASSO penalty $ \sum_{j=1}^{p}\left( \sum_{k=1}^{K}\tau_{jk}^{2} \right)^{1/2} $ takes $\{\tau_{jk}\}_{k=1}^{K}$ for the $j$-th variable across all treatments as a group, and $ \lambda_{\sA} $ is a tuning parameter and can be chosen using cross validation.

In observational studies, the propensity scores can be vulnerable to model misspecification. Another approach for estimating $ p_{\sA} $ is to consider flexible nonparametric regression using the regression forest \citep{athey2019generalized}. Specifically, for each $ 1 \le k \le K $, 	we run a regression forest using $ \bbone(A = k) $ as the response and $ \bX $ as the covariates. Then each fitted regression forest provides a prediction for $ \bbE[\bbone(A=k)|\bX] $. The final estimate of $ p_{\sA}(k|\bX) $ is the prediction after normalization such that the summation over $ k = 1, \cdots, K $ is one.

\subsubsection{Estimating the Treatment-Free Effect Function}\label{sec:tf}
Similar to Section \ref{sec:prop}, the treatment-free effect function $ \mu_{0} $ can be estimated from a parametric model or nonparametric regression. For parametric estimation, we consider the linear working model $ \widecheck{\mu}_{0}(\bX;\bmeta) = \bmeta^{\intercal}\bX $. In this case, the outcome mean model in \eqref{eq:model} is fully parametrized. For example, if $ \gamma(\bX,A;\sfB) = (1-1/K)\langle \bomega_{A},\sfB^{\intercal}\bX \rangle $, then we can consider the following joint penalized inverse-probability weighted least-squares problem with the $ \ell_{1} $-penalty:
\[ \min_{\bmeta \in \bbR^{p},\sfB \in \bbR^{p\times(K-1)}} \left\{ \bbE_{n}\left[ {1 \over \widehat{p}_{\sA,n}(A|\bX)}\left( Y - \bmeta^{\intercal}\bX - \left( 1 - {1 \over K} \right)\langle \bomega_{A},\sfB^{\intercal}\bX \rangle \right)^{2} \right] + \lambda_{\mu_{0}}\left( \| \bmeta \|_{1} + \| \sfB \|_{1} \right) \right\}, \]
where $ \widehat{p}_{\sA,n} $ is the estimated treatment assignment probability, $ \lambda_{\mu_{0}} $ is a tuning parameter and can be chosen using cross validation. Here, if $ \widehat{p}_{\sA,n}(A|\bX) $ is the correct treatment assignment probability, then the above estimate for $ \bmeta $ can be consistent even if the model for the interaction effect $ \gamma(\bX,A;\bbeta) $ is incorrect. If the model for the interaction effect $ \gamma(\bX,A;\bbeta) $ is correct, then the above estimate for $ \bmeta $ can also be consistent for any arbitrary $ \widehat{p}_{\sA,n} $ besides the correct one. 

For nonparametric regression, we first divide the data into $ K $ subsets according to the received treatments. For each $ 1 \le k \le K $, we use $ Y $ as the response and $ \bX $ as the covariates to fit a regression forest on the data subset $ \{ (\bX_{i},Y_{i}): A_{i} = k \} $. Then each fitted regression forest corresponds to the prediction of $ \bbE(Y|\bX,A = k) $. We average the predictions over $ k = 1,\cdots,K $ to obtain the treatment-free effect estimate.

\subsubsection{Estimating the Variance Function}\label{sec:var}
Suppose $ \widehat{e} $ is the working residual in \textbf{Step 2}. In order to estimate the variance function, we specifically consider the regression forest using $ \widehat{e}^{2} $ as the response and $ (\bX,A) $ as the covariates. Then $ \widehat{\sigma}_{n}^{2}(\bX,k) $ is the regression forest prediction at $ (\bX,k) $ for $ 1 \le k \le K $. 

In the simulation study in Section \ref{sec:simulation_multiple}, we also study another two nonparametric regression methods, the \textit{Multivariate Adaptive Regression Splines (MARS)} \citep{friedman1991multivariate} and the \textit{COmponent Selection and Smoothing Operator (COSSO)} \citep{lin2006component}. Here, the COSSO estimate of the working variance function is based on the following \textit{Smoothing Spline ANalysis Of VAriance (SS-ANOVA)} model: $ \bbE(\widehat{e}^{2}|\bX,A) = \nu_{0} + \sum_{j=1}^{p}f_{j}(X_{j}) + \sum_{k=1}^{K}\alpha_{k} + \sum_{j=1}^{p}\sum_{k=1}^{K}f_{jk}(X_{j}) + u $, where $ \nu_{0} $ is the global main effect, $ \{ f_{j}(X_{j}) \}_{j=1}^{p} $ are the covariate main effects, $ \{ \alpha_{k} \}_{k=1}^{K} $ are the treatment main effects, $ \{ f_{jk}(X_{j}) \}_{1 \le j \le p, 1 \le k \le K} $ are the covariate-treatment interaction effects, and $ u $ is the remainder term that is not modeled.

\subsubsection{Solving the Regularized E-Learning Estimating Equation}\label{sec:reg}
In this section, we consider further regularization $ J(\bbeta) $ on the parameters of interest. One example from \citet{qi2020multi} is to consider the linear angle-based decision function $ \vec{\bbf}(\bX;\sfB) = \sfB^{\intercal}\bX $ in Lemma \ref{lem:ang}, where the covariate vector $ \bX \in \bbR^{p} $ can be high-dimensional. They introduced the row-wise group-LASSO penalty on the matrix coefficient $ \sfB = [\beta_{jk}]_{p\times(K-1)} \in \bbR^{p\times(K-1)} $ as $ J(\sfB) := \|\sfB\|_{2,1} = \sum_{j=1}^{p}(\sum_{k=1}^{K-1}\beta_{jk}^{2})^{1/2} $, which encourages sparsity among input covariates. Another example can be the extension to nonlinear modeling of the decision function $ \vec{\bbf}(\bX) $, where a functional penalty $ J(\vec{\bbf}) $ is applied.

To incorporate regularization in E-Learning from \eqref{eq:eLearn}, we solve the penalized estimating equations \citep{johnson2008penalized}:
\begin{align}
	\min_{\bbeta \in \bbR^{p}}\left\{ {1 \over 2}\left\| \bbE_{n}[\bphi_{\eff}(\bbeta;\widehat{\mu}_{0,n},\widehat{p}_{\sA,n},\widehat{\sigma}_{n}^{2})] \right \|^{2}_{\sfW} + \lambda J(\bbeta) \right\},\label{eq:eLearn_pen}
\end{align}
where $ \|\bx\|_{\sfW}^{2} := \bx^{\intercal}\sfW\bx $ with some weighting matrix $ \sfW \in \bbR^{p\times p} $. A typical choice of $ \sfW $ can be $ \sfI_{p \times p} $ or the inverse of the empirical information matrix $ \left\{ \bbE_{n}[(\partial/\partial \bbeta^{\intercal})\bphi_{\eff}(\bbeta;\widehat{\mu}_{0,n},\widehat{p}_{\sA,n},\widehat{\sigma}_{n}^{2})]  \right\}^{-1} $.
Problem \eqref{eq:eLearn_pen} can be solved by the accelerated proximal gradient method \citep{nesterov2013gradient} with the gradient $ \bbeta \mapsto \bbE_{n}[\bphi_{\eff}(\bbeta;\widehat{\mu}_{0,n},\widehat{p}_{\sA,n},\widehat{\sigma}_{n}^{2})] $. A comprehensive lists of the proximal operators on various penalties $ J(\bbeta) $ can be found in \citet{mo2021supervised}.
For a fixed $ \lambda $, the estimation procedure follows the three steps in Section \ref{sec:eLearn}. The parameter $ \lambda $ can be further tuned by cross validation. The IPWE of the value function is used as the tuning criteria. Denote $ \widehat{\bbeta}_{n}(\lambda) $ as the solution to \eqref{eq:eLearn_pen}. The corresponding ITR becomes $ \widehat{d}_{n}(\bX;\lambda) := \argmax_{1 \le k \le K}\gamma(\bX,k;\widehat{\bbeta}_{n}(\lambda)) $. Let $ \{ (\bX_{i},A_{i},Y_{i}) \}_{i=1}^{n_{\rm valid}} $ be the validation dataset. Then the criteria for $ \lambda $ is $ {1 \over n_{\rm valid}}\sum_{i=1}^{n_{\rm valid}}{\bbone[\widehat{d}_{n}(\bX_{i};\lambda) = A_{i}] \over \widehat{p}_{\sA,n}(A_{i}|\bX_{i})}Y_{i} $, which is larger the better.

More implementation details for E-Learning are discussed in Sections \ref{sec:implement_detail} and \ref{sec:var_cosso} in the Supplementary Material. 

\section{Connections to Existing Literature}\label{sec:connect}
In this section, we discuss the connection of the E-Learning estimating function \eqref{eq:estim_fun} to several methods in the existing literature. It can be shown that with more assumptions in addition to Model \eqref{eq:model}, several existing methods can be equivalent to \eqref{eq:estim_fun}. That is, E-Learning can incorporate these methods as special cases. The motivating example in Section \ref{sec:eg} is such a special case. In Sections \ref{sec:binary} and \ref{sec:multiple}, we discuss the equivalence and the specific additional assumptions.  In Section \ref{sec:compare}, we further provide the general comparisons for these methods and some other nonparametric methods in the literature.

\subsection{Binary Treatment}\label{sec:binary}
We first consider the binary treatment case $ K = 2 $ and relate the efficient estimating function (\ref{eq:estim_fun}) to some existing methods. We follow the convention to denote $ \cA = \{ 0,1 \} $. Then we have one-dimensional coding for two treatment arms as $ \omega_{0}$, $\omega_{1} $, which satisfies $ c_{0}\omega_{0} + c_{1}\omega_{1} = 0 $  if and only if $ c_{0} = c_{1} $. Then we have $ \omega_{1} = - \omega_{0} $. Without loss of generality, we can assume that $ \omega_{1} = 1 $ and $ \omega_{0} = -1 $, which become the sign coding of treatments. Then $ \Omega^{\intercal}\Omega = 1 $.

The variance matrix from Proposition \ref{prop:eff_ang} becomes a scalar: $ v_{\epsilon}(\bX) := {\sigma^{2}(\bX,1) \over p_{\sA}(1|\bX)} + {\sigma^{2}(\bX,0) \over p_{\sA}(0|\bX)} $. The decision function $ f(\bX;\bbeta) $ is $ \bbR $-valued, such that $ \gamma(\bX,A;\bbeta) = (1/2)\omega_{A}f(\bX;\bbeta) $. Then the E-Learning efficient estimating function (\ref{eq:estim_fun}) becomes
\begin{align}
	\bphi_{\eff}(\bbeta;\widecheck{\mu}_{0},\widecheck{p}_{\sA},\widecheck{\sigma}^{2}) = [Y - \widecheck{\mu}_{0}(\bX) - (1/2)\omega_{A}f(\bX;\bbeta)]{\widecheck{v}_{\epsilon}^{-1}(\bX)\omega_{A} \over \widecheck{p}_{\sA}(A|\bX)}\dot{\fb}(\bX;\bbeta),\label{eq:eLearn_binary}
\end{align}
where $ \dot{\fb}(\bX;\bbeta) := (\partial/\partial\bbeta)f(\bX;\bbeta) \in \bbR^{p} $. Moreover, \eqref{eq:eLearn_binary} is also equivalent to the following weighed least-squares problem:
\begin{align}
	\min_{\bbeta \in \bbR^{p}}\bbE_{n}\left\{ {\widecheck{v}_{\epsilon}^{-1}(\bX) \over \widecheck{p}_{\sA}(A|\bX)}[Y - \widecheck{\mu}_{0}(\bX) - (1/2)\omega_{A}f(\bX;\bbeta)]^{2} \right\}.\label{eq:eLearn_wls}
\end{align}
There are some connections for this formulation to several methods in the existing literature.

\paragraph{Q-Learning}
Consider the additional assumptions: (a) homoscedasticity $ \sigma^{2}(\bX,1) = \sigma^{2}(\bX,0) = \sigma^{2} $; and (b) complete-at-random treatment assignment $ p_{\sA}(1|\bX) = p_{\sA}(0|\bX) = 1/2 $. Then E-Learning (\ref{eq:eLearn_wls}) reduces to an OLS problem. If we also assume that: (c) the treatment-free effect satisfies $ \mu_{0}(\bX) = \bX^{\intercal}(\bmeta + \bbeta/2) $, where $ (\bbeta,\bmeta) $ are jointly estimated, then E-Learning (\ref{eq:eLearn_wls}) can be equivalent to the standard \textit{Q-Learning} \citep{watkins1989learning} in this case:
\[ \min_{\bmeta,\bbeta\in \bbR^{p}}\bbE_{n}(Y - \bX^{\intercal}\bmeta - A\bX^{\intercal}\bbeta)^{2}. \]

\paragraph{G-Estimation, A-Learning and dWOLS}
Consider the additional assumption: (a) homoscedasticity $ \sigma^{2}(\bX,1) = \sigma^{2}(\bX,0) = \sigma^{2} $. Then $ v_{\epsilon}^{-1}(\bX) = \sigma^{-2}p_{\sA}(1|\bX)p_{\sA}(0|\bX) $. Without loss of generality, we can further assume that $ \sigma^{2} = 1 $. Denote $ \pi_{\sA}(\bX) := p_{\sA}(1|\bX) = \bbE(A|\bX) $. Then we have $ {v_{\epsilon}^{-1}(\bX)\omega_{A} \over p_{\sA}(A|\bX)} = A - \pi_{\sA}(\bX) = |A - \pi_{\sA}(\bX)|\omega_{A} $ and $ {v_{\epsilon}^{-1}(\bX) \over p_{\sA}(A|\bX)} = |A - \pi_{\sA}(\bX)| $.

\citet{robins2004optimal} proposed the \textit{G-Estimation} strategy for dynamic treatment regimes, which is equivalent to the standard \textit{A-Learning} \citep{murphy2003optimal} in the single-stage setting. In particular, G-Estimation solves the estimating equation
\[ \bbE_{n}\Big\{ \left[Y - \widehat{\mu}_{0,n}(\bX) - A\bX^{\intercal}\bbeta\right]\left[A-\widehat{\pi}_{\sA,n}(\bX)\right]\bX \Big\} = \bzero, \]
while A-Learning is equivalent to the estimating equation
\[ \bbE_{n}\Big\{ \left[Y - \widehat{m}_{0,n}(\bX) - \big(A-\widehat{\pi}_{\sA,n}(\bX)\big)\bX^{\intercal}\bbeta\right]\left[A-\widehat{\pi}_{\sA,n}(\bX)\right]\bX \Big\} = \bzero. \]
Then G-Estimation and A-Learning are  equivalent to E-Learning \eqref{eq:eLearn_binary} in this case up to reparametrization, where $ \widehat{\mu}_{0,n}(\bX) $ is replaced by $ \widehat{m}_{0,n}(\bX) - \widehat{\pi}_{\sA,n}(\bX)\bX^{\intercal}\bbeta $.

\citet{wallace2015doubly} proposed the dWOLS method. In the single-stage setting, they considered the following weighted least-squares problem:
\[ \min_{\bmeta,\bbeta \in \bbR^{p}}\bbE_{n}\Big\{ w(\bX,A)(Y - \bX^{\intercal}\bmeta - A\bX^{\intercal}\bbeta)^{2} \Big\}, \]
where $ w(\bX,A) $ satisfies the balancing condition $ \pi_{\sA}(\bX)w(\bX,1) = [1-\pi_{\sA}(\bX)]w(\bX,0) $. Note that $ w(\bX,A) = |A - \pi_{\sA}(\bX)| $ meets this balancing condition. Assume that: (b) the treatment assignment probability $ \pi_{\sA}(\bX) = p_{\sA}(1|\bX) $ is known; and (c) the treatment-free effect satisfies $ \mu_{0}(\bX) = \bX^{\intercal}(\bmeta + \bbeta/2) $, where $ (\bbeta,\bmeta) $ are jointly estimated. Then dWOLS with $ w(\bX,A) = |A - \pi_{\sA}(\bX)| $ is equivalent to E-Learning \eqref{eq:eLearn_wls}:
\[ \min_{\bmeta,\bbeta \in \bbR^{p}}\bbE_{n}\Big\{ |A-\pi_{\sA}(\bX)|(Y - \bX^{\intercal}\bmeta - A \bX^{\intercal}\bbeta)^{2} \Big\}. \]

\paragraph{Subgroup Identification, D-Learning and RD-Learning}

Consider the additional assumptions: (a) the variance function satisfies $ v_{\epsilon}(\bX) = {\sigma^{2}(\bX,1) \over p_{\sA}(1|\bX)} + {\sigma^{2}(\bX,0) \over p_{\sA}(0|\bX)} = v_{\epsilon} $, which is a constant; (b) the treatment assignment probability $ p_{\sA}(A|\bX) $ is known; and (c) the treatment-free effect satisfies $ \mu_{0}(\bX) = 0 $. Then E-Learning (\ref{eq:eLearn_wls}) is equivalent to the standard \textit{Subgroup Identification} \citep{tian2014simple,chen2017general} and the binary \textit{D-Learning} \citep{qi2018d}:
\[ \min_{\bbeta \in \bbR^{p}}\bbE_{n}\left\{ {1 \over p_{\sA}(A|\bX)}[Y - (1/2)\omega_{A}\bX^{\intercal}\bbeta]^{2} \right\}. \]
If both (b) and (c) are relaxed, then E-Learning (\ref{eq:eLearn_wls}) is equivalent to the augmented Subgroup Identification \citep[Web Appendix B]{chen2017general} and the binary \textit{RD-Learning} \citep{meng2020robust}:
\[ \min_{\bbeta \in \bbR^{p}}\bbE_{n}\left\{ {1 \over \widehat{p}_{\sA,n}(A|\bX)}[Y - \widehat{\mu}_{0,n}(\bX) - (1/2)\omega_{A}\bX^{\intercal}\bbeta]^{2} \right\}. \]

\subsection{Multiple Treatments and Partially Linear Model}\label{sec:multiple}

For general $ K \ge 3 $, we consider the linear decision function $ \vec{\bbf}(\bX;\sfB) = \sfB^{\intercal}\bX $, where $ \sfB \in \bbR^{p \times (K-1)} $ is a parameter matrix. By Lemma \ref{lem:ang}, Model (\ref{eq:model}) becomes
\begin{align}
	Y = \mu_{0}(\bX) + \left( 1 - {1 \over K} \right)\langle \bomega_{A}, \sfB^{\intercal}\bX \rangle + \epsilon; \quad \bbE(\epsilon|\bX,A) = 0; \quad \sigma^{2}(\bX,A) = \bbE(\epsilon^{2}|\bX,A) < +\infty,\label{eq:plm}
\end{align}
which is a \textit{Heteroscedasticitic Partially Linear Model (HPLM)} \citep{ma2006efficient}.

Denote $ \sfvec(\sfB) \in \bbR^{p(K-1)} $ as the vectorization of $ \sfB $. The we further have $ \vec{\bbf}(\bX;\sfB) = (\Ib_{(K-1)\times(K-1)} \otimes \bX)^{\intercal}\sfvec(\sfB) $ and $ \dot{\sfF}(\bX;\sfB) = [\partial/\partial \sfvec(\sfB)^{\intercal}]\vec{\bbf}(\bX;\bbeta) = (\Ib_{(K-1)\times(K-1)} \otimes \bX)^{\intercal} $, where $ \otimes $ denotes the Kronecker product. 
The E-Learning efficient estimating function \eqref{eq:estim_fun} becomes
\begin{small}
	\begin{align}
		\bphi_{\eff}(\sfB;\widecheck{\mu}_{0},\widecheck{p}_{\sA},\widecheck{\sigma}^{2}) = \underbrace{[(\Omega^{\intercal}\Omega)\otimes \Ib_{d\times d}]}_{\text{constant matrix}} \times \left[ Y - \widecheck{\mu}_{0}(\bX) - \left( 1 - {1 \over K} \right)\langle \bomega_{A},\sfB^{\intercal}\bX \rangle \right] \times{\widecheck{\sfV}_{\epsilon}(\bX)^{-1}\bomega_{A} \over \widecheck{p}_{\sA}(A|\bX)} \otimes \bX,\label{eq:eLearn_multiple}
	\end{align}
\end{small}%
where $ \widecheck{\sfV}_{\epsilon}(\bX) := \sum_{k=1}^{K}{\widecheck{\sigma}^{2}(\bX,k)\bomega_{k}^{\otimes 2} \over \widecheck{p}_{\sA}(k|\bX)} $, and $ \widecheck{\sfV}_{\epsilon}(\bX)^{-1} $ denotes the generalized inverse if not invertible.

Consider the additional assumption: (a) the variance function satisfies $ \sfV_{\epsilon}(\bX) = \sum_{k=1}^{K}{\sigma^{2}(\bX,k)\bomega_{k}^{\otimes 2} \over p_{\sA}(k|\bX)} = \sfV_{\epsilon} $, which is a constant matrix. Then E-Learning \eqref{eq:eLearn_multiple} is equivalent to the multi-arm RD-Learning:
\[ \min_{\sfB \in \bbR^{p\times(K-1)}}\bbE_{n}\left\{ {1 \over \widehat{p}_{\sA,n}(A|\bX)}\left[ Y - \widehat{\mu}_{0,n}(\bX) - \left( 1 - {1 \over K} \right)\langle \bomega_{A},\sfB^{\intercal}\bX \rangle \right]^{2} \right\}. \]

Notice that the multi-arm D-Learning \citep{qi2020multi} cannot be equivalent to E-Learning. In fact, D-Learning solves the following vectorized least-squares problem:
\begin{align}
	\min_{\sfB \in \bbR^{p\times(K-1)}}\bbE_{n}\left\{ {1 \over 2Kp_{\sA}(A|\bX)}\left\|KY\bomega_{A} - \sfB^{\intercal}\bX\right\|_{2}^{2} \right\}.\label{eq:dLearn}
\end{align}
The estimating function of \eqref{eq:dLearn} is
\[ \underbrace{\left[ Y - \left( 1 - {1 \over K} \right)\langle \bomega_{A},\sfB^{\intercal}\bX \rangle \right] \times {\bomega_{A} \over p_{\sA}(A|\bX)} \otimes \bX}_{\text{efficient estimating function if (a) and $ \mu_{0}(\bX) = 0 $}} + \underbrace{{1 \over p_{\sA}(A|\bX)} \left[ \left( 1 - {1 \over K} \right)\bomega_{A}^{\otimes 2} - {1 \over K}\sfI_{(K-1)\times(K-1)} \right]\sfvec(\bX^{\otimes 2}\sfB)}_{:= \bphi_{\rm D}(\bX,A)}. \]
Note that $ \bbE[\bphi_{\rm D}(\bX,A)|\bX] = \bzero $ and $ \bbE[\bphi_{\rm D}(\bX,A)^{\otimes 2}] $ is strictly positive definite, which contributes an extra term to the $ \sqrt{n} $-asymptotic variance of the D-Learning estimate. This suggests that when $ K \ge 3 $, the D-Learning estimate can generally have a larger asymptotic variance than E-Learning.

\subsection{General Comparisons}\label{sec:compare}

In Table \ref{tab:compare}, we provide the comparisons of the methods discussed in Sections \ref{sec:binary} and \ref{sec:multiple}. We also compare several popular nonparametric approaches including \textit{Outcome Weighted Learning (OWL)} \citep{zhao2012estimating}, \textit{Residual Weighted Learning (RWL)} \citep{zhou2017residual,liu2018augmented}, \textit{Efficient Augmentation and Relaxation Learning (EARL)} \citep{zhao2019efficient}, and Policy Learning \citep{athey2021policy,zhou2018offline}. In particular, EARL and Policy Learning utilize the AIPWE of the value function, which incorporates the outcome and propensity score models and is doubly robust. The listed methods are also compared in the simulation studies in Section \ref{sec:simulation_binary}.

\renewcommand{\arraystretch}{1}
\begin{table}[!ht]
	\centering
	\caption[Comparison]{Comparisons of E-Learning with Several Existing Methods in the Literature}\label{tab:compare}
	\begin{threeparttable}
		\begin{tabular}{@{}c@{}|@{}c@{}||>{\centering}m{1.5cm}|>{\centering}m{1.5cm}||>{\centering}m{1.5cm}|>{\centering}m{1.5cm}|>{\centering}m{1.5cm}|>{\centering}m{1.5cm}|>{\centering\arraybackslash}m{1.5cm}@{}}
			\hline\hline
			\multicolumn{2}{c||}{\multirow{3}{*}{\textbf{Method}}} & \multicolumn{2}{c||}{\textbf{Nuisance Models}} & \multirow{3}{1.5cm}{\makecell[c]{\textbf{Doubly} \\ \textbf{Robust}}} & \multicolumn{3}{c|}{\textbf{\footnotesize Assumptions for Being Optimal}} & \multirow{3}{1.5cm}{\makecell[c]{\textbf{Allow} \\ $ K \ge 3 $}} \\
			\cline{3-4}\cline{6-8}
			\multicolumn{2}{c||}{} & \textit{Outcome} & \textit{\footnotesize Propensity} &  & \textit{\scriptsize Treatment-Free Effect} & \textit{\footnotesize Propensity} & \textit{Variance} & \\
			\hline\hline
			\multicolumn{2}{c||}{\textit{E-Learning}} & Yes & Yes & Yes & Arbitrary & Correct & Hetero. & Yes \\
			\hline
			\multicolumn{2}{c||}{\textit{Q-Learning}} & Yes & No & No & Correct & $ 1/K $ & Homo. & Yes \\
			\hline
			\multicolumn{2}{c||}{\textit{G-Estimation}} & Yes & Yes & Yes & \multirow{3}{1.5cm}{\makecell[c]{Correct}} & \multirow{3}{1.5cm}{\makecell[c]{Correct}} & \multirow{3}{1.5cm}{\makecell[c]{Homo.}} & No \\
			\cline{1-5}\cline{9-9}
			\multicolumn{2}{c||}{\textit{A-Learning}} & Yes & Yes & Yes &  &  & & No \\
			\cline{1-5}\cline{9-9}
			\multicolumn{2}{c||}{\textit{dWOLS}} & Yes & Yes & Yes & & & & No \\
			\hline
			\multirow{2}{*}{\makecell[c]{\textit{\footnotesize Subgroup} \\ \textit{\footnotesize Identification}}}& \textit{\footnotesize Std.} & No & Yes & No & 0 & Known & Const. & Yes \\
			\cline{2-9}
			& \textit{\footnotesize Aug.}& Yes & Yes & No & Correct & Known & Const. & Yes \\
			\hline
			\multicolumn{2}{c||}{\textit{RD-Learning}} & Yes & Yes & Yes & Correct & Correct & Const. & Yes \\
			\hline
			\multirow{2}{*}{\makecell[c]{\textit{\footnotesize D-Learning}}} & {\tiny $ K=2 $} & \multirow{2}{*}{\makecell[c]{No}} & \multirow{2}{*}{\makecell[c]{Yes}} & \multirow{2}{*}{\makecell[c]{No}} & 0 & Known & Const. & \multirow{2}{*}{\makecell[c]{Yes}} \\
			\cline{2-2}\cline{6-8}
			& {\tiny $ K\ge 3 $} &  & & & \multicolumn{3}{c|}{\multirow{5}{4.5cm}{\makecell[c]{N/A}}} &  \\
			\cline{1-5}\cline{9-9}
			\multicolumn{2}{c||}{\textit{OWL}} & No & Yes & No & \multicolumn{3}{c|}{\multirow{4}{4.5cm}{}} & No \\
			\cline{1-5}\cline{9-9}
			\multicolumn{2}{c||}{\textit{RWL}} & Yes & Yes & No & \multicolumn{3}{c|}{\multirow{4}{4.5cm}{}} & No \\
			\cline{1-5}\cline{9-9}
			\multicolumn{2}{c||}{\textit{EARL}} & Yes & Yes & Yes & \multicolumn{3}{c|}{\multirow{4}{4.5cm}{}} & No \\
			\cline{1-5}\cline{9-9}
			\multicolumn{2}{c||}{\textit{\footnotesize Policy Learning}}
			& Yes & Yes & Yes & \multicolumn{3}{c|}{\multirow{4}{4.5cm}{}} & Yes \\
			\hline\hline
		\end{tabular}
		\begin{tablenotes}
			\fontsize{8}{5}\selectfont
			\item[1] ``Being optimal'' is defined as the estimate of $ \bbeta $ in Model \eqref{eq:model} achieves the smallest $ \sqrt{n} $-asymptotic variance among the class of estimates in Definition \ref{dfn:regular}. 
			\item[2] Methods of Subgroup Identification include the standard (\texttt{std.}) and augmented (\texttt{aug.}) versions. 
			\item[3] Variance assumptions are: \texttt{homo.} $ \Leftrightarrow $ constant $ \sigma^{2}(\bX,A) $; \texttt{hetero.} $ \Leftrightarrow $ general $ \sigma^{2}(\bX,A) $; \texttt{const.} $ \Leftrightarrow $ $ \sfV_{\epsilon}(\bX)  = \sum_{k=1}^{K}{\sigma^{2}(\bX,k)\bomega_{k}^{\otimes 2} \over p_{\sA}(k|\bX)} $ is a constant matrix.
		\end{tablenotes}
	\end{threeparttable}
\end{table}

We also discuss the estimation optimality for $ \bbeta $ in Table \ref{tab:compare}. Note that the nonparametric methods do not assume Model \eqref{eq:model}. Therefore, the estimation optimality for $ \bbeta $ is not available. In Theorem \ref{thm:eff_opt} in Section \ref{sec:asymptotic_misspecified}, we establish that the E-Learning estimate of $ \bbeta $ achieves the smallest $ \sqrt{n} $-asymptotic variance among the class of estimates in Definition \ref{dfn:regular}. This is also referred as ``being optimal'' in Table \ref{tab:compare}. Since the methods discussed in Sections \ref{sec:binary} and \ref{sec:multiple}, except for D-Learning with $ K \ge 3 $, are equivalent to E-Learning under specific additional assumptions, this also implies that the equivalent methods are optimal under those specific additional assumptions. However, this is not true for the general case. In contrast, our proposed E-Learning remains optimal under the most general scenario among all these methods. 

\section{Theoretical Properties}\label{sec:theory}
We investigate some theoretical properties of E-Learning. In particular, in Section \ref{sec:asymptotic}, we establish estimation properties based on the efficient estimating function (\ref{eq:estim_fun}). In Section \ref{sec:regret}, we further relate the asymptotic properties to the regret bound of the estimated ITR.

\subsection{Asymptotic Properties}\label{sec:asymptotic}
We first focus on estimation properties of the proposed E-Learning. In Proposition \ref{prop:dr}, we show the double robustness property of the estimating function (\ref{eq:estim_fun}). 

\begin{prop}[Double Robustness]\label{prop:dr}
	Consider Model (\ref{eq:model}) and the estimating function (\ref{eq:estim_fun}). Suppose $ \widecheck{\mu}_{0}(\bX) $, $ \widecheck{p}_{\sA}(A|\bX) $ and $ \widecheck{\sigma}^{2}(\bX,A) $ are arbitrary nuisance functions. Then we have
	\[ \bbE[ \bphi_{\eff}(\bbeta;\widecheck{\mu}_{0},p_{\sA},\widecheck{\sigma}^{2})] = \bbE[\bphi_{\eff}(\bbeta;\mu_{0},\widecheck{p}_{\sA},\widecheck{\sigma}^{2})] = \bzero. \]
\end{prop}

If either $ \widecheck{\mu}_{0} = \mu_{0} $ or $ \widecheck{p}_{\sA} = p_{\sA} $, then $ \bbE[ \bphi_{\eff}(\bbeta;\widecheck{\mu}_{0},\widecheck{p}_{\sA},\widecheck{\sigma}^{2})] = \bzero $ at the true parameter $ \bbeta $ in Model \eqref{eq:model}. By assuming the positivity of the information matrix at $ \bbeta $ (Assumption \ref{enum:curv}), the consistency of $ \widehat{\bbeta}_{n} \in \argmin_{\bbeta \in \sB} {1 \over 2}\left\| \bbE_{n}[\bphi_{\eff}(\widehat{\bbeta}_{n};\widecheck{\mu}_{0},\widecheck{p}_{\sA},\widecheck{\sigma}^{2})] \right\|_{2}^{2} $ can be established by the consistency of an M-estimator \citep[Corollary 3.2.3]{van1996weak}. This implies the doubly robust property of $ \widehat{\bbeta}_{n} $. If $ (\widecheck{\mu}_{0},\widecheck{p}_{\sA},\widecheck{\sigma}^{2}) $ are replaced by their finite-sample estimate $ (\widehat{\mu}_{0,n},\widehat{p}_{\sA,n},\widehat{\sigma}_{n}^{2}) $, then Lemma \ref{lem:plug-in} can be further applied to obtain consistency. Based on the connections from Section \ref{sec:connect}, Proposition \ref{prop:dr} provides a more general framework to explain the double robustness property discussed in \citet{robins2004optimal,lu2013variable,wallace2015doubly,meng2020robust}.

Our next goal is to study how model specifications can affect estimation efficiency. In Section \ref{sec:asymptotic_correct}, we study the asymptotic properties of the parameter estimate under correctly specified models. In Section \ref{sec:asymptotic_misspecified}, we further consider the case of misspecified treatment-free effect, and show that there exists an optimal choice of the working variance function for efficiency improvement.

\subsubsection{Correctly Specified Models}\label{sec:asymptotic_correct}

For simplicity, we assume that the treatment assignment probability $ p_{\sA} $ is known, so that the estimating function is consistent due to Proposition \ref{prop:dr}. This assumption can be relaxed to assuming a consistent estimate $ \widehat{p}_{\sA,n} $ of $ p_{\sA} $, and the theoretical results can be extended following the cross-fitting argument in \citet{ertefaie2021robust}. For example, we can assume a correctly specified parametric model for $ p_{\sA} $. 

We make additional assumptions on the squared integrability of Model (\ref{eq:model}) and the convergence of the plug-in treatment-free effect and variance function estimates. The estimated variance function $ \widehat{\sigma}_{n}^{2}(\bX,A) $ is furthered assumed uniformly bounded away from 0 to ensure that the smallest eigenvalue of $ \widehat{\sfV}_{\epsilon,n}(\bX) = \sum_{k=1}^{K}[\widehat{\sigma}_{n}^{2}(\bX,k)/p_{\sA}(k|\bX)]\bomega_{k}^{\otimes 2} $ is uniformly bounded away from 0, so that the largest eigenvalue of $ \widehat{\sfV}_{\epsilon,n}(\bX)^{-1} $ can be bounded from above. This can also be relaxed by considering a specific generalized inverse of $ \widehat{\sfV}_{\epsilon,n}(\bX) $ to extend the theoretical results.

\begin{asm}[Treatment Assignment Probability]\label{asm:prop}
	The treatment assignment probability $ p_{\sA} $ is known, such that for some $ \ubar{p}_{\sA} > 0 $, we have $ p_{\sA}(a|\bx) \ge \ubar{p}_{\sA} $ for all $ \bx \in \cX $ and $ a \in \cA $.
\end{asm}

\begin{asm}[Squared Integrability]\label{asm:int}
	Consider Model (\ref{eq:model}) and the angle-based decision function $ \vec{\bbf}(\bX;\bbeta) $ in Lemma \ref{lem:ang}. We assume the following:
	
	\begin{itemize}
		\item $ \bbE[\mu(\bX)^{2}] < +\infty $; 
		
		\item $ \bbE\sup_{\widecheck{\bbeta} \in \sB}\gamma(\bX,A;\widecheck{\bbeta})^{2} < +\infty $; 
		
		\item $ \bbE(\epsilon^{2}) = \bbE\sigma^{2}(\bX,A) < +\infty $; 
		
		\item $ \dot{\sfF}(\bX;\widecheck{\bbeta}) = (\partial/\partial\bbeta^{\intercal})\vec{\bbf}(\bX;\widecheck{\bbeta}) \in \bbR^{(K-1)\times p} $ exists for $ \widecheck{\bbeta} \in \sB $, and $ 
		\bbE\sup_{\widecheck{\bbeta} \in \sB}\| \dot{\sfF}(\bX;\widecheck{\bbeta}) \|_{2}^{2} < +\infty $, where $ \| \cdot \|_{2} $ is the spectral norm on $ \bbR^{(K-1) \times p} $.
	\end{itemize}
\end{asm}

\begin{asm}[Convergence of Plug-in Estimates]\label{asm:plug-in}
	\item 
	\begin{itemize}
		\item There exists some $ \widecheck{\mu}_{0}: \cX \to \bbR $, such that $ \bbE[\widecheck{\mu}_{0}(\bX)^{2}]  < +\infty $ and $ (1/n)\sum_{i=1}^{n}[\widehat{\mu}_{0,n}(\bX_{i}) - \widecheck{\mu}_{0}(\bX_{i})]^{2} = \smallcO_{\bbP}(n^{-1}) $.
		
		\item There exists some $ 0 < \ubar{\sigma}^{2}\le \widebar{\sigma}^{2} < +\infty $ and $ \widecheck{\sigma}^{2}:\cX\times\cA \to \bbR_{+} $, such that $ \ubar{\sigma}^{2} \le \widehat{\sigma}_{n}^{2}(\bx,a),\widecheck{\sigma}^{2}(\bx,a) \le \widebar{\sigma}^{2} $, and $ \| \widehat{\sigma}_{n}^{2} - \widecheck{\sigma}^{2} \|_{\infty} = \sup_{\bx \in \cX,a \in \sA}|\widehat{\sigma}_{n}^{2}(\bx,a) - \widecheck{\sigma}^{2}(\bx,a)| = \smallcO_{\bbP}(n^{-1/2}) $.
	\end{itemize}
\end{asm}

\noindent Given Assumptions \ref{asm:prop}-\ref{asm:plug-in}, we show in Lemma \ref{lem:plug-in} that the plug-in estimating equation associated with (\ref{eq:estim_fun}) is $ \sqrt{n} $-asymptotically equivalent to the limiting estimating equation. 
\begin{lem}[Plug-in Estimating Equation]\label{lem:plug-in}
	Consider Model (\ref{eq:model}) and the estimating function (\ref{eq:estim_fun}). Under Assumptions \ref{asm:prop}-\ref{asm:plug-in}, we have
	\[ \sup_{\widecheck{\bbeta} \in \sB} \bbE_{n}\left\| \bphi_{\eff}(\widecheck{\bbeta};\widehat{\mu}_{0,n},p_{\sA},\widehat{\sigma}_{n}^{2}) - \bphi_{\eff}(\widecheck{\bbeta};\widecheck{\mu}_{0},p_{\sA},\widecheck{\sigma}^{2}) \right\|_{2} = \smallcO_{\bbP}(n^{-1/2}). \]
\end{lem}

\noindent Lemma \ref{lem:plug-in} implies that the plug-in estimates $ (\widehat{\mu}_{0,n},\widehat{\sigma}_{n}^{2}) $ do not affect the $ \sqrt{n} $-asymptotic properties of the estimating function (\ref{eq:estim_fun}). Then we can show the asymptotic normality of $ \widehat{\bbeta}_{\eff,n} $ as the solution to $ \bbE_{n}[\bphi_{\eff}(\widehat{\bbeta}_{\eff,n};\widehat{\mu}_{0,n},p_{\sA},\widehat{\sigma}_{n}^{2})] = \bzero $ following the argument in \citet{newey1994asymptotic}. Moreover, if the treatment-free effect and the variance function are correctly specified, \textit{i.e.}, $ (\widecheck{\mu}_{0},\widecheck{\sigma}^{2}) = (\mu_{0},\sigma^{2}) $ in Model (\ref{eq:model}), then $ \widehat{\bbeta}_{\eff,n} $ is semiparametric efficient, in the sense that its $ \sqrt{n} $-asymptotic variance achieves the semiparametric variance lower bound. We summarize the regularity conditions in Assumption \ref{asm:regular}.

\begin{asm}[Regularity Conditions]\label{asm:regular}
	Consider Model (\ref{eq:model}) and the angle-based representation in Lemma \ref{lem:ang}. We assume the following:
	\begin{enumerate}[label=\ref{asm:regular}.\arabic*]
		\item $ \sB $ is a compact subset in $ \bbR^{p} $ and the true parameter $ \bbeta \in \mathring{\sB} $, where $ \mathring{\sB} $ is the interior of $ \sB $;
		
		\item \label{enum:int} $ \ddot{\sfF}(\bx;\widecheck{\bbeta}) = (\partial/\partial\bbeta)\dot{\sfF}(\bx;\widecheck{\bbeta}) \in \bbR^{(K-1) \times p \times p} $ exists and satisfies $ \bbE\sup_{\widecheck{\bbeta} \in \sB}\| \ddot{\sfF}(\bX;\widecheck{\bbeta}) \|_{2}^{2} < +\infty $, where $ \| \cdot \|_{2} $ is the operator norm of $ (\bbR^{p},\| \cdot \|_{2}) \to (\bbR^{(K-1)\times p},\| \cdot \|_{2}) $;
		
		\item \label{enum:curv} Define $ \widecheck{\sfV}_{\epsilon}(\bX) := \sum_{k=1}^{K}{\widecheck{\sigma}^{2}(\bX,k)\bomega_{k}^{\otimes 2} \over \widecheck{p}_{\sA}(k|\bX)} $ and $ \widecheck{\cI}(\bbeta) := \bbE\left[ \dot{\sfF}(\bX;\bbeta)^{\intercal}\Omega^{\intercal}\Omega\widecheck{\sfV}_{\epsilon}(\bX)^{-1}\Omega^{\intercal}\Omega\dot{\sfF}(\bX;\bbeta) \right] $, where $ \widecheck{\sfV}_{\epsilon}(\bX)^{-1} $ denotes the generalized inverse if not invertible. Assume $ \widecheck{\cI}(\bbeta) $ is positive definite;
		
		\item \label{enum:unique} The true parameter $ \bbeta $ is a unique solution to $ \bbE[\bphi_{\eff}(\bbeta;\widecheck{\mu}_{0},p_{\sA},\widecheck{\sigma}^{2})] = \bzero $.
	\end{enumerate}
\end{asm}
Note that the definition of $ \widecheck{\cI}(\bbeta) $ only depends on the working variance function $ \widecheck{\sigma}^{2} $ through $ \widecheck{\sfV}_{\epsilon}(\bX) $. We denote $ \widecheck{\cI} $ to reflect that it depends on $ \widecheck{\sigma}^{2} $. It can be shown that for any $ \widecheck{\bbeta} \in \sB $, we have $ \widecheck{\cI}(\widecheck{\bbeta}) = \bbE[-(\partial/\partial \bbeta^{\intercal}) \bphi_{\eff}(\widecheck{\bbeta};\widecheck{\mu}_{0},p_{\sA},\widecheck{\sigma}^{2})] $. In Theorem \ref{thm:eff}, we establish the semiparametric efficiency of E-Learning. For symmetric matrices $ \sfA $ and $ \sfB $, the matrix inequality $ \sfA \le \sfB $ means that $ \sfB - \sfA $ is positive semi-definite.

\begin{thm}[Semiparametric Efficiency under Correct Specification]\label{thm:eff}
	Consider Model (\ref{eq:model}) and the angle-based representation in Lemma \ref{lem:ang}. Suppose $ \widehat{\bbeta}_{\eff,n} $ is the solution to the estimating equation $ \bbE_{n}[\bphi_{\eff}(\widehat{\bbeta}_{\eff,n};\widehat{\mu}_{n},p_{\sA},\widehat{\sigma}_{n}^{2})] = \bzero $ from (\ref{eq:eLearn}). Then under Assumptions \ref{asm:prop}-\ref{asm:regular}, we have
	\[ \widehat{\bbeta}_{\eff,n} - \bbeta = \widecheck{\cI}(\bbeta)^{-1}\bbE_{n}[\bphi_{\eff}(\bbeta;\widecheck{\mu}_{0},p_{\sA},\widecheck{\sigma}^{2})] + \smallcO_{\bbP}(n^{-1/2}). \]
	Moreover, if $ (\widecheck{\mu}_{0},\widecheck{\sigma}^{2}) = (\mu_{0},\sigma^{2}) $, then $ \widehat{\bbeta}_{\eff,n} $ is semiparametric efficient, in the sense that for any other \textit{Regular and Asymptotic Linear (RAL)} estimate $ \widehat{\bbeta}_{n} $, we have
	\[ \cI(\bbeta)^{-1} = \lim_{n\to\infty}n\sfvar(\widehat{\bbeta}_{\eff,n}) \le \lim_{n\to\infty}n\sfvar(\widehat{\bbeta}_{n}), \]
	where $ \cI(\bbeta) := \bbE\left[ \dot{\sfF}(\bX;\bbeta)^{\intercal}\Omega^{\intercal}\Omega\sfV_{\epsilon}(\bX)^{-1}\Omega^{\intercal}\Omega\dot{\sfF}(\bX;\bbeta) \right] $ is the semiparametric information matrix.
\end{thm}

For specific parametric models of $ \vec{f}(\bX;\bbeta) $, the information matrix can be simplified. In the binary treatment case discussed in Section \ref{sec:binary}, we have $ v_{\epsilon}(\bX) = {\sigma^{2}(\bX,1) \over p_{\sA}(1|\bX)} + {\sigma^{2}(\bX,0) \over p_{\sA}(0|\bX)} $, which becomes a scalar weight. It is shown that E-Learning is equivalent to the generalized least-squares problem \eqref{eq:eLearn_wls} with the weight $ v_{\epsilon}^{-1}(\bX)p_{\sA}(A|\bX)^{-1} $, which is also the overlap weight under heteroscedasticity \citep{crump2006moving,li2019propensity}. Then the information is $ \cI(\bbeta) = \bbE[v_{\epsilon}(\bX)^{-1}\dot{\fb}(\bX;\bbeta)^{\otimes 2}] $, where $ \dot{\fb}(\bX;\bbeta) := (\partial/\partial\bbeta)f(\bX;\bbeta) $. For HPLM \eqref{eq:plm} in the multiple treatment case (Section \ref{sec:multiple}), the information matrix becomes $ \cI(\sfB) = \bbE\left[ \sfV_{\epsilon}(\bX)^{-1} \otimes \bX^{\otimes 2} \right] $.

In Theorem \ref{thm:eff}, if $ \widecheck{\mu}_{0} \ne \mu_{0} $, then $ \widehat{\bbeta}_{\eff,n} $ is not semiparametric efficient. A natural question is to ask whether there exists some $ \widecheck{\sigma}^{2} $ such that $ \widehat{\bbeta}_{\eff,n} $ is still ``optimal'' in certain sense. This motivates our discussion in Section \ref{sec:asymptotic_misspecified}.

\subsubsection{Misspecified Treatment-Free Effect Model}\label{sec:asymptotic_misspecified}

Going beyond the double robustness and semiparametric efficiency of the estimating function (\ref{eq:estim_fun}), we are further interested in certain optimality when misspecified treatment-free effect happens. Specifically, we first define the \textit{regular} class of semiparametric estimates of $ \bbeta $.

\begin{dfn}[Regular Class of Semiparametric Estimates]\label{dfn:regular}
	Denote $ \widehat{\bbeta}_{n}=\widehat{\bbeta}_{n}(\widecheck{\mu}_{0}) $ as an estimate based on $ n $ observations independent and identically distributed from Model (\ref{eq:model}), and take the working treatment-free effect function $ \widecheck{\mu}_{0} $ as its input. We define a regular class of semiparametric estimates $ \cB_{n}(\widecheck{\mu}_{0}) $ as follows. For any $ \widehat{\bbeta}_{n}(\widecheck{\mu}_{0}) \in \cB_{n}(\widecheck{\mu}_{0}) $, there exists some $ \bh:\cX \times \cA \to \bbR^{p} $, which can depend on $ (\bbeta,\widecheck{\eta}) $, such that:
	\begin{itemize}
		\item The estimate $ \widehat{\bbeta}_{n}(\widecheck{\mu}_{0}) $ corresponds to the estimating function
		\[ \bphi(\bbeta;\widecheck{\mu}_{0}) = [Y - \widecheck{\mu}_{0}(\bX) - \gamma(\bX,A;\bbeta)]\bh(\bX,A;\bbeta,\widecheck{\eta}). \]
		That is, $ \bbE_{n}[\bphi(\widehat{\bbeta}_{n}(\widecheck{\mu}_{0});\widecheck{\eta})] = \bzero $;
		
		\item (Consistency) $ \bbE[\bh(\bX,A;\bbeta,\widecheck{\eta})|\bX] = \bzero $.
	\end{itemize}
\end{dfn}

Note that the consistency condition is equivalent to $ \bbE[\bphi(\bbeta;\widecheck{\mu}_{0})] = \bzero $ for any $ \widecheck{\mu}_{0}:\cX \to \bbR $. This can be concluded from that $ \bbE[\bh(\bX,A;\bbeta,\widecheck{\eta})|\bX] = \bzero $ if and only if for any $ \widecheck{\mu}_{0}$, we have $ \bzero = \bbE[\bphi(\bbeta;\widecheck{\mu}_{0})] = \bbE\Big\{ [\mu_{0}(\bX) - \widecheck{\mu}_{0}(\bX)]\bbE[\bh(\bX,A;\bbeta,\widecheck{\eta})|\bX] \Big\} $. The consistency can be met by any doubly robust estimates with a correct propensity score, such as G-Estimation, A-Learning, dWOLS, and RD-Learning.

If $ \widecheck{\mu}_{0} $ is the true treatment-free effect $ \mu_{0} $ in Model (\ref{eq:model}), then by \citet[Theorem 4.2]{tsiatis2007semiparametric}, any semiparametric RAL estimate of $ \bbeta $ must have an influence function in the form of $ \bphi(\bbeta;\mu_{0}) $ in Definition \ref{dfn:regular}. That is, for any RAL estimate $ \widetilde{\bbeta}_{n} $, there exists some $ \widehat{\bbeta}_{n}(\mu_{0}) \in \cB_{n}(\mu_{0}) $, such that $ \widetilde{\bbeta}_{n} = \widehat{\bbeta}_{n}(\mu_{0})+ \smallcO_{\bbP}(n^{-1/2}) $ under Model (\ref{eq:model}). Therefore, $ \cB_{n}(\mu_{0}) $ can represent the equivalent classes of RAL estimates, where two RAL estimates are ``equivalent'' if and only if their $ \sqrt{n} $-asymptotic variances are the same. In particular, $ \cB_{n}(\mu_{0}) $ consists of the ``regular versions'' such that their estimating functions coincide with their IFs. 

Definition \ref{dfn:regular} provides a useful class of estimates with a specific form of dependency on the working treatment-free effect function $ \widecheck{\mu}_{0} $. In fact, the following Theorem \ref{thm:eff_opt} shows that, given a working treatment-free effect function $ \widecheck{\mu}_{0} $, there exists some optimal RAL estimate among the regular class $ \cB_{n}(\widecheck{\mu}_{0}) $, in the sense that its $ \sqrt{n} $-asymptotic variance is the smallest.

\begin{thm}[Optimal Efficiency Improvement under Misspecification]\label{thm:eff_opt}
	Given a working treatment-free effect function $ \widecheck{\mu}_{0}:\cX \to \bbR $, consider Model (\ref{eq:model}) and the regular class of semiparametric estimates $ \cB_{n}(\widecheck{\mu}_{0}) $ in Definition \ref{dfn:regular}. Define 
	\[ \sigma_{\opt}^{2}(\bX,A;\widecheck{\mu}_{0}) := [\widecheck{\mu}_{0}(\bX) - \mu_{0}(\bX)]^{2} + \sigma^{2}(\bX,A), \]
	and $ \widehat{\bbeta}_{\eff,n}(\widecheck{\mu}_{0}) \in \cB_{n}(\widecheck{\mu}_{0}) $ as the solution to $ \bbE_{n}[\bphi_{\eff}(\widehat{\bbeta}_{\eff,n}(\widecheck{\mu}_{0});\widecheck{\mu}_{0},p_{\sA},\sigma_{\opt}^{2})] = \bzero $ from (\ref{eq:eLearn}). Then under Assumptions \ref{asm:prop}-\ref{asm:regular}, we have
	\[ \cI(\bbeta;\widecheck{\mu}_{0})^{-1} = \lim_{n\to\infty}n\sfvar[\widehat{\bbeta}_{\eff,n}(\widecheck{\mu}_{0})] \le \lim_{n\to\infty}n\sfvar[\widehat{\bbeta}_{n}(\widecheck{\mu}_{0})]; \quad \forall \widehat{\bbeta}_{n}(\widecheck{\mu}_{0})\in \cB_{n}(\widecheck{\mu}_{0}), \]
	where $ \sfV_{\epsilon}(\bX;\widecheck{\mu}_{0}) := \sum_{k=1}^{K}{\sigma_{\opt}^{2}(\bX,A;\widecheck{\mu}_{0})\bomega_{k}^{\otimes 2} \over p_{\sA}(k|\bX)} $ and  $ \cI(\bbeta;\widecheck{\mu}_{0}) := \bbE\left[ \dot{\sfF}(\bX;\bbeta)^{\intercal}\Omega^{\intercal}\Omega\sfV_{\epsilon}(\bX;\widecheck{\mu}_{0})^{-1}\Omega^{\intercal}\Omega\dot{\sfF}(\bX;\bbeta) \right] $.
\end{thm}

Note that Theorem \ref{thm:eff_opt} can be more general than the semiparametric efficiency in Theorem \ref{thm:eff}, in the sense that the optimality in Theorem \ref{thm:eff_opt} is for a general working treatment-free effect function. Specifically, if $ \widecheck{\mu}_{0} = \mu_{0} $, then $ \cB_{n}(\mu_{0}) $ in Definition \ref{dfn:regular} represents the equivalent classes of RAL estimates with the $ \sqrt{n} $-asymptotic variance as the equivalence relationship. In that case, Theorem \ref{thm:eff_opt} recovers Theorem \ref{thm:eff} that $ \widehat{\bbeta}_{\eff,n} $ has the smallest $ \sqrt{n} $-asymptotic variance. As a remark, we would like to point out that	Theorem \ref{thm:eff_opt} can be extended to the estimating equation $ \bbE_{n}[\bphi_{\eff}(\widehat{\bbeta}_{\eff,n};\widehat{\mu}_{0,n},p_{\sA},\widehat{\sigma}_{n}^{2})] = \bzero $ with plug-in nuisance function estimates $ (\widehat{\mu}_{0,n},\widehat{\sigma}_{n}^{2}) $. The argument is similar to Theorem \ref{thm:eff}, and we omit the details here.

If the working treatment-free effect function $ \widecheck{\mu}_{0} $ is not identical to the true treatment-free effect function $ \mu_{0} $ in Model (\ref{eq:model}), then Theorem \ref{thm:eff_opt} suggests an optimal variance function $ \sigma_{\opt}^{2}(\bX,A;\widecheck{\mu}_{0}) $. For the binary treatment case, the optimal working variance function can correspond to
\[ v_{\epsilon}(\bx;\widecheck{\mu}_{0}) = {\sigma_{\opt}^{2}(\bx,1;\widecheck{\mu}_{0}) \over p_{\sA}(1|\bx)} + {\sigma_{\opt}^{2}(\bx,0;\widecheck{\mu}_{0}) \over p_{\sA}(0|\bx)} = {\sigma^{2}(\bx,1) \over p_{\sA}(1|\bx)} + {\sigma^{2}(\bx,0) \over p_{\sA}(0|\bx)} + {[\widecheck{\mu}_{0}(\bx) - \mu_{0}(\bx)]^{2} \over p_{\sA}(1|\bx)p_{\sA}(0|\bx)}. \]
The corresponding generalized least-squares estimate from (\ref{eq:eLearn_wls}) can achieve the smallest $ \sqrt{n} $-asymptotic variance among the regular class of estimates $ \cB_{n}(\widecheck{\mu}_{0}) $. The motivating example in Section \ref{sec:eg} is a special case when we further assume $ p_{\sA}(1|\bX) = p_{\sA}(0|\bX) = 1/2 $.

\begin{rmk}[General Asymptotic Variance]
	It can be useful to compute the $ \sqrt{n} $-asymptotic variance for arbitrary working treatment-free effect and variance function $ (\widecheck{\mu}_{0},\widecheck{\sigma}^{2}) $. Suppose $ \widehat{\bbeta}_{n} $ is the solution to $ \bbE[\bphi_{\eff}(\widehat{\bbeta}_{n};\widecheck{\mu}_{0},p_{\sA},\widecheck{\sigma}^{2})] = \bzero $. Then we have
	\begin{align}
		\bbE[\bphi_{\eff}(\bbeta;\widecheck{\mu}_{0},p_{\sA},\widecheck{\sigma}^{2})^{\otimes 2}] = \bbE\Big[ \dot{\sfF}(\bX;\bbeta)^{\intercal}\Omega^{\intercal}\Omega\widecheck{\sfV}_{\epsilon}(\bX)^{-1} \sfV_{\epsilon}(\bX;\widecheck{\mu}_{0}) \widecheck{\sfV}_{\epsilon}(\bX)^{-1}\Omega^{\intercal}\Omega\dot{\sfF}(\bX;\bbeta) \Big],\label{eq:var_misspecified}
	\end{align}
	and $ \bbE[-(\partial/\partial\bbeta^{\intercal})\bphi_{\eff}(\bbeta;\widecheck{\mu}_{0},p_{\sA},\widecheck{\sigma}^{2})] = \widecheck{\cI}(\bbeta) $. The $ \sqrt{n} $-asymptotic variance is given by the sandwich form $ \lim_{n\to\infty}n\sfvar(\widehat{\bbeta}_{n}) = \widecheck{\cI}(\bbeta)^{-1}[(\ref{eq:var_misspecified})]\widecheck{\cI}(\bbeta)^{-1} $.
\end{rmk}

\begin{rmk}[Incorrect Propensity Score] 
	In our theoretical analysis, we assume that the propensity score is known or can be consistently estimated. In Section \ref{sec:prop_mis} in the Supplementary Material, we further discuss the case when the propensity score is incorrect. Although the optimality of Theorem \ref{thm:eff_opt} cannot be recovered in this case, the covariate-dependent variance adjustment for the optimal working variance function $ \widecheck{\sigma}_{\rm opt}^{2}(\bX,A) $ can still be helpful. We demonstrate in our simulation studies (Section \ref{sec:simulation}) that E-Learning still outperforms other methods even with incorrect propensity.
\end{rmk}

In Theorem \ref{thm:eff_opt}, we establish the optimality of using the working variance function $ \sigma_{\opt}^{2}(\bX,A;\widecheck{\mu}_{0}) $ in the proposed E-Learning. As discussed in Section \ref{sec:eLearn}, the optimal working variance function can be identified by the expectation of the squared working residual. This can confirm the optimality of the E-Learning estimate.

\subsection{Regret Bound}\label{sec:regret}
In this section, we relate the theoretical results for estimation in Section \ref{sec:asymptotic} to the regret bound for the estimated ITR. Recall from Theorem \ref{thm:regret_est} that the estimation error of the interaction effect can dominate the regret. We further make compactness assumption on covariates to establish the regret bound.

\begin{asm}[Compact Covariate Domain]\label{asm:cmpt}
	The support of the distribution $ p_{\sX}(\bx) $ is compact.
\end{asm}

\begin{thm}[Regret Bound for RAL Estimate]\label{thm:regret_par}
	Consider Model (\ref{eq:model}) and the angle-based representation in Lemma \ref{lem:ang}. Suppose $ \sqrt{n}(\widehat{\bbeta}_{n} - \bbeta) \stackrel{\cD}{\to} \cN_{p}(\bzero,\Sigma) $ for some $ \Sigma > 0 $. Define $ \widehat{d}_{n}(\bx) := \argmax_{1 \le k \le K}\langle \bomega_{k},\vec{\bbf}(\bx;\widehat{\bbeta}_{n}) \rangle $ and $ d^{\star}(\bx) := \argmax_{1 \le k \le K}\langle \bomega_{k},\vec{\bbf}(\bx;\bbeta) \rangle $. Then under Assumptions \ref{asm:int}, \ref{enum:int} and \ref{asm:cmpt}, we have
	\begin{align*}
		\limsup_{n\to\infty}\sqrt{n}[\cV(d^{\star}) - \bbE\cV(\widehat{d}_{n})] &\le 2\lim_{n\to\infty}\left\{ n\sum_{k=1}^{K} \bbE[\gamma(\bX,k;\widehat{\bbeta}_{n}) - \gamma(\bX,k;\bbeta)]^{2} \right\}^{1/2}\\
		&= 2\left( 1 - {1 \over K} \right)^{1/2}\tr\left\{ \bbE\left[ \dot{\sfF}(\bX;\bbeta)^{\intercal}\Omega^{\intercal}\Omega\dot{\sfF}(\bX;\bbeta) \right]\Sigma \right\}^{1/2}.
	\end{align*}
\end{thm}

The regret bound in Theorem \ref{thm:regret_par} can be tight compared to Theorem \ref{thm:regret_est}, since Theorem \ref{thm:regret_par} only relaxes the absolute estimation error to the squared estimation error, and the maximization to the summation. Theorem \ref{thm:regret_par} further implies that the regret bound and the estimation error are both in $ \sqrt{n} $-order, where the leading constant depends on the $ \sqrt{n} $-asymptotic variance $ \Sigma $ of the estimated parameters $ \widehat{\bbeta}_{n} $. In particular, denote $ \| \cdot \|_{\rm F} $ as the Frobenius norm. Then we have
\[ \tr\left\{ \bbE\left[ \dot{\sfF}(\bX;\bbeta)^{\intercal}\Omega^{\intercal}\Omega\dot{\sfF}(\bX;\bbeta) \right]\Sigma \right\} \le \left \| \bbE\left[ \dot{\sfF}(\bX;\bbeta)^{\intercal}\Omega^{\intercal}\Omega\dot{\sfF}(\bX;\bbeta) \right] \right \|_{\rm F} \times \| \Sigma \|_{\rm F}, \]
with equality if $ \dot{\sfF}(\bX;\bbeta) $ contains $ \bX $ and we take supremum over all possible covariate distribution $ p_{\sX} $ on both sides. This  suggests that an RAL estimate with the smallest $ \sqrt{n} $-asymptotic variance $ \Sigma $ can achieve the minimal regret bound. This complements the theoretical results in Sections \ref{sec:asymptotic_correct} and \ref{sec:asymptotic_misspecified} that establish the optimality of E-Learning estimate of $ \bbeta $ in terms of the $ \sqrt{n} $-asymptotic variance. In particular, if we use the efficient estimate $ \widehat{\bbeta}_{\eff,n}(\widecheck{\mu}_{0}) $ with the optimal choice of working variance function $ \sigma_{\opt}^{2}(\bX,A;\widecheck{\mu}_{0}) $, then the $ \sqrt{n} $-asymptotic variance $ \Sigma $ becomes $ \cI(\bbeta;\widecheck{\mu}_{0})^{-1} $, and the regret bound above is the smallest among all RAL estimates in $ \cB(\widecheck{\mu}_{0}) $.

To conclude this section, we have established that E-Learning is doubly robust and optimal with the smallest $\sqrt{n}$-asymptotic variance among the class of regular semiparametric estimates in Definition \ref{dfn:regular}, which can allow multiple treatments, heteroscedasticity and misspecified treatment-free effect. The corresponding regret bound can also have an optimal leading constant in the $n^{-1/2}$-order.

%

\section{Simulation Study}\label{sec:simulation}

We consider several simulation studies to compare the proposed E-Learning with existing methods from the literature and demonstrate the superiority of E-Learning. In particular, the data generation based on the heteroscedasticitic partially linear model \eqref{eq:plm} is assumed throughout this section, where the interaction effect is linear. In the Supplementary Material Section \ref{sec:simulation_nonlinear}, we consider additional simulation setups with nonlinear interaction effects, and explore the competitive performance of E-Learning with the linear or cubic polynomial basis.

\subsection{Data Generating Process and Model Specifications}\label{sec:DGP}
The synthetic data generation process is as follows. Let $ n \in \{100,200,400,800,1600\} $ be the training sample size, $ p \in \{ 10,50,100 \} $ be the number of variables, and $ K \in \{2,3,5,7\} $ be the number of treatments. First, we generate the coefficients of the treatment-covariate-interaction effect by $ (\widetilde{\beta}_{0k},\widetilde{\beta}_{1k},\widetilde{\beta}_{2k},\widetilde{\beta}_{3k},\widetilde{\beta}_{4k},\widetilde{\beta}_{5k}) \sim \unif\{ \bu \in \bbR^{6}: \|\bu\|_{2} = 1 \} $ independently for $ 1 \le k \le K $, $ \beta_{jk} := \widetilde{\beta}_{jk} - {1 \over K}\sum_{k'=1}^{K}\widetilde{\beta}_{jk'} $ for $ 0 \le j \le 5 $, and $ \bbeta_{k} := (\beta_{0k},\beta_{1k},\beta_{2k},\beta_{3k},\beta_{4k},\beta_{5k},\overbrace{0,\cdots,0}^{p-5})^{\intercal} $ for $ 1 \le k \le K $. Then we generate the data from:
\[ \bX \sim \cN_{p}(\bzero,\Ib_{p\times p}); \quad \bbP(A=k|\bX) = \underbrace{p_{\sA}(k|\bX)}_{\text{propensity score}} = {e(\bX,k) \over \sum_{k'=1}^{K}e(\bX,k')}; \quad 1 \le k \le K; \]
\[ Y|(\bX,A) = \underbrace{\mu_{0}(\bX) - \bbE[\mu_{0}(\bX)]}_{\text{treatment-free effect}} + \underbrace{\bbeta_{A}^{\intercal}(1,\bX^{\intercal})^{\intercal}}_{\text{interaction effect}}+ \underbrace{\sigma(\bX,A)}_{\text{variance function}} \times \cN(0,1). \]
For the coefficient vectors $ \{\bbeta_{k}\}_{k=1}^{K} $, the optimal ITR is $ d^{\star}(\bx) = \argmax_{1 \le k \le K}\bbeta_{k}^{\intercal}(1,\bx^{\intercal})^{\intercal} $. Here, the true treatment-free effect function $ \mu_{0}(\bx) $, variance functions $ \{\sigma^{2}(\bx,k)\}_{k=1}^{K} $ and the propensity score functions $ \{ e(\bx,k) \}_{k=1}^{K} $ are defined according to Table \ref{tab:model}. 

\renewcommand{\arraystretch}{1.5}
\begin{table}[!ht]
	\centering
	\caption[Simulation, Specification]{True Models and the Implying Model Specifications in the Simulation Studies}
	\label{tab:model}
	\begin{threeparttable}
		\begin{tabular}{c||r|>{\centering}m{1.2cm}|>{\centering}m{1.2cm}|>{\centering}m{1.2cm}|>{\centering}m{1.2cm}|l||>{\raggedright}m{1cm}|l}
			\hline\hline
			\multicolumn{2}{c||}{} & \multicolumn{2}{c|}{\textit{Correctly Specified}} & \multicolumn{2}{c|}{\textit{Misspecified}} & \multicolumn{3}{l}{\textbf{\textit{Treatment-Free Effect}}} \\
			\cline{3-9}
			\multicolumn{2}{c||}{} & \textit{\footnotesize Homo-scedastic} & \textit{\footnotesize Hetero-scedastic} & \textit{\footnotesize Homo-scedastic} & \textit{\footnotesize Hetero-scedastic} & \multicolumn{3}{l}{\textbf{\textit{Variance}}} \\
			\hline\hline
			\parbox[t]{2mm}{\multirow{4}{*}{\rotatebox[origin=c]{90}{\textbf{Truth}}}} & $ \mu_{0}(\bx) = $ &
			\multicolumn{2}{c|}{$ {1 \over \sqrt{K}}\sum_{k'=1}^{K}x_{k'}  $} & \multicolumn{2}{c||}{$ {1 \over K}\sum_{k'=1}^{K}e^{\sqrt{2}x_{k'}}$} & \multirow{2}{*}{$ e(\bx,k) = e^{x_{k}/2}  $} & \parbox[t]{2mm}{\multirow{2}{*}{\rotatebox[origin=c]{-90}{\makecell[c]{\textit{\scriptsize Correctly} \\ \textit{\small Specified}}}}} & \parbox[t]{2mm}{\multirow{4}{*}{\rotatebox[origin=c]{-90}{\textbf{\scriptsize \textit{Propensity Score}}}}} \\
			\cline{3-6}
			& $ \sigma^{2}(\bx,k) = $ & $ 1 $ & $  e^{2\sqrt{2}x_{k}} $ & $ 1 $ & \multicolumn{1}{c||}{$ e^{2\sqrt{2}x_{k}} $} & & & \\
			\cline{2-8} 
			& $ \mu_{0}(\bx) = $ &
			\multicolumn{2}{c|}{$ {1 \over \sqrt{K}}\sum_{k'=1}^{K}x_{k'} $} & \multicolumn{2}{c||}{$ {1 \over K}\sum_{k'=1}^{K}e^{\sqrt{2}x_{k'}} $} & \multirow{2}{*}{$ e(\bx,k) = |x_{k}|^{1/2} $} & \parbox[t]{2mm}{\multirow{2}{*}{\rotatebox[origin=c]{-90}{\makecell[c]{\textit{\small Mis-} \\ \textit{\small specified}}}}} & \\
			\cline{3-6}
			& $ \sigma^{2}(\bx,k) = $ & $ 1 $ & $ e^{2\sqrt{2}x_{k}} $ & $ 1 $ & \multicolumn{1}{c||}{$ e^{2\sqrt{2}x_{k}} $} & & & \\
			\hline\hline
		\end{tabular}
		\begin{tablenotes}
			\fontsize{8}{5}\selectfont
			\item[1] The treatment-free effect is estimated by a linear working model. 
			\item[2] The propensity score is estimated by a multinomial logistic working model.
		\end{tablenotes}
	\end{threeparttable}
\end{table}

When estimating the treatment-free effect $ \mu_{0}(\bX) - \bbE[\mu_{0}(\bX)] $, we consider a linear working model $ \widecheck{\mu}_{0}(\bX;\eta) = \bmeta^{\intercal}(1,\bX^{\intercal})^{\intercal} $. Then the treatment-free effect model is correctly specified if the truth is $ \mu_{0}(\bx) = {1 \over \sqrt{K}}\sum_{k'=1}^{K}x_{k'} $, while misspecified if the truth is $ \mu_{0}(\bx) = {1 \over K}\sum_{k'=1}^{K}e^{\sqrt{2}x_{k'}}$.  In Figure \ref{fig:tf} in the Supplementary Material, we provide the fitted treatment-free effect plots when the model is correctly and incorrectly specified. It shows that the estimated treatment-free effect is consistent if correctly specified, and deviates from the truth if misspecified. When estimating the propensity score functions $ \{p_{\sA}(k|\bX)\}_{k=1}^{K} $, we consider a multinomial logistic working model $ \widecheck{p}_{\sA}(k|\bX;\btau_{1},\btau_{2},\cdots,\btau_{K}) = {\exp[\btau_{k}^{\intercal}(1,\bX^{\intercal})^{\intercal}] \over \sum_{k'=1}^{K}\exp[\btau_{k'}^{\intercal}(1,\bX^{\intercal})^{\intercal}]} $. Then the propensity score model is correctly specified if the truth is generated from $ e(\bx,k) = e^{x_{k}/2} $, while misspecified if the truth is generated from $ e(\bx,k) = |x_{k}|^{1/2} $. In Figure \ref{fig:prop} in the Supplementary Material, we provide the fitted propensity score plots when the model is correctly and incorrectly speceified, and demonstrate how the misspecified model affects the fitted propensity scores.  As discussed in Section \ref{sec:eLearn}, if one of or both misspecified treatment-free effect model and heteroscedasticity exist, the squared residuals can depend on $ (\bX,A) $. In Figures \ref{fig:resid2_1}-\ref{fig:resid2_4}, we provide the residual plots in all these cases to demonstrate such dependencies.

\subsection{Binary Treatments}\label{sec:simulation_binary}
In this section, we consider the binary treatment case ($ K = 2 $) and compare E-Learning with existing methods from literature discussed in Table \ref{tab:compare} in Section \ref{sec:compare}. The implementation details of these methods are provided in Section \ref{sec:implement_detail} in the Supplementary Material.

For the implementation of E-Learning, we consider HPLM \eqref{eq:plm} and solve the regularized estimating equation in Section \ref{sec:reg} with the row-wise group-LASSO penalty. We follow the implementation in Section \ref{sec:implement} for the estimation of the treatment-free effect with the linear working model, the propensity score with the multinomial logistic working model, and the variance function with regression forest. The tuning parameter $ \lambda $ is chosen based on 10-fold cross validation. We consider the oracle working variance function $ \sigma_{\opt}^{2}(\bX,A) = [\widecheck{\mu}_{0}(\bX) - \mu_{0}(\bX)]^{2} + \sigma^{2}(\bX,A) $ and the estimated one from the regression forest using the squared residual as the response and $ (\bX,A) $ as the covariates. At the testing stage, a testing covariate sample $ \{ \bX_{i} \}_{i=1}^{n_{\rm test}=10000} \stackrel{\rm i.i.d.}{\sim}\cN_{p}(\bzero,\Ib_{p \times p}) $ is generated, and the testing value of an estimated ITR $ \widehat{d} $ is computed as $ \widehat{\cV}_{\rm test}(\widehat{\bd}) = {1 \over n_{\rm test}}\sum_{i=1}^{n_{\rm test}}\sum_{k=1}^{K}\bbeta_{k}^{\intercal}(1,\bX_{i}^{\intercal})^{\intercal}\bbone[\widehat{d}(\bX_{i}) = k] $. Recall that the optimal ITR is $ d^{\star}(\bx) = \argmax_{1 \le k \le K}\bbeta_{k}^{\intercal}(1,\bx^{\intercal})^{\intercal} $. Then we report the testing regret, $ \widehat{\cV}_{\rm test}(d^{\star}) - \widehat{\cV}_{\rm test}(\widehat{d}) $, and the testing misclassification rate, $ {1 \over n_{\rm test}}\sum_{i=1}^{n_{\rm test}}\bbone[\widehat{d}(\bX_{i}) \ne d^{\star}(\bX_{i})] $. The training-testing process is replicated for 100 times for each of the model specification scenarios in Table \ref{tab:model}. 

\begin{figure}[!ht]
	\centering
	\includegraphics[width=\linewidth]{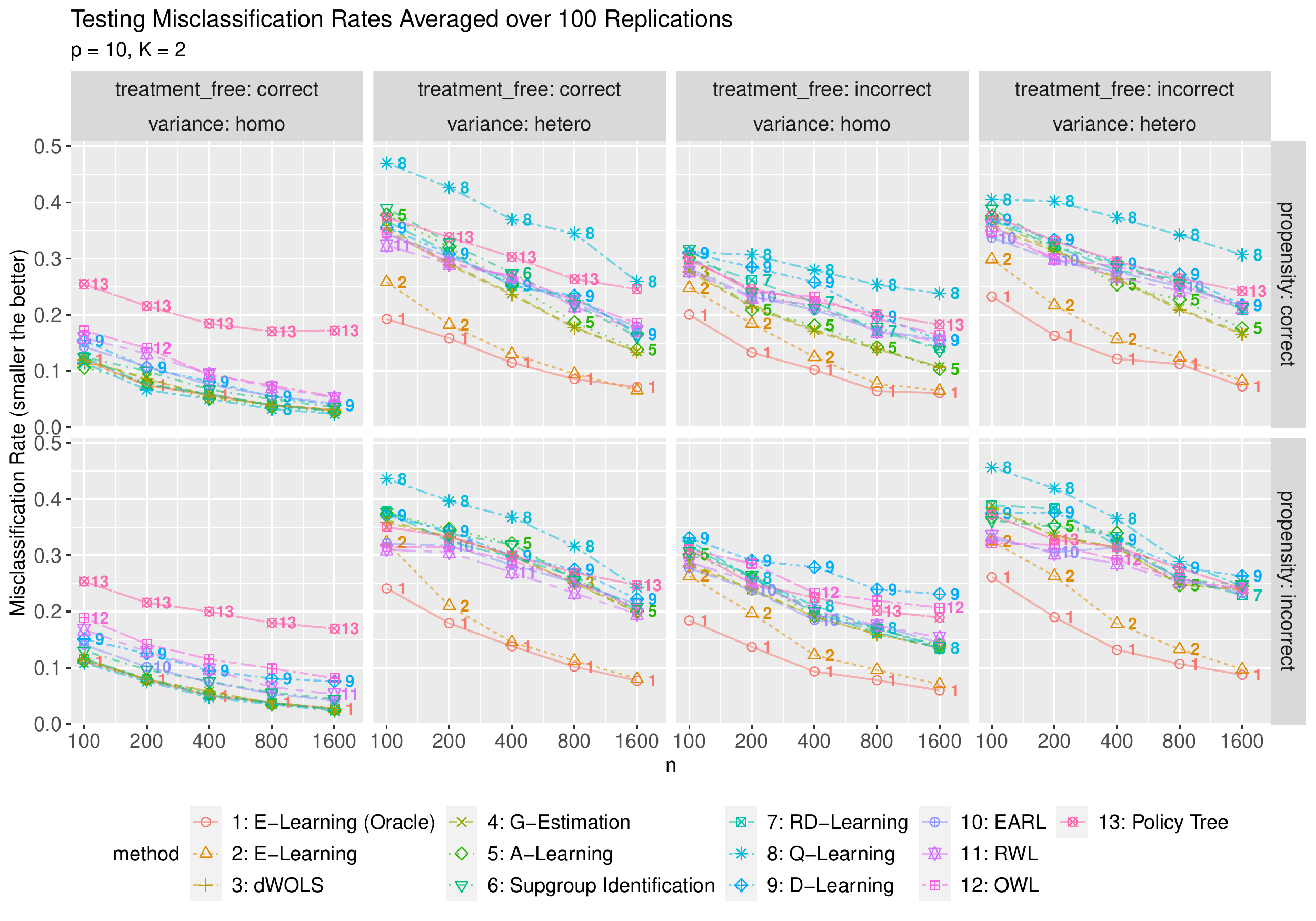}
	\caption[Simulation, Misclassification Rates]{Testing misclassification rates (smaller the better) for $ n \in \{ 100,200,400,800,1600 \} $, $ p = 10 $, $ K = 2 $ and each of the model specification scenarios in Table \ref{tab:model}. Methods in Table \ref{tab:compare} are compared, where \textit{E-Learning (Oracle)} corresponds to E-Learning with the oracle working variance function, and \textit{Policy Tree} corresponds to \textit{Policy Learning} with decision trees.}
	\label{fig:all_misclass}
\end{figure}

\begin{figure}[!ht]
	\centering
	\includegraphics[width=\linewidth]{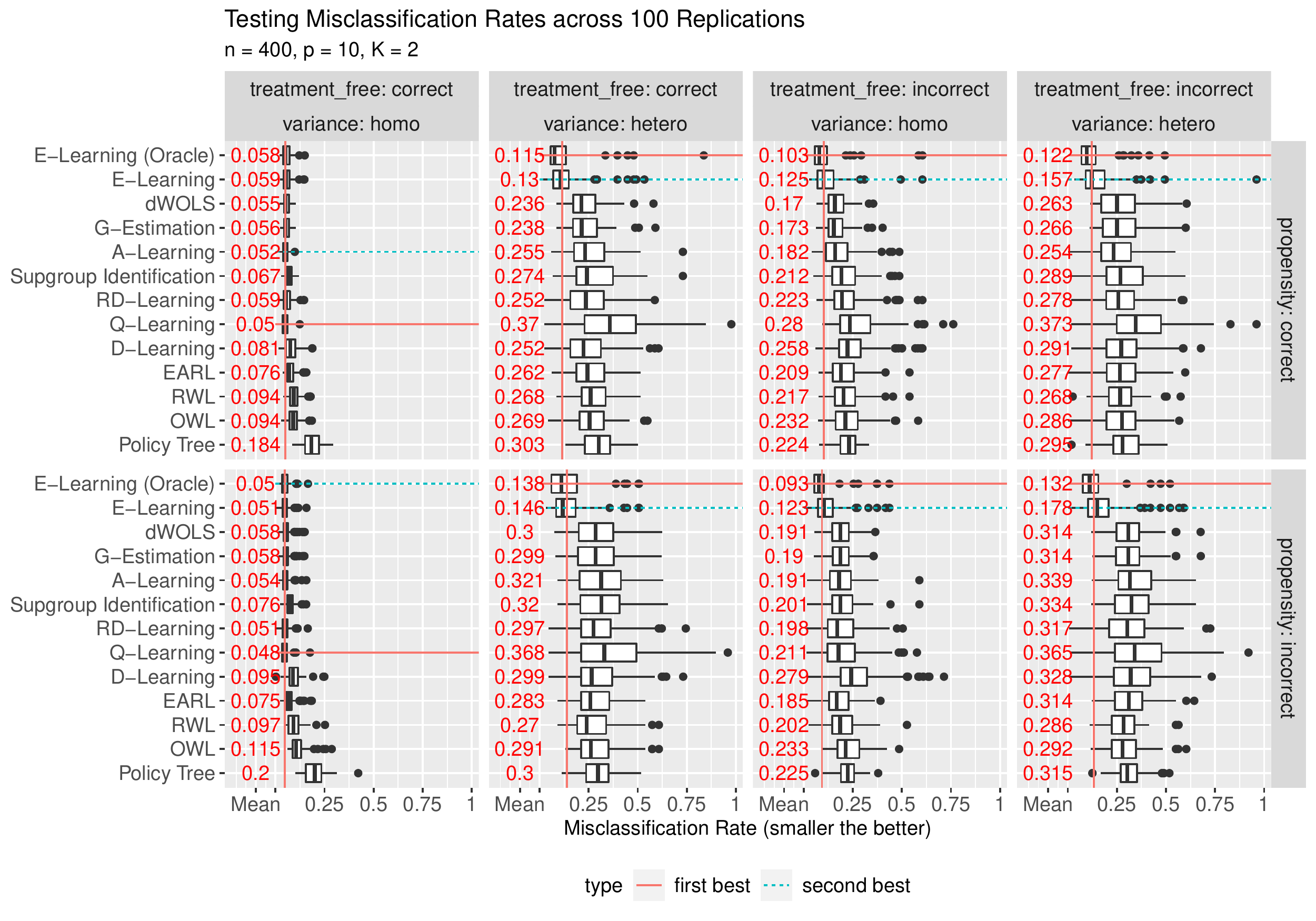}
	\caption[Simulation, Misclassification Rates, $ n = 400 $]{Testing misclassification rates (smaller the better) for $ n = 400 $, $ p = 10 $, $ K = 2 $ and each of the model specification scenarios in Table \ref{tab:model}. Methods in Table \ref{tab:compare} are compared, where \textit{E-Learning (Oracle)} corresponds to E-Learning with the oracle working variance function, and \textit{Policy Tree} corresponds to \textit{Policy Learning} with decision trees. First and second best methods in terms of the averaged misclassification rates are annotated in horizontal lines. The minimal averaged misclassification rate is shown by the vertical line.}
	\label{fig:all_n=400_misclass}
\end{figure}

We first consider the low-dimensional setting ($ p = 10 $). Figure \ref{fig:all_misclass} reports the testing misclassification rates for the training sample sizes $ n \in \{100,200,400,800,1600\} $ and each of the specification scenarios listed in Table \ref{tab:model}, while Figure \ref{fig:all_n=400_misclass} provides more details for $ n = 400 $. In the case of correctly specified treatment-free effect, correctly specified propensity score, and homoscedasticity (upper-left panel of the plots), E-Learning, Q-Learning, G-Estimation, A-Learning, RD-Learning, dWOLS and Subgroup Identification have similar testing performance, since all of them leverage the correct parametric model assumption. Here, although Subgroup Identification does not rely on a specific parametric model assumption, it is equivalent to RD-Learning in this case as discussed in Section \ref{sec:binary}. Therefore, it can enjoy similar performance as other model-based methods. In contrast, D-Learning, OWL, RWL, EARL and Policy Tree are based on nonparametric models, and can have inferior performance in this case. When one of or both misspecified treatment-free effect and heteroscedasticity happen (columns 2-4 of the plots), the E-Learning procedures with the oracle and estimated working variance function both demonstrate the best performance among all methods. In particular, the advantages of E-Learning are more evident as $ n $ increases. Such a superiority can still maintain even if the propensity score model is misspecified (second rows of the plots). This suggests that incorrect propensity score can have relatively small impacts.

In the Supplementary Material, we further provide more plots of misclassification rates for $ n \in \{ 100,200,800,1600 \} $ (Figures \ref{fig:all_n=100_misclass}-\ref{fig:all_n=1600_misclass}) and testing regrets (Figures \ref{fig:all_regret}-\ref{fig:all_n=1600_regret}). All of them show the same patterns as in Figures \ref{fig:all_misclass} and \ref{fig:all_n=400_misclass}.
In order to further demonstrate the superiority of E-Learning in presence of moderately large number of variables, we also study the case of $ p = 50 $ and report the testing performance in Figures \ref{fig:all_p_misclass} and \ref{fig:all_p_regret}. They show that even though the increase in $ p $ can result in worse performance of all methods, the efficiency gain in sufficiently large samples $ (n = 200,400,800,1600) $ of E-Learning remains.

\subsection{Multiple Treatments}\label{sec:simulation_multiple}
We consider the multiple treatment case ($ K = 3 $) and compare E-Learning with model-based methods that can allow multiple treatments (Q-Learning, D-Learning, RD-Learning). In particular, we are interested in the following questions:
\begin{itemize}
	\item [(\Romannum{1})] Efficiency of different methods as $ n $ increases across all model specifications in Table \ref{tab:model};
	
	\item [(\Romannum{2})] The impacts of increase in the number of variables $ p $;
	
	\item [(\Romannum{3})] The impacts of increase in the number of treatments $ K $;
	
	\item [(\Romannum{4})] Effects of different nonparametric estimation methods for variance function on the performance of E-Learning.
\end{itemize}

For Question \Romannum{1}, we consider the same setup as in Section \ref{sec:simulation_binary} but with $ K = 3 $. The testing results are provided in Figure \ref{fig:n} in the Supplementary Material. In particular, E-Learning shows the same superiority over Q-Learning, D-Learning, RD-Learning as in the binary case.
For Question \Romannum{2}, we consider $ K = 3 $ and varying $ p \in \{10,50,100\} $ (Figures \ref{fig:p_misclass} and \ref{fig:p_regret}). As the number of variables $ p $ increases, the performance of all methods become worse. For $ p=50,100 $, Q-Learning, D-Learning and RD-Learning have much worse performance when one of or both treatment-free effect misspecification and heteroscedasticity happen, even with the sample size $ n = 1600 $. The misclassification rates of these methods are 0.562, 0.429 and 0.433 respectively for incorrectly specified treatment-free effect and heteroscedasticity with $ n = 1600 $ and $ p = 100 $. In contrast, for sufficiently large sample sizes ($ n = 400, 800, 1600 $), the number of variables $ p $ has less impacts on E-Learning with the oracle working variance function, while it requires 
sizes ($ n = 800, 1600 $) for E-Learning with the estimated working variance function to have comparable performance across $ p $'s. The reason for requiring larger sample sizes is due to the challenge of the high-dimensional nonparametric estimation of the working variance function. The misclassification rates of E-Learning for incorrectly specified treatment-free effect and heteroscedasticity with $ n = 1600 $ and $ p = 100 $ are 0.167 for the oracle working variance function and 0.248 for the estimated working variance function respectively. These results can confirm the superiority of E-Learning even when the number of variables increases to 100.

In order to study Question \Romannum{3}, we consider $ p = 10 $ and varying $ K \in \{2,3,5,7\} $. Notice that increasing the number of treatments can have two folds of effects. On one hand, the effective dimensionality generally increases in $ K $. For HPLM \eqref{eq:plm}, the interaction effect $ \gamma(\bX,A;\sfB) = (1-1/K)\langle \bomega_{A},\sfB^{\intercal}\bX \rangle $ is indexed by the matrix-valued parameter $ \sfB \in \bbR^{p\times(K-1)} $. The effective dimension is $ p(K-1) $ and increases with $ K $. Moreover, the number of variance functions $ \{\sigma_{\opt}^{2}(\bX,k)\}_{k=1}^{K} $ also increases in $ K $, which means more nuisance functions to be nonparametrically estimated. On the other hand, more treatments can lead to a harder classification problem. In particular, the misclassification rate of a random treatment rule $ d_{\rm rand} $ with $ \bbP[d_{\rm rand}(\bX) = k] = 1/K $ for $ 1 \le k \le K $ is $ 1 - 1/K $. Then the misclassification rate of the random treatment rule increases in $ K $, which suggests that the difficulty of the learning problem is also increasing. In Figures \ref{fig:K_misclass} and \ref{fig:K_regret} in the Supplementary Material, Q-Learning, D-Learning and RD-Learning have poor performance in presence of one of or both treatment-free effect misspecification and heteroscedasticity. When both treatment-free effect misspecification and heteroscedasticity exist, the misclassification rates of these methods with $ n = 1600 $ and $ K = 7 $ are $ 0.811 $, $ 0.648 $ and $ 0.645 $ respectively. Notice that the misclassification rate of the random treatment rule in this case is $ 1 - 1/7 = 0.857 $, which suggests that the performance of Q-Learning is close to the random treatment rule. In contrast, the E-Learning procedures with oracle working variance and estimated working variance have misclassification rates 0.299 and 0.424 in this case, which significantly outperform other methods. 

Finally, for Question \Romannum{4}, we consider $ p \in \{ 10, 50, 100 \} $, $ K = 3 $ and the comparisons among E-Learning procedures with the oracle optimal working variance function, the working variance function estimated by regression forest, MARS and COSSO. The numerical results in Figures \ref{fig:var_misclass} and \ref{fig:var_regret} suggest that E-Learning with regression forest can have better performance than E-Learning with MARS or COSSO, and the superiority remains even for $ p = 50,100 $. Therefore, we recommend using regression forest for the working variance function estimation in E-Learning.

\section{Application to a Type 2 Diabetes Mellitus (T2DM) Study}\label{sec:t2dm}

We consider a T2DM dataset from an observational study based on the \textit{Clinical Practice Research Datalink (CPRD)} \citep{herrett2015data,chen2018estimating}. The study population comprises T2DM patients of age $ \ge $ 21 years (registered 
at a CPRD practice) who received at least one of the long-acting insulins (Glargine or Detemir), the intermediate-acting insulins, the short-acting insulins, and the \textit{Glucagon-Like Peptide 1 Receptor Agonists (GLP-1 RAs)} of Exenatide and Liraglutide during 01/01/2012 - 12/31/2013. The treatment exposure $ A $ is defined as: 1) the long-acting insulins alone (with no addition of any short or intermediate-acting insulin within 60 days); 2) the intermediate-acting insulins alone (with no addition of any short or long-acting insulins within 60 days); 3) any insulin regimens including a short-acting insulin (the short-acting insulins either alone or in combinations with any long or intermediate-acting insulin); 4) the GLP-1 RAs alone. Here, for patients who received one of the insulins as well as the GLP-1 RAs, the corresponding treatment is defined as the earliest received one. 

The primary outcome $ Y $ of this study is the change of the \textit{Hemoglobin A1c (HbA1c)} lab value (\%, smaller the better) between Day 182 and Day 1 (defined as the first treatment date). The following individual covariates $ \bX $ are measured: age, gender, ethnicity, weight, height, \textit{Body Mass Index (BMI)}, \textit{High Density Lipoprotein (HDL)}, \textit{Low Density Lipoprotein (LDL)}, baseline HbA1c, smoking status, and comorbidities (any of angina, congestive heart failure, myocardial infarction, stroke, retinopathy, macular edema, renal status, neuropathy, and lower extremity amputation). The total number of records from this study is 1139, with the primary outcome available for 591 records and missing for the rest. Among the 591 observations, there is a large proportion of missingness in HDL and LDL. Therefore, for HDL and LDL, we first discretize the available observations into two levels: if the observation is above the median, then set as \texttt{high}; otherwise, set as \texttt{low}. Then we code the missing observations as \texttt{n/a}. Consequently,  all possible levels of LDL and HDL become: \texttt{high}, \texttt{low}, and \texttt{n/a}. For  categorical variables (gender, ethnicity, smoking status and comorbidities), we also code the missing observations as \texttt{n/a} and combine it with the original levels of these variables. Finally, the remaining numerical variables (age, weight, height, BMI, baseline HbA1c) have mild missingness, and we remove the records that contain any missing entries among these variables. After pre-processing the dataset as above, there remains 430 records for further analysis.

Next, we estimate the propensity scores from the dataset using the regression forest estimator in Section \ref{sec:prop}. Then we are ready to apply E-Learning, RD-Learning, D-Learning, Q-Learning and Policy Tree to the analysis of this dataset. In order to estimate the expected change of HbA1c under the fitted ITRs, we randomly sample two disjoint subsets from the dataset for training and testing. We choose various training sample sizes as $ n \in \{100,200,300\} $, and a testing sample size $ n_{\rm test} = 100 $. On the training set, we consider estimation of the propensity scores based on regression forest in the same way as that on the full dataset. We also apply different estimation methods of the treatment-free effect for RD-Learning, including: 1) the linear model on $ \bX $ with the $ \ell_{1} $-penalty (fitted by \texttt{glmnet}) as in Section \ref{sec:simulation}, 2) the regression forest on $ \bX $, 3) fitted treatment-free effect as the mean of the primary outcome on the training set, and 4) fitted treatment-free effect as 0. We find that the fitted treatment-free effect as 0 can result in better testing performance for RD-Learning. Therefore, we also use 0 as the estimated treatment-free effect in E-Learning. Since a smaller outcome is better for this problem, we negate the outcome before fitting all models. Other implementation details of all these methods remain the same as in Section \ref{sec:simulation}. On the testing dataset, we use the IPWE $ {1 \over n_{\rm test}}\sum_{i=1}^{n_{\rm test}}{\bbone[\widehat{d}(\bX_{i}) = A_{i}] \over \widehat{p}_{\sA}(A_{i}|\bX_{i})}Y_{i} $ to estimate the expected change in HbA1c under the estimated ITR $ \widehat{d} $. The training-testing process is repeated for 500 times on this dataset. 

\begin{figure}[!ht]
	\centering
	\includegraphics[width=\linewidth]{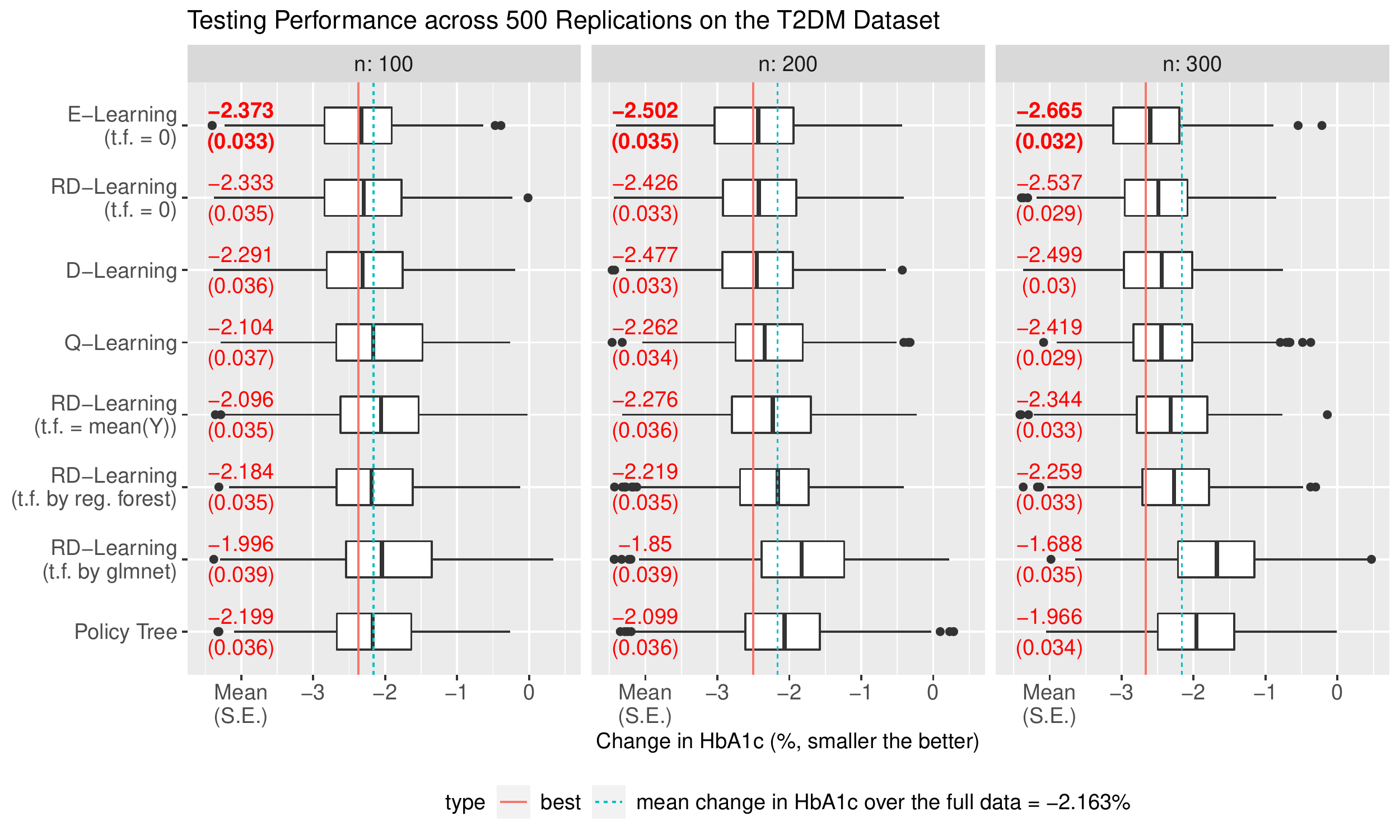}
	\caption[T2DM, Testing]{Testing changes in HbA1c (\%, smaller the better) for training sample sizes $ n \in \{100,200,300\} $ on the T2DM dataset. Here, \textit{t.f.} represents the fitted treatment-free effect, and \textit{reg. forest} corresponds to the regression forest.}
	\label{fig:T2DM}
\end{figure}

The testing results are reported in Figure \ref{fig:T2DM}. E-Learning enjoys the best testing performance among all training sample sizes. As the training sample size $ n $ increases, the advantage of E-Learning is more evident compared with other methods. This can confirm the efficiency improvement of E-Learning by using an optimal working variance function on this dataset. Among patients in the T2DM dataset, E-Learning recommends 19.77\% for long-acting insulins, 18.14\% for intermediate-acting insulins, 30.23\% for short-acting insulins, and 31.86\% for GLP-1 RAs. The fitted E-Learning coefficients are reported in Table \ref{tab:T2DM_coef}. In particular, short-acting insulins (\texttt{A = 3}) is recommended for the patients with average covariates. Patients as former smokers are more recommended for the short-acting insulins than other patients. The general benefits of short-acting insulins are consistent with the results in \citet{chen2018estimating,meng2020near}. Moreover, it can be observed that the coefficients for \texttt{baseline HbA1c} in Table \ref{tab:T2DM_coef} increase in the treatment arm. In fact, the averaged baseline HbA1c values among recommended treatments $ A = 1,2,3,4 $ are 7.35\%, 10.67\%, 10.91\% and 11.18\% respectively.
This suggests that patients with worse baseline HbA1c are recommended for faster-acting therapies, where the GLP-1 RAs (\texttt{A = 4}) can be regarded as an alternative for the rapid-acting insulin \citep{ostroff2016glp}. Such a phenomenon is also consistent with the recommended treatment ordinality pointed out by \citet{chen2018estimating}.

\renewcommand{\arraystretch}{1}
\begin{table}
	\centering
	\fontsize{8}{10}\selectfont
	\caption{E-Learning Coefficients on the T2DM Dataset}\label{tab:T2DM_coef}
	
\begin{tabular}{>{}l|>{\raggedleft\arraybackslash}p{1cm}>{\raggedleft\arraybackslash}p{1cm}>{\raggedleft\arraybackslash}p{1cm}>{\raggedleft\arraybackslash}p{1cm}>{}p{1cm}>{}p{1cm}>{}p{1cm}>{}p{1cm}>{}p{1cm}>{}p{1cm}>{}p{1cm}>{}p{1cm}>{}p{1cm}>{}p{1cm}>{}p{1cm}>{}p{1cm}>{}p{1cm}}
\toprule
\ttfamily{ } & \ttfamily{A = 1} & \ttfamily{A = 2} & \ttfamily{A = 3} & \ttfamily{A = 4}\\
\midrule
\ttfamily{Intercept} & \textbf{-0.164} & -0.053 & \textbf{0.168} & 0.048\\
\ttfamily{gender (male)} & -0.008 & -0.014 & 0.029 & -0.007\\
\ttfamily{ethnic (others)} &  &  &  & \\
\ttfamily{ethnic (white)} & -0.023 & 0.081 & -0.029 & -0.03\\
\ttfamily{smoke (former)} & -0.001 & \textbf{-0.174} & \textbf{0.361} & \textbf{-0.186}\\
\ttfamily{smoke (no)} &  &  &  & \\
\ttfamily{smoke (yes)} &  &  &  & \\
\ttfamily{comorbidity (yes)} &  &  &  & \\
\ttfamily{HDL (low)} & 0.033 & -0.009 & -0.053 & 0.029\\
\ttfamily{HDL (high)} &  &  &  & \\
\ttfamily{LDL (low)} & 0.004 & \textbf{0.115} & \textbf{-0.122} & 0.003\\
\ttfamily{LDL (high)} & 0.007 & -0.002 & -0.023 & 0.018\\
\ttfamily{baseline HbA1c} & \textbf{-0.492} & \textbf{0.139} & \textbf{0.168} & \textbf{0.185}\\
\ttfamily{age} & -0.058 & \textbf{0.169} & \textbf{0.106} & \textbf{-0.217}\\
\ttfamily{weight} &  &  &  & \\
\ttfamily{height} &  &  &  & \\
\ttfamily{BMI} &  &  &  & \\
\bottomrule
\multicolumn{5}{l}{\rule{0pt}{1em}\underline{\textit{Note:}}}\\
\multicolumn{5}{l}{\rule{0pt}{1em}Larger coefficients encourage better outcome.}\\
\multicolumn{5}{l}{\rule{0pt}{1em}Coefficients are fitted at standardized scales of covariates.}\\
\multicolumn{5}{l}{\rule{0pt}{1em}Blank coefficients are 0's. Absolute value $>$ 0.1 are bolded.}\\
\end{tabular}

\end{table}

\section{Discussion}\label{sec:discuss}

In this paper, we propose E-Learning for learning an optimal ITR under heteroscedasticity or misspecified treatment-free effect. In particular, E-Learning is developed from semiparametric efficient estimation in the multi-armed treatment setting. When nuisance models are correctly specified, even if heteroscedasticity exists, the $ \sqrt{n} $-asymptotic variance of the estimated parameters achieve the semiparametric variance lower bound. When the treatment-free effect model is misspecified, E-Learning targets the optimal working variance function, so that the $ \sqrt{n} $-asymptotic variance of the estimated parameters is still the smallest among the class of regular semiparametric estimates. In summary, E-Learning extends the optimality of existing model-based methods to allow multiple treatments, heteroscedasticity and treatment-free effect misspecification. The efficiency gain of E-Learning is demonstrated by our simulation studies when either of or both heteroscedasticity and misspecified treatment-free effect happen, where existing methods can have much worse performance. This also can be consistent with \citet{kang2007demystifying}'s finding that the misspecified treatment-free effect can have severe consequences.
	

Our proposed E-Learning requires the parametric assumption on the interaction effect. While we mainly consider the linear interaction effect in our simulation studies and the data applications for illustration purpose, a nonlinear interaction effect can also be allowed. In that case, we may consider the basis expansion under this framework. E-Learning with the cubic polynomial basis can be a concrete example. Our nonlinear simulation examples in the Supplementary Material Section \ref{sec:simulation_nonlinear} further demonstrate the advantage of E-Learning with the cubic polynomial basis.

For a more flexible interaction effect model, the kernel extension of E-Learning can also be possible. In particular, under certain conditions, it can be shown that an extension of the weighted least-squares problem based on \citet{kennedy2020optimal} can be $ \sqrt{n} $-asymptotically equivalent to our E-Learning estimating equations \eqref{eq:eLearn}. However, the implementation of such an extension can be very different from our Section \ref{sec:eLearn}, and there remains some theoretical challenges to address along this line. Because of the competitive performance of E-learning using basis expansion, we believe kernel E-learning has the potential to work well and leave possible kernel extensions  as a future work.

Another direction of future work can be the high-dimensional problem. In Section \ref{sec:reg}, we propose to solve the regularized estimating equation, which can handle high-dimensional parameter estimation. However, the nonparametric estimation of the working variance function is also a potential challenge when the dimension is growing. In our simulation study, our proposed E-Learning with estimated working variance function requires larger sample sizes in presence of increasing numbers of variables and treatments. In the literature, there exists three possible strategies to accommodate this challenge: 1) considering index models for the variance function that can allow dimension reduction \citep{zhu2013semiparametric,lian2015variance}; 2) estimating the central variance subspace for sufficient dimension reduction \citep{zhu2009dimension,luo2014efficient,ma2019semiparametric}; 3) performing simultaneous nonlinear variable selection during nonparametric regression \citep{lin2006component,lafferty2008rodeo,zhang2011linear,allen2013automatic}. These can have potential for further improvement of E-Learning.

\section*{Acknowledgments}

The authors would like to thank the Editor, the Associate Editor, and reviewers, whose helpful comments and suggestions led to a much improved presentation. The research was supported in part by NSF grants DMS-1821231 and  DMS-2100729, and NIH grant R01GM126550.

\singlespacing
\bibliography{bibfile.bib}
\bibliographystyle{asa}
\end{document}


\pagenumbering{arabic}
\maketitle
\tableofcontents
\listoftables
\listoffigures

\newpage 

\renewcommand{\thesection}{S}
\section{Supplementary Materials}

\subsection{Nonlinear Simulation Studies}\label{sec:simulation_nonlinear}
This section focuses on nonlinear data generation and the learning methods based on polynomial basis. Specifically,
We consider the same data generating process as in Section \ref{sec:DGP}, but with the treatment-free effect $ \mu_{0}(\bx) = {1 \over \sqrt{K}}\sum_{k'=1}^{K}h(x_{k'}) $ and the interaction effect $ \bbeta_{A}^{\intercal}\big(1,h(X_{1}),\cdots,h(X_{p})\big)^{\intercal} $. Here, both $ h(x) = x^{3} $ and $ h(x) = e^{\sqrt{2}x} $ are studied. The propensity score is $ p_{\sA}(A|\bX) = {e^{X_{A}/2} \over \sum_{k=1}^{K}e^{X_{k}/2}} $, while both the homoscedastic variance function $ \sigma^{2}(\bX,A) = 1 $ and the heteroscedastic variance function $ \sigma^{2}(\bX,A) = e^{2\sqrt{2} X_{A}} $ are considered.

For implementation, we consider the same linear Q-Learning, D-Learning, RD-Learning, and E-Learning as in Section \ref{sec:implement_detail}, and the cubic versions with $ \bX^{(3)} = (X_{1},X_{1}^{2},X_{1}^{3};\cdots;X_{p},X_{p}^{2},X_{p}^{3})^{\intercal} $ as the covariate vector. If $ h(x) = x^{3} $, then the linear methods misspecify both the treatment-free effect and interaction effect models, while the cubic methods correctly specify these models. If $ h(x) = e^{\sqrt{2}x} $, then all these methods misspecify the models.

\begin{figure}[p]
	\centering
	\includegraphics[width=\linewidth]{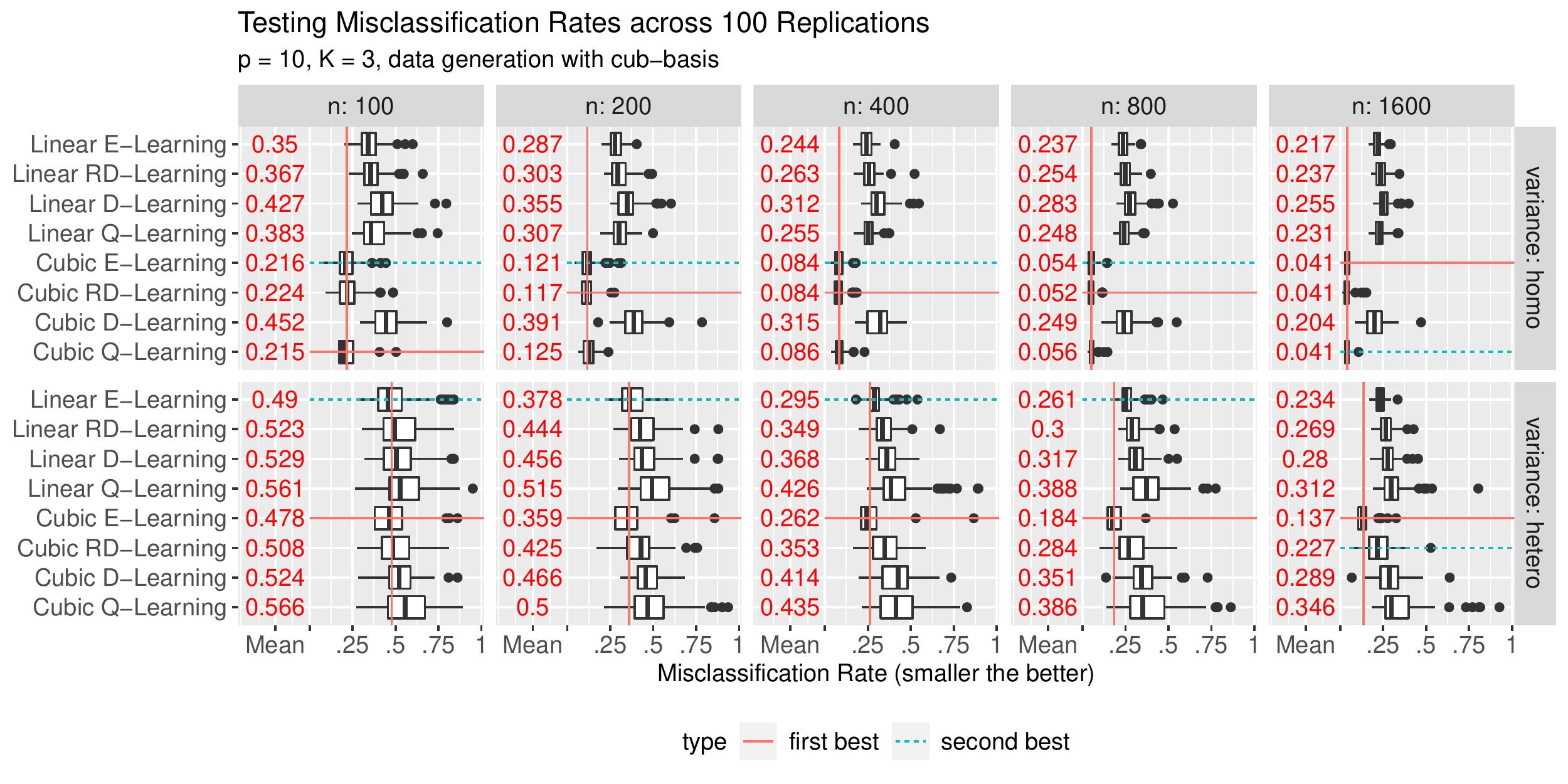}
	\includegraphics[width=\linewidth]{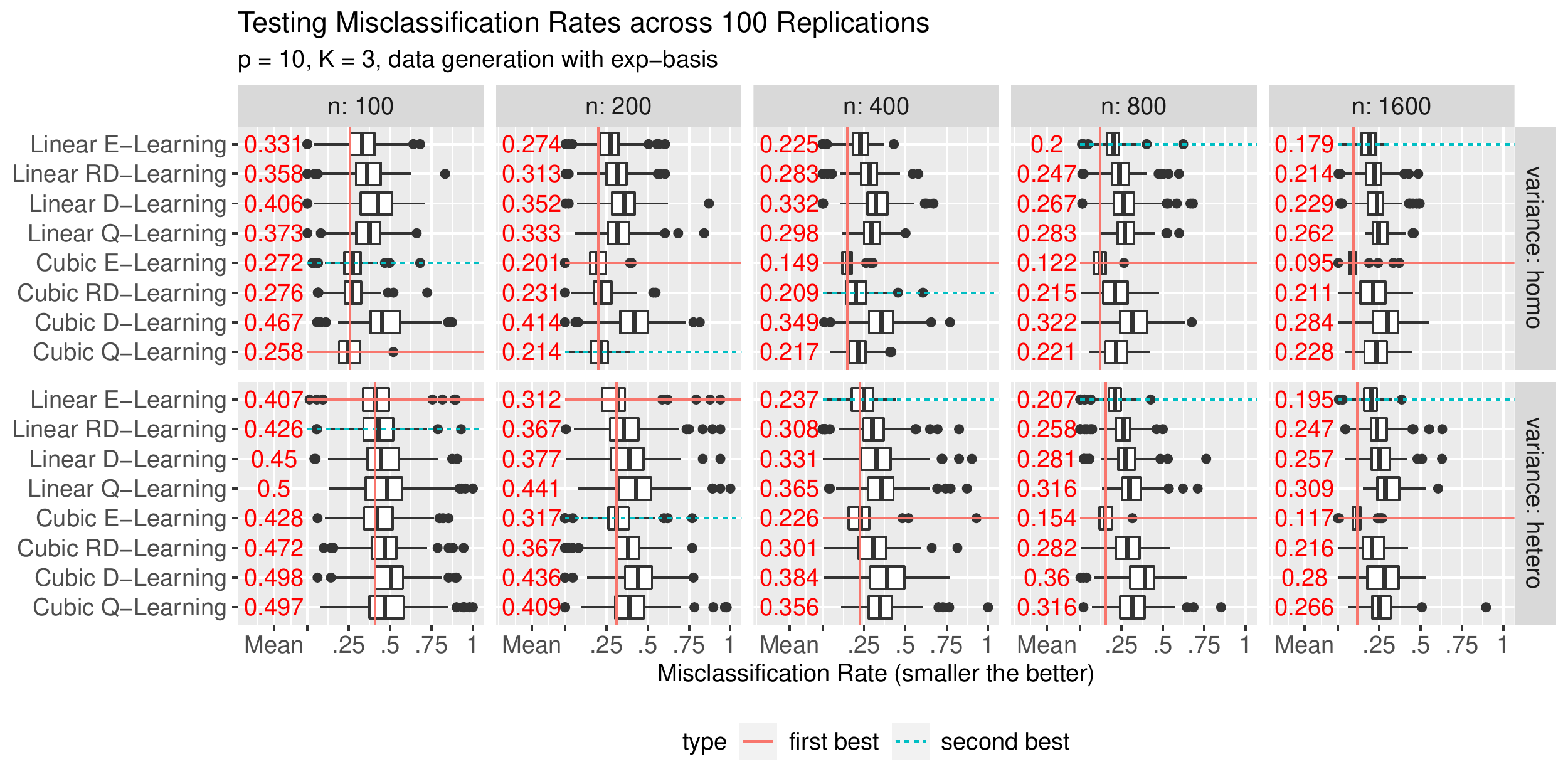}
	\caption[Nonlinear Simulation, Misclassification Rates]{\footnotesize Testing misclassification rates (smaller the better) for $ n = \{100,200,400,800,1600\}  $, $ p = 10 $, $ K = 3 $ under homoscedasticity or heteroscedasticity. The upper panel corresponds to the data generation with $ h(x) = x^{3} $, and the lower panel corresponds to the data generation with $ h(x) = e^{\sqrt{2}x} $ in Section \ref{sec:simulation_nonlinear}.}
	\label{fig:nonlinear_misclass}
\end{figure}

\begin{figure}[p]
	\centering
	\includegraphics[width=\linewidth]{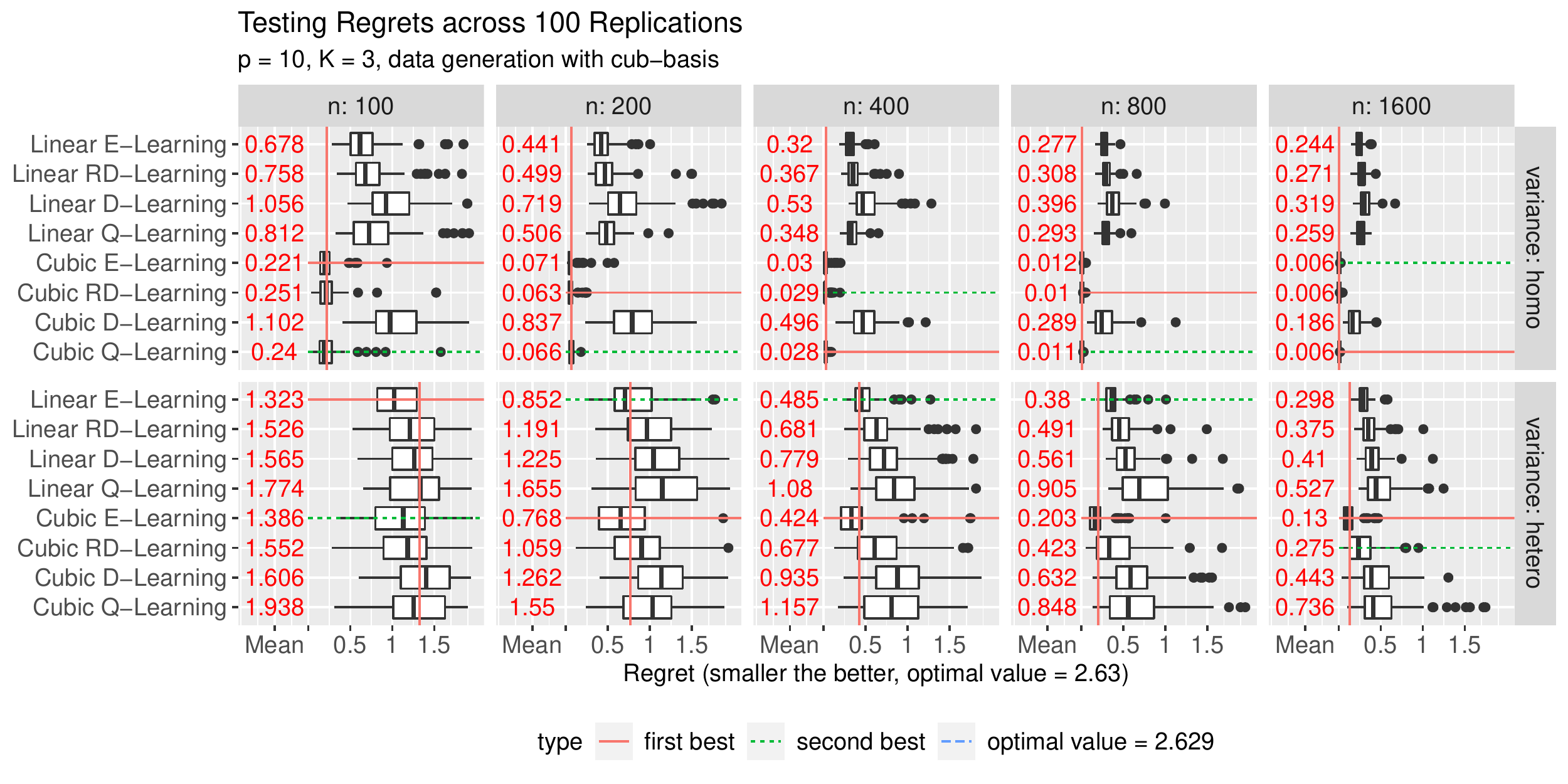}
	\includegraphics[width=\linewidth]{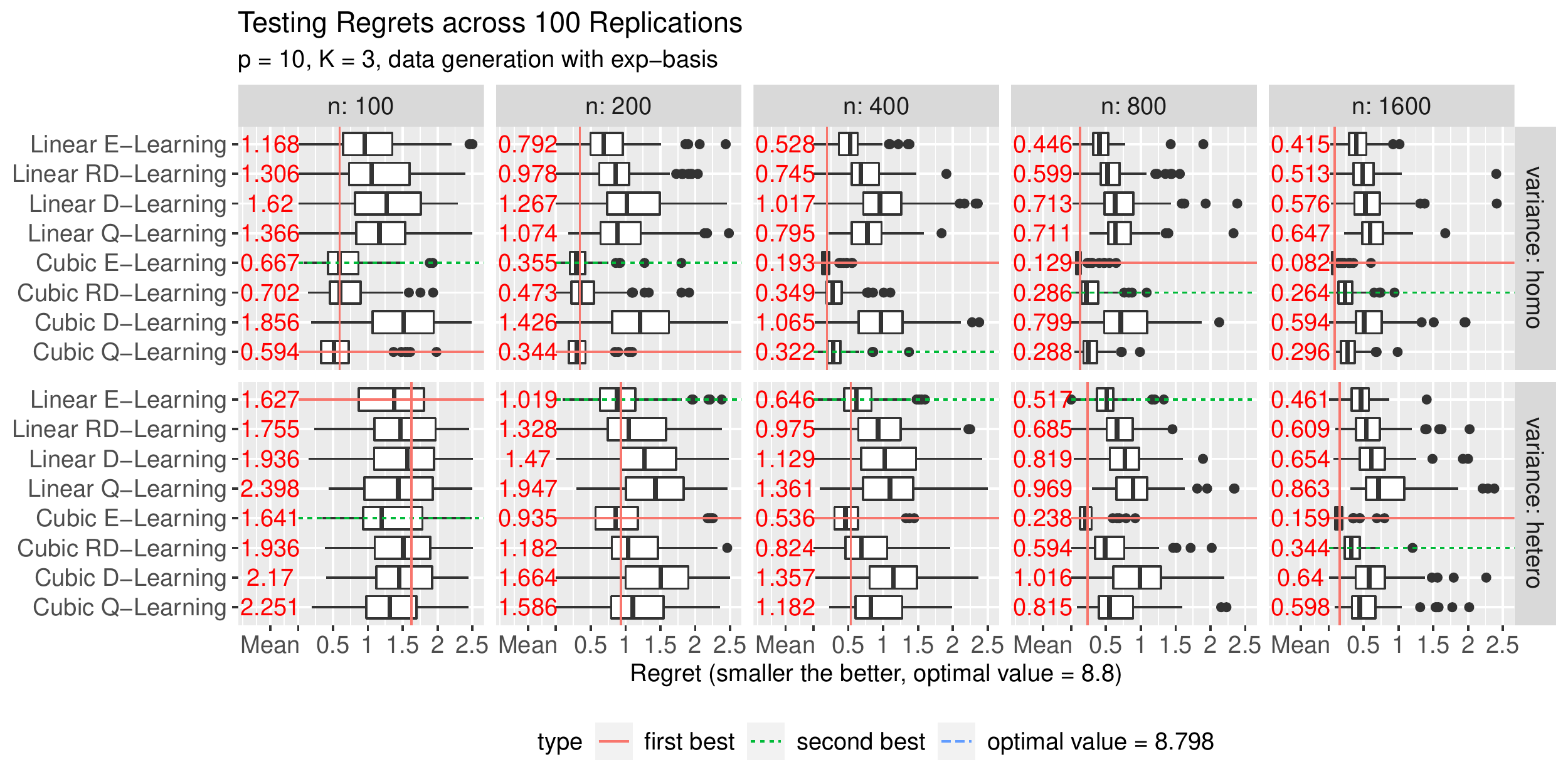}
	\caption[Nonlinear Simulation, Regrets]{\footnotesize Testing regrets (smaller the better) for $ n = \{100,200,400,800,1600\}  $, $ p = 10 $, $ K = 3 $ under homoscedasticity or heteroscedasticity. The upper panel corresponds to the data generation with $ h(x) = x^{3} $, and the lower panel corresponds to the data generation with $ h(x) = e^{\sqrt{2}x} $ in Section \ref{sec:simulation_nonlinear}.}
	\label{fig:nonlinear_regret}
\end{figure}

The testing misclassification rates and regrets based on 100 replications are reported in Figures \ref{fig:nonlinear_misclass} and \ref{fig:nonlinear_regret} respectively. 
Linear 	E-Learning can maintain the best performance among the linear methods, even though the advantage is less evident due to the misspecified interaction effect. This suggests that the efficiency gain of E-Learning towards the misspecified treatment-free effect and heteroscedasticity can be maintained even in presence of the misspecified interaction effect.

The cubic methods generally have better testing performance than the linear methods, since the cubic versions are more flexible to approximate the true treatment-free and interaction effects. In particular, for the data generation with $ h(x) = x^{3} $, the treatment-free and interaction effect models in the cubic methods are correctly specified. In that case, the superiority of cubic E-Learning among the cubic methods is the same as in Section \ref{sec:simulation}. For the data generation with $ h(x) = e^{\sqrt{2}x} $, cubic E-Learning can also be superior to the other cubic methods with a strong advantage, even when the interaction effect is misspecified. In particular, the misclassification rate of cubic E-Learning when $ n = 1600 $ is 0.117, while that of cubic RD-Learning, D-Learning and Q-Learning are 0.216, 0.28 and 0.266 respectively. This can confirm the effectiveness of cubic E-Learning for the nonlinear problems.
\newpage

\subsection{Analysis of the ACTG 175 Trial Data}\label{sec:ACTG175}
We evaluate the effectiveness of our proposed E-Learning on a clinical trial dataset from the ``AIDS clinical trial group study 175" \citep{hammer1996trial}. The goal of this study was to compare four treatment arms among 2,139 randomly assigned subjects with human immunodeficiency virus type 1 (HIV-1), whose CD4 counts were 200-500 cells/mm$ ^{3} $. The four treatment options of $ A $ are the zidovudine (ZDV) monotherapy, the didanosine (ddI) monotherapy, the ZDV combined with ddI, and the ZDV combined with zalcitabine (ZAL). 

The primary outcome $ Y $ of our interest is the difference between the CD4 cell counts at early stage (20$ \pm $5 weeks from baseline) and the CD4 counts at baseline, which is larger the better. We follow the analyses in \citet{lu2013variable,qi2020multi,meng2020robust} and consider 12 selected baseline covariates $ \bX $. There are 5 continuous covariates: age (year), weight (kg, coded as \texttt{wtkg}), CD4 count (cells/mm$ ^{3} $) at baseline, Karnofsky score (scale of 0-100, coded as \texttt{karnof}), CD8 count (cells/mm$ ^{3} $) at baseline. They are centered and scaled before further analysis. In addition, there are 7 binary variables: gender (1 = male, 0 = female), homosexual activity (\texttt{homo}, 1 = yes, 0 = no), race (1 = nonwhite, 0 = white), history of intravenous drug use (\texttt{drug}, 1 = yes, 0 = no), symptomatic status (\texttt{symptom}, 1 = symptomatic, 0 = asymptomatic), antiretroviral history (\texttt{str2}, 1 = experienced, 0 = naive) and hemophilia (\texttt{hemo}, 1 = yes, 0 = no). 

\begin{figure}[p]
	\centering
	\includegraphics[width=0.95\linewidth]{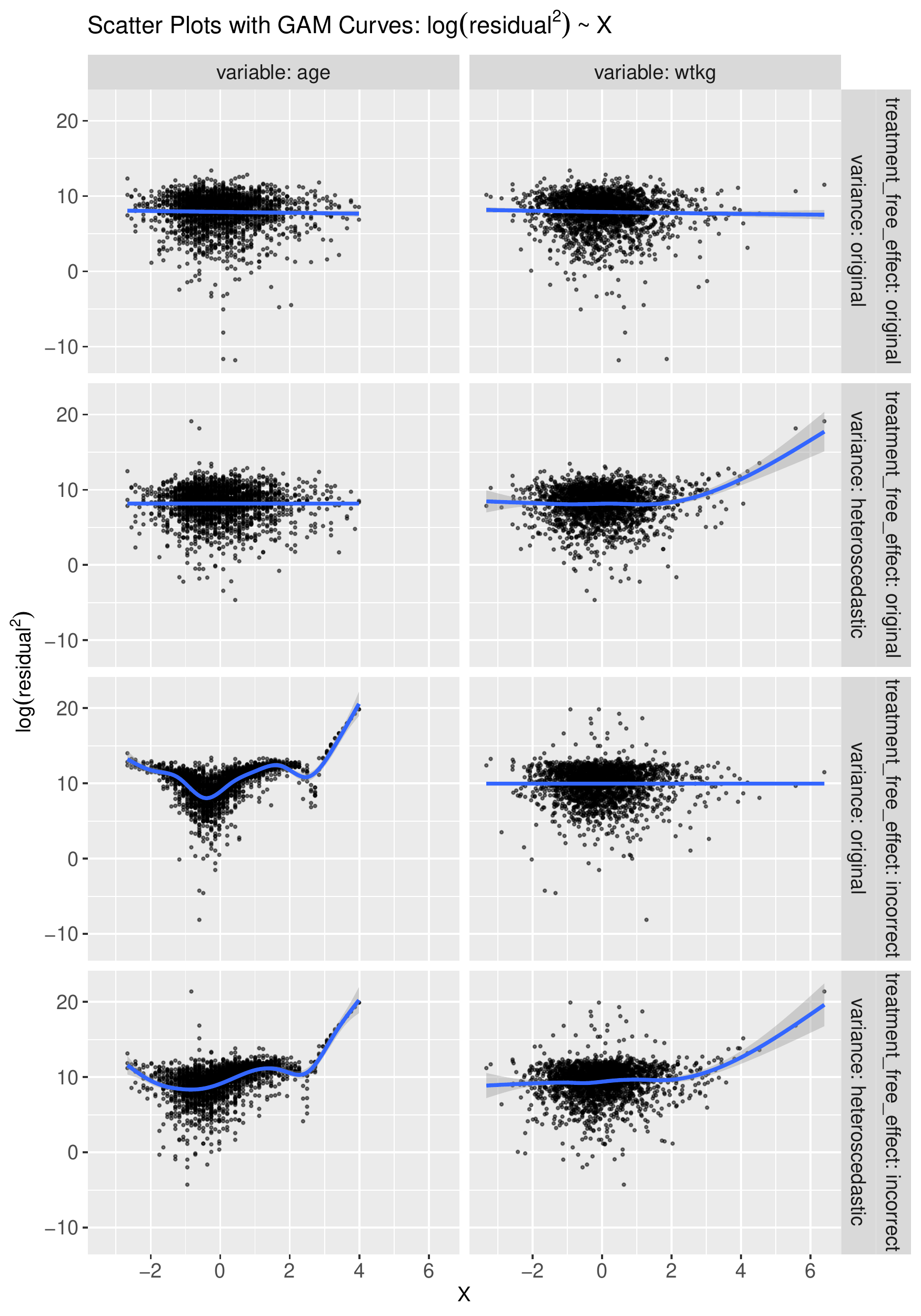}
	\caption[ACTG 175, Residual Plots]{\footnotesize Residual plots with respect to \texttt{age} and \texttt{wtkg} on the ACTG175 dataset (Section \ref{sec:ACTG175}). Curves are fitted by the \textit{Generalized Additive Model (GAM)} of cubic spline. Residuals are computed from the fitted E-Learning on each modified dataset according to Table \ref{tab:ACTG175}, and averaged over 10 replications.}
	\label{fig:ACTG175_resid2}
\end{figure}

We consider the training sample size $ n \in \{100,200,400,800,1600\} $ and the testing sample size $ n_{\rm test} = 400 $. The full  dataset is randomly split into training and testing according to the given sample sizes. Since the dataset is obtained from a randomized controlled trial, the propensity score function is known to be $ p_{\sA}(k|\bX) = 1/4 $ for $ k = 1,2,3,4 $. For the treatment-free effect estimation, we consider a linear working model with the $ \ell_{1} $-penalty throughout the analysis, which will be different from the implementation in \citet{meng2020robust}. For this real-world data application, the underlying truth is unknown to us. We cannot verify whether any of misspecified treatment-free effect and heteroscedasticity on the original dataset exist. Nevertheless, after modifying the dataset according to the following Table \ref{tab:ACTG175}, the treatment-free effect misspecification and heteroscedasticity can be anticipated. Note that the unmodified cases can also have treatment-free effect misspecification and heteroscedasticity as well. Our modification can enlarge such effects. The goal of our analysis is to demonstrate the efficiency improvement of E-Learning in presence of heavy treatment-free effect misspecification and heteroscedasticity. We further provide the residual plots in Figure \ref{fig:ACTG175_resid2} in Section \ref{sec:tab_fig}. Residuals are computed from the fitted E-Learning on each modified dataset according to Table \ref{tab:ACTG175}, and averaged over 10 replications. It confirms that the modifications can result in the squared residuals heavily depending on the variables \texttt{age} and \texttt{wtkg}.

\renewcommand{\arraystretch}{1}
\begin{table}[!ht]
	\centering
	\caption[ACTG 175, Modifications]{Modifications on the ACTG 175 Dataset and the Implying Model Specifications}
	\label{tab:ACTG175}
	\begin{threeparttable}
		\begin{tabular}{l|cc}
			\hline\hline
			\textbf{Training Outcome Modification} & \textbf{Treatment-Free Effect} & \textbf{Variance Function} \\
			\hline
			Original $ Y $ & \textit{unmodified} & \textit{unmodified} \\
			$ Y \gets Y + e^{2.5\times \texttt{age}} $ & \textbf{\textit{misspecified}} & \textit{unmodified} \\
			$ Y \gets Y + 5e^{1.5\times\texttt{wtkg}} \times \xi $ & \textit{unmodified} & \textbf{\textit{heteroscedastic}} \\
			$ Y \gets Y + e^{2.5\times \texttt{age}} + 5e^{1.5\times\texttt{wtkg}} \times \xi $ & \textbf{\textit{misspecified}} & \textbf{\textit{heteroscedastic}} \\
			\hline\hline
		\end{tabular}
		\begin{tablenotes}
			\fontsize{8}{5}\selectfont
			\item[1] The variables \texttt{age} and \texttt{wtkg} are centered and scaled at the data preparation stage. 
			\item[2] The additional noise $ \xi $ is randomly generated from $ \bbP(\xi = 1) = \bbP(\xi = -1) = 1/2 $ independent of $ \bX,A,Y $. 
			\item[3] We further round the modified outcomes to their nearest integers to respect the integer nature of $ Y $. 
			\item[4] The treatment-free effect is estimated by a linear working model with the $ \ell_{1} $-penalty.
		\end{tablenotes}
	\end{threeparttable}
\end{table}

\begin{figure}[!ht]
	\centering
	\includegraphics[width=\linewidth]{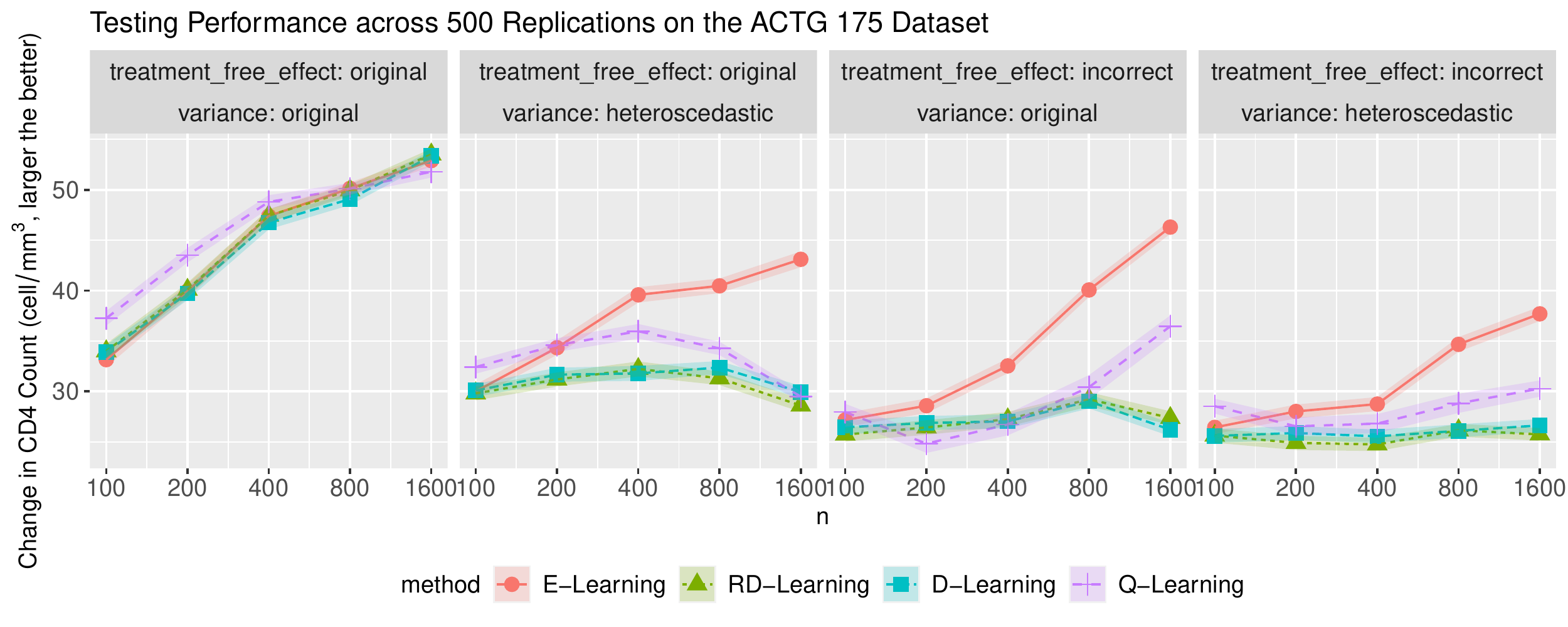}
	\caption[ACTG 175, Testing]{Testing changes in CD4 count (cell/mm$ {}^{3} $, larger the better) on the ACTG 175 dataset.}
	\label{fig:ACTG175}
\end{figure}

On the training sample, we implement the same procedures as in Section \ref{sec:simulation} to fit Q-Learning, D-Learning, RD-Learning, and our proposed E-Learning. On the testing dataset, we evaluate the an estimated ITR $ \widehat{d} $ by the IPWE $ (1/n_{\rm test})\sum_{i=1}^{n_{\rm test}}Y_{i}\bbone[A_{i} = \widehat{d}(\bX_{i})]/(1/4) $, which is larger the better. Here, the testing outcome $ Y_{i} $ is unmodified in contrast to the training outcome to ensure comparable testing evaluation. Testing results based on 500 repeated training and testing for each of the four cases in Table \ref{tab:ACTG175} are reported in Figure \ref{fig:ACTG175}.

When we use the original training outcome $ Y $, the testing CD4 count improvements of all methods close to each other. In particular, Q-Learning demonstrates slightly better performance for $ n = 100,200,400 $, but all methods have similar performance when $ n =800 $ and 1600. All these methods have improving testing performance  as $ n $ increases. When we modify $ Y $ to incorporate heavy heteroscedasticity or/and treatment-free effect misspecification, E-Learning can maintain the improvements as $ n $ increases, while other methods can have much poorer performance. In particular, other methods can even get worse as $ n $ increases in presence of heavy heteroscedasticity. We shall anticipate that the scientific findings for the analysis with original outcome $ Y $ will not be disturbed when we introduce additional treatment-free effect misspecification or/and heteroscedasticity. The results in Figure \ref{fig:ACTG175} show that E-Learning maintains the testing performance well during these modifications, while RD-Learning, D-Learning and Q-Learning are heavily affected. In this way, E-Learning demonstrates its superiority of efficiency gain in presence of misspecified treatment-free effect or/and heteroscedasticity.

\begin{figure}[p]
	\centering
	\includegraphics[width=\linewidth]{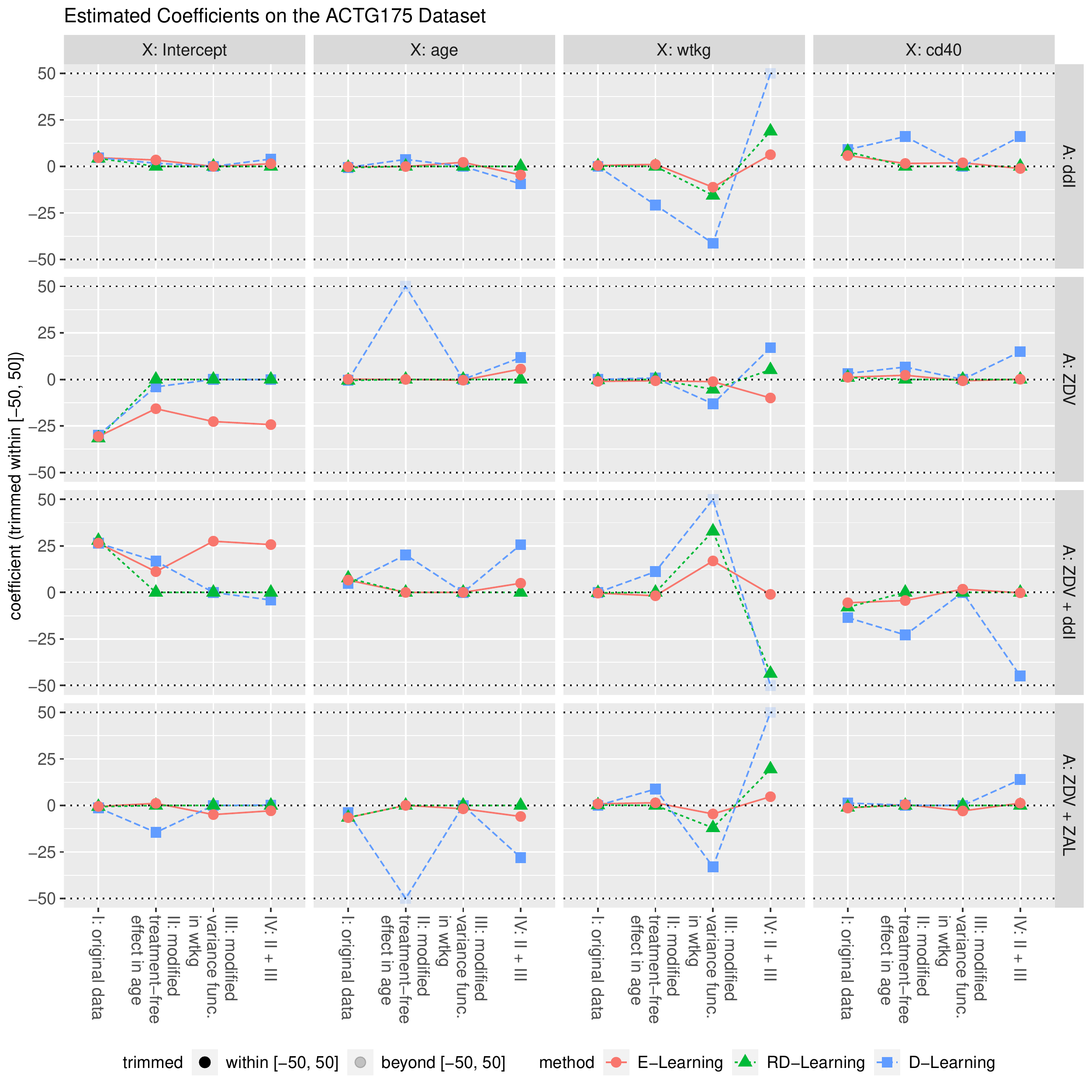}
	\caption[ACTG 175, Coefficients]{Fitted coefficients on each modified ACTG 175 dataset according to Table \ref{tab:ACTG175}, and averaged over 10 replications.}
	\label{fig:ACTG175_coef}
\end{figure}

We further report the estimated coefficients for D-Learning, RD-Learning and E-Learning in Table \ref{tab:eLearn_ACTG175_coef} and Figure \ref{fig:ACTG175_coef} in Section \ref{sec:tab_fig}. The fitted coefficients on the original data are consistent with existing literature. Specifically, \texttt{Intercept}, \texttt{age} and \texttt{cd40} are common important covariates that were frequently reported in the literature \citep{lu2013variable,qi2020multi,meng2020robust}. When we incorporate heavy treatment-free effect misspecification or/and heteroscedasticity in cases \Romannum{2}, \Romannum{3} and \Romannum{4}, the fitted coefficients of D-Learning and RD-Learning become highly unstable with many extreme coefficients. In contrast, the fitted E-Learning coefficients are relatively stable across these cases. This suggests that the E-Learning estimate can be more resilient to the training outcome modifications in Table \ref{tab:ACTG175} compared with the other methods.

\newpage

\subsection{Optimal Estimating Function under Misspecified Propensity Score Model}\label{sec:prop_mis}

Let $ \widecheck{p}_{\sA}(a|\bx) $ be an arbitrary propensity score function. For any $ \sfH: \cX \to \bbR^{p\times (K-1)} $, which can depend on $ \bbeta $, consider the following estimating function:
\[ \bphi(\bbeta;\widecheck{p}_{\sA}) = [Y - \mu_{0}(\bX) - \gamma(\bX,A;\bbeta)]{\sfH(\bX)\bomega_{A} \over \widecheck{p}_{\sA}(A|\bX)}. \]
By Proposition \ref{prop:dr}, since the working treatment-free effect function $ \mu_{0} $ is true, for any working propensity score function $ \widecheck{p}_{\sA} $, we have $ \bbE[\bphi(\bbeta;\widecheck{p}_{\sA})] = \bzero $ at the true $ \bbeta $. Our goal is to find the optimal $ \sfH(\bX) $ for a given working propensity score function $ \widecheck{p}_{\sA} $. 

The following derivations are analogous to the proof of Theorem \ref{thm:eff_opt}:
\begin{align*}
	\bbE[\bphi(\bbeta;\widecheck{p}_{\sA})^{\otimes 2}] &= \bbE\left[ \sfH(\bX)\left( {\bomega_{A}^{\otimes 2}\epsilon^{2} \over \widecheck{p}_{\sA}(A|\bX)^{2}} \right)\sfH(\bX)^{\intercal} \right]\\
	&= \bbE\left[ \sfH(\bX) \sfV_{\epsilon}(\bX;\widecheck{p}_{\sA}) \sfH(\bX)^{\intercal} \right];\\
	\bbE \left[-{\partial\widecheck{\bphi}(\bbeta) \over \partial \bbeta^{\intercal}}\right] &= \bbE\left\{ \sfH(\bX)\left[ \left( 1 - {1 \over K} \right){\bomega_{A}^{\otimes 2} \over \widecheck{p}_{\sA}(A|\bX)} \right]\dot{\sfF}(\bX;\bbeta) \right\} - \underbrace{\bbE\left[ \dot{\sfH}(\bX;\bbeta){\bomega_{A}\epsilon \over \widecheck{p}_{\sA}(A|\bX)} \right]}_{= 0}\\
	&= \bbE\left[ \sfH(\bX)\sfV_{\sA}(\bX;\widecheck{p}_{\sA})\dot{\sfF}(\bX;\bbeta) \right].
\end{align*}
In the second equality, we define
\[ \sfV_{\epsilon}(\bX;\widecheck{p}_{\sA}) := \sum_{k=1}^{K}{p_{\sA}(k|\bX)\sigma^{2}(\bX,k)\bomega_{k}^{\otimes 2} \over \widecheck{p}_{\sA}(k|\bX)^{2}}. \]
In the forth equality, it follows from that $ \bbE(\epsilon|\bX,A) = 0 $ and we define
\[ \sfV_{\sA}(\bX;\widecheck{p}_{\sA}) := \left( 1 - {1 \over K} \right)\sum_{k=1}^{K}{p_{\sA}(k|\bX)\bomega_{k}^{\otimes 2} \over \widecheck{p}_{\sA}(k|\bX)}. \]

Let $ \widehat{\bbeta}_{n}(\widecheck{p}_{\sA}) $ be the solution to $ \bbE_{n}[\bphi(\bbeta;\widecheck{p}_{\sA})] = \bzero $. Then under the same regularity conditions as in Theorems \ref{thm:eff} and \ref{thm:eff_opt}, we have
\begin{align*}
	&\lim_{n\to\infty}n\sfvar[\widehat{\bbeta}_{n}(\widecheck{p}_{\sA})] \\
	=& \left\{ \bbE \left[-{\partial \bphi(\bbeta;\widecheck{p}_{\sA})\over \partial \bbeta^{\intercal}}\right]\right\}^{-1}\bbE [\widecheck{\bphi}(\bbeta)^{\otimes 2}]\left\{ \bbE \left[-{\partial \bphi(\bbeta;\widecheck{p}_{\sA})^{\intercal}\over \partial \bbeta}\right]\right\}^{-1} \\
	=& \widetilde{\sfB}^{-1}\widetilde{\sfA}\widetilde{\sfB}^{-\intercal} = (\widetilde{\sfB}^{\intercal}\widetilde{\sfA}^{-1}\widetilde{\sfB})^{-1} \ge \widetilde{\sfC}^{-1},
\end{align*}
where, analogous to Lemma \ref{lem:sandwich}, we define
\begin{align*}
	\widetilde{\sfA} &:= \bbE[\sfH(\bX)\sfV_{\epsilon}(\bX;\widecheck{p}_{\sA})\sfH(\bX)^{\intercal}];\\
	\widetilde{\sfB} &:= \bbE[\sfH(\bX)\sfV_{\sA}(\bX;\widecheck{p}_{\sA})\dot{\sfF}(\bX;\bbeta)];\\
	\widetilde{\sfC} &:= \bbE[\dot{\sfF}(\bX;\bbeta)^{\intercal}\sfV_{\sA}(\bX;\widecheck{p}_{\sA})\sfV_{\epsilon}(\bX;\widecheck{\mu}_{0})^{-1}\sfV_{\sA}(\bX;\widecheck{p}_{\sA})\dot{\sfF}(\bX;\bbeta)],
\end{align*}
with equality if and only if exists some non-singular constant matrix $ \sfH_{0} \in \bbR^{p \times p} $ such that
\[ \sfH(\bX) = \sfH_{0}\dot{\sfF}(\bX;\bbeta)^{\intercal}\sfV_{\sA}(\bX;\widecheck{p}_{\sA})\sfV_{\epsilon}(\bX;\widecheck{p}_{\sA})^{-1}. \]
Therefore, the optimal estimating function under the working propensity score function $ \widecheck{p}_{\sA} $ is
\begin{align*}
	\bphi_{\eff}(\bbeta;\widecheck{p}_{\sA}) :=& [Y - \mu_{0}(\bX) - \gamma(\bX,A;\bbeta)] \times\\
	& \underbrace{\dot{\sfF}(\bX;\bbeta)^{\intercal}\overbrace{\left[ \left( 1- {1 \over K} \right)\sum_{k=1}^{K}{p_{\sA}(k|\bX)\bomega_{k}^{\otimes 2} \over \widecheck{p}_{\sA}(k|\bX)} \right]}^{:=\sfV_{\sA}(\bX;\widecheck{p}_{\sA})}\Bigg[\overbrace{\sum_{k=1}^{K}{p_{\sA}(k|\bX)\sigma^{2}(\bX,k)\bomega_{k}^{\otimes 2} \over \widecheck{p}_{\sA}(k|\bX)^{2}}}^{:=\sfV_{\epsilon}(\bX;\widecheck{p}_{\sA})}\Bigg]^{-1}{\bomega_{A} \over \widecheck{p}_{\sA}(A|\bX)}}_{\text{optimal instrument}}.
\end{align*}
Here, different from the case of misspecified treatment-free effect model discussed in Section \ref{sec:asymptotic_misspecified}, the optimal estimating function cannot be defined from an optimal working variance as in Theorem \ref{thm:eff_opt}. It require different strategies to estimate the optimal variance components $ \sfV_{\sA}(\bX;\widecheck{p}_{\sA}) $ and $ \sfV_{\epsilon}(\bX;\widecheck{p}_{\sA}) $. One potential strategy is to identified them from
\[ \sfV_{\sA}(\bX;\widecheck{p}_{\sA}) = \bbE\left[ \left( 1-{1 \over K} \right){\bomega_{A}^{\otimes 2} \over \widecheck{p}_{\sA}(A|\bX)}\middle|\bX \right]; \quad \sfV_{\epsilon}(\bX;\widecheck{p}_{\sA}) = \bbE\left( {\bomega_{A}^{\otimes 2} \epsilon^{2} \over \widecheck{p}_{\sA}(A|\bX)^{2}}\middle|\bX \right), \]
where $ \epsilon $ is replaced by the working residual $ e = Y - \mu_{0}(\bX) - \gamma(\bX,A;\bbeta) $. Therefore, we can perform nonparametric regression on the $ \bbR^{(K-1)\times(K-1)} $-valued matrices $ \left( 1 - {1 \over K} \right){\bomega_{A}^{\otimes 2} \over \widecheck{p}_{\sA}(A|\bX)} $ and $ {\bomega_{A}^{\otimes 2}e^{2} \over \widecheck{p}_{\sA}(A|\bX)} $ on $ \bX $. However, such a strategy will be much different from the methodology proposed in this paper. In particular, in this paper, we only need to estimate an $ \bbR^{K} $-valued function $ \big( \widecheck{\sigma}_{\opt}^{2}(\bX,k): 1 \le k \le K \big) $, while the optimal estimating function under misspecified propensity score model require the estimation of two $ \bbR^{(K-1)\times(K-1)} $-valued functions $ \sfV_{\sA}(\bX;\widecheck{p}_{\sA}) $ and $ \sfV_{\epsilon}(\bX;\widecheck{p}_{\sA}) $.

\newpage

\subsection{More Implementation Details}\label{sec:implement_detail}
\begin{itemize}[leftmargin=*]
	\item \textbf{E-Learning} (general $ K $): When fitting the treatment-free effect and the propensity score functions, we consider the 10-fold cross-fitting strategy as in \citet{chernozhukov2018double,zhao2019efficient,athey2021policy}. Specifically, the training sample is randomly divided into 10 folds. For the $ k $-th fold fitting, we utilize the data other the $ k $-th fold to fit a treatment-free effect or propensity score model, and then predict the treatment-free effects or the propensity scores for the $ k $-th fold data. 
	
	We use regression forest to estimate variance function from the squared working residual. The \texttt{regression\_forest} function from the \texttt{grf} package in \texttt{R} is called. We also fit the MARS and COSSO estimates of variance function. The \texttt{earth} function from the \texttt{earth} package and a modified program based on the \texttt{cosso.Gaussian} function from the \texttt{cosso} package are applied. More details on the COSSO model and program are discussed in Section \ref{sec:var_cosso}.
	
	Before fitting E-Learning, we first center and scale each variables to ensure $ (1/n)\sum_{i=1}^{n}X_{ij} = 0 $ and $ (1/n)\sum_{i=1}^{n}X_{ij}^{2} = 1 $. When solving the penalized minimization problem \eqref{eq:eLearn_pen} by the accelerated proximal gradient descent, we call the \texttt{apg} function in \texttt{R} to perform optimization. When determining the tuning sequence of $ \lambda $'s, we take the strategy analog to \texttt{glmnet} \citep{friedman2010regularization}.
	
	\item \textbf{Q-Learning} (general $ K $): We consider the linear model using $ Y $ as the response and $ \big( 1,\bX^{\intercal},\vec{\bA}^{\intercal}, \vec{\bA}^{\intercal} \otimes \bX^{\intercal} \big)^{\intercal} $ as the covariates with the $ \ell_{1} $-penalty. Here, $ \vec{\bA} = \big( \bbone(A=2),\bbone(A=3),\cdots,\bbone(A=K) \big)^{\intercal} $, and $ \otimes $ denotes the Kronecker product. The method is also known as the \textit{$ \ell_{1} $-Penalized Least Square ($ \ell_{1} $-PLS)} \citep{qian2011performance}, and implemented by the \texttt{glmnet} function in \texttt{R}.
	
	\item \textbf{G-Estimation, dWOLS} ($ K=2 $, $ A \in \{0,1\} $): The \texttt{DTRreg} function from the \texttt{DTRreg} package \citep{wallace2017dynamic} in \texttt{R} is called to fit G-Estimation (\texttt{method = "gest"}) and dWOLS (\texttt{method = "dwols"}). The treatment-free effect model (\texttt{tf.mod}) is specified as linear in $ (1,\bX^{\intercal})^{\intercal} $. The propensity score model (\texttt{treat.mod}) is specified as the logistic model of $ A $ with respect to $ (1,\bX^{\intercal}) ^{\intercal} $. The interaction effect model (\texttt{blip.mod}) is specified as linear in $ (1,\bX^{\intercal})^{\intercal} $. For dWOLS, the weight function $ w(\bX,A) = |A - \widehat{\pi}_{\sA,n}(\bX)| $ in Section \ref{sec:binary} is used.
	
	\item \textbf{A-Learning, Subgroup Identification} ($ K = 2 $, $ A \in \{-1,1\} $): The \texttt{fit.subgroup} function from the \texttt{personalized} package \citep{huling2018subgroup} in \texttt{R} is called to fit A-Learning and Subgroup Identification with the $ \ell_{1} $-penalty (\texttt{method = "a\_learning"} and \texttt{method = "weighting"} respectively, \texttt{loss = "sq\_loss\_lasso"}). The propensity score model (\texttt{propensity.func}) is specified as the logistic model of $ A $ with respect to $ (1,\bX^{\intercal})^{\intercal} $ with the $ \ell_{1} $-penalty, which is fitted by \texttt{glmnet}. The treatment-free effect, also known as the augmentation function (\texttt{augment.func}), is specified as: for A-Learning, the linear model of $ Y $ with respect to $ (1,\bX^{\intercal})^{\intercal} $ with the $ \ell_{1} $-penalty, fitted by \texttt{glmnet}; and for Subgroup Identification, the linear model of $ Y $ with respect to $ (1,\bX^{\intercal},A,A\bX^{\intercal})^{\intercal} $ with the $ \ell_{1} $-penalty, fitted by \texttt{glmnet}, and outputting the arithmetic average of  predictions at $ A = 1 $ and $ A = -1 $.
	
	\item \textbf{D-Learning, RD-Learning:} (general $ K $) We consider the class of linear functions with the row-wise grouped LASSO penalty. The training process is performed  by the accelerated proximal gradient descent using the \texttt{apg} function. The estimation of the treatment-free effect and propensity score functions, the fitting details and the tuning strategy are the same as in E-Learning.
	
	\item \textbf{OWL, RWL, EARL} ($ K = 2 $, $ A \in \{0,1\} $): The \texttt{owl}, \texttt{rwl} and \texttt{earl} functions are called from the \texttt{R} package \texttt{DynTxRegime} to fit OWL, RWL and EARL respectively. The propensity score model (\texttt{moPropen}) is specified as the logistic model of $ A $ with respect to $ (1,\bX^{\intercal})^{\intercal} $ with the $ \ell_{1} $-penalty, which is fitted by \texttt{glmnet}. The outcome models, including the main effect model (\texttt{moMain}, used in \texttt{rwl}) and the contrast model (\texttt{moCont}, used in \texttt{rwl} and \texttt{earl}), are both specified as linear in $ (1,\bX^{\intercal})^{\intercal} $ with the $ \ell_{1} $-penalty, which are fitted by \texttt{glmnet}. The corresponding outcome mean model is $ \bbE(Y|\bX, A) = \texttt{moMain}(\bX) + A \times  \texttt{moCont}(\bX) $. These methods are fitted with linear decision functions (\texttt{kernel = "linear"}). For \texttt{owl} and \texttt{earl}, the hinge surrogate loss is used (\texttt{surrogate = "hinge"}). For \texttt{rwl}, the surrogate loss is the smoothed ramp loss \citep{zhou2017residual}. The tuning parameter $ \lambda $ for all methods is determined by 5-fold cross validation (\texttt{cvFolds = 5}). The sequence of $ \lambda $'s for tuning is determined analog to \texttt{glmnet} \citep{friedman2010regularization}.
	
	\item \textbf{Policy Learning:} (general $ K $) We use the \texttt{policy\_tree} function from the \texttt{policytree} package \citep{sverdrup2020policytree} in \texttt{R} to fit policy learning with decision trees. The outcome mean function and the propensity score function are both fitted by \texttt{regression\_forest} from the \texttt{grf} package.
\end{itemize}

\newpage

\subsection{COSSO Estimate of the Working Variance Function}\label{sec:var_cosso}

In this section, we consider the implementation details of estimating $ \sigma_{\opt}^{2}(\bx,A;\widehat{\mu}_{0,n}) $ from COSSO \citep{lin2006component}. Specifically, we perform nonparametric regression using the squared working residual $ \widehat{e}^{2} $ as the response and $ (\bX,A) $ as the covariates. 

First of all, we discuss an SS-ANOVA model in terms of the covariate vector $ \bX=(X_{1},\cdots,X_{p})^{\intercal} \in \bbR^{p} $ and the treatment variable $ A \in \{1,2,\cdots,K\} $. If the $ j $-th variable $ X_{j} $ is continuously ranged, then the we first scale the domain of $ X_{j} $ to $ [0,1] $, and consider the $ j $-th covariate function space as $ \cH_{j} = \cS_{2} $, where $ (\cS_{2},\| \cdot \|_{\cS_{2}}) $ is the second order Sobolev Hilbert space:
\begin{align*}
	\cS_{2} &= \left\{ f:[0,1] \to \bbR~\middle|~\text{$ f $ and $ f' $ are absolutely continuous}, ~ \int f''(x)^{2} \rd x < +\infty \right\};\\
	\| f \|_{\cS_{2}}^{2} &= \left( \int_{0}^{1}f(x)\rd x \right)^{2} + \left( \int_{0}^{1}f'(x)\rd x \right)^{2} + \int_{0}^{1}f''(x)^{2}\rd x.
\end{align*}
In particular, $ \cH_{j} = \cS_{2} $ can be decomposed as $ \{1\} \oplus \widebar{\cH}_{j} $ \citep[Equation (2.26)]{gu2013smoothing}, where $ \widebar{\cH}_{j} $ is the \textit{reproducing kernel Hilbert space (RKHS)} corresponding to the kernel function
\[ \widebar{\kappa}_{j}(x,x') = k_{1}(x)k_{1}(x') + k_{2}(x)k_{2}(x') - k_{4}(|x-x'|); \quad x,x' \in [0,1]. \]
Here, $ k_{1}(x) = x - 0.5 $, $ k_{2}(x) = (1/2)[k_{1}(x)^{2} - 1/12] $, and $ k_{4}(x) = (1/24)[k_{1}(x)^{4} - k_{1}(x)^{2}/2 + 7/240] $. If $ X_{j} $ takes finitely many values in $ \{ 1,2,\cdots,L_{j} \} $, then we consider $ \cH_{j} = \bbR^{L_{j}} $, which can be further decomposed as $ \{1\} \oplus \widebar{\cH}_{j} $. Here, $ \widebar{\cH}_{j} = \{ (\alpha_{1},\alpha_{2},\cdots,\alpha_{L_{j}})^{\intercal} \in \bbR^{L_{j}}: \sum_{l=1}^{L_{j}}\alpha_{l} = 0 \} $, and can be regarded as an $ L_{j} $-dimensional RKHS corresponding to the kernel matrix $ [\widebar{\kappa}_{j}(x,x')]_{x,x'=1}^{L_{j}} = \sfI_{L_{j}\times L_{j}} - (1/L_{j})\vec{\bone}_{L_{j}}\vec{\bone}_{L_{j}}^{\intercal} $. Similarly, since the treatment variable $ A $ is valued in $ \{1,2,\cdots,K\} $, we also consider the treatment function space $ \cH_{\sA} = \bbR^{K} $ with the decomposition $ \cH_{\sA} = \{1\} \oplus \widebar{\cH}_{\sA} $, where $ \widebar{\cH}_{\sA} $ is the subspace of $ \bbR^{K} $ with the sum-to-zero constraint and corresponds to the kernel matrix $ [\widebar{\kappa}_{\sA}(a,a')]_{a,a'=1}^{K} = \sfI_{K\times K} - (1/K)\vec{\bone}_{K}\vec{\bone}_{K}^{\intercal} $.

The SS-ANOVA model is based on the following tensor-product RKHS \citep[Section 2.4.1]{gu2013smoothing}:
\begin{align*}
	\cH := \left[ \bigotimes_{j=1}^{d}\cH_{j} \right] \otimes \cH_{\sA} &= \left[ \bigotimes_{j=1}^{d}\left( \{1\} \oplus \widebar{\cH}_{j}\right) \right] \otimes \left(\{1\} \oplus \widebar{\cH}_{\sA} \right) \\
	&= \underbrace{\{1\}}_{\text{global main effect}} \oplus \underbrace{\left[ \bigoplus_{j=1}^{d}\widebar{\cH}_{j} \right]}_{\text{covariate main effects}} \oplus \underbrace{\widebar{\cH}_{\sA}}_{\text{treatment main effect}} \oplus \underbrace{\left[ \bigoplus_{j=1}^{d}\left(\widebar{\cH}_{j}\otimes \widebar{\cH}_{\sA}\right) \right]}_{\text{covariate-treatment interaction effect}}\\
	&\quad \oplus \underbrace{\left\{ \bigoplus_{\substack{\sJ \subseteq \{1,2,\cdots,d\} \\ |\sJ| \ge 2}} \left[ \left( \bigotimes_{j \in \sJ}\widebar{\cH}_{j} \right) \oplus \left( \bigotimes_{j\in \sJ}\widebar{\cH}_{j}\otimes\widebar{\cH}_{\sA} \right) \right] \right\}}_{\text{higher-order interaction effects}}.
\end{align*}
Here, we only consider first four effects from the above tensor-sum decomposition and ignore the higher-order interaction effects. Then the SS-ANOVA model is
\[ \bbE(\widehat{e}^{2}|\bX,A) = \underbrace{\nu_{0}}_{\text{global main effect}} + \underbrace{\sum_{j=1}^{p}f_{j}(X_{j})}_{\text{covariate main effect}} + \underbrace{\sum_{k=1}^{K}\alpha_{k}}_{\text{treatment main effect}} + \underbrace{\sum_{j=1}^{K}\sum_{k=1}^{K}f_{jk}(X_{j})}_{\text{covariate-treatment interaction effect}} + \underbrace{u}_{\text{remainder}}. \]
In particular, the tensor-product RKHS $ \widebar{\cH}_{j} \otimes \widebar{\cH}_{\sA} $, which models the covariate-treatment interaction effect, corresponds to the kernel function 
\[ (\widebar{\kappa}_{j} \otimes \widebar{\kappa}_{\sA})\Big( (x_{j},a)^{\intercal},(x_{j}',a')^{\intercal} \Big) = \widebar{\kappa}_{j}(x_{j},x_{j}')\widebar{\kappa}_{\sA}(a,a'). \]
Then the COSSO estimate $ \widehat{\sigma}_{n}^{2}(\bX,A) $ of the working variance function is obtained by solving:
\[ \min_{f \in \cH}\left\{ {1 \over n}\sum_{i=1}^{n}[e_{i}^{2} - f(\bX_{i},A_{i})]^{2} + \lambda_{\sigma^{2}}\left( \sum_{j=1}^{d}\| f \|_{\widebar{\cH}_{j}} + \| f \|_{\widebar{\cH}_{\sA}} + \sum_{j=1}^{d}\| f \|_{\widebar{\cH}_{j}\otimes \widebar{\cH}_{\sA}} \right) \right\}. \]
Here, $ \| \cdot \|_{\widebar{\cH}_{j}} $, $ \| \cdot \|_{\widebar{\cH}_{\sA}} $ and $ \| \cdot \|_{\widebar{\cH}_{j}\otimes\widebar{\cH}_{\sA}} $ are the RKHS-norms corresponding to the associated component spaces, and $ \lambda_{\sigma^{2}} $ is a tuning parameter.

For implementation, we define the empirical kernel matrices $ \widebar{\sfK}_{j} := [\widebar{\kappa}_{j}(X_{ij},X_{i'j})]_{i,i'=1}^{n} $ and $ \widebar{\sfK}_{\sA} := [\widebar{\kappa}_{\sA}(A_{i},A_{i'})]_{i,i'=1}^{n} $. Then $ \widebar{\sfK}_{j} $, $ \widebar{\sfK}_{\sA} $ and $ \widebar{\sfK}_{j,\sA} := \widebar{\sfK}_{j} \odot \widebar{\sfK}_{\sA} $ are the empirical kernel matrices on $ \widebar{\cH}_{j} $, $ \widebar{\cH}_{\sA} $ and $ \widebar{\cH}_{j} \otimes \widebar{\cH}_{\sA} $ respectively, where $ \widebar{\sfK}_{j} \odot \widebar{\sfK}_{\sA} $ is the elementwise product of $ \widebar{\sfK}_{j} $ and $ \widebar{\sfK}_{\sA} $. For a vector $ \btheta := (\theta_{1},\cdots,\theta_{d};\theta_{\sA};\theta_{1,\sA},\cdots,\theta_{d,\sA})^{\intercal} \in \bbR_{+}^{2d+1} $ of kernel weights, we write $ \widebar{\sfK}_{\btheta} := \sum_{j=1}^{d}\theta_{j}\widebar{\sfK}_{j} + \theta_{\sA}\widebar{\sfK}_{\sA} + \sum_{j=1}^{d}\theta_{j,\sA}\widebar{\sfK}_{j,\sA} \in \bbR^{n\times n} $ as the weighted sum of the empirical kernel matrices. For a vector $ \balpha \in \bbR^{n} $ of representer coefficients, we write $ \sfG_{\balpha} := [\widebar{\sfK}_{1}\balpha,\cdots,\widebar{\sfK}_{d}\balpha;\widebar{\sfK}_{\sA}\balpha;\widebar{\sfK}_{1,\sA}\balpha,\cdots,\widebar{\sfK}_{d,\sA}\balpha] \in \bbR^{n\times (2d+1)} $ as the gram matrix of the componentwise prediction values. We also denote $ \vec{\be}^{2} := (e_{1}^{2},\cdots,e_{n}^{2})^{\intercal} $ as the empirical squared residual vector. Then we fit a COSSO model by calling the \texttt{R} function \texttt{cosso::cosso.Gaussian} with the aforementioned kernel matrices and the squared residual vector as inputs. In particular, a random subset of sample points with size $ \max\{40,\lceil 12n^{2/9} \rceil\} $ is used for representers. The following two steps are alternatively implemented:
\begin{itemize}
	\item For a given kernel weight vector $ \btheta $, we solve
	\[ \min_{b,\balpha}\left\{ {1 \over n}\left\| \vec{\be}^{2} - b\vec{\bone}_{n} - \widebar{\sfK}_{\btheta}\balpha \right\|_{2}^{2} + \lambda_{0}\balpha^{\intercal}\widebar{\sfK}_{\btheta}\balpha\right\} \]
	for the representer coefficient vector $ (b,\balpha^{\intercal})^{\intercal} $;
	
	\item For a given representer coefficient vector $ (b,\balpha^{\intercal})^{\intercal} $, we solve
	\[ \min_{\btheta}\left\{ {1 \over n}\left\| \vec{\be}^{2} - b\vec{\bone}_{n} - \sfG_{\balpha}\btheta \right\|_{2}^{2} + \lambda_{0}\balpha^{\intercal}\sfG_{\balpha}\btheta \quad \text{subject to} \quad \btheta \in \bbR_{+}^{2d+1}, ~ \vec{\bone}_{2d+1}^{\intercal}\btheta \le M \right\}, \]
	for the kernel weight vector $ \btheta $.
\end{itemize}
Here, the tuning parameters $ (\lambda_{0},M) $ are chosen according to \citet[Section 6]{lin2006component}. 

\newpage

\subsection{Technical Proofs}\label{sec:proof}

\subsubsection{Proof of Theorem \ref{thm:regret_est}}

\begin{proof}[Proof of Theorem \ref{thm:regret_est}]
	\begin{align*}
	\cV(d^{\star}) - \cV(\widehat{d}_{n}) =& \bbE\left\{ \sum_{k=1}^{K}\gamma(\bX,k)\{ \bbone[d^{\star}(\bX) = k] - \bbone[\widehat{d}_{n}(\bX) = k]\}  \right\}\\
	\le&\bbE\left\{ \sum_{k \ne k'}|\gamma(\bX,k) - \gamma(\bX,k')|;d^{\star}(\bX) = k, ~ \widehat{d}_{n}(\bX) = k' \right\}\\
	=&{1 \over 2}\bbE\Bigg\{ \sum_{k \ne k'}|\gamma(\bX,k) - \gamma(\bX,k')|; \\
	&\qquad d^{\star}(\bX) = k ~ \text{and} ~ \widehat{d}_{n}(\bX) = k' ~ \text{or} ~ d^{\star}(\bX) = k' ~ \text{and} ~ \widehat{d}_{n}(\bX) = k \Bigg\}\\
	\le &{1 \over 2}\bbE\Bigg\{ \sum_{k \ne k'}|\gamma(\bX,k) - \gamma(\bX,k')| + |\widehat{\gamma}_{n}(\bX,k) - \widehat{\gamma}_{n}(\bX,k')|; \\
	&\qquad d^{\star}(\bX) = k ~ \text{and} ~ \widehat{d}_{n}(\bX) = k' ~ \text{or} ~ d^{\star}(\bX) = k' ~ \text{and} ~ \widehat{d}_{n}(\bX) = k \Bigg\}\\
	=&{1 \over 2}\bbE\Bigg\{ \sum_{k \ne k'}\left |[\gamma(\bX,k) - \gamma(\bX,k')]-[\widehat{\gamma}_{n}(\bX,k) - \widehat{\gamma}_{n}(\bX,k')]\right |; \\
	&\qquad d^{\star}(\bX) = k ~ \text{and} ~ \widehat{d}_{n}(\bX) = k' ~ \text{or} ~ d^{\star}(\bX) = k' ~ \text{and} ~ \widehat{d}_{n}(\bX) = k \Bigg\} \tag{$ \star $}\label{eq:regret_est} \\
	\le &{1 \over 2}\bbE\Bigg\{ \sum_{k \ne k'}|\gamma(\bX,k) - \widehat{\gamma}_{n}(\bX,k)| + |\gamma(\bX,k') - \widehat{\gamma}_{n}(\bX,k')|; \\
	&\qquad d^{\star}(\bX) = k ~ \text{and} ~ \widehat{d}_{n}(\bX) = k' ~ \text{or} ~ d^{\star}(\bX) = k' ~ \text{and} ~ \widehat{d}_{n}(\bX) = k \Bigg\}\\
	\le&\bbE\left |\gamma(\bX,d^{\star}) - \widehat{\gamma}_{n}(\bX,d^{\star})\right | + \bbE\left |\gamma(\bX,\widehat{d}_{n}) - \widehat{\gamma}_{n}(\bX,\widehat{d}_{n})\right | \\
	\le & 2\max_{1 \le k \le K}\bbE\left |\gamma(\bX,k) - \widehat{\gamma}_{n}(\bX,k)\right |.
	\end{align*}
	The equality (\ref{eq:regret_est}) holds since the event $ d^{\star}(\bX) = k ~ \text{and} ~ \widehat{d}_{n}(\bX) = k' ~ \text{or} ~ d^{\star}(\bX) = k' ~ \text{and} ~ \widehat{d}_{n}(\bX) = k  $ implies that
	\[ [\gamma(\bX,k) - \gamma(\bX,k')][\widehat{\gamma}_{n}(\bX,k) - \widehat{\gamma}_{n}(\bX,k')] < 0. \]
\end{proof}

\subsubsection{Proof of Lemma \ref{lem:tangent}}

\begin{proof}[Proof of Lemma \ref{lem:tangent}] Denote 
	\[ \widetilde{\Lambda} := \Big\{ \bH \in \cH: \bbE(\bH\epsilon|\bX,A) = \bbE(\bH\epsilon|\bX)\Big\}. \]
	Consider the following family of parametric submodels of (\ref{eq:model}):
	\[ \sP_{\bbeta,\alpha} := \Big\{ (\bX,A,Y) \sim p_{\sX}(\bX;\balpha_{\sX}) \times p_{\sA}(A|\bX;\balpha_{\sA}) \times p_{\epsilon}\big( \overbrace{Y - \mu_{0}(\bX;\balpha_{\mu_{0}}) - \gamma(\bX,A;\bbeta)}^{\epsilon}\big| \bX, A; \balpha_{\epsilon} \big) \Big\}, \]
	where $ \balpha = \balpha_{\sX} \oplus \balpha_{\sA} \oplus \balpha_{\epsilon} \oplus \balpha_{\mu_{0}} $ are finite-dimensional parameters for the nuisance components $ \eta = (p_{\sX},p_{\sA},p_{\epsilon},\mu_{0}) $. Then the nuisance score vector is
	\[ \bS_{\balpha} = \bordermatrix{
		~ & \cr
		\bS_{\sX} & {\partial \log p_{\sX}(\bX;\balpha_{\sX}) \over \partial\balpha_{\sX}} \cr 
		\bS_{\sA} & {\partial \log p_{\sA}(A|\bX;\balpha_{\sA}) \over \partial\balpha_{\sA}} \cr
		\bS_{\epsilon} & {\partial \log p_{\epsilon}(\epsilon|\bX,A;\balpha_{\epsilon}) \over \partial\balpha_{\epsilon}} \cr 
		\bS_{\mu_{0}} & -{\partial \log p_{\epsilon}(\epsilon|\bX,A;\balpha_{\epsilon}) \over \partial\epsilon}{\partial \mu_{0}(\bX;\balpha_{\mu_{0}}) \over \partial \balpha_{\mu_{0}}}
	}.
	\]
	Then the nuisance tangent space of the family of submodels $ \sP_{\bbeta,\balpha} $ is defined as \citep{tsiatis2007semiparametric}:
	\begin{align*}
		\Lambda_{\balpha} =& \left\{ \sfB \bS_{\balpha}: \sfB \in \bbR^{p \times \dim(\balpha)} \right\} \\
		=& \Big\{ \sfB_{\sX}\bS_{\sX} + \sfB_{\sA}\bS_{\sA} + \sfB_{\epsilon}\bS_{\epsilon} + \sfB_{\mu_{0}}\bS_{\mu_{0}}: \\
		&\quad \sfB_{\sX} \in \bbR^{p \times \dim(\balpha_{\sX})}, ~ \sfB_{\sA} \in \bbR^{p \times \dim(\balpha_{\sA})}, \sfB_{\epsilon} \in \bbR^{p \times \dim(\balpha_{\epsilon})}, \sfB_{\mu_{0}} \in \bbR^{p \times \dim(\balpha_{\mu_{0}})} \Big\}.
	\end{align*}
	We aim to show that $ \Lambda_{\balpha} \subseteq \widetilde{\Lambda} $. By $ \widetilde{\Lambda} $ is a linear space, it is equivalent to show that $ \bS_{\sX},\bS_{\sA},\bS_{\epsilon},\bS_{\mu_{0}} \in \widetilde{\Lambda} $. Denote $ \bbE_{\bbeta,\balpha} $ as the expectation under the submodel parametrized by $ (\bbeta,\balpha) $. 
	\begin{itemize}
		\item Since $ \bS_{\sX} $ is a function of $ \bX $, we have that 
		\[ \bbE_{\bbeta,\balpha}(\bS_{\sX}\epsilon|\bX,A) = \bS_{\sX}\bbE_{\bbeta,\balpha}(\epsilon|\bX,A) = \bzero = \bS_{\sX}\bbE_{\bbeta,\balpha}(\epsilon|\bX) = \bbE_{\bbeta,\balpha}(\bS_{\sX}\epsilon|\bX). \]
		That is, $ \bS_{\sX} \in \widetilde{\Lambda} $.
		
		\item Since $ \bS_{\sA} $ is a function of $ (\bX,A) $, we have that
		\[ \bbE_{\bbeta,\balpha}(\bS_{\sA}\epsilon|\bX,A) = \bS_{\sA}\bbE_{\bbeta,\balpha}(\epsilon|\bX,A) = \bzero. \]
		On the other hand, by $ \bbE_{\bbeta,\balpha}(\epsilon|\bX) = 0 $ for any $ \balpha $, we have
		\begin{align*}
			\bzero &= {\partial \over \balpha_{\sA}}\bbE_{\bbeta,\balpha}(\epsilon|\bX) \\
			&= {\partial \over \balpha_{\sA}}\int \epsilon p_{\sA}(a|\bX;\balpha_{\sA})p_{\epsilon}(\epsilon|\bX,a;\balpha_{\epsilon})\rd a\rd \epsilon\\
			&= \int \epsilon \times {{(\partial/\partial\balpha_{\sA})}p_{\sA}(a|\bX;\balpha_{\sA}) \over p_{\sA}(a|\bX;\balpha_{\sA})} \times p_{\sA}(a|\bX;\balpha_{\sA})p_{\epsilon}(\epsilon|\bX,a;\balpha_{\epsilon})\rd a\rd \epsilon \\
			&= \int \epsilon \times {\partial \log p_{\sA}(a|\bX;\balpha_{\sA}) \over \partial \balpha_{\sA}} \times p_{\sA}(a|\bX;\balpha_{\sA})p_{\epsilon}(\epsilon|\bX,a;\balpha_{\epsilon})\rd a\rd \epsilon \\
			&= \bbE_{\bbeta,\balpha}(\bS_{\sA}\epsilon|\bX).
		\end{align*}
		That is, $ \bbE_{\bbeta,\balpha}(\bS_{\sA}\epsilon|\bX,A) = \bzero = \bbE_{\bbeta,\balpha}(\bS_{\sA}\epsilon|\bX) $, and $ \bS_{\sA} \in \widetilde{\Lambda} $.
		
		\item By $ \bbE_{\bbeta,\balpha}(\epsilon|\bX,A) = 0 $ for any $ \balpha $, we have
		\begin{align*}
			\bzero &= {\partial \over \balpha_{\epsilon}}\bbE_{\bbeta,\balpha}(\epsilon|\bX,A) \\
			&= {\partial \over \balpha_{\epsilon}}\int \epsilon p_{\epsilon}(\epsilon|\bX,A;\balpha_{\epsilon})\rd \epsilon\\
			&= \int \epsilon \times {{(\partial/\partial\balpha_{\epsilon})}p_{\epsilon}(\epsilon|\bX,A;\balpha_{\epsilon}) \over p_{\epsilon}(\epsilon|\bX,A;\balpha_{\epsilon})} \times p_{\epsilon}(\epsilon|\bX,A;\balpha_{\epsilon})\rd \epsilon\\
			&= \int \epsilon \times {\partial \log p_{\epsilon}(\epsilon|\bX,A;\balpha_{\epsilon}) \over \partial \balpha_{\epsilon}} \times p_{\epsilon}(\epsilon|\bX,A;\balpha_{\epsilon})\rd \epsilon \\
			&= \bbE_{\bbeta,\balpha}(\bS_{\epsilon}\epsilon|\bX,A).
		\end{align*}
		That is, $ \bbE_{\bbeta,\balpha}(\bS_{\epsilon}\epsilon|\bX,A) = \bzero = \bbE_{\bbeta,\balpha}(\bS_{\epsilon}\epsilon|\bX) $, and $ \bS_{\epsilon} \in \widetilde{\Lambda} $.
		
		\item Note that
		\[ \bS_{\mu_{0}}\epsilon = - {\epsilon\partial \log p_{\epsilon}(\epsilon|\bX,A;\balpha_{\epsilon}) \over \partial\epsilon}\times\underbrace{\partial \mu_{0}(\bX;\balpha_{\mu_{0}}) \over \partial \balpha_{\mu_{0}}}_{\text{function of $ \bX $}}. \]
		
		\begin{lem}\label{lem:score}
			Suppose $ X \sim p(x) $ where $ p(x) $ is a probability density function with respect to the Lebesgue measure on $ \bbR $, such that $ |x|p(x) \to 0 $ as $ |x| \to +\infty $. Then
			\[  \bbE\left( {X \rd \log p(X) \over \rd x} \right) = -1. \]
		\end{lem}
	
		By Lemma \ref{lem:score}, we have
		\begin{align}
			\bbE\left( {\epsilon\partial\log p_{\epsilon}(\epsilon|\bX,A;\balpha_{\epsilon}) \over \partial \epsilon}\middle|\bX,A \right) = -1.\label{eq:tangent}
		\end{align}
		Then
		\[ \bbE_{\bbeta,\balpha}(\bS_{\mu_{0}}\epsilon|\bX,A) = - \bbE\left( {\epsilon\log p_{\epsilon}(\epsilon|\bX,A;\balpha_{\epsilon}) \over \partial \epsilon}\middle|\bX,A \right){\partial \mu_{0}(\bX;\balpha_{\mu_{0}}) \over \partial\balpha_{\mu_{0}}} = {\partial \mu_{0}(\bX;\balpha_{\mu_{0}}) \over \partial\balpha_{\mu_{0}}}, \]
		which is a function of $ \bX $. Consequently, $ \bbE_{\bbeta,\balpha}(\bS_{\mu_{0}}\epsilon|\bX,A) = \bbE_{\bbeta,\balpha}(\bS_{\mu_{0}}\epsilon|\bX) $, and $ \bS_{\mu_{0}} \in \widetilde{\Lambda} $.
	\end{itemize}
	Therefore, we can conclude that $ \Lambda_{\balpha} \subseteq \widetilde{\Lambda} $. By the definition of nuisance tangent space of the semiparametric model, we have that $ \Lambda \subseteq \widetilde{\Lambda} $.
	
	Next, we aim to justify $ \Lambda \supseteq \widetilde{\Lambda} $. Fix an $ \bH = \bh(\bX,A,\epsilon) \in \widetilde{\Lambda} $, \textit{i.e.},
	\[ \bbE(\bH) = \bzero; \quad \bbE(\bH\epsilon|\bX,A) = \bbE(\bH\epsilon|\bX). \]
	We need to construct a parametric submodel whose nuisance score vector is $ \bH $. Consider the following orthogonal decompositions:
	\[ \bH = \underbrace{\bbE(\bH|\bX)}_{:= \bH_{\sX}} + \underbrace{\bbE(\bH|\bX,A) - \bbE(\bH|\bX)}_{:= \bH_{\sA}} + \underbrace{\bH - \bbE(\bH|\bX,A)}_{:=\bH_{\epsilon}}. \]
	Without loss of generality, we assume that $ \| \bH \|_{2} \le M < +\infty $. Define
	\begin{align*}
		p_{\sX}(\bX;\balpha_{\sX}) &:= p_{\sX}(\bX)(1 + \balpha_{\sX}^{\intercal}\bH_{\sX}); && \| \balpha_{\sX} \|_{2} \le 1/M;\\
		p_{\sA}(A|\bX;\balpha_{\sA}) &:= p_{\sA}(A|\bX)(1 + \balpha_{\sA}^{\intercal}\bH_{\sA}); && \| \balpha_{\sA} \|_{2} \le 1/M;\\
		p_{\widetilde{\epsilon}}(\widetilde{\epsilon}|\bX,A;\balpha_{\epsilon}) &:= p_{\epsilon}(\widetilde{\epsilon}|\bX,A)(1 + \balpha_{\epsilon}^{\intercal}\bH_{\epsilon}); && \| \balpha_{\epsilon} \|_{2} \le 1/M.
	\end{align*}
	Here, $ p_{\sX}(\bX) $, $ p_{\sA}(A|\bX) $ and $ p_{\epsilon}(\epsilon|\bX,A) $ are the densities of Model (\ref{eq:model}). Then we consider the following data generating process (DGP) parametrized by $ (\bbeta,\balpha) $ where $ \balpha = \balpha_{\sX} \oplus \balpha_{\sA} \oplus \balpha_{\epsilon} $:
	\begin{align*}
		&Y = \mu_{0}(\bX) + \gamma(\bX,A;\bbeta) + \widetilde{\epsilon};\\
		\text{subject to} \quad &\sum_{k=1}^{K}\gamma(\bX,k;\bbeta) = 0; \quad \bbE_{\bbeta,\balpha}(\widetilde{\epsilon}|\bX,A) = \bbE_{\bbeta,\balpha}(\widetilde{\epsilon}|\bX);\\
		&(\bX,A,\widetilde{\epsilon}) \sim p_{\sX}(\bx;\balpha_{\sX})p_{\sA}(a|\bx;\balpha_{\sA})p_{\widetilde{\epsilon}}(\widetilde{\epsilon}|\bx,a;\balpha_{\epsilon}).
	\end{align*}
	Notice that the above DGP can be transformed to the form of (\ref{eq:model}) by setting $ \mu_{0}(\bX;\balpha_{\epsilon}) := \mu_{0}(\bX) + \bbE_{\bbeta,\balpha}(\widetilde{\epsilon}|\bX) $ and $ \epsilon := \widetilde{\epsilon} - \mu_{0}(\bX;\balpha_{\epsilon}) $. Next, we verify that the above DGP is well defined and corresponds to the nuisance score vectors $ \bH_{\sX} $, $ \bH_{\sA} $ and $ \bH_{\epsilon} $.
	\begin{itemize}
		\item By 
		\[ |\balpha^{\intercal}_{\sX}\bH_{\sX}| \le \|\balpha_{\sX}\|_{2}\|\bH\|_{2} \le 1;\quad |\balpha^{\intercal}_{\sA}\bH_{\sA}| \le \|\balpha_{\sA}\|_{2}\|\bH\|_{2} \le 1;\quad|\balpha^{\intercal}_{\epsilon}\bH_{\epsilon}| \le \|\balpha_{\epsilon}\|_{2}\|\bH\|_{2} \le 1, \]
		we have $ p_{\sX}(\bX;\balpha_{\sX}),p_{\sA}(A|\bX;\balpha_{\sA}),p_{\widetilde{\epsilon}}(\widetilde{\epsilon}|\bX,A;\balpha_{\epsilon}) \ge 0 $.
		
		\item By $ \bbE(\bH_{\sX}) = \bzero $, $ \bbE(\bH_{\sA}|\bX) = \bzero $ and $ \bbE(\bH_{\epsilon}|\bX,A) = \bzero $, we have
		\begin{align*}
			\int p_{\sX}(\bx;\balpha_{\sX})\rd \bx &= \int p_{\sX}(\bx)\rd \bx + \balpha^{\intercal}_{\sX}\bbE(\bH_{\sX}) = 1;\\
			\int p_{\sA}(a|\bX;\balpha_{\sA})\rd a &= \int p_{\sA}(a|\bX)\rd a + \balpha_{\sA}^{\intercal}\bbE(\bH_{\sA}|\bX) = 1; \\
			\int p_{\widetilde{\epsilon}}(\widetilde{\epsilon}|\bX,A;\balpha_{\epsilon}) \rd \widetilde{\epsilon} &= \int p_{\epsilon}(\widetilde{\epsilon}|\bX,A) \rd \widetilde{\epsilon} + \balpha_{\epsilon}^{\intercal}\bbE(\bH_{\epsilon}|\bX,A) = 1.
		\end{align*}
		Therefore, $ p_{\sX}(\bx;\balpha_{\sX}) $, $ p_{\sX}(a|\bX;\balpha_{\sA}) $ and $ p_{\widetilde{\epsilon}}(\widetilde{\epsilon}|\bX,A;\balpha_{\epsilon}) $ are probability density functions.
		
		\item The conditional mean restriction becomes
		\begin{align*}
			\bbE_{\bbeta,\balpha}(\widetilde{\epsilon}|\bX,A) &= \int \widetilde{\epsilon}(1 + \balpha_{\epsilon}^{\intercal}\bH_{\epsilon})p_{\epsilon}(\widetilde{\epsilon}|\bX,A)\rd \widetilde{\epsilon} \\
			&= \bbE(\epsilon|\bX,A) + \balpha_{\epsilon}^{\intercal}\bbE(\bH_{\epsilon }\epsilon|\bX,A)\\
			&= \balpha_{\epsilon}^{\intercal}\bbE(\bH_{\epsilon }\epsilon|\bX,A),
		\end{align*}
		which is a function of $ \bX $, by the fact that
		\[ \bbE(\bH_{\epsilon}\epsilon|\bX,A) = \bbE(\bH\epsilon|\bX,A) - \underbrace{\bbE(\bH|\bX,A)\bbE(\epsilon|\bX,A)}_{= \bzero} = \bbE(\bH\epsilon|\bX) - \underbrace{\bbE(\bH|\bX,A)\bbE(\epsilon|\bX)}_{= \bzero} = \bbE(\bH_{\epsilon}\epsilon|\bX). \]
		Then it can be clear that
		\[ \bbE_{\bbeta,\balpha}(\widetilde{\epsilon}|\bX,A) = \bbE_{\bbeta,\balpha}(\widetilde{\epsilon}|\bX). \]
		
		\item The nuisance score vectors are
		\[ \begin{array}{rclclcl}
			\bS_{\sX} &=& \left[ {\partial \log p_{\sX}(\bX;\balpha_{\sX}) \over \partial \balpha_{\sX}} \right]_{\balpha_{\sX} = \bzero} &=& \left[ {\bH_{\sX} \over 1 + \balpha_{\sX}^{\intercal}\bH_{\sX}} \right]_{\balpha_{\sX} = \bzero} &=& \bH_{\sX};\\
			\bS_{\sA} &=& \left[ {\partial \log p_{\sA}(A|\bX;\balpha_{\sA}) \over \partial \balpha_{\sA}} \right]_{\balpha_{\sA} = \bzero} &=& \left[ {\bH_{\sA} \over 1 + \balpha_{\sX}^{\intercal}\bH_{\sA}} \right]_{\balpha_{\sA} = \bzero} &=& \bH_{\sA};\\
			\bS_{\epsilon} &=& \left[ {\partial \log p_{\widetilde{\epsilon}}(\widetilde{\epsilon}|\bX,A;\balpha_{\epsilon}) \over \partial \balpha_{\epsilon}} \right]_{\balpha_{\epsilon} = \bzero} &=& \left[ {\bH_{\epsilon} \over 1 + \balpha_{\epsilon}^{\intercal}\bH_{\epsilon}} \right]_{\balpha_{\epsilon} = \bzero} &=& \bH_{\epsilon}.
		\end{array} \]
	\end{itemize}
	Therefore, the DGP above corresponds to the nuisance score vectors $ \bH $, so that $ \Lambda \supseteq \widetilde{\Lambda} $. That is,
	\[ \Lambda = \Big\{ \bH \in \cH: \bbE(\bH\epsilon|\bX,A) = \bbE(\bH\epsilon|\bX)\Big\}. \]
\end{proof}

\begin{proof}[Proof of Lemma \ref{lem:score}]
	\begin{align*}
		\bbE\left( {X \partial \log p(X) \over \partial x} \right) &= \int {x \partial \rd p(x) \over 
			\rd x}p(x)\rd x \\
		&= \int x {\rd p(x) \over \rd x} \times {1 \over p(x)} \times p(x) \rd x\\
		&= \int x \rd p(x)\\
		&= \Big[ xp(x) \Big]_{x \to -\infty}^{x \to +\infty} - \int p(x)\rd x\\
		&= -1.
	\end{align*}
\end{proof}

\subsubsection{Proof of Lemmas \ref{lem:equiang} and \ref{lem:ang}}

\begin{proof}[Proof of Lemma \ref{lem:equiang}]
	Since the column space of $ \Omega $ is the orthogonal complement of the space spanned by $ \bone_{K} $, we have
	\[ \Omega(\Omega^{\intercal}\Omega)^{-1}\Omega^{\intercal} = \Ib_{K\times K} - {1 \over K}\bone_{K\times 1}^{\otimes 2}. \]
	In particular, for $ 1 \le k,k'\le K $, we have5
	\[ \langle  (\Omega^{\intercal}\Omega)^{-1/2}\bomega_{k},(\Omega^{\intercal}\Omega)^{-1/2}\bomega_{k'}  \rangle = \left(1-{1 \over K}\right)^{-1} [\Omega(\Omega^{\intercal}\Omega)^{-1}\Omega^{\intercal}]_{kk'} =  \bbone(k = k') - {1 \over K-1}\bbone(k \ne k'). \]
	Therefore, $ \{ \bomega_{k} \}_{k=1}^{K} $ are unit vectors and equiangular.
\end{proof}

\begin{proof}[Proof of Lemma \ref{lem:ang}]
	By Lemma \ref{lem:equiang}, we have
	\[ \Omega(\Omega^{\intercal}\Omega)^{-1}\Omega^{\intercal} = \Ib_{K\times K} - (1/K)\vec{\bone}_{K\times 1}^{\otimes 2}. \]
	Denote $ \vec{\bgamma}(\bx;\bbeta) := \big( \gamma(\bx,1;\bbeta),\gamma(\bx,2;\bbeta),\cdots,\gamma(\bx,K;\bbeta) \big)^{\intercal} $. Then
	\begin{align*}
		\sqrt{1 - 1/K}\Omega \vec{\bbf}(\bx;\bbeta) &= \Omega(\Omega^{\intercal}\Omega)^{-1}\Omega^{\intercal}\vec{\bgamma}(\bx;\bbeta)\\
		&= \vec{\bgamma}(\bx;\bbeta) - \underbrace{{1 \over K}\sum_{k=1}^{K}\gamma(\bx,k;\bbeta)}_{= 0}\vec{\bone}_{K}\\
		&= \vec{\bgamma}(\bx;\bbeta).
	\end{align*}
	That is,
	\[ \gamma(\bx,k;\bbeta) = \left( 1 - {1 \over K} \right)\langle \bomega_{k},\vec{\bbf}(\bx;\bbeta) \rangle; \quad 1 \le k \le K. \]
\end{proof}

\subsubsection{Proof of Lemma \ref{lem:tangent_proj}}

\begin{proof}[Proof of Lemma \ref{lem:tangent_proj}]
	Suppose $ \bH \in \Lambda $ and $ \sfH: \cX \to \bbR^{p \times (K-1)} $. Then
	\[ \bbE\left[ \bH\left({\sfH(\bX)\bomega_{A}\epsilon \over p_{\sA}(A|\bX)}\right)^{\intercal} \right] = \bbE\left[ \bbE\left( {\bH\bomega_{A}^{\intercal}\epsilon \over p_{\sA}(A|\bX)} \middle|\bX \right)\sfH(\bX)^{\intercal} \right] = \sfO_{p\times p}. \]
	That is,
	\[ \Lambda \perp \left \{{\sfH(\bX)\bomega_{A}\epsilon \over p_{\sA}(A|\bX)}\middle|\sfH:\cX\to \bbR^{p\times(K-1)} \right \}. \]
	Now suppose $ \bH \in \cH $. Define
	\[ \bH_{1} := \underbrace{\bbE\left\{ {\bH\bomega_{A}^{\intercal}\epsilon \over p_{\sA}(A|\bX)}\middle|\bX \right\}\sfV_{\epsilon}(\bX)^{-1}}_{\text{$ \bbR^{p\times(K-1)} $-valued function of $ \bX $}}{\bomega_{A}\epsilon \over p_{\sA}(A|\bX)}; \quad \bH_{2} := \bH - \bH_{1}. \]
	Then the following shows that $ \bH_{2} \in \Lambda $:
	\[ \bbE\left( {\bH_{2}\bomega_{A}^{\intercal}\epsilon \over p_{\sA}(A|\bX)}\middle|\bX \right) = \bbE\left( {\bH\bomega_{A}^{\intercal}\epsilon \over p_{\sA}(A|\bX)}\middle|\bX \right) - \bbE\left\{ {\bH\bomega_{A}^{\intercal}\epsilon \over p_{\sA}(A|\bX)}\middle|\bX \right\}\sfV_{\epsilon}(\bX)^{-1}\bbE\left( {\bomega_{A}^{\otimes 2}\epsilon^{2} \over p_{\sA}(A|\bX)^{2}}\middle|\bX \right) = \sfO_{p\times(K-1)}, \]
	where
	\[ \bbE\left( {\bomega_{A}^{\otimes 2}\epsilon^{2} \over p_{\sA}(A|\bX)^{2}}\middle|\bX \right) = \bbE\left( {\bomega_{A}^{\otimes 2}\sigma^{2}(\bX,A) \over p_{\sA}(A|\bX)^{2}}\middle|\bX \right) = \sum_{k=1}^{K}{\sigma^{2}(\bX,A)\bomega_{k}^{\otimes 2} \over p_{\sA}(k|\bX)} = \sfV_{\epsilon}(\bX). \]
	Therefore, 
	\[ \Lambda^{\perp} = \left \{{\sfH(\bX)\bomega_{A}\epsilon \over p_{\sA}(A|\bX)}\middle|\sfH:\cX\to \bbR^{p\times(K-1)} \right \}. \]
\end{proof}

\subsubsection{Proof of Proposition \ref{prop:eff_ang}}

\begin{proof}[Proof of Proposition \ref{prop:eff_ang}]
	The score vector of Model (\ref{eq:model}) is defined as \citep{tsiatis2007semiparametric}
	\[ \bS_{\bbeta} = {\partial\log p_{\epsilon}\Big( Y - \mu_{0}(\bX) - \gamma(\bX,A;\bbeta)\Big|\bX,A\Big) \over \partial\bbeta} = -\dot{\gamma}(\bX,A;\bbeta){\partial \log p_{\epsilon}(\epsilon|\bX,A) \over \partial\epsilon}. \]
	In particular,
	\[ \bbE\left\{ {\bS_{\bbeta}\bomega_{A}^{\intercal}\epsilon\over p_{\sA}(A|\bX)}\middle|\bX \right\} = \bbE\left\{ {\dot{\gamma}(\bX,A;\bbeta)\bomega_{A}^{\intercal}\over p_{\sA}(A|\bX)} \left[ - {\epsilon\partial\log p_{\epsilon}(\epsilon|\bX,A) \over \partial\epsilon} \right] \middle|\bX \right\} \stackrel{(\ref{eq:tangent})}{=} \bbE\left\{ {\dot{\gamma}(\bX,A;\bbeta)\bomega_{A}^{\intercal}\over p_{\sA}(A|\bX)} \middle|\bX \right\}. \]
	Then by Lemma \ref{lem:tangent_proj}, the efficient score \citep{tsiatis2007semiparametric} is 
	\[ \bS_{\eff} = \bbE(\bS_{\bbeta}|\Lambda^{\perp}) = \bbE\left\{ {\dot{\gamma}(\bX,A;\bbeta)\bomega_{A}^{\intercal}\over p_{\sA}(A|\bX)} \middle|\bX \right\}\sfV_{\epsilon}(\bX)^{-1}{\bomega_{A}\epsilon \over p_{\sA}(A|\bX)}. \]
	Consider the angle-based decision function (\ref{eq:ang_opt}). We have
	\[ \dot{\gamma}(\bX,A;\bbeta) = \left( 1 - {1 \over K} \right)\dot{\sfF}(\bX;\bbeta)^{\intercal}\bomega_{A}. \]
	Then
	\[ \bbE\left\{ {\dot{\gamma}(\bX,A;\bbeta)\bomega_{A}^{\intercal}\over p_{\sA}(A|\bX)} \middle|\bX \right\} = \left( 1 - {1 \over K} \right)\bbE\left\{ {\dot{\sfF}(\bX;\bbeta)^{\intercal}\bomega_{A}^{\otimes 2} \over p_{\sA}(A|\bX)}\middle|\bX \right\} = \dot{\sfF}(\bX;\bbeta)^{\intercal}\left( 1 - {1 \over K} \right)\sum_{k=1}^{K}\bomega_{k}^{\otimes 2} = \dot{\sfF}(\bX;\bbeta)^{\intercal}\Omega^{\intercal}\Omega. \]
	Therefore,
	\[ \bS_{\eff} = \dot{\sfF}(\bX;\bbeta)^{\intercal}\Omega^{\intercal}\Omega\sfV_{\epsilon}(\bX)^{-1}{\bomega_{A}\epsilon \over p_{\sA}(A|\bX)}. \]
\end{proof}

\subsubsection{Proof of Proposition \ref{prop:dr}}

\begin{proof}[Proof of Proposition \ref{prop:dr}]
	It follows from direct calculation that
	\[ \bbE[\bphi_{\eff}(\bbeta;\widecheck{\mu}_{0},\widecheck{p}_{\sA},\widecheck{\sigma}^{2})|\bX,A] = [\mu_{0}(\bX) - \widecheck{\mu}_{0}(\bX)]\sfH(\bX){p_{\sA}(A|\bX) \over \widecheck{p}_{\sA}(A|\bX)}{\bomega_{A} \over p_{\sA}(A|\bX)}. \]
	If $ \widecheck{\mu}_{0} = \mu_{0} $, then the above is $ \bzero $. If $ \widecheck{p}_{\sA} = p_{\sA} $, the above becomes 
	\[ \bbE[\bphi_{\eff}(\bbeta;\widecheck{\mu}_{0},\widecheck{p}_{\sA},\widecheck{\sigma}^{2})|\bX,A] = [\mu_{0}(\bX) - \widecheck{\mu}_{0}(\bX)]\sfH(\bX){\bomega_{A} \over p_{\sA}(A|\bX)}. \]
	Then we have
	\[ \bbE[\bphi_{\eff}(\bbeta;\widecheck{\mu}_{0},\widecheck{p}_{\sA},\widecheck{\sigma}^{2})|\bX] = [\mu_{0}(\bX) - \widecheck{\mu}_{0}(\bX)]\sfH(\bX)\bbE\left(  {\bomega_{A} \over p_{\sA}(A|\bX)} \middle|\bX \right) = \bzero. \]
\end{proof}

\subsubsection{Proof of Lemma \ref{lem:plug-in}}\label{sec:lem_plug-in}

\begin{proof}[Proof of Lemma \ref{lem:plug-in}]
	Denote $ \dot{F}(\bX) := \sup_{\widecheck{\bbeta} \in \sB}\|\dot{\sfF}(\bX;\widecheck{\bbeta})\|_{2} $. Define
	\[ \widehat{\sfV}_{\epsilon,n}(\bX) := \sum_{k=1}^{K}{\widehat{\sigma}_{n}^{2}(\bX,A)\bomega_{k}^{\otimes 2} \over p_{\sA}(A|\bX)}, \]
	and
	\begin{align*}
		\widehat{\bh}_{\eff,n}(\bX,A;\widecheck{\bbeta}) &:= \dot{\sfF}(\bX;\widecheck{\bbeta})^{\intercal}\Omega^{\intercal}\Omega\widehat{\sfV}_{\epsilon,n}(\bX)^{-1}{\bomega_{A} \over p_{\sA}(A|\bX)};\\
		\widecheck{\bh}_{\eff}(\bX,A;\widecheck{\bbeta}) &:= \dot{\sfF}(\bX;\widecheck{\bbeta})^{\intercal}\Omega^{\intercal}\Omega\widecheck{\sfV}_{\epsilon}(\bX)^{-1}{\bomega_{A} \over p_{\sA}(A|\bX)}.
	\end{align*}
	Then
	\begin{align}
		&\bphi_{\eff}(\widecheck{\bbeta};\widehat{\mu}_{0,n},p_{\sA},\widehat{\sigma}_{n}^{2}) - \bphi_{\eff}(\widecheck{\bbeta};\widecheck{\mu}_{0},p_{\sA},\widecheck{\sigma}^{2}) \nonumber \\
		=& \quad [\underbrace{Y - \widecheck{\mu}_{0}(\bX) - \gamma(\bX,A;\widecheck{\bbeta})}_{:= \widecheck{e}(\widecheck{\bbeta})}][\widehat{\bh}_{\eff,n}(\bX,A;\widecheck{\bbeta}) - \widecheck{\bh}_{\eff}(\bX,A;\widecheck{\bbeta})] \label{eq:plug-in-1} \\
		& - [\widehat{\mu}_{0,n}(\bX) - \widecheck{\mu}_{0}(\bX)]\widecheck{\bh}_{\eff,n}(\bX,A;\widecheck{\bbeta}) \label{eq:plug-in-2}\\
		& - [\widehat{\mu}_{0,n}(\bX) - \widecheck{\mu}_{0}(\bX)][\widehat{\bh}_{\eff,n}(\bX,A,\widecheck{\bbeta}) - \widecheck{\bh}_{\eff}(\bX,A;\widecheck{\bbeta})]. \label{eq:plug-in-3}
	\end{align}
	
	\begin{itemize}[leftmargin=*]
		\item We first relate $ \widehat{\bh}_{\eff,n} - \widecheck{\bh}_{\eff} $ to $ \widehat{\sigma}_{n}^{2} - \widecheck{\sigma}^{2} $. Note that 
		\begin{align*}
			\left\| \widehat{\bh}_{\eff,n}(\bX,A;\widecheck{\bbeta}) - \widecheck{\bh}_{\eff}(\bX,A;\widecheck{\bbeta}) \right\|_{2} &\le \left \| \dot{\sfF}(\bX;\widecheck{\bbeta}) \right \|_{2} \times \| \Omega \|_{2}^{2} \times \left \| \widehat{\sfV}_{\epsilon,n}(\bX)^{-1} - \widecheck{\sfV}_{\epsilon}(\bX)^{-1} \right \|_{2} \times {\|\bomega_{A}\|_{2} \over p_{\sA}(A|\bX)}\\
			&\le \left( 1 - {1 \over K} \right){\| \Omega \|_{2}^{2} \|\Omega\|_{\rm F} \over \ubar{p}_{\sA}} \times \dot{F}(\bX) \times \left \| \widehat{\sfV}_{\epsilon,n}(\bX)^{-1} - \widecheck{\sfV}_{\epsilon}(\bX)^{-1} \right \|_{2}.
		\end{align*}
		Here, $ \| \cdot \|_{\rm F} $ is the Frobenius norm. By $ \ubar{\sigma}^{2}\le \widehat{\sigma}_{n}^{2}(\bX,k),\widecheck{\sigma}^{2}(\bX,k) \le \widebar{\sigma}^{2} $ for $ 1 \le k \le K $, we further have
		\[ \widehat{\sfV}_{\epsilon,n}(\bX), ~ \widecheck{\sfV}_{\epsilon}(\bX) \ge \ubar{\sigma}^{2}\sum_{k=1}^{K}\bomega_{k}^{\otimes 2} = \underbrace{\ubar{\sigma}^{2}\left( 1 - {1 \over K} \right)^{-1}\Omega^{\intercal}\Omega}_{:= \ubar{\sfV}_{\epsilon}} > 0. \]
		Here, ``$ \sfA \ge \sfB $'' means that $ \sfA - \sfB $ is positive semi-definite, with strict inequality if $ \sfA - \sfB $ is positive definite. Then
		\begin{align}
			\left\|  \widehat{\sfV}_{\epsilon,n}(\bX)^{-1}-\widecheck{\sfV}_{\epsilon}(\bX)^{-1} \right\|_{2} &\le \| \ubar{\sfV}_{\epsilon}^{-2} \|_{2} \times \left\|  \widehat{\sfV}_{\epsilon,n}(\bX)-\widecheck{\sfV}_{\epsilon}(\bX) \right\|_{2} \nonumber \\
			&= {1 \over \lambda_{\min}(\ubar{\sfV}_{\epsilon})^{2}} \times \left\| \sum_{k=1}^{K}{[\widehat{\sigma}_{n}^{2}(\bX,k) - \widecheck{\sigma}^{2}(\bX,k)]\bomega_{k}^{\otimes 2} \over p_{\sA}(k|\bX)} \right\|_{2} \nonumber \\
			&\le {1 \over \ubar{p}_{\sA}\lambda_{\min}(\ubar{\sfV}_{\epsilon})^{2}} \Bigg\|\underbrace{\sum_{k=1}^{K}\bomega_{k}^{\otimes 2}}_{= (1-1/K)^{-1}\Omega^{\intercal}\Omega}\Bigg\|_{2} \times \| \widehat{\sigma}_{n}^{2} - \widecheck{\sigma}^{2} \|_{\infty}. \label{eq:var_bdd}
		\end{align}
		Therefore,
		\[ \left\| \widehat{\bh}_{\eff,n}(\bX,A;\widecheck{\bbeta}) - \widecheck{\bh}_{\eff}(\bX,A;\widecheck{\bbeta}) \right\|_{2} \le \text{constant} \times \dot{F}(\bX) \times \| \widehat{\sigma}_{n}^{2} - \widecheck{\sigma}^{2} \|_{\infty}. \]
		
		\item (Bound for Residual) First note that
		\begin{align*}
			\bbE_{n}[\widecheck{e}(\widecheck{\bbeta})^{2}] &= \bbE[\widecheck{e}(\widecheck{\bbeta})^{2}] + \smallcO_{\bbP}(1) &&(\text{by SLLN})\\
			&\lesssim 5\bbE\Big\{ \widecheck{\mu}_{0}(\bX)^{2} + \mu_{0}(\bX)^{2} + \gamma(\bX,A;\bbeta)^{2} + \gamma(\bX,A;\widecheck{\bbeta})^{2} + \epsilon^{2} \Big\} + \smallcO_{\bbP}(1),
		\end{align*}
		which is bounded by Assumption \ref{asm:int}.
		
		\item (Convergence of (\ref{eq:plug-in-1}))
		\begin{align*}
			[\bbE_{n}\|(\ref{eq:plug-in-1})\|_{2}]^{2} &\le \text{constant} \times \bbE_{n}\left\{ \left| \widecheck{e}(\widecheck{\bbeta}) \right| \dot{F}(\bX) \right\}^{2} \| \widehat{\sigma}_{n}^{2} - \widecheck{\sigma}^{2} \|_{\infty}^{2}\\
			&\lesssim \text{constant} \times \left\{ \bbE[\widecheck{e}(\widecheck{\bbeta})^{2}]\bbE[\dot{F}(\bX)^{2}] + \smallcO_{\bbP}(1) \right\} \times \| \widehat{\sigma}_{n}^{2} - \widecheck{\sigma}^{2} \|_{\infty}^{2} \\
			&= \smallcO_{\bbP}(n^{-1}).
		\end{align*}
		
		\item (Convergence of (\ref{eq:plug-in-2})) First note that
		\begin{align}
			\left \| \widehat{\bh}_{\eff,n}(\bX,A;\widecheck{\bbeta}) \right \|_{2} &\le \left \| \dot{\sfF}(\bX;\widecheck{\bbeta}) \right \|_{2} \times \| \Omega \|_{2}^{2} \times \left \| \widecheck{\sfV}_{\epsilon}(\bX)^{-1} \right \|_{2} \times {\| \bomega_{A} \|_{2} \over p_{\sA}(A|\bX)} \nonumber \\
			&\le \left( 1 - {1 \over K} \right)^{-1/2}{\| \Omega \|_{2}^{2}\| \Omega \|_{\rm F} \over \ubar{p}_{\sA} \lambda_{\min}(\ubar{\sfV}_{\epsilon})} \times \dot{F}(\bX).\label{eq:instrument_bdd}
		\end{align}
		Here, $ \lambda_{\min}(\cdot) $ is the smallest eigenvalue of a matrix. Then
		\begin{align*}
			[\bbE_{n}\| (\ref{eq:plug-in-2}) \|_{2}]^{2} &\le \text{constant} \times \bbE_{n}[\dot{F}(\bX)^{2}] \times \bbE_{n}[\widehat{\mu}_{0,n}(\bX) - \widecheck{\mu}_{0}(\bX)]^{2} \\
			&\lesssim \text{constant} \times \left\{  \bbE[\dot{F}(\bX)^{2}] + \smallcO_{\bbP}(1) \right\} \times \bbE_{n}[\widehat{\mu}_{0,n}(\bX) - \widecheck{\mu}_{0}(\bX)]^{2} \\
			&= \smallcO_{\bbP}(n^{-1}).
		\end{align*}
		
		\item (Convergence of (\ref{eq:plug-in-3}))
		\begin{align*}
			[\bbE_{n}\| (\ref{eq:plug-in-3}) \|_{2}]^{2} &\le \text{constant} \times \bbE_{n}\left\{ \left| \widehat{\mu}_{0,n}(\bX) - \widecheck{\mu}_{0}(\bX) \right| \times \dot{F}(\bX) \right\}^{2} \times \| \widehat{\sigma}_{n}^{2} - \widecheck{\sigma}^{2} \|_{\infty}^{2} \\
			&\le \text{constant} \times \left\{ \bbE[\dot{F}(\bX)^{2}] + \smallcO_{\bbP}(1) \right\} \times \bbE_{n}[\widehat{\mu}_{0,n}(\bX) - \widecheck{\mu}_{0}(\bX)]^{2} \times \| \widehat{\sigma}_{n}^{2} - \widecheck{\sigma}^{2} \|_{\infty}^{2} \\
			&= \smallcO_{\bbP}(n^{-2}).
		\end{align*}
	\end{itemize}
	Therefore,
	\begin{align*}
		& \sup_{\widecheck{\bbeta} \in \sB}\bbE_{n}\| \bphi_{\eff}(\widecheck{\bbeta};\widehat{\mu}_{0,n},p_{\sA},\widehat{\sigma}_{n}^{2}) - \bphi_{\eff}(\widecheck{\bbeta};\widecheck{\mu}_{0},p_{\sA},\widecheck{\sigma}^{2}) \|_{2}\\
		\le & \bbE_{n}\| (\ref{eq:plug-in-1}) \|_{2} + \bbE_{n}\| (\ref{eq:plug-in-2}) \|_{2} + \bbE_{n}\| (\ref{eq:plug-in-3}) \|_{2}\\
		\lesssim & \smallcO_{\bbP}(n^{-1/2}) + \smallcO_{\bbP}(n^{-1/2}) + \smallcO_{\bbP}(n^{-1})\\
		=& \smallcO_{\bbP}(n^{-1/2}).
	\end{align*}
\end{proof}

\subsubsection{Proof of Theorem \ref{thm:eff}}

\begin{proof}[Proof of Theorem \ref{thm:eff}]
	We follow \citet[Lemmas 5.1-5.3]{newey1994asymptotic} to establish the asymptotic linear representation. 
	
	\begin{enumerate}[label=\textbf{Step \Roman*:},leftmargin=0cm]
		\item (Asymptotic Linear Representation) Our Lemma \ref{lem:plug-in} can imply the asymptotic linear representation of the plug-in estimating function as in \citet[Lemma 5.1]{newey1994asymptotic}:
		\[ \sqrt{n}\bbE_{n}[\bphi_{\eff}(\bbeta;\widehat{\mu}_{0,n},p_{\sA},\widehat{\sigma}_{n}^{2})] = \sqrt{n}\bbE_{n}[\bphi_{\eff}(\bbeta;\widecheck{\mu}_{0},p_{\sA},\widecheck{\sigma}^{2})] + \smallcO_{\bbP}(n^{-1/2}). \]
		
		\item (Uniform Convergence and Consistency) We aim to establish the convergence of $ \bbE_{n}[\bphi_{\eff}(\widecheck{\bbeta};\widehat{\mu}_{0,n},p_{\sA},\widehat{\sigma}_{n}^{2})] $ and $ \bbE_{n}[(\partial/\partial\bbeta^{\intercal})\bphi_{\eff}(\widecheck{\bbeta};\widehat{\mu}_{0,n},p_{\sA},\widehat{\sigma}_{n}^{2})] $ uniform for $ \widecheck{\bbeta} \in \sB $, and the consistency of $ \widehat{\bbeta}_{\eff,n} $.
		
		Recall from Lemma \ref{lem:plug-in} that:
		\[ \sup_{\widecheck{\bbeta} \in \sB} \bbE_{n}\left\| \bphi_{\eff}(\widecheck{\bbeta};\widehat{\mu}_{0,n},p_{\sA},\widehat{\sigma}_{n}^{2}) - \bphi_{\eff}(\widecheck{\bbeta};\widecheck{\mu}_{0},p_{\sA},\widecheck{\sigma}^{2}) \right\|_{2} = \smallcO_{\bbP}(n^{-1/2}). \]
		The same conclusion can be drawn for $ (\partial/\partial\bbeta^{\intercal})\bphi_{\eff} $ following the same argument.
		
		\begin{lem}\label{lem:plug-in_grad}
			Consider Model (\ref{eq:model}) and the estimating function (\ref{eq:estim_fun}). Under Assumptions \ref{asm:prop}-\ref{asm:plug-in} and \ref{enum:int}, we have
			\[ \sup_{\widecheck{\bbeta} \in \sB} \bbE_{n}\left\| {\partial \bphi_{\eff}(\widecheck{\bbeta};\widehat{\mu}_{0,n},p_{\sA},\widehat{\sigma}_{n}^{2})\over \partial \bbeta^{\intercal}} - {\partial \bphi_{\eff}(\widecheck{\bbeta};\widecheck{\mu}_{0},p_{\sA},\widecheck{\sigma}^{2})\over \partial \bbeta^{\intercal}} \right\|_{2} = \smallcO_{\bbP}(n^{-1/2}). \]
		\end{lem}
		
		Next, we apply Glivenko-Cantelli Theorem to replace $ \bbE_{n} $ by $ \bbE $. We establish the conditions in Lemma \ref{lem:estim_fun}.
		
		\begin{lem}\label{lem:estim_fun}
			Consider Model (\ref{eq:model}) and the estimating function (\ref{eq:estim_fun}). Under Assumptions \ref{asm:prop}-\ref{asm:plug-in} and \ref{enum:int}, we have:
			\begin{enumerate}[label=(\Roman*)]
				\item There exists $ L: \cX \times\cA\times\bbR \to \bbR_{+} $ such that $ \bbE L(\bX,A,\epsilon) < \infty $, and for any $ \widecheck{\bbeta}_{1}, \widecheck{\bbeta}_{2} \in \sB $, we have
				\[ \left\|  \bphi_{\eff}(\widecheck{\bbeta}_{1};\widecheck{\mu}_{0},p_{\sA},\widecheck{\sigma}^{2}) - \bphi_{\eff}(\widecheck{\bbeta}_{2};\widecheck{\mu}_{0},p_{\sA},\widecheck{\sigma}^{2}) \right \|_{2} \le L(\bX,A,\epsilon) \left\|\widecheck{\bbeta}_{1} - \widecheck{\bbeta}_{2}\right \|_{2}. \]
				
				\item \label{enum:estim_fun_grad} There exists $ \widetilde{L}: \cX \times\cA\times\bbR \to \bbR_{+} $ such that $ \bbE \widetilde{L}(\bX,A,\epsilon) < \infty $, and for any $ \widecheck{\bbeta}_{1}, \widecheck{\bbeta}_{2} \in \sB $, we have
				\[ \left\|  {\partial\bphi_{\eff}(\widecheck{\bbeta}_{1};\widecheck{\mu}_{0},p_{\sA},\widecheck{\sigma}^{2}) \over \partial \bbeta^{\intercal}} - {\partial \bphi_{\eff}(\widecheck{\bbeta}_{2};\widecheck{\mu}_{0},p_{\sA},\widecheck{\sigma}^{2})\over \partial \bbeta^{\intercal}} \right \|_{2} \le \widetilde{L}(\bX,A,\epsilon) \left\|\widecheck{\bbeta}_{1} - \widecheck{\bbeta}_{2}\right \|_{2}. \]
				
				\item $ \bbE\sup_{\widecheck{\bbeta} \in \sB}\|\bphi_{\eff}(\widecheck{\bbeta};\widecheck{\mu}_{0},p_{\sA},\widecheck{\sigma}^{2})\|_{2} < +\infty $, $ \bbE\sup_{\widecheck{\bbeta} \in \sB}\|(\partial/\partial\bbeta^{\intercal})\bphi_{\eff}(\widecheck{\bbeta};\widecheck{\mu}_{0},p_{\sA},\widecheck{\sigma}^{2})\|_{2} < +\infty $.
			\end{enumerate}
		\end{lem}
		
		By $ \sB $ is compact and Lemma \ref{lem:estim_fun}, we can conclude that 
		\[ \left \{ \bphi_{\eff}(\widecheck{\bbeta};\widecheck{\mu}_{0},p_{\sA},\widecheck{\sigma}^{2}): \widecheck{\bbeta} \in \sB  \right \} \quad \text{and} \quad \left \{ (\partial/\partial\bbeta^{\intercal})\bphi_{\eff}(\widecheck{\bbeta};\widecheck{\mu}_{0},p_{\sA},\widecheck{\sigma}^{2}): \widecheck{\bbeta} \in \sB  \right \} \]
		are both $ \bbP $-Glivenko-Cantelli. Then, by Glivenko-Cantelli Theorem, we have
		\begin{align}
			\sup_{\widehat{\bbeta} \in \sB}(\bbP_{n} - \bbP)\left\| \bphi_{\eff}(\widecheck{\bbeta};\widecheck{\mu}_{0},p_{\sA},\widecheck{\sigma}^{2})\right\|_{2}, ~ \sup_{\widehat{\bbeta} \in \sB}(\bbP_{n} - \bbP)\left\| {\partial \bphi_{\eff}(\widecheck{\bbeta};\widecheck{\mu}_{0},p_{\sA},\widecheck{\sigma}^{2}) \over \partial \bbeta^{\intercal}} \right\|_{2} \stackrel{\bbP}{\longrightarrow} 0.\label{eq:gc}
		\end{align}
		\begin{rmk}
			From the proof of Lemma \ref{lem:estim_fun}, we have 
			\begin{align*}
				&\bbE\left[ -{\partial \bphi_{\eff}(\widecheck{\bbeta};\widecheck{\mu}_{0},p_{\sA},\widecheck{\sigma}^{2}) \over \partial \bbeta^{\intercal}} \right] \\
				=& \bbE(\ref{eq:eff_grad-1}) + \bbE(\ref{eq:eff_grad-2})\\
				=& \bbE\bigg\{ \dot{\sfF}(\bX;\widecheck{\bbeta})^{\intercal}\Omega^{\intercal}\Omega\widecheck{\sfV}_{\epsilon}(\bX)^{-1}\underbrace{\bbE\left[ \left( 1 - {1 \over K} \right){\bomega_{A}^{\otimes 2} \over p_{\sA}(A|\bX)}\middle|\bX \right]}_{= \Omega^{\intercal}\Omega }\dot{\sfF}(\bX;\widecheck{\bbeta}) \bigg\} \\
				& -\bbE\bigg\{  \ddot{\sfF}(\bX;\widecheck{\bbeta})^{\intercal}\Omega^{\intercal}\Omega\widecheck{\sfV}_{\epsilon}(\bX)^{-1}\underbrace{\bbE\left(  {\bomega_{A} \over p_{\sA}(A|\bX)}\middle|\bX \right)}_{= \bzero}\widecheck{e}(\widecheck{\bbeta}) \bigg\}\\
				=& \widecheck{\cI}(\widecheck{\bbeta}).
			\end{align*}
			By Lemma \ref{lem:estim_fun} \ref{enum:estim_fun_grad} and (\ref{eq:gc}), we further have that $ \widecheck{\bbeta} \to \widecheck{\cI}(\widecheck{\bbeta}) $ is continuous.
		\end{rmk}
	
		Combining Lemmas \ref{lem:plug-in}, \ref{lem:plug-in_grad} and (\ref{eq:gc}), we have
		\begin{align}
			&\sup_{\widehat{\bbeta} \in \sB}\left\| \bbE_{n}[\bphi_{\eff}(\widecheck{\bbeta};\widehat{\mu}_{0,n},p_{\sA},\widehat{\sigma}_{n}^{2})] - \bbE[\bphi_{\eff}(\widecheck{\bbeta};\widecheck{\mu}_{0},p_{\sA},\widecheck{\sigma}^{2})] \right\|_{2}\stackrel{\bbP}{\longrightarrow} 0; \label{eq:cvg}	\\
			&\sup_{\widecheck{\bbeta} \in \sB} \left\| \bbE_{n}\left[ {\partial \bphi_{\eff}(\widecheck{\bbeta};\widehat{\mu}_{0,n},p_{\sA},\widehat{\sigma}_{n}^{2})\over \partial \bbeta^{\intercal}} \right] - \bbE\left[ {\partial \bphi_{\eff}(\widecheck{\bbeta};\widecheck{\mu}_{0},p_{\sA},\widecheck{\sigma}^{2})\over \partial \bbeta^{\intercal}} \right] \right\|_{2} \stackrel{\bbP}{\longrightarrow} 0. \label{eq:cvg_grad}
		\end{align}
		The consistency of $ \widehat{\bbeta}_{\eff,n} $ follows from that $ \sB $ is compact,
		\[ \widehat{\bbeta}_{\eff,n} \in \argmax_{\widecheck{\bbeta} \in \sB}\left\| \bbE_{n}[\bphi_{\eff}(\widecheck{\bbeta};\widehat{\mu}_{0,n},p_{\sA},\widehat{\sigma}_{n}^{2})] \right\|_{2}^{2}; \quad \bbeta \in \argmax_{\widecheck{\bbeta} \in \sB}\left\| \bbE[\bphi_{\eff}(\widecheck{\bbeta};\widecheck{\mu}_{0},p_{\sA},\widecheck{\sigma}^{2})] \right\|_{2}^{2}, \]
		and the uniform convergence in probability in (\ref{eq:cvg}).
		
		\item By Mean Value Theorem, there exists some $ \alpha_{n} \in [0,1] $ and $ \widetilde{\bbeta}_{n} = (1-\alpha_{n})\widehat{\bbeta}_{\eff,n} + \alpha_{n}\bbeta $, such that
		\begin{align*}
			\bzero &= \bbE_{n}[\bphi_{\eff}(\widehat{\bbeta}_{\eff,n};\widehat{\mu}_{0,n},p_{\sA},\widehat{\sigma}_{n}^{2})]\\
			&= \bbE_{n}[\bphi_{\eff}(\bbeta;\widehat{\mu}_{0,n},p_{\sA},\widehat{\sigma}_{n}^{2})] + \bbE_{n}\left[ {\partial \bphi_{\eff}(\widetilde{\bbeta}_{n};\widehat{\mu}_{0,n},p_{\sA},\widehat{\sigma}_{n}^{2}) \over \partial \bbeta^{\intercal}} \right](\widehat{\bbeta}_{\eff,n} - \bbeta).
		\end{align*}
		That is,
		\begin{align*}
			\widehat{\bbeta}_{\eff,n} - \bbeta &= \left\{ \bbE_{n}\left[ -{\partial \bphi_{\eff}(\widetilde{\bbeta}_{n};\widehat{\mu}_{0,n},p_{\sA},\widehat{\sigma}_{n}^{2}) \over \partial \bbeta^{\intercal}} \right] \right\}^{-1}\bbE_{n}[\bphi_{\eff}(\bbeta;\widehat{\mu}_{0,n},p_{\sA},\widehat{\sigma}_{n}^{2})]\\
			&= \left\{ \widecheck{\cI}(\bbeta) + \smallcO_{\bbP}(1) \right\}^{-1}\times\left\{ \bbE_{n}[\bphi_{\eff}(\bbeta;\widecheck{\mu}_{0},p_{\sA},\widecheck{\sigma}^{2})] + \smallcO_{\bbP}(n^{-1/2}) \right\}\\
			&= \widecheck{\cI}(\bbeta)^{-1}\bbE_{n}[\bphi_{\eff}(\bbeta;\widecheck{\mu}_{0},p_{\sA},\widecheck{\sigma}^{2})] + \smallcO_{\bbP}(n^{-1/2}).
		\end{align*}
		Here, the second equality follows from that
		\begin{align*}
			\bbE_{n}\left[ -{\partial \bphi_{\eff}(\widetilde{\bbeta}_{n};\widehat{\mu}_{0,n},p_{\sA},\widehat{\sigma}_{n}^{2}) \over \partial \bbeta^{\intercal}}\right] &= \widecheck{\cI}(\widetilde{\bbeta}_{n}) + \smallcO_{\bbP}(1) &&(\text{by (\ref{eq:cvg_grad})})\\
			&= \widecheck{\cI}(\bbeta) + \smallcO_{\bbP}(1). &&(\text{by $ \widehat{\bbeta}_{\eff,n} \stackrel{\bbP}{\to} \bbeta $ and the continuity of  $ \widecheck{\cI} $})
		\end{align*}
		\item (Semiparametric Efficiency) If $ (\widecheck{\mu}_{0},\widecheck{\sigma}^{2}) = (\mu_{0},\sigma^{2}) $, then $ \bphi_{\eff}(\bbeta;\widecheck{\mu}_{0},p_{\sA},\widecheck{\sigma}^{2}) = \bS_{\eff}(\bbeta) $. Moreover,
		\begin{align*}
			\bbE[\bS_{\eff}(\bbeta)^{\otimes 2}] &= \bbE\left[ \dot{\sfF}(\bX;\bbeta)^{\intercal}\Omega^{\intercal}\Omega\sfV_{\epsilon}(\bX)^{-1}{\bomega_{A}^{\otimes 2}\epsilon^{2} \over p_{\sA}(A|\bX)^{2}} \sfV_{\epsilon}(\bX)^{-1}\Omega^{\intercal}\Omega\dot{\sfF}(\bX;\bbeta) \right]\\
			&=\bbE\Bigg[ \dot{\sfF}(\bX;\bbeta)^{\intercal}\Omega^{\intercal}\Omega\sfV_{\epsilon}(\bX)^{-1}\underbrace{\bbE\left(  {\bomega_{A}^{\otimes 2}\bbE(\epsilon^{2}|\bX,A) \over p_{\sA}(A|\bX)^{2}}\middle|\bX \right)}_{= \sfV_{\epsilon}(\bX)} \sfV_{\epsilon}(\bX)^{-1}\Omega^{\intercal}\Omega\dot{\sfF}(\bX;\bbeta) \Bigg]\\
			&= \cI(\bbeta).
		\end{align*}
		That is, $ \cI(\bbeta) $ defined in Theorem \ref{thm:eff} is the semiparametric Fisher information matrix. We further have
		\[ \sqrt{n}(\widehat{\bbeta}_{\eff,n}-\bbeta) = \sqrt{n}\cI(\bbeta)^{-1}\bbE_{n}[\bS_{\eff}(\bbeta)] + \smallcO_{\bbP}(1) \stackrel{\cD}{\longrightarrow} \cN_{p}\big(\bzero, \cI(\bbeta)^{-1} \big). \]
		Therefore, $ \widehat{\bbeta}_{\eff,n} $ is semiparametric efficient.
	\end{enumerate}
\end{proof}

\begin{proof}[Proof of Lemma \ref{lem:estim_fun}]\label{sec:lem_estim_fun}
	We follow the notations in Section \ref{sec:lem_plug-in}. Denote $ \ddot{F}(\bX) := \sup_{\widecheck{\bbeta} \in \sB}\|\ddot{\sfF}(\bX;\widecheck{\bbeta})\|_{2} $.
	\begin{enumerate}[label=(\Roman*),leftmargin=0cm]
		\item Fix $ \widecheck{\bbeta}_{1},\widecheck{\bbeta}_{2} \in \sB $. Then
		\begin{align}
			&\bphi_{\eff}(\widecheck{\bbeta}_{1};\widecheck{\mu}_{0},p_{\sA},\widecheck{\sigma}^{2}) - \bphi_{\eff}(\widecheck{\bbeta}_{2};\widecheck{\mu}_{0},p_{\sA},\widecheck{\sigma}^{2}) \nonumber \\
			=& -\left( 1 - {1 \over K} \right)\bomega_{A}^{\intercal}[\vec{\bbf}(\bX;\widecheck{\bbeta}_{1}) - \vec{\bbf}(\bX;\widecheck{\bbeta}_{2})]\bh_{\eff}(\bX,A;\widecheck{\bbeta}_{1}) \label{eq:eff-1} \\
			& + \widecheck{e}(\widecheck{\bbeta}_{2})[\widecheck{\bh}_{\eff}(\bX,A;\widecheck{\bbeta}_{1}) - \widecheck{\bh}_{\eff}(\bX,A;\widecheck{\bbeta}_{2})]. \label{eq:eff-2}
		\end{align}
		\begin{itemize}
			\item (Lipschitz Bound on (\ref{eq:eff-1}))
			\begin{itemize}
				\item (Lipschitz Bound on $ \vec{\bbf} $)
				\begin{align*}
					\left \| \vec{\bbf}(\bX;\widecheck{\bbeta}_{1}) - \vec{\bbf}(\bX;\widecheck{\bbeta}_{2}) \right \|_{2} &= \left\| \dot{\sfF}(\bX;\widetilde{\bbeta})[\widecheck{\bbeta}_{1} - \widecheck{\bbeta}_{2}]\right\|_{2} &&\left (\text{for some $ \widetilde{\bbeta} = (1-\alpha)\widecheck{\bbeta}_{1} + \alpha\widecheck{\bbeta}_{2} $}\right ) \\
					&\le \left \| \dot{\sfF}(\bX;\widetilde{\bbeta}) \right \|_{2} \times \left \|\widecheck{\bbeta}_{1} - \widecheck{\bbeta}_{2}\right \|_{2}\\
					&\le \dot{F}(\bX) \times\left \|\widecheck{\bbeta}_{1} - \widecheck{\bbeta}_{2}\right \|_{2}.
				\end{align*}
				
				\item Then we have
				\begin{align*}
					\| (\ref{eq:eff-1}) \|_{2} &= \underbrace{\left( 1 - {1 \over K} \right)\| \bomega_{A} \|_{2}}_{\le (1-1/K)^{1/2}\| \Omega \|_{\rm F}}\left \| \vec{\bbf}(\bX;\widecheck{\bbeta}_{1}) - \vec{\bbf}(\bX;\widecheck{\bbeta}_{2}) \right \|_{2}\underbrace{\left \| \widecheck{\bh}_{\eff}(\bX,A;\widecheck{\bbeta}_{1}) \right \|_{2}}_{\le (\ref{eq:instrument_bdd})}\\
					&\le \text{constant} \times \dot{F}(\bX)^{2} \times \left \| \widecheck{\bbeta}_{1} - \widecheck{\bbeta}_{2} \right \|_{2}.
				\end{align*}
			\end{itemize}
			
			\item (Lipschitz Bound on (\ref{eq:eff-2}))
			\begin{itemize}
				\item  Note that
				\[ \widecheck{\bh}_{\eff}(\bX,A;\widecheck{\bbeta}_{1}) - \widecheck{\bh}_{\eff}(\bX,A;\widecheck{\bbeta}_{2}) = [\dot{\sfF}(\bX;\widecheck{\bbeta}_{1}) - \dot{\sfF}(\bX;\widecheck{\bbeta}_{2})]^{\intercal}\Omega^{\intercal}\Omega\widecheck{\sfV}_{\epsilon}(\bX)^{-1}{\bomega_{A} \over p_{\sA}(A|\bX)}. \]
				
				\item (Lipschitz Bound on $ \dot{\sfF} $)
				\[ \left \|\dot{\sfF}(\bX;\widecheck{\bbeta}_{1}) - \dot{\sfF}(\bX;\widecheck{\bbeta}_{2})\right \|_{2} \le \ddot{F}(\bX) \times \left \| \widecheck{\bbeta}_{1} - \widecheck{\bbeta}_{2} \right \|_{2}. \]
				
				\item Then we have
				\begin{align*}
					\left \| \widecheck{\bh}_{\eff}(\bX,A;\widecheck{\bbeta}_{1}) - \widecheck{\bh}_{\eff}(\bX,A;\widecheck{\bbeta}_{2}) \right \|_{2} &= \left \|\dot{\sfF}(\bX;\widecheck{\bbeta}_{1}) - \dot{\sfF}(\bX;\widecheck{\bbeta}_{2})\right \|_{2}\underbrace{\| \Omega \|_{2}^{2}\left \| \widecheck{\sfV}_{\epsilon}(\bX)^{-1} \right \|_{2}{\| \bomega_{A} \|_{2} \over p_{\sA}(A|\bX)}}_{\le (1-1/K)^{-1/2}\| \Omega \|_{2}^{2}\| \Omega \|_{\rm F}/[\ubar{p}_{\sA}\lambda_{\min}(\ubar{\sfV}_{\epsilon})]} \\
					&\le \text{constant} \times \ddot{F}(\bX) \times \left \| \widecheck{\bbeta}_{1} - \widecheck{\bbeta}_{2} \right \|_{2}.
				\end{align*}
			
				\item The residual $ \widecheck{e}(\widecheck{\bbeta}_{2}) $ in (\ref{eq:eff-2}) can be bounded by
				\begin{align}
					\sup_{\widecheck{\bbeta} \in \sB}\left| \widecheck{e}(\widecheck{\bbeta})  \right| &\le | \widecheck{\mu}_{0}(\bX) - \mu_{0}(\bX)| + \left( 1 - {1 \over K} \right)\sup_{\widecheck{\bbeta} \in \sB}\left|  \bomega_{A}^{\intercal}[\vec{\bbf}(\bX;\widecheck{\bbeta}) - \vec{\bbf}(\bX;\bbeta)] \right| + | \epsilon | \nonumber \\
					&\le |\widecheck{\mu}_{0}(\bX)| + |\mu_{0}(\bX)| + \left( 1 - {1 \over K} \right)^{1/2}\| \Omega \|_{\rm F} \sup_{\widecheck{\bbeta} \in \sB}\left\| \vec{\bbf}(\bX;\widecheck{\bbeta}) - \vec{\bbf}(\bX;\bbeta) \right\|_{2} + | \epsilon| \nonumber\\
					&\le |\widecheck{\mu}_{0}(\bX)| + |\mu_{0}(\bX)| + \text{constant} \times \dot{F}(\bX) \times \underbrace{\sup_{\widecheck{\bbeta} \in \sB}\left \| \widecheck{\bbeta} - \bbeta \right \|_{2}}_{\le \mathsf{diam}(\sB)} + |\epsilon| \nonumber\\
					&\le |\widecheck{\mu}_{0}(\bX)| + |\mu_{0}(\bX)| + \text{constant} \times \dot{F}(\bX) + |\epsilon|. \label{eq:resid_bdd}
				\end{align}
			
				\item Then we have
				\[ \| (\ref{eq:eff-2}) \|_{2} \le \text{constant} \times \left[ |\widecheck{\mu}_{0}(\bX)| + |\mu_{0}(\bX)| + \dot{F}(\bX) + |\epsilon| \right] \times \ddot{F}(\bX) \times \left \| \widecheck{\bbeta}_{1} - \widecheck{\bbeta}_{2} \right \|_{2} \]
			\end{itemize}
		\end{itemize}
		Combining the Lipschitz bounds on (\ref{eq:eff-1}) and (\ref{eq:eff-2}), we have
		\begin{align*}
			&\left \| \bphi_{\eff}(\widecheck{\bbeta}_{1};\widecheck{\mu}_{0},p_{\sA},\widecheck{\sigma}^{2}) - \bphi_{\eff}(\widecheck{\bbeta}_{2};\widecheck{\mu}_{0},p_{\sA},\widecheck{\sigma}^{2}) \right \|_{2} \\
			\le& \underbrace{\text{constant} \times \left\{  \dot{F}(\bX)^{2} + \left[ |\widecheck{\mu}_{0}(\bX)| + |\mu_{0}(\bX)| + \dot{F}(\bX) + |\epsilon| \right] \times \ddot{F}(\bX) \right\}}_{:= L(\bX,A,\epsilon)} \times \left \| \widecheck{\bbeta}_{1} - \widecheck{\bbeta}_{2} \right \|_{2}.
		\end{align*}
		In particular,
		\begin{align*}
			&\bbE L(\bX,A,\epsilon) \\
			\le & \text{constant} \times \left [\bbE[\dot{F}(\bX)^{2}] +\left(  \left\{ \bbE[\widecheck{\mu}_{0}(\bX)^{2}] \right\}^{1/2} + \left\{ \bbE[\mu_{0}(\bX)^{2}] \right\}^{1/2} + \left\{ \bbE[\dot{F}(\bX)^{2}] \right\}^{1/2} + \left\{ \bbE \epsilon^{2} \right\}^{1/2} \right) \left\{ \bbE[\ddot{F}(\bX)^{2}] \right\}^{1/2}\right ],
		\end{align*}
		which is finite by Assumptions \ref{asm:int} and \ref{enum:int}.
		
		\item Note that
		\begin{align}
			-{\partial \bphi_{\eff}(\widecheck{\bbeta};\widecheck{\mu}_{0},p_{\sA},\widecheck{\sigma}^{2}) \over \partial \bbeta^{\intercal}} &= \dot{\sfF}(\bX;\widecheck{\bbeta})^{\intercal}\Omega^{\intercal}\Omega\widecheck{\sfV}_{\epsilon}(\bX)^{-1}\left[ \left( 1 - {1 \over K} \right){\bomega_{A}^{\otimes 2} \over p_{\sA}(A|\bX)} \right]\dot{\sfF}(\bX;\widecheck{\bbeta}) \label{eq:eff_grad-1}\\
			&\quad -\ddot{\sfF}(\bX;\widecheck{\bbeta})^{\intercal}\Omega^{\intercal}\Omega\widecheck{\sfV}_{\epsilon}(\bX)^{-1}{\bomega_{A}\widecheck{e}(\widecheck{\bbeta}) \over p_{\sA}(A|\bX)}.\label{eq:eff_grad-2}
		\end{align}
		Fix $ \widecheck{\bbeta}_{1},\widecheck{\bbeta}_{2} \in \sB $. Denote $ (\ref{eq:eff_grad-1})(\widecheck{\bbeta}_{1}) $ and $ (\ref{eq:eff_grad-1})(\widecheck{\bbeta}_{2}) $ as (\ref{eq:eff_grad-1}) with $ \widecheck{\bbeta} $ replaced by $ \widecheck{\bbeta}_{1} $ and $ \widecheck{\bbeta}_{2} $ respectively.
		\begin{align*}
			(\ref{eq:eff_grad-1})(\widecheck{\bbeta}_{1}) - (\ref{eq:eff_grad-1})(\widecheck{\bbeta}_{2}) =& \dot{\sfF}(\bX;\widecheck{\bbeta}_{1})^{\intercal}\Omega^{\intercal}\Omega\widecheck{\sfV}_{\epsilon}(\bX)^{-1}\left[ \left( 1 - {1 \over K} \right){\bomega_{A}^{\otimes 2} \over p_{\sA}(A|\bX)} \right][\dot{\sfF}(\bX;\widecheck{\bbeta}_{1}) - \dot{\sfF}(\bX;\widecheck{\bbeta}_{2})] + \\
			& [\dot{\sfF}(\bX;\widecheck{\bbeta}_{1}) - \dot{\sfF}(\bX;\widecheck{\bbeta}_{2})]^{\intercal}\Omega^{\intercal}\Omega\widecheck{\sfV}_{\epsilon}(\bX)^{-1}\left[ \left( 1 - {1 \over K} \right){\bomega_{A}^{\otimes 2} \over p_{\sA}(A|\bX)} \right]\dot{\sfF}(\bX;\widecheck{\bbeta}_{2}).
		\end{align*}
		Then
		\begin{align*}
			\left \| (\ref{eq:eff_grad-1})(\widecheck{\bbeta}_{1}) - (\ref{eq:eff_grad-1})(\widecheck{\bbeta}_{2}) \right \|_{2} &\le {2\|\Omega \|_{2}^{2}\| \Omega \|_{\rm F}^{2} \over \ubar{p}_{\sA}\lambda_{\min}(\ubar{\sfV}_{\epsilon})}\times \dot{F}(\bX) \times \left\| \dot{\sfF}(\bX;\widecheck{\bbeta}_{1}) - \dot{\sfF}(\bX;\widecheck{\bbeta}_{2}) \right\|_{2}\\
			&\le \text{constant} \times \dot{F}(\bX)\ddot{F}(\bX) \times \left\|\widecheck{\bbeta}_{1} - \widecheck{\bbeta}_{2}\right\|_{2}.
		\end{align*}
		Denote $ (\ref{eq:eff_grad-2})(\widecheck{\bbeta}_{1}) $ and $ (\ref{eq:eff_grad-2})(\widecheck{\bbeta}_{2}) $ as (\ref{eq:eff_grad-2}) with $ \widecheck{\bbeta} $ replaced by $ \widecheck{\bbeta}_{1} $ and $ \widecheck{\bbeta}_{2} $ respectively.
		\begin{align}
			(\ref{eq:eff_grad-2})(\widecheck{\bbeta}_{1}) - (\ref{eq:eff_grad-2})(\widecheck{\bbeta}_{2}) =&
			-[\dot{\sfF}(\bX;\widecheck{\bbeta}_{1})-\dot{\sfF}(\bX;\widecheck{\bbeta}_{2})]^{\intercal}\Omega^{\intercal}\Omega\widecheck{\sfV}_{\epsilon}(\bX)^{-1}{\bomega_{A}\widecheck{e}(\widecheck{\bbeta}_{1}) \over p_{\sA}(A|\bX)} \label{eq:eff_grad-2-1} \\
			& - \ddot{\sfF}(\bX;\widecheck{\bbeta}_{2})^{\intercal}\Omega^{\intercal}\Omega\widecheck{\sfV}_{\epsilon}(\bX)^{-1}{\bomega_{A} \over p_{\sA}(A|\bX)}[\widecheck{e}(\widecheck{\bbeta}_{1}) - \widecheck{e}(\widecheck{\bbeta}_{2})]. \label{eq:eff_grad-2-2}
		\end{align}
		Here,
		\begin{align*}
			\left| \widecheck{e}(\widecheck{\bbeta}_{1}) - \widecheck{e}(\widecheck{\bbeta}_{2}) \right| &= \left( 1 - {1 \over K} \right)\left| \bomega_{A}^{\intercal}[\vec{\bbf}(\bX;\widecheck{\bbeta}_{1}) - \vec{\bbf}(\bX;\widecheck{\bbeta}_{2})] \right|\\
			&\le \left( 1 - {1 \over K} \right)^{1/2}\| \Omega \|_{\rm F} \times \dot{F}(\bX) \times \left\| \widecheck{\bbeta}_{1} - \widecheck{\bbeta}_{2} \right\|_{2}.
		\end{align*}
		Then
		\begin{align*}
			\left \| (\ref{eq:eff_grad-2-1}) \right \|_{2} &\le \left( 1 - {1 \over K} \right)^{-1/2}{\| \Omega \|_{2}^{2}\| \Omega \|_{\rm F} \over \ubar{p}_{\sA}\lambda_{\min}(\ubar{\sfV}_{\epsilon})} \times \left[ |\widecheck{\mu}_{0}(\bX)| + |\mu_{0}(\bX)| + \text{constant} \times \dot{F}(\bX) + |\epsilon| \right] \times \left\| \dot{\sfF}(\bX;\widecheck{\bbeta}_{1}) - \dot{\sfF}(\bX;\widecheck{\bbeta}_{2}) \right\|_{2}\\
			&\le \text{constant} \times \left[ |\widecheck{\mu}_{0}(\bX)| + |\mu_{0}(\bX)| + \dot{F}(\bX) + |\epsilon| \right] \times \ddot{F}(\bX) \times \left \| \widecheck{\bbeta}_{1} - \widecheck{\bbeta}_{2} \right \|_{2};\\
			\left \| (\ref{eq:eff_grad-2-2}) \right \|_{2} &\le \left( 1 - {1 \over K} \right)^{-1/2}{\| \Omega \|_{2}^{2}\| \Omega \|_{\rm F} \over \ubar{p}_{\sA}\lambda_{\min}(\ubar{\sfV}_{\epsilon})}\times \ddot{F}(\bX) \times \left| \widecheck{e}(\widecheck{\bbeta}_{1}) - \widecheck{e}(\widecheck{\bbeta}_{2}) \right|\\
			&\le \text{constant} \times \dot{F}(\bX)\ddot{F}(\bX) \times \left\| \widecheck{\bbeta}_{1} - \widecheck{\bbeta}_{2} \right\|_{2}.
		\end{align*}
		And the Lipschitz bound for (\ref{eq:eff_grad-2}) is
		\begin{align*}
			\left \| (\ref{eq:eff_grad-2})(\widecheck{\bbeta}_{1}) - (\ref{eq:eff_grad-2})(\widecheck{\bbeta}_{2}) \right \|_{2} &\le \| (\ref{eq:eff_grad-2-1}) \|_{2} + \| (\ref{eq:eff_grad-2-2}) \|_{2} \\
			&\le \text{constant} \times \left[ |\widecheck{\mu}_{0}(\bX)| + |\mu_{0}(\bX)| + \dot{F}(\bX) + |\epsilon| \right] \times \ddot{F}(\bX) \times \left \| \widecheck{\bbeta}_{1} - \widecheck{\bbeta}_{2} \right \|_{2}
		\end{align*}
		Combining the Lipschitz bounds for (\ref{eq:eff_grad-1}) and (\ref{eq:eff_grad-2}), we have
		\begin{align*}
			&\left\|  {\partial\bphi_{\eff}(\widecheck{\bbeta}_{1};\widecheck{\mu}_{0},p_{\sA},\widecheck{\sigma}^{2}) \over \partial \bbeta^{\intercal}} - {\partial \bphi_{\eff}(\widecheck{\bbeta}_{2};\widecheck{\mu}_{0},p_{\sA},\widecheck{\sigma}^{2})\over \partial \bbeta^{\intercal}} \right \|_{2}\\
			\le & \underbrace{\text{constant} \times \left[ |\widecheck{\mu}_{0}(\bX)| + |\mu_{0}(\bX)| + \dot{F}(\bX) + |\epsilon| \right] \times \ddot{F}(\bX)}_{:= \widetilde{L}(\bX,A,\epsilon)} \times \left \| \widecheck{\bbeta}_{1} - \widecheck{\bbeta}_{2} \right \|_{2}
		\end{align*}
		In particular,
		\[ \bbE\widetilde{L}(\bX,A,\epsilon) \le \text{constant} \times \left(  \left\{ \bbE[\widecheck{\mu}_{0}(\bX)^{2}] \right\}^{1/2} + \left\{ \bbE[\mu_{0}(\bX)^{2}] \right\}^{1/2} + \left\{ \bbE[\dot{F}(\bX)^{2}] \right\}^{1/2} + \left\{ \bbE \epsilon^{2} \right\}^{1/2} \right) \left\{ \bbE[\ddot{F}(\bX)^{2}] \right\}^{1/2}, \]
		which is finite by Assumptions \ref{asm:int} and \ref{enum:int}.
		
		\item Note that
		\begin{align*}
			\left\| \bphi_{\eff}(\widecheck{\bbeta};\widecheck{\mu}_{0},p_{\sA},\widecheck{\sigma}^{2}) \right\|_{2} &= \underbrace{\left| \widecheck{e}(\widecheck{\bbeta}) \right|}_{\le (\ref{eq:resid_bdd})}\times
			\underbrace{\left\| \widecheck{\bh}_{\eff}(\bX,A;\widecheck{\bbeta}) \right\|_{2}}_{\le (\ref{eq:instrument_bdd})}\\
			&\le \text{constant} \times \left[  |\widecheck{\mu}_{0}(\bX)| + |\mu_{0}(\bX)| + \dot{F}(\bX) + |\epsilon| \right] \times \dot{F}(\bX).
		\end{align*}
		Therefore,
		\begin{align*}
			&\bbE\sup_{\widecheck{\bbeta} \in \sB}\left\| \bphi_{\eff}(\widecheck{\bbeta};\widecheck{\mu}_{0},p_{\sA},\widecheck{\sigma}^{2}) \right\|_{2} \\
			\le& \text{constant} \times \left(  \left\{ \bbE[\widecheck{\mu}_{0}(\bX)^{2}] \right\}^{1/2} + \left\{ \bbE[\mu_{0}(\bX)^{2}] \right\}^{1/2} + \left\{ \bbE[\dot{F}(\bX)^{2}] \right\}^{1/2} + \left\{ \bbE \epsilon^{2} \right\}^{1/2} \right) \left\{ \bbE[\dot{F}(\bX)^{2}] \right\}^{1/2},
		\end{align*}
		which is finite by Assumptions \ref{asm:int} and \ref{enum:int}. Next, we consider bounds for (\ref{eq:eff_grad-1}) and (\ref{eq:eff_grad-2}).
		\begin{align*}
			\| (\ref{eq:eff_grad-1}) \|_{2} &\le {\| \Omega \|_{2}^{2}\| \Omega \|_{\rm F}^{2} \over \ubar{p}_{\sA}\lambda_{\min}(\ubar{\sfV}_{\epsilon})}\times \dot{F}(\bX)^{2};\\
			\| (\ref{eq:eff_grad-2}) \|_{2}
			&\le \left( 1 - {1 \over K} \right)^{-1/2}{\| \Omega \|_{2}^{2}\| \Omega \|_{\rm F} \over \ubar{p}_{\sA}\lambda_{\min}(\ubar{\sfV}_{\epsilon})}\times \ddot{F}(\bX) \times \underbrace{\left| \widecheck{e}(\widecheck{\bbeta}) \right|}_{\le (\ref{eq:resid_bdd})}\\
			&\le \text{constant} \times \left[ |\widecheck{\mu}_{0}(\bX)| + |\mu_{0}(\bX)| + \dot{F}(\bX) + |\epsilon| \right] \times \ddot{F}(\bX).
		\end{align*}
		Then we have
		\begin{align*}
			&\bbE\left\| {\partial \bphi_{\eff}(\widecheck{\bbeta};\widecheck{\mu}_{0},p_{\sA},\widecheck{\sigma}^{2}) \over \partial \bbeta^{\intercal}} \right\|_{2} \le \bbE\| (\ref{eq:eff_grad-1}) \|_{2} + \bbE\| (\ref{eq:eff_grad-2}) \|_{2}\\
			\le& \text{constant} \times \left [\bbE[\dot{F}(\bX)^{2}] + \left(  \left\{ \bbE[\widecheck{\mu}_{0}(\bX)^{2}] \right\}^{1/2} + \left\{ \bbE[\mu_{0}(\bX)^{2}] \right\}^{1/2} + \left\{ \bbE[\dot{F}(\bX)^{2}] \right\}^{1/2} + \left\{ \bbE \epsilon^{2} \right\}^{1/2} \right) \left\{ \bbE[\ddot{F}(\bX)^{2}] \right\}^{1/2}\right ],
		\end{align*}
		which is finite by Assumptions by \ref{asm:int} and \ref{enum:int}.
	\end{enumerate}
\end{proof}

\begin{proof}[Proof of Lemma \ref{lem:plug-in_grad}]
	We follow the notations in Section \ref{sec:lem_plug-in} and the proof of Lemma \ref{lem:estim_fun}. Note that
	\begin{align}
		&- {\partial \bphi_{\eff}(\widecheck{\bbeta};\widehat{\mu}_{0,n},p_{\sA},\widehat{\sigma}_{n}^{2})\over \partial \bbeta^{\intercal}} + {\partial \bphi_{\eff}(\widecheck{\bbeta};\widecheck{\mu}_{0},p_{\sA},\widecheck{\sigma}^{2})\over \partial \bbeta^{\intercal}} \nonumber \\
		=&\quad~ \dot{\sfF}(\bX;\widecheck{\bbeta})^{\intercal}\Omega^{\intercal}\Omega[\widehat{\sfV}_{\epsilon,n}(\bX)^{-1} - \widecheck{\sfV}_{\epsilon}(\bX)^{-1}]\left[ \left( 1 - {1 \over K} \right) {\bomega_{A}^{\otimes 2}\over p_{\sA}(A|\bX)} \right]\dot{\sfF}(\bX;\widecheck{\bbeta})\label{eq:plug-in_grad-1}\\
		& +\ddot{\sfF}(\bX;\widecheck{\bbeta})^{\intercal}\Omega^{\intercal}\Omega\widecheck{\sfV}_{\epsilon}(\bX)^{-1}{\bomega_{A} \over p_{\sA}(A|\bX)}[\widehat{\mu}_{0,n}(\bX) - \widecheck{\mu}_{0}(\bX)] \label{eq:plug-in_grad-2}\\
		& - \ddot{\sfF}(\bX;\widecheck{\bbeta})^{\intercal}\Omega^{\intercal}\Omega[\widehat{\sfV}_{\epsilon,n}(\bX)^{-1} - \widecheck{\sfV}_{\epsilon}(\bX)^{-1}]{\bomega_{A}\widecheck{e}(\widecheck{\bbeta}) \over p_{\sA}(A|\bX)} \label{eq:plug-in_grad-3}\\
		& + \ddot{\sfF}(\bX;\widecheck{\bbeta})^{\intercal}\Omega^{\intercal}\Omega[\widehat{\sfV}_{\epsilon,n}(\bX)^{-1} - \widecheck{\sfV}_{\epsilon}(\bX)^{-1}]{\bomega_{A} \over p_{\sA}(A|\bX)}[\widehat{\mu}_{0,n}(\bX) - \widecheck{\mu}_{0}(\bX)]. \label{eq:plug-in_grad-4}
	\end{align}
	Then
	\begin{align*}
		\| (\ref{eq:plug-in_grad-1}) \|_{2} &\le {\| \Omega \|_{2}^{2}\| \Omega \|_{\rm F}^{2} \over \ubar{p}_{\sA}} \times \dot{F}(\bX)^{2} \times \underbrace{\left \| \widehat{\sfV}_{\epsilon,n}(\bX)^{-1} - \widecheck{\sfV}_{\epsilon}(\bX)^{-1} \right \|_{2}}_{\le (\ref{eq:var_bdd})}\\
		&\le \text{constant} \times \dot{F}(\bX)^{2} \times \| \widehat{\sigma}_{n}^{2} - \widecheck{\sigma}^{2} \|_{\infty};\\
		\| (\ref{eq:plug-in_grad-2}) \|_{2} &\le \left( 1 - {1 \over K} \right)^{-1/2}{\|\Omega\|_{2}^{2}\| \Omega \|_{\rm F} \over \ubar{p}_{\sA}\lambda_{\min}(\ubar{\sfV}_{\epsilon})} \times \ddot{F}(\bX) \times \left| \widehat{\mu}_{0,n}(\bX) - \widecheck{\mu}_{0}(\bX) \right|;\\
		\| (\ref{eq:plug-in_grad-3}) \|_{2} &\le \left( 1 - {1 \over K} \right)^{-1/2}{\| \Omega \|_{2}^{2}\| \Omega \|_{\rm F} \over \ubar{p}_{\sA}} \times \ddot{F}(\bX) \times \underbrace{\left| \widecheck{e}(\widecheck{\bbeta}) \right|}_{\le (\ref{eq:resid_bdd})} \times \underbrace{\left \| \widehat{\sfV}_{\epsilon,n}(\bX)^{-1} - \widecheck{\sfV}_{\epsilon}(\bX)^{-1} \right \|_{2}}_{\le (\ref{eq:var_bdd})}\\
		&\le \text{constant} \times \left[ |\widecheck{\mu}_{0}(\bX)| + |\mu_{0}(\bX)| + \dot{F}(\bX) + |\epsilon| \right] \times \ddot{F}(\bX) \times \| \widehat{\sigma}_{n}^{2} - \widecheck{\sigma}^{2} \|_{\infty};\\
		\| (\ref{eq:plug-in_grad-4}) \|_{2} &\le \left( 1 - {1 \over K} \right)^{-1/2}{\| \Omega \|_{2}^{2}\| \Omega \|_{\rm F} \over \ubar{p}_{\sA}} \times \ddot{F}(\bX) \times \underbrace{\left \| \widehat{\sfV}_{\epsilon,n}(\bX)^{-1} - \widecheck{\sfV}_{\epsilon}(\bX)^{-1} \right \|_{2}}_{\le (\ref{eq:var_bdd})} \times \left| \widehat{\mu}_{0,n}(\bX) - \widecheck{\mu}_{0}(\bX) \right|\\
		&\le \text{constant} \times \ddot{F}(\bX) \times \| \widehat{\sigma}_{n}^{2} - \widecheck{\sigma}^{2} \|_{\infty} \times \left| \widehat{\mu}_{0,n}(\bX) - \widecheck{\mu}_{0}(\bX) \right|.
	\end{align*}
	Then we have
	\begin{align*}
		&\bbE_{n}\left\| {\partial \bphi_{\eff}(\widecheck{\bbeta};\widehat{\mu}_{0,n},p_{\sA},\widehat{\sigma}_{n}^{2})\over \partial \bbeta^{\intercal}} - {\partial \bphi_{\eff}(\widecheck{\bbeta};\widecheck{\mu}_{0},p_{\sA},\widecheck{\sigma}^{2})\over \partial \bbeta^{\intercal}} \right\|_{2} \\
		\le& \bbE_{n}\| (\ref{eq:plug-in_grad-1}) \|_{2} + \bbE_{n}\| (\ref{eq:plug-in_grad-2}) \|_{2} + \bbE_{n}\| (\ref{eq:plug-in_grad-3}) \|_{2} + \bbE_{n}\| (\ref{eq:plug-in_grad-4}) \|_{2}\\
		\lesssim& \smallcO_{\bbP}(n^{-1/2}) + \smallcO_{\bbP}(n^{-1/2}) + \smallcO_{\bbP}(n^{-1/2}) \smallcO_{\bbP}(n^{-1/2}) + \smallcO_{\bbP}(n^{-1})\\
		=& \smallcO_{\bbP}(n^{-1/2}).
	\end{align*}
\end{proof}

\subsubsection{Proof of Theorem \ref{thm:eff_opt}}

\begin{proof}[Proof of Theorem \ref{thm:eff_opt}]
	By Theorem \ref{thm:eff}, we have
	\[ \widehat{\bbeta}_{\eff,n}(\widecheck{\mu}_{0}) - \bbeta = \cI(\bbeta;\widecheck{\mu}_{0})^{-1}\bbE_{n}[\bphi_{\eff}(\bbeta;\widecheck{\mu}_{0},p_{\sA},\sigma_{\opt}^{2})] + \smallcO_{\bbP}(n^{-1/2}). \]
	Therefore, it suffices to study the asymptotic variance
	First of all, we derive the $ \sqrt{n} $-asymptotic variance of $ \widehat{\bbeta}_{\eff,n}(\widecheck{\mu}_{0}) $. Denote
	\begin{align*}
		\widecheck{\bphi}_{\eff}(\bbeta) &:= \bphi_{\eff}(\bbeta;\widecheck{\mu}_{0},p_{\sA},\sigma_{\opt}^{2})\\
		&= \left [Y - \widecheck{\mu}_{0}(\bX) - \left( 1 - {1 \over K} \right) \langle \bomega_{A},\vec{\bbf}(\bX;\bbeta) \rangle \right]\bh_{\eff}(\bX,A;\bbeta,\widecheck{\mu}_{0})\\
		&= [\underbrace{\mu_{0}(\bX) - \widecheck{\mu}_{0}(\bX) + \epsilon}_{=\widecheck{e}(\bbeta)}]\bh_{\eff}(\bX,A;\bbeta,\widecheck{\mu}_{0});\\
		\bh_{\eff}(\bX,A;\bbeta,\widecheck{\mu}_{0}) &:= \dot{\sfF}(\bX;\bbeta)^{\intercal}\Omega^{\intercal}\Omega\sfV_{\epsilon}(\bX;\widecheck{\mu}_{0})^{-1}{\bomega_{A} \over p_{\sA}(A|\bX)}.
	\end{align*}
	First notice that
	\begin{align*}
	\bbE[\widecheck{\bphi}_{\eff}(\bbeta)^{\otimes 2}] &= \bbE\left[ \dot{\sfF}(\bX;\bbeta)^{\intercal}\Omega^{\intercal}\Omega \sfV_{\epsilon}(\bX;\widecheck{\mu}_{0})^{-1}\left( {\bomega_{A}^{\otimes 2}\widecheck{e}(\bbeta)^{2} \over p_{\sA}(A|\bX)^{2}} \right)\sfV_{\epsilon}(\bX;\widecheck{\mu}_{0})^{-1} \Omega^{\intercal}\Omega\dot{\sfF}(\bX;\bbeta) \right]\\
	&=\underbrace{\bbE\left[ \dot{\sfF}(\bX;\bbeta)^{\intercal}\Omega^{\intercal}\Omega \sfV_{\epsilon}(\bX;\widecheck{\mu}_{0})^{-1} \Omega^{\intercal}\Omega\dot{\sfF}(\bX;\bbeta) \right]}_{= \cI(\bbeta;\widecheck{\mu}_{0})};\\
	\bbE\left[ -{\partial \widecheck{\bphi}_{\eff}(\bbeta) \over \partial \bbeta^{\intercal}} \right] &=\bbE\left\{ \dot{\sfF}(\bX;\bbeta)^{\intercal}\Omega^{\intercal}\Omega \sfV_{\epsilon}(\bX;\widecheck{\mu}_{0})^{-1} \left[ \left( 1 - {1 \over K} \right) {\bomega_{A}^{\otimes 2} \over p_{\sA}(A|\bX)} \right]\dot{\sfF}(\bX;\bbeta) \right\}\\
	& \quad - \underbrace{\bbE\left\{ \ddot{\sfF}(\bX;\bbeta)^{\intercal}\Omega^{\intercal}\Omega\sfV_{\epsilon}(\bX;\widecheck{\mu}_{0})^{-1}{[\mu_{0}(\bX) - \widecheck{\mu}_{0}(\bX) + \epsilon]\bomega_{A} \over p_{\sA}(A|\bX)} \right\}}_{= 0}\\
	&= \underbrace{\bbE\left[ \dot{\sfF}(\bX;\bbeta)^{\intercal}\Omega^{\intercal}\Omega \sfV_{\epsilon}(\bX;\widecheck{\mu}_{0})^{-1} \Omega^{\intercal}\Omega\dot{\sfF}(\bX;\bbeta) \right]}_{= \cI(\bbeta;\widecheck{\mu}_{0})}.
	\end{align*}
	Here, the second equality follows from that
	\begin{align*}
		&\bbE[\widecheck{e}(\bbeta)^{2}|\bX,A] = [\widecheck{\mu}_{0}(\bX) - \mu_{0}(\bX)]^{2} + \sigma^{2}(\bX,A) = \sigma_{\opt}^{2}(\bX,A;\widecheck{\mu}_{0});\\
		&\bbE\left( {\bomega_{A}^{\otimes 2}\widecheck{e}(\bbeta)^{2} \over p_{\sA}(A|\bX)^{2}}\middle|\bX \right) = \bbE\left( {\bomega_{A}^{\otimes 2}\sigma_{\opt}^{2}(\bX,A;\widecheck{\mu}_{0}) \over p_{\sA}(A|\bX)^{2}}\middle|\bX \right) = \sum_{k=1}^{K}{\sigma_{\opt}^{2}(\bX,k;\widecheck{\mu}_{0})\bomega_{k}^{\otimes 2} \over p_{\sA}(k|\bX)} = \sfV_{\epsilon}(\bX;\widecheck{\mu}_{0}).
	\end{align*}
	The forth equality follows from that $ \bbE[\bomega_{A}/p_{\sA}(A|\bX)|\bX] = \bzero $, $ \bbE(\epsilon|\bX,A) = 0 $ and
	\[ \left( 1 - {1 \over K} \right)\bbE\left( {\bomega_{A}^{\otimes 2} \over p_{\sA}(A|\bX)}\middle|\bX \right) = \left( 1 - {1 \over K} \right)\sum_{k=1}^{K}\bomega_{k}^{\otimes 2} = \Omega^{\intercal}\Omega. \]
	Then
	\[ \lim_{n\to\infty}n\sfvar[\widehat{\bbeta}_{\eff,n}(\widecheck{\mu}_{0})] = \left\{ \bbE \left[-{\partial \widecheck{\bphi}_{\eff}(\bbeta) \over \partial \bbeta^{\intercal}}\right]\right\}^{-1}\bbE[\widecheck{\bphi}_{\eff}(\bbeta)^{\otimes 2}]\left\{ \bbE \left[-{\partial \widecheck{\bphi}_{\eff}(\bbeta)^{\intercal} \over \partial \bbeta}\right]\right\}^{-1} = \cI(\bbeta;\widecheck{\mu}_{0})^{-1}. \]
	
	Next, we study the $ \sqrt{n} $-asymptotic variance of estimates from the regular class $ \cB_{n}(\widecheck{\mu}_{0}) $. Fix $ \widehat{\bbeta}_{n}(\widecheck{\mu}_{0}) \in \cB_{n}(\widecheck{\mu}_{0}) $ that corresponds to the estimating function $ \bphi(\bbeta;\widecheck{\mu}_{0}) $, and denote $ \widecheck{\bphi}(\bbeta) := \bphi(\bbeta;\widecheck{\mu}_{0}) $. There exists $ \bh:\cX\times A\to \bbR^{p} $, which can depend on $ (\bbeta,\widecheck{\mu}_{0}) $, such that $ \bbE[\bh(\bX,A)|\bX] = \bzero $ and
	\[ \widecheck{\bphi}(\bbeta) = \bigg[\underbrace{Y - \widecheck{\mu}_{0}(\bX) - \left( 1 - {1 \over K} \right) \langle \bomega_{A},\vec{\bbf}(\bX;\bbeta) \rangle}_{= \widecheck{e}(\bbeta)} \bigg]\bh(\bX,A). \]
	Here, we suppress the potential dependency of $ \bh $ on $ (\bbeta,\widecheck{\mu}_{0}) $ and only mention it if necessary. Note that $ \bbE[\bh(\bX,A)|\bX] = \bzero $ informs the representation
	\[ \bh(\bX,A) = {\sfH(\bX)\bomega_{A} \over p_{\sA}(A|\bX)}; \quad \sfH(\bX) := \left( 1 - {1 \over K} \right)\sum_{k=1}^{K}p_{\sA}(k|\bX)\bh(\bX,k)\bomega_{k}^{\intercal}(\Omega^{\intercal}\Omega)^{-1} \in \bbR^{p\times (K-1)}, \]
	since by $ \Omega^{\intercal}(\Omega^{\intercal}\Omega)^{-1}\Omega^{\intercal} = \Ib_{K\times K} - (1/K)\vec{\bone}_{K}^{\otimes 2} $, we have
	\[ \sfH(\bX)\bomega_{A} = p_{\sA}(A|\bX)\bh(\bX,A) - {1 \over K}\underbrace{\sum_{k = 1}^{K}p_{\sA}(k|\bX)\bh(\bX,k)}_{= \bbE[\bh(\bX,A)|\bX] = \bzero} = p_{\sA}(A|\bX)\bh(\bX,A). \]
	Then
	\begin{align*}
		\bbE[\widecheck{\bphi}(\bbeta)^{\otimes 2}] &= \bbE\left[ \sfH(\bX)\left( {\bomega_{A}^{\otimes 2}\widecheck{e}(\bbeta)^{2} \over p_{\sA}(A|\bX)^{2}} \right)\sfH(\bX)^{\intercal} \right]\\
		&= \bbE\left[ \sfH(\bX) \sfV_{\epsilon}(\bX;\widecheck{\mu}_{0}) \sfH(\bX)^{\intercal} \right];\\
		\bbE \left[-{\partial\widecheck{\bphi}(\bbeta) \over \partial \bbeta^{\intercal}}\right] &= \bbE\left\{ \sfH(\bX)\left[ \left( 1 - {1 \over K} \right){\bomega_{A}^{\otimes 2} \over p_{\sA}(A|\bX)} \right]\dot{\sfF}(\bX;\bbeta) \right\} - \underbrace{\bbE\left\{ \dot{\sfH}(\bX;\bbeta,\widecheck{\mu}_{0}){[\mu_{0}(\bX) - \widecheck{\mu}_{0}(\bX) + \epsilon]\bomega_{A} \over p_{\sA}(A|\bX)} \right\}}_{= 0}\\
		&= \bbE\left[ \sfH(\bX)\Omega^{\intercal}\Omega\dot{\sfF}(\bX;\bbeta) \right].
	\end{align*}
	Here, the second equality follows from that $ \bbE[\widecheck{e}(\bbeta)^{2}|\bX,A] = \sigma_{\opt}^{2}(\bX,A) $ and
	\[ \bbE\left( {\bomega_{A}^{\otimes 2}\widecheck{e}(\bbeta)^{2} \over p_{\sA}(A|\bX)^{2}}\middle|\bX \right) = \bbE\left( {\bomega_{A}^{\otimes 2}\sigma_{\opt}^{2}(\bX,A) \over p_{\sA}(A|\bX)^{2}}\middle|\bX \right) = \sum_{k=1}^{K}{\sigma^{2}_{\opt}(\bX,A)\bomega_{k}^{\otimes 2} \over p_{\sA}(k|\bX)} = \sfV_{\epsilon}(\bX;\widecheck{\mu}_{0}). \]
	The forth equality follows from that $ \bbE[\bomega_{A}/p_{\sA}(A|\bX)|\bX] = \bzero $, $ \bbE(\epsilon|\bX,A) = 0 $ and
	\[ \left( 1 - {1 \over K} \right)\bbE\left( {\bomega_{A}^{\otimes 2} \over p_{\sA}(A|\bX)}\middle|\bX \right) = \left( 1 - {1 \over K} \right)\sum_{k=1}^{K}\bomega_{k}^{\otimes 2} = \Omega^{\intercal}\Omega. \]
	\begin{lem}[Sandwich Variance Inequality]\label{lem:sandwich}
		Define
		\begin{align*}
			\sfA &:= \bbE[\sfH(\bX)\sfV_{\epsilon}(\bX;\widecheck{\mu}_{0})\sfH(\bX)^{\intercal}];\\
			\sfB &:= \bbE[\sfH(\bX)\Omega^{\intercal}\Omega\dot{\sfF}(\bX;\bbeta)];\\
			\sfC &:= \bbE[\dot{\sfF}(\bX;\bbeta)^{\intercal}\Omega^{\intercal}\Omega\sfV_{\epsilon}(\bX;\widecheck{\mu}_{0})^{-1}\Omega^{\intercal}\Omega\dot{\sfF}(\bX;\bbeta)];\\
			\sfX &:= \begin{pmatrix}
				\sfA & \sfB\\
				\sfB^{\intercal} & \sfC
			\end{pmatrix}.
		\end{align*}
		Then
		\[ \sfX \ge 0 \quad \Leftrightarrow \quad \overbrace{\sfX/\sfA = \sfC - \sfB^{\intercal}\sfA^{-1}\sfB}^{\text{Schur complement}} \ge 0, \]
		with equality if and only if there exists some non-singular constant matrix $ \sfH_{0} \in \bbR^{p \times p} $ such that
		\[ \sfH(\bX) = \sfH_{0}\dot{\sfF}(\bX;\bbeta)^{\intercal}\Omega^{\intercal}\Omega\sfV_{\epsilon}(\bX;\widecheck{\mu}_{0})^{-1}. \]
	\end{lem}

	\noindent Following the notations in Lemma \ref{lem:sandwich}, we have
	\begin{align*}
		&\lim_{n\to\infty}n\sfvar[\widehat{\bbeta}_{n}(\widecheck{\mu}_{0})] \\
		=& \left\{ \bbE \left[-{\partial \widecheck{\bphi}(\bbeta)\over \partial \bbeta^{\intercal}}\right]\right\}^{-1}\bbE [\widecheck{\bphi}(\bbeta)^{\otimes 2}]\left\{ \bbE \left[-{\partial \widecheck{\bphi}(\bbeta)^{\intercal}\over \partial \bbeta}\right]\right\}^{-1} \\
		=& \sfB^{-1}\sfA\sfB^{-\intercal} = (\sfB^{\intercal}\sfA^{-1}\sfB)^{-1} \ge \sfC^{-1} &&(\text{by Lemma \ref{lem:sandwich}}) \\
		=& \cI(\bbeta;\widecheck{\mu}_{0})^{-1},
	\end{align*}
	with equality attained at
	\[ \sfH(\bX) = \sfH_{0}\dot{\sfF}(\bX;\bbeta)^{\intercal}\Omega^{\intercal}\Omega\sfV_{\epsilon}(\bX;\widecheck{\mu}_{0})^{-1}, \]
	for some non-singular constant matrix $ \sfH_{0} \in \bbR^{p \times p} $.
\end{proof}

\begin{proof}[Proof of Lemma \ref{lem:sandwich}]
	For any $ \bu,\bv \in \bbR^{p} $, define
	\begin{align*}
		\bU &:= \sfV_{\epsilon}(\bX;\widecheck{\mu}_{0})^{1/2}\sfH(\bX)^{\intercal}\bu;\\
		\bV &:= \sfV_{\epsilon}(\bX;\widecheck{\mu}_{0})^{-1/2}\Omega^{\intercal}\Omega\dot{\sfF}(\bX;\bbeta)\bv.
	\end{align*}
	Then
	\begin{align*}
		(\bu^{\intercal},\bv^{\intercal})\sfX\begin{pmatrix}
			\bu \\ \bv
		\end{pmatrix} &= \bu^{\intercal}\sfA\bu + 2\bv^{\intercal}\sfB^{\intercal}\bu + \bv^{\intercal}\sfC\bv\\
		&= \bbE(\bU^{\intercal}\bU + 2\bV^{\intercal}\bU + \bV^{\intercal}\bV)\\
		&= \bbE\|\bU + \bV\|_{2}^{2}\\
		&\ge 0,
	\end{align*}
	with equality if and only if 
	\[ \bU = -\bV \quad \Leftrightarrow \quad \sfV_{\epsilon}(\bX;\widecheck{\mu}_{0})^{1/2}\sfH(\bX)^{\intercal}\bu = - \sfV_{\epsilon}(\bX;\widecheck{\mu}_{0})^{-1/2}\Omega^{\intercal}\Omega\dot{\sfF}(\bX;\bbeta)\bv \quad a.s. \]
	That is, for some constant matrix $ \sfK \in \bbR^{p \times p} $, we have $ \bu = -\sfK^{\intercal}\bv $, and
	\begin{align*}
		\sfV_{\epsilon}(\bX;\widecheck{\mu}_{0})^{1/2}\sfH(\bX)^{\intercal}\sfK^{\intercal} = \sfV_{\epsilon}(\bX;\widecheck{\mu}_{0})^{-1/2}\Omega^{\intercal}\Omega\dot{\sfF}(\bX;\bbeta) \quad \Leftrightarrow \quad \sfK\sfH(\bX) = \dot{\sfF}(\bX;\bbeta)^{\intercal}\Omega^{\intercal}\Omega\sfV_{\epsilon}(\bX;\widecheck{\mu}_{0})^{-1} \quad a.s.
	\end{align*}
	This proves $ \sfX \ge 0 $ and the equality condition. Finally, by $ \sfA > 0 $, we have
	\[ \bv^{\intercal}(\sfX/\sfA)\bv = \bv^{\intercal}\sfC\bv - \bv^{\intercal}\sfB^{\intercal}\sfA^{-1}\sfB\bv = \left[ (\bu^{\intercal},\bv^{\intercal})\sfX\begin{pmatrix}
		\bu \\ \bv
	\end{pmatrix} \right]_{\bu = -\sfA^{-1}\sfB\bv} \ge 0. \]
	That is, $ \sfX/\sfA \ge 0 $ with equality if and only if for $ \sfK = \sfB^{\intercal}\sfA $ and $ \sfH_{0} := \sfK^{-1} = \sfA^{-1}\sfB^{-\intercal} $, we have
	\[ \sfH(\bX) = \sfH_{0}\dot{\sfF}(\bX;\bbeta)^{\intercal}\Omega^{\intercal}\Omega\sfV_{\epsilon}(\bX;\widecheck{\mu}_{0})^{-1} \quad a.s. \]
\end{proof}

\subsubsection{Proof of Theorem \ref{thm:regret_par}}

\begin{proof}[Proof of Theorem \ref{thm:regret_par}] By Assumption \ref{asm:cmpt}, the distribution of $ \bX $ has compact support. Without loss of generality, assume that $ \cX $ is compact. Recall from the proof of Lemma \ref{lem:plug-in_grad} that $ \ddot{F} $ is the envelop function of $ \ddot{\sfF} $. Then by Assumption \ref{enum:int} that $ \bbE \ddot{F}(\bX) +\infty $, we further have that $ \| \ddot{F} \|_{\infty} = \sup_{\bx \in \cX}\ddot{F}(\bx) < +\infty $.
	
	By $ \widehat{\bbeta}_{n} = \bbeta + \cO_{\bbP}(n^{-1/2}) $, we have
	\[ \sup_{\bx \in \cX}\left \| \dot{\sfF}(\bx,\widehat{\bbeta}_{n}) - \dot{\sfF}(\bx,\bbeta) \right \|_{2} \le \|\ddot{F}\|_{\infty} \times \| \widehat{\bbeta}_{n} - \bbeta \|_{2} = \cO_{\bbP}(n^{-1/2}). \]
	Fix $ \bx \in \cX $. By Mean Value Theorem, there exists some $ \alpha_{n} \in [0,1] $ and $ \widetilde{\bbeta}_{n} = (1 - \alpha_{n})\widehat{\bbeta}_{n} + \alpha_{n}\bbeta $, such that
	\begin{align*}
		\vec{\bbf}(\bx;\widehat{\bbeta}_{n}) - \vec{\bbf}(\bx;\bbeta)
		&= \dot{\sfF}(\bx,\widetilde{\bbeta}_{n})(\widehat{\bbeta}_{n} - \bbeta)\\
		&= [\dot{\sfF}(\bx,\bbeta) + \cO_{\bbP}(n^{-1/2})](\widehat{\bbeta}_{n} - \bbeta) \\
		&= \dot{\sfF}(\bx,\bbeta)(\widehat{\bbeta}_{n} - \bbeta) + \cO_{\bbP}(n^{-1}).
	\end{align*}
	By $ \lim_{n\to\infty}n\sfvar(\widehat{\bbeta}_{n}) = \Sigma $, we have
	\[ \lim_{n\to\infty}n\sfvar[\vec{\bbf}(\bx;\widehat{\bbeta}_{n})] = \dot{\sfF}(\bx;\bbeta)\Sigma\dot{\sfF}(\bx;\bbeta)^{\intercal}. \]
	Suppose $ \bX \sim p_{\sX}(\bx) $ and $ \bX \indep \widehat{\bbeta}_{n} $. Then
	\begin{align*}
		&\limsup_{n\to\infty}n\sum_{k=1}^{K}\bbE[\gamma(\bX,k;\widehat{\bbeta}_{n}) - \gamma(\bX,k;\bbeta)]^{2} \\
		=& \limsup_{n\to\infty}\left( 1 - {1 \over K} \right)^{2}\sum_{k=1}^{K}\bomega_{k}^{\intercal}\bbE  [\vec{\bbf}(\bX;\widehat{\bbeta}_{n}) - \vec{\bbf}(\bX;\bbeta)]^{\otimes 2}\bomega_{k} &&(\text{by Lemma \ref{lem:ang}}) \\
		=& \left( 1 - {1 \over K} \right)^{2}\sum_{k=1}^{K}\bomega_{k}^{\intercal}\bbE\bigg[\lim_{n\to\infty}n\underbrace{\bbE\left\{  [\vec{\bbf}(\bX;\widehat{\bbeta}_{n}) - \vec{\bbf}(\bX;\bbeta)]^{\otimes 2}\middle|\bX \right\}}_{= \sfvar[\vec{\bbf}(\bx;\widehat{\bbeta}_{n})]|_{\bx = \bX}}\bigg]\bomega_{k}  &&\text{(by DCT)}\\
		=& \left( 1 - {1 \over K} \right)^{2}\sum_{k=1}^{K}\bomega_{k}^{\intercal} \bbE[\dot{\sfF}(\bX;\bbeta)\Sigma\dot{\sfF}(\bX;\bbeta)^{\intercal}] \bomega_{k}\\
		=&\left( 1 - {1 \over K} \right)^{2}\tr\left\{  \sum_{k=1}^{K}\bomega_{k}^{\intercal} \bbE[\dot{\sfF}(\bX;\bbeta)\Sigma\dot{\sfF}(\bX;\bbeta)^{\intercal}]\bomega_{k} \right\} &&(\text{trace of a scalar}) \\
		=&\left( 1 - {1 \over K} \right)\tr\left\{ \bbE[\dot{\sfF}(\bX;\bbeta)^{\intercal}\Omega^{\intercal}\Omega \dot{\sfF}(\bX;\bbeta)]\Sigma \right\}. &&(\text{commutativity under trace})
	\end{align*}
	Finally, by Theorem \ref{thm:regret_est}, we have
	\begin{align*}
		\limsup_{n\to\infty}\sqrt{n}[\cV(d^{\star}) - \bbE\cV(\widehat{d}_{n})] &\le 2\limsup_{n\to\infty} \left\{\sqrt{n} \max_{1 \le k \le K}\bbE\left| \gamma(\bX,k;\widehat{\bbeta}_{n}) - \gamma(\bX,k;\bbeta) \right| \right\}\\
		&\le 2\limsup_{n\to\infty}\left\{ n\max_{1 \le k \le K} \bbE[\gamma(\bX,k;\widehat{\bbeta}_{n}) - \gamma(\bX,k;\bbeta)]^{2} \right\}^{1/2} \\
		&\le 2\lim_{n\to\infty}\left\{ n\sum_{k=1}^{K} \bbE[\gamma(\bX,k;\widehat{\bbeta}_{n}) - \gamma(\bX,k;\bbeta)]^{2} \right\}^{1/2}\\
		&= 2\left( 1 - {1 \over K} \right)^{1/2}\tr\left\{ \bbE[\dot{\sfF}(\bX;\bbeta)^{\intercal}\Omega^{\intercal}\Omega \dot{\sfF}(\bX;\bbeta)]\Sigma \right\}^{1/2}.
	\end{align*}
	Here, $ \bbE $ is taken over $ (\bX,\widehat{\bbeta}_{n}) $.
\end{proof}


\newpage

\subsection{Additional Tables and Figures}\label{sec:tab_fig}
\renewcommand{\baselinestretch}{1}
\renewcommand{\arraystretch}{0.99}
\begin{table}[H]

\caption[ACTG 175, Coefficients]{\label{tab:eLearn_ACTG175_coef}Estimated Coefficients on the ACTG175 Dataset (averaged over 10 replications)}
\centering
\fontsize{7}{9}\selectfont
\begin{tabular}[t]{>{}r|rrr>{}r|rrr>{}r|rrrr}
\toprule
\multicolumn{1}{c}{\textbf{variable}} & \multicolumn{4}{c}{\textbf{D-Learning}} & \multicolumn{4}{c}{\textbf{RD-Learning}} & \multicolumn{4}{c}{\textbf{E-Learning}} \\
\cmidrule(l{3pt}r{3pt}){1-1} \cmidrule(l{3pt}r{3pt}){2-5} \cmidrule(l{3pt}r{3pt}){6-9} \cmidrule(l{3pt}r{3pt}){10-13}
\textbf{} & \textbf{\texttt{ddI}} & \textbf{\texttt{ZDV}} & \textbf{\texttt{ZDV+ddI}} & \textbf{\texttt{ZDV+ZAL}} & \textbf{\texttt{ddI}} & \textbf{\texttt{ZDV}} & \textbf{\texttt{ZDV+ddI}} & \textbf{\texttt{ZDV+ZAL}} & \textbf{\texttt{ddI}} & \textbf{\texttt{ZDV}} & \textbf{\texttt{ZDV+ddI}} & \textbf{\texttt{ZDV+ZAL}}\\
\midrule
\addlinespace[0.3em]
\multicolumn{13}{l}{\textbf{I: original data}}\\
\ttfamily{\hspace{1em}Intercept} & \bgroup\fontsize{3}{5}\selectfont 4.72\egroup{} & \bgroup\fontsize{6}{8}\selectfont \textbf{-29.9}\egroup{} & \bgroup\fontsize{6}{8}\selectfont \textbf{26.45}\egroup{} & \bgroup\fontsize{3}{5}\selectfont -1.27\egroup{} & \bgroup\fontsize{3}{5}\selectfont 4.33\egroup{} & \bgroup\fontsize{6}{8}\selectfont \textbf{-31.54}\egroup{} & \bgroup\fontsize{6}{8}\selectfont \textbf{27.87}\egroup{} & \bgroup\fontsize{3}{5}\selectfont -0.66\egroup{} & \bgroup\fontsize{3}{5}\selectfont 4.67\egroup{} & \bgroup\fontsize{6}{8}\selectfont \textbf{-30.67}\egroup{} & \bgroup\fontsize{6}{8}\selectfont \textbf{26.52}\egroup{} & \bgroup\fontsize{3}{5}\selectfont -0.52\egroup{}\\
\ttfamily{\hspace{1em}gender} &  &  &  &  &  &  &  &  & \bgroup\fontsize{3}{5}\selectfont -0.25\egroup{} & \bgroup\fontsize{3}{5}\selectfont 0.36\egroup{} & \bgroup\fontsize{3}{5}\selectfont 0.05\egroup{} & \bgroup\fontsize{3}{5}\selectfont -0.16\egroup{}\\
\ttfamily{\hspace{1em}homo} & \bgroup\fontsize{3}{5}\selectfont 2.2\egroup{} & \bgroup\fontsize{3}{5}\selectfont -0.12\egroup{} & \bgroup\fontsize{3}{5}\selectfont -2.74\egroup{} & \bgroup\fontsize{3}{5}\selectfont 0.65\egroup{} & \bgroup\fontsize{3}{5}\selectfont 3.63\egroup{} & \bgroup\fontsize{3}{5}\selectfont -0.53\egroup{} & \bgroup\fontsize{3}{5}\selectfont -4.55\egroup{} & \bgroup\fontsize{3}{5}\selectfont 1.45\egroup{} & \bgroup\fontsize{3}{5}\selectfont 3.26\egroup{} & \bgroup\fontsize{3}{5}\selectfont -0.4\egroup{} & \bgroup\fontsize{3}{5}\selectfont -4.36\egroup{} & \bgroup\fontsize{3}{5}\selectfont 1.49\egroup{}\\
\ttfamily{\hspace{1em}race} & \bgroup\fontsize{3}{5}\selectfont -0.12\egroup{} & \bgroup\fontsize{3}{5}\selectfont 0.4\egroup{} & \bgroup\fontsize{3}{5}\selectfont -0.41\egroup{} & \bgroup\fontsize{3}{5}\selectfont 0.13\egroup{} & \bgroup\fontsize{3}{5}\selectfont -0.41\egroup{} & \bgroup\fontsize{3}{5}\selectfont 1.04\egroup{} & \bgroup\fontsize{3}{5}\selectfont -1.08\egroup{} & \bgroup\fontsize{3}{5}\selectfont 0.45\egroup{} & \bgroup\fontsize{3}{5}\selectfont -0.57\egroup{} & \bgroup\fontsize{3}{5}\selectfont 2.56\egroup{} & \bgroup\fontsize{3}{5}\selectfont -2.67\egroup{} & \bgroup\fontsize{3}{5}\selectfont 0.69\egroup{}\\
\ttfamily{\hspace{1em}drugs} & \bgroup\fontsize{3}{5}\selectfont -3.29\egroup{} & \bgroup\fontsize{3}{5}\selectfont -0.96\egroup{} & \bgroup\fontsize{3}{5}\selectfont 2.77\egroup{} & \bgroup\fontsize{3}{5}\selectfont 1.49\egroup{} & \bgroup\fontsize{3}{5}\selectfont -3.3\egroup{} & \bgroup\fontsize{3}{5}\selectfont -1.12\egroup{} & \bgroup\fontsize{3}{5}\selectfont 2.48\egroup{} & \bgroup\fontsize{3}{5}\selectfont 1.94\egroup{} & \bgroup\fontsize{3}{5}\selectfont -4.26\egroup{} & \bgroup\fontsize{3}{5}\selectfont -2.68\egroup{} & \bgroup\fontsize{3}{5}\selectfont 3.13\egroup{} & \bgroup\fontsize{3}{5}\selectfont 3.81\egroup{}\\
\ttfamily{\hspace{1em}symptom} &  &  &  &  &  &  &  &  & \bgroup\fontsize{3}{5}\selectfont -0.08\egroup{} & \bgroup\fontsize{3}{5}\selectfont 0.08\egroup{} & \bgroup\fontsize{3}{5}\selectfont -0.06\egroup{} & \bgroup\fontsize{3}{5}\selectfont 0.06\egroup{}\\
\ttfamily{\hspace{1em}str2} &  &  &  &  &  &  &  &  & \bgroup\fontsize{3}{5}\selectfont 0.11\egroup{} & \bgroup\fontsize{3}{5}\selectfont 0.07\egroup{} & \bgroup\fontsize{3}{5}\selectfont -0.26\egroup{} & \bgroup\fontsize{3}{5}\selectfont 0.09\egroup{}\\
\ttfamily{\hspace{1em}hemo} &  &  &  &  &  &  &  &  & \bgroup\fontsize{3}{5}\selectfont 0.25\egroup{} & \bgroup\fontsize{3}{5}\selectfont -0.5\egroup{} & \bgroup\fontsize{3}{5}\selectfont -0.01\egroup{} & \bgroup\fontsize{3}{5}\selectfont 0.26\egroup{}\\
\ttfamily{\hspace{1em}age} & \bgroup\fontsize{3}{5}\selectfont -0.54\egroup{} & \bgroup\fontsize{3}{5}\selectfont -0.48\egroup{} & \bgroup\fontsize{3}{5}\selectfont 4.79\egroup{} & \bgroup\fontsize{3}{5}\selectfont -3.77\egroup{} & \bgroup\fontsize{3}{5}\selectfont -0.64\egroup{} & \bgroup\fontsize{3}{5}\selectfont -0.55\egroup{} & \bgroup\fontsize{6}{8}\selectfont \textbf{7.67}\egroup{} & \bgroup\fontsize{6}{8}\selectfont \textbf{-6.48}\egroup{} & \bgroup\fontsize{3}{5}\selectfont -0.2\egroup{} & \bgroup\fontsize{3}{5}\selectfont 0.1\egroup{} & \bgroup\fontsize{6}{8}\selectfont \textbf{6.65}\egroup{} & \bgroup\fontsize{6}{8}\selectfont \textbf{-6.55}\egroup{}\\
\ttfamily{\hspace{1em}wtkg} &  &  &  &  & \bgroup\fontsize{3}{5}\selectfont 0.24\egroup{} & \bgroup\fontsize{3}{5}\selectfont -0.26\egroup{} & \bgroup\fontsize{3}{5}\selectfont -0.19\egroup{} & \bgroup\fontsize{3}{5}\selectfont 0.21\egroup{} & \bgroup\fontsize{3}{5}\selectfont 0.5\egroup{} & \bgroup\fontsize{3}{5}\selectfont -1.03\egroup{} & \bgroup\fontsize{3}{5}\selectfont -0.36\egroup{} & \bgroup\fontsize{3}{5}\selectfont 0.9\egroup{}\\
\ttfamily{\hspace{1em}cd40} & \bgroup\fontsize{6}{8}\selectfont \textbf{9.02}\egroup{} & \bgroup\fontsize{3}{5}\selectfont 3.22\egroup{} & \bgroup\fontsize{6}{8}\selectfont \textbf{-13.54}\egroup{} & \bgroup\fontsize{3}{5}\selectfont 1.29\egroup{} & \bgroup\fontsize{6}{8}\selectfont \textbf{7.94}\egroup{} & \bgroup\fontsize{3}{5}\selectfont 1.02\egroup{} & \bgroup\fontsize{6}{8}\selectfont \textbf{-7.94}\egroup{} & \bgroup\fontsize{3}{5}\selectfont -1.03\egroup{} & \bgroup\fontsize{6}{8}\selectfont \textbf{5.82}\egroup{} & \bgroup\fontsize{3}{5}\selectfont 1.15\egroup{} & \bgroup\fontsize{6}{8}\selectfont \textbf{-5.56}\egroup{} & \bgroup\fontsize{3}{5}\selectfont -1.41\egroup{}\\
\ttfamily{\hspace{1em}karnof} & \bgroup\fontsize{3}{5}\selectfont -0.02\egroup{} & \bgroup\fontsize{3}{5}\selectfont -0.08\egroup{} & \bgroup\fontsize{3}{5}\selectfont 0.1\egroup{} & \bgroup\fontsize{3}{5}\selectfont 0\egroup{} &  &  &  &  & \bgroup\fontsize{3}{5}\selectfont 0.15\egroup{} & \bgroup\fontsize{3}{5}\selectfont -0.25\egroup{} & \bgroup\fontsize{3}{5}\selectfont -0.01\egroup{} & \bgroup\fontsize{3}{5}\selectfont 0.1\egroup{}\\
\ttfamily{\hspace{1em}cd80} &  &  &  &  &  &  &  &  & \bgroup\fontsize{3}{5}\selectfont -0.22\egroup{} & \bgroup\fontsize{3}{5}\selectfont -0.1\egroup{} & \bgroup\fontsize{3}{5}\selectfont -0.03\egroup{} & \bgroup\fontsize{3}{5}\selectfont 0.35\egroup{}\\
\midrule
\addlinespace[0.3em]
\multicolumn{13}{l}{\textbf{II: modified treatment-free effect in \texttt{age}}}\\
\ttfamily{\hspace{1em}Intercept} & \bgroup\fontsize{3}{5}\selectfont 1.69\egroup{} & \bgroup\fontsize{3}{5}\selectfont -3.98\egroup{} & \bgroup\fontsize{6}{8}\selectfont \textbf{16.8}\egroup{} & \bgroup\fontsize{6}{8}\selectfont \textbf{-14.5}\egroup{} &  &  &  &  & \bgroup\fontsize{3}{5}\selectfont 3.44\egroup{} & \bgroup\fontsize{6}{8}\selectfont \textbf{-15.74}\egroup{} & \bgroup\fontsize{6}{8}\selectfont \textbf{11.17}\egroup{} & \bgroup\fontsize{3}{5}\selectfont 1.12\egroup{}\\
\ttfamily{\hspace{1em}gender} & \bgroup\fontsize{3}{5}\selectfont -0.5\egroup{} & \bgroup\fontsize{6}{8}\selectfont \textbf{5.59}\egroup{} & \bgroup\fontsize{6}{8}\selectfont \textbf{-6.06}\egroup{} & \bgroup\fontsize{3}{5}\selectfont 0.97\egroup{} &  &  &  &  & \bgroup\fontsize{3}{5}\selectfont 0.95\egroup{} & \bgroup\fontsize{3}{5}\selectfont -0.28\egroup{} & \bgroup\fontsize{3}{5}\selectfont -1\egroup{} & \bgroup\fontsize{3}{5}\selectfont 0.34\egroup{}\\
\ttfamily{\hspace{1em}homo} & \bgroup\fontsize{3}{5}\selectfont 0.16\egroup{} & \bgroup\fontsize{3}{5}\selectfont -0.19\egroup{} & \bgroup\fontsize{3}{5}\selectfont 0.81\egroup{} & \bgroup\fontsize{3}{5}\selectfont -0.78\egroup{} &  &  &  &  & \bgroup\fontsize{3}{5}\selectfont 0.84\egroup{} & \bgroup\fontsize{3}{5}\selectfont -0.31\egroup{} & \bgroup\fontsize{3}{5}\selectfont -0.77\egroup{} & \bgroup\fontsize{3}{5}\selectfont 0.24\egroup{}\\
\ttfamily{\hspace{1em}race} & \bgroup\fontsize{3}{5}\selectfont -3.52\egroup{} & \bgroup\fontsize{3}{5}\selectfont 2.19\egroup{} & \bgroup\fontsize{3}{5}\selectfont -4.69\egroup{} & \bgroup\fontsize{6}{8}\selectfont \textbf{6.03}\egroup{} &  &  &  &  & \bgroup\fontsize{3}{5}\selectfont -0.76\egroup{} & \bgroup\fontsize{3}{5}\selectfont 0.46\egroup{} & \bgroup\fontsize{3}{5}\selectfont -1.87\egroup{} & \bgroup\fontsize{3}{5}\selectfont 2.18\egroup{}\\
\ttfamily{\hspace{1em}drugs} & \bgroup\fontsize{6}{8}\selectfont \textbf{-12.55}\egroup{} & \bgroup\fontsize{6}{8}\selectfont \textbf{-16.86}\egroup{} & \bgroup\fontsize{6}{8}\selectfont \textbf{24.37}\egroup{} & \bgroup\fontsize{6}{8}\selectfont \textbf{5.04}\egroup{} &  &  &  &  & \bgroup\fontsize{3}{5}\selectfont -0.64\egroup{} & \bgroup\fontsize{3}{5}\selectfont -0.4\egroup{} & \bgroup\fontsize{3}{5}\selectfont 0.49\egroup{} & \bgroup\fontsize{3}{5}\selectfont 0.55\egroup{}\\
\ttfamily{\hspace{1em}symptom} & \bgroup\fontsize{6}{8}\selectfont \textbf{13.4}\egroup{} & \bgroup\fontsize{6}{8}\selectfont \textbf{-8.08}\egroup{} & \bgroup\fontsize{6}{8}\selectfont \textbf{-9.37}\egroup{} & \bgroup\fontsize{3}{5}\selectfont 4.05\egroup{} &  &  &  &  &  &  &  & \\
\ttfamily{\hspace{1em}str2} & \bgroup\fontsize{6}{8}\selectfont \textbf{-14.12}\egroup{} & \bgroup\fontsize{3}{5}\selectfont -2.69\egroup{} & \bgroup\fontsize{6}{8}\selectfont \textbf{9.25}\egroup{} & \bgroup\fontsize{6}{8}\selectfont \textbf{7.56}\egroup{} &  &  &  &  & \bgroup\fontsize{3}{5}\selectfont -0.01\egroup{} & \bgroup\fontsize{3}{5}\selectfont -0.09\egroup{} & \bgroup\fontsize{3}{5}\selectfont -0.06\egroup{} & \bgroup\fontsize{3}{5}\selectfont 0.16\egroup{}\\
\ttfamily{\hspace{1em}hemo} & \bgroup\fontsize{3}{5}\selectfont -0.19\egroup{} & \bgroup\fontsize{3}{5}\selectfont 1.34\egroup{} & \bgroup\fontsize{3}{5}\selectfont 1.08\egroup{} & \bgroup\fontsize{3}{5}\selectfont -2.23\egroup{} &  &  &  &  & \bgroup\fontsize{3}{5}\selectfont 0.05\egroup{} & \bgroup\fontsize{3}{5}\selectfont -0.56\egroup{} & \bgroup\fontsize{3}{5}\selectfont -0.91\egroup{} & \bgroup\fontsize{3}{5}\selectfont 1.43\egroup{}\\
\ttfamily{\hspace{1em}age} & \bgroup\fontsize{3}{5}\selectfont 3.77\egroup{} & \bgroup\fontsize{6}{8}\selectfont \textbf{50.53}\egroup{} & \bgroup\fontsize{6}{8}\selectfont \textbf{20.2}\egroup{} & \bgroup\fontsize{6}{8}\selectfont \textbf{-74.49}\egroup{} &  &  &  &  & \bgroup\fontsize{3}{5}\selectfont -0.07\egroup{} & \bgroup\fontsize{3}{5}\selectfont 0.02\egroup{} & \bgroup\fontsize{3}{5}\selectfont 0.01\egroup{} & \bgroup\fontsize{3}{5}\selectfont 0.04\egroup{}\\
\ttfamily{\hspace{1em}wtkg} & \bgroup\fontsize{6}{8}\selectfont \textbf{-20.77}\egroup{} & \bgroup\fontsize{3}{5}\selectfont 0.78\egroup{} & \bgroup\fontsize{6}{8}\selectfont \textbf{11.18}\egroup{} & \bgroup\fontsize{6}{8}\selectfont \textbf{8.81}\egroup{} &  &  &  &  & \bgroup\fontsize{3}{5}\selectfont 1.08\egroup{} & \bgroup\fontsize{3}{5}\selectfont -0.71\egroup{} & \bgroup\fontsize{3}{5}\selectfont -1.79\egroup{} & \bgroup\fontsize{3}{5}\selectfont 1.42\egroup{}\\
\ttfamily{\hspace{1em}cd40} & \bgroup\fontsize{6}{8}\selectfont \textbf{16.05}\egroup{} & \bgroup\fontsize{6}{8}\selectfont \textbf{6.65}\egroup{} & \bgroup\fontsize{6}{8}\selectfont \textbf{-22.85}\egroup{} & \bgroup\fontsize{3}{5}\selectfont 0.15\egroup{} &  &  &  &  & \bgroup\fontsize{3}{5}\selectfont 1.58\egroup{} & \bgroup\fontsize{3}{5}\selectfont 2.24\egroup{} & \bgroup\fontsize{3}{5}\selectfont -4.34\egroup{} & \bgroup\fontsize{3}{5}\selectfont 0.53\egroup{}\\
\ttfamily{\hspace{1em}karnof} & \bgroup\fontsize{6}{8}\selectfont \textbf{-9.9}\egroup{} & \bgroup\fontsize{3}{5}\selectfont -1.56\egroup{} & \bgroup\fontsize{6}{8}\selectfont \textbf{9.31}\egroup{} & \bgroup\fontsize{3}{5}\selectfont 2.14\egroup{} &  &  &  &  & \bgroup\fontsize{3}{5}\selectfont -0.01\egroup{} & \bgroup\fontsize{3}{5}\selectfont 0\egroup{} & \bgroup\fontsize{3}{5}\selectfont 0.01\egroup{} & \bgroup\fontsize{3}{5}\selectfont -0.01\egroup{}\\
\ttfamily{\hspace{1em}cd80} & \bgroup\fontsize{6}{8}\selectfont \textbf{-18.92}\egroup{} & \bgroup\fontsize{6}{8}\selectfont \textbf{7.55}\egroup{} & \bgroup\fontsize{6}{8}\selectfont \textbf{17.75}\egroup{} & \bgroup\fontsize{6}{8}\selectfont \textbf{-6.37}\egroup{} &  &  &  &  & \bgroup\fontsize{3}{5}\selectfont -0.18\egroup{} & \bgroup\fontsize{3}{5}\selectfont 0.13\egroup{} & \bgroup\fontsize{3}{5}\selectfont -0.24\egroup{} & \bgroup\fontsize{3}{5}\selectfont 0.28\egroup{}\\
\midrule
\addlinespace[0.3em]
\multicolumn{13}{l}{\textbf{III: modified variance function in \texttt{wtkg}}}\\
\ttfamily{\hspace{1em}Intercept} &  &  &  &  &  &  &  &  & \bgroup\fontsize{3}{5}\selectfont 0.02\egroup{} & \bgroup\fontsize{6}{8}\selectfont \textbf{-22.65}\egroup{} & \bgroup\fontsize{6}{8}\selectfont \textbf{27.54}\egroup{} & \bgroup\fontsize{3}{5}\selectfont -4.91\egroup{}\\
\ttfamily{\hspace{1em}gender} &  &  &  &  &  &  &  &  & \bgroup\fontsize{3}{5}\selectfont 0.74\egroup{} & \bgroup\fontsize{3}{5}\selectfont 3.24\egroup{} & \bgroup\fontsize{6}{8}\selectfont \textbf{-8.76}\egroup{} & \bgroup\fontsize{3}{5}\selectfont 4.78\egroup{}\\
\ttfamily{\hspace{1em}homo} &  &  &  &  &  &  &  &  & \bgroup\fontsize{3}{5}\selectfont 4.36\egroup{} & \bgroup\fontsize{3}{5}\selectfont -1.31\egroup{} & \bgroup\fontsize{3}{5}\selectfont 1.44\egroup{} & \bgroup\fontsize{3}{5}\selectfont -4.49\egroup{}\\
\ttfamily{\hspace{1em}race} &  &  &  &  &  &  &  &  & \bgroup\fontsize{3}{5}\selectfont -2.17\egroup{} & \bgroup\fontsize{3}{5}\selectfont -0.28\egroup{} & \bgroup\fontsize{3}{5}\selectfont 3.72\egroup{} & \bgroup\fontsize{3}{5}\selectfont -1.28\egroup{}\\
\ttfamily{\hspace{1em}drugs} &  &  &  &  &  &  &  &  & \bgroup\fontsize{3}{5}\selectfont -1.77\egroup{} & \bgroup\fontsize{3}{5}\selectfont -0.9\egroup{} & \bgroup\fontsize{3}{5}\selectfont -0.51\egroup{} & \bgroup\fontsize{3}{5}\selectfont 3.18\egroup{}\\
\ttfamily{\hspace{1em}symptom} &  &  &  &  &  &  &  &  & \bgroup\fontsize{3}{5}\selectfont -1.06\egroup{} & \bgroup\fontsize{3}{5}\selectfont -0.78\egroup{} & \bgroup\fontsize{3}{5}\selectfont -1.01\egroup{} & \bgroup\fontsize{3}{5}\selectfont 2.85\egroup{}\\
\ttfamily{\hspace{1em}str2} &  &  &  &  &  &  &  &  & \bgroup\fontsize{3}{5}\selectfont 0.72\egroup{} & \bgroup\fontsize{3}{5}\selectfont -1.68\egroup{} & \bgroup\fontsize{3}{5}\selectfont -1.18\egroup{} & \bgroup\fontsize{3}{5}\selectfont 2.14\egroup{}\\
\ttfamily{\hspace{1em}hemo} &  &  &  &  &  &  &  &  & \bgroup\fontsize{3}{5}\selectfont 1.6\egroup{} & \bgroup\fontsize{3}{5}\selectfont -1.41\egroup{} & \bgroup\fontsize{3}{5}\selectfont 2.57\egroup{} & \bgroup\fontsize{3}{5}\selectfont -2.76\egroup{}\\
\ttfamily{\hspace{1em}age} &  &  &  &  &  &  &  &  & \bgroup\fontsize{3}{5}\selectfont 2.2\egroup{} & \bgroup\fontsize{3}{5}\selectfont -0.41\egroup{} & \bgroup\fontsize{3}{5}\selectfont 0.02\egroup{} & \bgroup\fontsize{3}{5}\selectfont -1.8\egroup{}\\
\ttfamily{\hspace{1em}wtkg} & \bgroup\fontsize{6}{8}\selectfont \textbf{-41.32}\egroup{} & \bgroup\fontsize{6}{8}\selectfont \textbf{-13.14}\egroup{} & \bgroup\fontsize{6}{8}\selectfont \textbf{87.41}\egroup{} & \bgroup\fontsize{6}{8}\selectfont \textbf{-32.96}\egroup{} & \bgroup\fontsize{6}{8}\selectfont \textbf{-15.54}\egroup{} & \bgroup\fontsize{6}{8}\selectfont \textbf{-5.29}\egroup{} & \bgroup\fontsize{6}{8}\selectfont \textbf{32.88}\egroup{} & \bgroup\fontsize{6}{8}\selectfont \textbf{-12.04}\egroup{} & \bgroup\fontsize{6}{8}\selectfont \textbf{-11.18}\egroup{} & \bgroup\fontsize{3}{5}\selectfont -1.19\egroup{} & \bgroup\fontsize{6}{8}\selectfont \textbf{16.93}\egroup{} & \bgroup\fontsize{3}{5}\selectfont -4.56\egroup{}\\
\ttfamily{\hspace{1em}cd40} &  &  &  &  &  &  &  &  & \bgroup\fontsize{3}{5}\selectfont 1.94\egroup{} & \bgroup\fontsize{3}{5}\selectfont -0.72\egroup{} & \bgroup\fontsize{3}{5}\selectfont 1.74\egroup{} & \bgroup\fontsize{3}{5}\selectfont -2.96\egroup{}\\
\ttfamily{\hspace{1em}karnof} &  &  &  &  &  &  &  &  & \bgroup\fontsize{3}{5}\selectfont 2.84\egroup{} & \bgroup\fontsize{3}{5}\selectfont -1.02\egroup{} & \bgroup\fontsize{3}{5}\selectfont -3.35\egroup{} & \bgroup\fontsize{3}{5}\selectfont 1.53\egroup{}\\
\ttfamily{\hspace{1em}cd80} & \bgroup\fontsize{6}{8}\selectfont \textbf{-20.1}\egroup{} & \bgroup\fontsize{6}{8}\selectfont \textbf{-20.18}\egroup{} & \bgroup\fontsize{6}{8}\selectfont \textbf{61}\egroup{} & \bgroup\fontsize{6}{8}\selectfont \textbf{-20.73}\egroup{} & \bgroup\fontsize{6}{8}\selectfont \textbf{-7.14}\egroup{} & \bgroup\fontsize{6}{8}\selectfont \textbf{-8.31}\egroup{} & \bgroup\fontsize{6}{8}\selectfont \textbf{24.04}\egroup{} & \bgroup\fontsize{6}{8}\selectfont \textbf{-8.59}\egroup{} & \bgroup\fontsize{3}{5}\selectfont -3.51\egroup{} & \bgroup\fontsize{3}{5}\selectfont -4.5\egroup{} & \bgroup\fontsize{6}{8}\selectfont \textbf{11.98}\egroup{} & \bgroup\fontsize{3}{5}\selectfont -3.97\egroup{}\\
\midrule
\addlinespace[0.3em]
\multicolumn{13}{l}{\textbf{IV = II + III: modified treatment-free effect in \texttt{age} and modified variance function in \texttt{wtkg}}}\\
\ttfamily{\hspace{1em}Intercept} & \bgroup\fontsize{3}{5}\selectfont 3.97\egroup{} & \bgroup\fontsize{3}{5}\selectfont -0.11\egroup{} & \bgroup\fontsize{3}{5}\selectfont -4.08\egroup{} & \bgroup\fontsize{3}{5}\selectfont 0.22\egroup{} &  &  &  &  & \bgroup\fontsize{3}{5}\selectfont 1.43\egroup{} & \bgroup\fontsize{6}{8}\selectfont \textbf{-24.28}\egroup{} & \bgroup\fontsize{6}{8}\selectfont \textbf{25.71}\egroup{} & \bgroup\fontsize{3}{5}\selectfont -2.87\egroup{}\\
\ttfamily{\hspace{1em}gender} & \bgroup\fontsize{6}{8}\selectfont \textbf{-7.29}\egroup{} & \bgroup\fontsize{3}{5}\selectfont 1.49\egroup{} & \bgroup\fontsize{6}{8}\selectfont \textbf{11.15}\egroup{} & \bgroup\fontsize{6}{8}\selectfont \textbf{-5.35}\egroup{} &  &  &  &  & \bgroup\fontsize{6}{8}\selectfont \textbf{5.8}\egroup{} & \bgroup\fontsize{3}{5}\selectfont 2.22\egroup{} & \bgroup\fontsize{6}{8}\selectfont \textbf{-5.33}\egroup{} & \bgroup\fontsize{3}{5}\selectfont -2.68\egroup{}\\
\ttfamily{\hspace{1em}homo} & \bgroup\fontsize{6}{8}\selectfont \textbf{5.02}\egroup{} & \bgroup\fontsize{3}{5}\selectfont 1.7\egroup{} & \bgroup\fontsize{6}{8}\selectfont \textbf{-10.68}\egroup{} & \bgroup\fontsize{3}{5}\selectfont 3.96\egroup{} &  &  &  &  & \bgroup\fontsize{6}{8}\selectfont \textbf{9.9}\egroup{} & \bgroup\fontsize{6}{8}\selectfont \textbf{-5.34}\egroup{} & \bgroup\fontsize{6}{8}\selectfont \textbf{-10.71}\egroup{} & \bgroup\fontsize{6}{8}\selectfont \textbf{6.16}\egroup{}\\
\ttfamily{\hspace{1em}race} & \bgroup\fontsize{6}{8}\selectfont \textbf{6.09}\egroup{} & \bgroup\fontsize{6}{8}\selectfont \textbf{10.4}\egroup{} & \bgroup\fontsize{6}{8}\selectfont \textbf{-27.53}\egroup{} & \bgroup\fontsize{6}{8}\selectfont \textbf{11.04}\egroup{} &  &  &  &  & \bgroup\fontsize{3}{5}\selectfont 4.47\egroup{} & \bgroup\fontsize{3}{5}\selectfont -0.29\egroup{} & \bgroup\fontsize{6}{8}\selectfont \textbf{-13.69}\egroup{} & \bgroup\fontsize{6}{8}\selectfont \textbf{9.51}\egroup{}\\
\ttfamily{\hspace{1em}drugs} & \bgroup\fontsize{3}{5}\selectfont -3.77\egroup{} & \bgroup\fontsize{3}{5}\selectfont -3.03\egroup{} & \bgroup\fontsize{6}{8}\selectfont \textbf{9.26}\egroup{} & \bgroup\fontsize{3}{5}\selectfont -2.45\egroup{} &  &  &  &  & \bgroup\fontsize{3}{5}\selectfont -1.61\egroup{} & \bgroup\fontsize{3}{5}\selectfont -2.18\egroup{} & \bgroup\fontsize{3}{5}\selectfont -0.12\egroup{} & \bgroup\fontsize{3}{5}\selectfont 3.9\egroup{}\\
\ttfamily{\hspace{1em}symptom} & \bgroup\fontsize{3}{5}\selectfont 1.18\egroup{} & \bgroup\fontsize{3}{5}\selectfont -0.57\egroup{} & \bgroup\fontsize{3}{5}\selectfont 0.56\egroup{} & \bgroup\fontsize{3}{5}\selectfont -1.17\egroup{} &  &  &  &  & \bgroup\fontsize{3}{5}\selectfont 1.79\egroup{} & \bgroup\fontsize{3}{5}\selectfont -0.47\egroup{} & \bgroup\fontsize{3}{5}\selectfont -1.29\egroup{} & \bgroup\fontsize{3}{5}\selectfont -0.03\egroup{}\\
\ttfamily{\hspace{1em}str2} & \bgroup\fontsize{6}{8}\selectfont \textbf{-5.19}\egroup{} & \bgroup\fontsize{3}{5}\selectfont 0.95\egroup{} & \bgroup\fontsize{6}{8}\selectfont \textbf{5.4}\egroup{} & \bgroup\fontsize{3}{5}\selectfont -1.17\egroup{} &  &  &  &  & \bgroup\fontsize{3}{5}\selectfont 1.98\egroup{} & \bgroup\fontsize{6}{8}\selectfont \textbf{-10.15}\egroup{} & \bgroup\fontsize{6}{8}\selectfont \textbf{-5.35}\egroup{} & \bgroup\fontsize{6}{8}\selectfont \textbf{13.52}\egroup{}\\
\ttfamily{\hspace{1em}hemo} & \bgroup\fontsize{3}{5}\selectfont 2.88\egroup{} & \bgroup\fontsize{3}{5}\selectfont 1.93\egroup{} & \bgroup\fontsize{6}{8}\selectfont \textbf{-6.75}\egroup{} & \bgroup\fontsize{3}{5}\selectfont 1.94\egroup{} &  &  &  &  & \bgroup\fontsize{3}{5}\selectfont 4.35\egroup{} & \bgroup\fontsize{6}{8}\selectfont \textbf{-8.64}\egroup{} & \bgroup\fontsize{3}{5}\selectfont -4.36\egroup{} & \bgroup\fontsize{6}{8}\selectfont \textbf{8.64}\egroup{}\\
\ttfamily{\hspace{1em}age} & \bgroup\fontsize{6}{8}\selectfont \textbf{-9.42}\egroup{} & \bgroup\fontsize{6}{8}\selectfont \textbf{11.74}\egroup{} & \bgroup\fontsize{6}{8}\selectfont \textbf{25.67}\egroup{} & \bgroup\fontsize{6}{8}\selectfont \textbf{-27.99}\egroup{} &  &  &  &  & \bgroup\fontsize{3}{5}\selectfont -4.6\egroup{} & \bgroup\fontsize{6}{8}\selectfont \textbf{5.59}\egroup{} & \bgroup\fontsize{3}{5}\selectfont 4.94\egroup{} & \bgroup\fontsize{6}{8}\selectfont \textbf{-5.93}\egroup{}\\
\ttfamily{\hspace{1em}wtkg} & \bgroup\fontsize{6}{8}\selectfont \textbf{65.69}\egroup{} & \bgroup\fontsize{6}{8}\selectfont \textbf{17.05}\egroup{} & \bgroup\fontsize{6}{8}\selectfont \textbf{-153.12}\egroup{} & \bgroup\fontsize{6}{8}\selectfont \textbf{70.38}\egroup{} & \bgroup\fontsize{6}{8}\selectfont \textbf{18.9}\egroup{} & \bgroup\fontsize{6}{8}\selectfont \textbf{5.13}\egroup{} & \bgroup\fontsize{6}{8}\selectfont \textbf{-43.57}\egroup{} & \bgroup\fontsize{6}{8}\selectfont \textbf{19.53}\egroup{} & \bgroup\fontsize{6}{8}\selectfont \textbf{6.31}\egroup{} & \bgroup\fontsize{6}{8}\selectfont \textbf{-9.97}\egroup{} & \bgroup\fontsize{3}{5}\selectfont -1.03\egroup{} & \bgroup\fontsize{3}{5}\selectfont 4.7\egroup{}\\
\ttfamily{\hspace{1em}cd40} & \bgroup\fontsize{6}{8}\selectfont \textbf{16.05}\egroup{} & \bgroup\fontsize{6}{8}\selectfont \textbf{14.91}\egroup{} & \bgroup\fontsize{6}{8}\selectfont \textbf{-44.87}\egroup{} & \bgroup\fontsize{6}{8}\selectfont \textbf{13.91}\egroup{} &  &  &  &  & \bgroup\fontsize{3}{5}\selectfont -1.12\egroup{} & \bgroup\fontsize{3}{5}\selectfont 0.11\egroup{} & \bgroup\fontsize{3}{5}\selectfont -0.22\egroup{} & \bgroup\fontsize{3}{5}\selectfont 1.23\egroup{}\\
\ttfamily{\hspace{1em}karnof} & \bgroup\fontsize{6}{8}\selectfont \textbf{-9.03}\egroup{} & \bgroup\fontsize{3}{5}\selectfont -1.3\egroup{} & \bgroup\fontsize{6}{8}\selectfont \textbf{16.11}\egroup{} & \bgroup\fontsize{6}{8}\selectfont \textbf{-5.78}\egroup{} &  &  &  &  & \bgroup\fontsize{6}{8}\selectfont \textbf{-5.12}\egroup{} & \bgroup\fontsize{3}{5}\selectfont 0.38\egroup{} & \bgroup\fontsize{3}{5}\selectfont 0.97\egroup{} & \bgroup\fontsize{3}{5}\selectfont 3.77\egroup{}\\
\ttfamily{\hspace{1em}cd80} & \bgroup\fontsize{6}{8}\selectfont \textbf{26.31}\egroup{} & \bgroup\fontsize{6}{8}\selectfont \textbf{40.74}\egroup{} & \bgroup\fontsize{6}{8}\selectfont \textbf{-95.75}\egroup{} & \bgroup\fontsize{6}{8}\selectfont \textbf{28.7}\egroup{} & \bgroup\fontsize{6}{8}\selectfont \textbf{5.37}\egroup{} & \bgroup\fontsize{6}{8}\selectfont \textbf{6.4}\egroup{} & \bgroup\fontsize{6}{8}\selectfont \textbf{-16.73}\egroup{} & \bgroup\fontsize{3}{5}\selectfont 4.96\egroup{} & \bgroup\fontsize{3}{5}\selectfont 0.98\egroup{} & \bgroup\fontsize{3}{5}\selectfont -1.14\egroup{} & \bgroup\fontsize{6}{8}\selectfont \textbf{-7.01}\egroup{} & \bgroup\fontsize{6}{8}\selectfont \textbf{7.18}\egroup{}\\
\bottomrule
\multicolumn{13}{l}{\rule{0pt}{1em}\underline{\textit{Note: }}}\\
\multicolumn{13}{l}{\rule{0pt}{1em}Larger coefficients encourage better outcome.}\\
\multicolumn{13}{l}{\rule{0pt}{1em}Coefficients are fitted at standardized scales of covariates.}\\
\multicolumn{13}{l}{\rule{0pt}{1em}Coefficients at blank are 0's. Absolute values $>$ 5 are bolded.}\\
\end{tabular}
\end{table}

\newpage


\begin{figure}[p]
	\centering
	\includegraphics[width=0.95\linewidth]{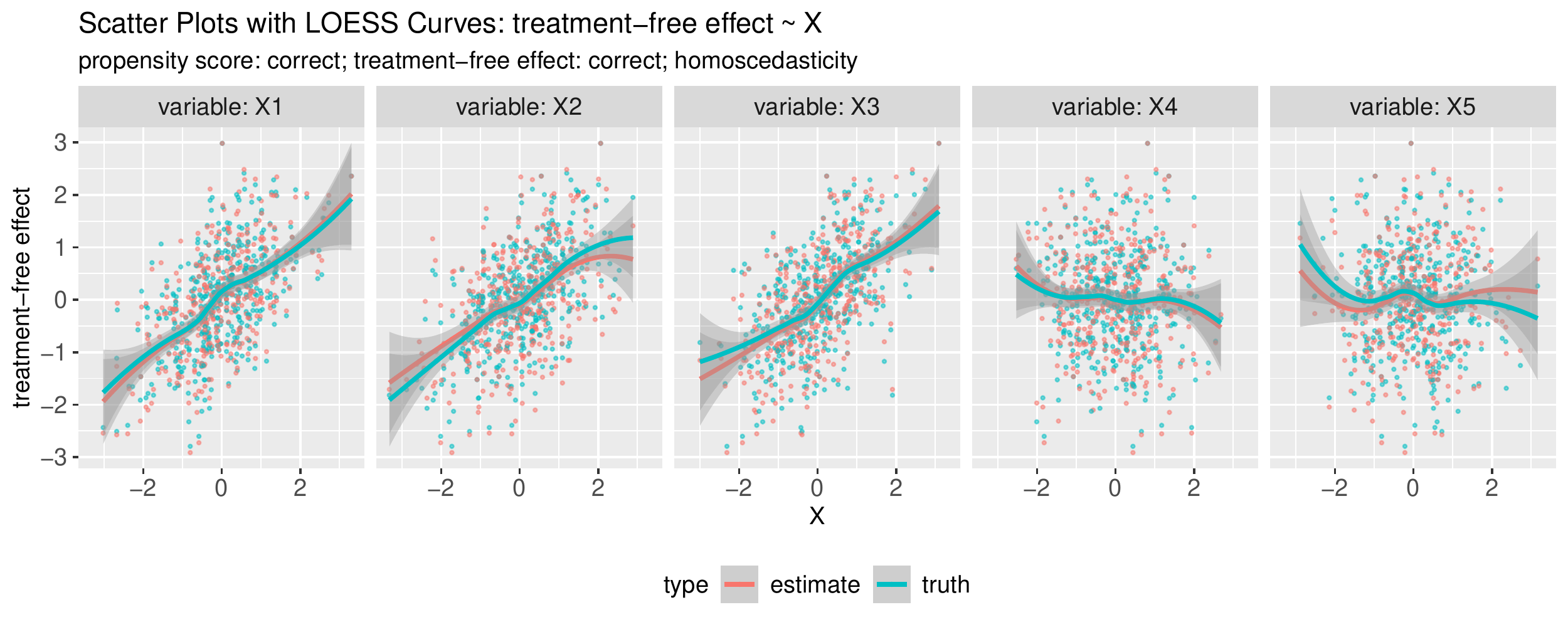}
	\includegraphics[width=0.95\linewidth]{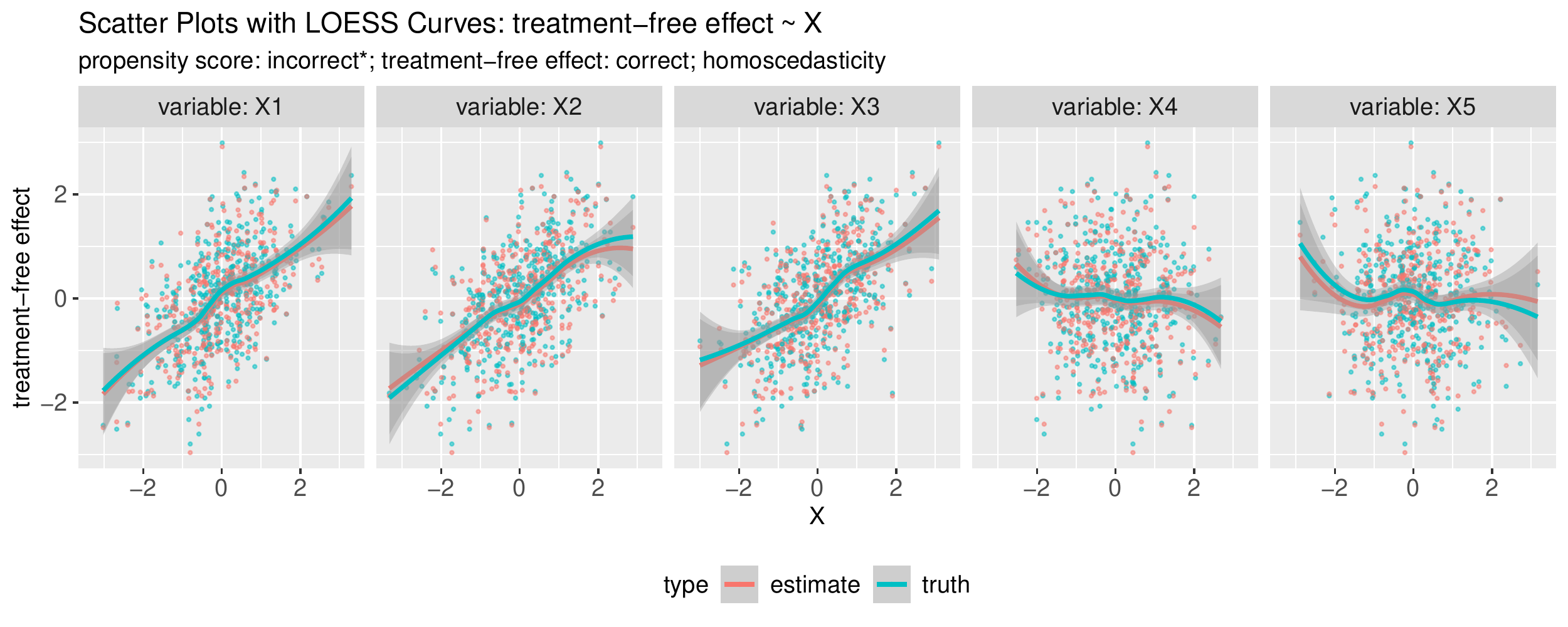}
	\includegraphics[width=0.95\linewidth]{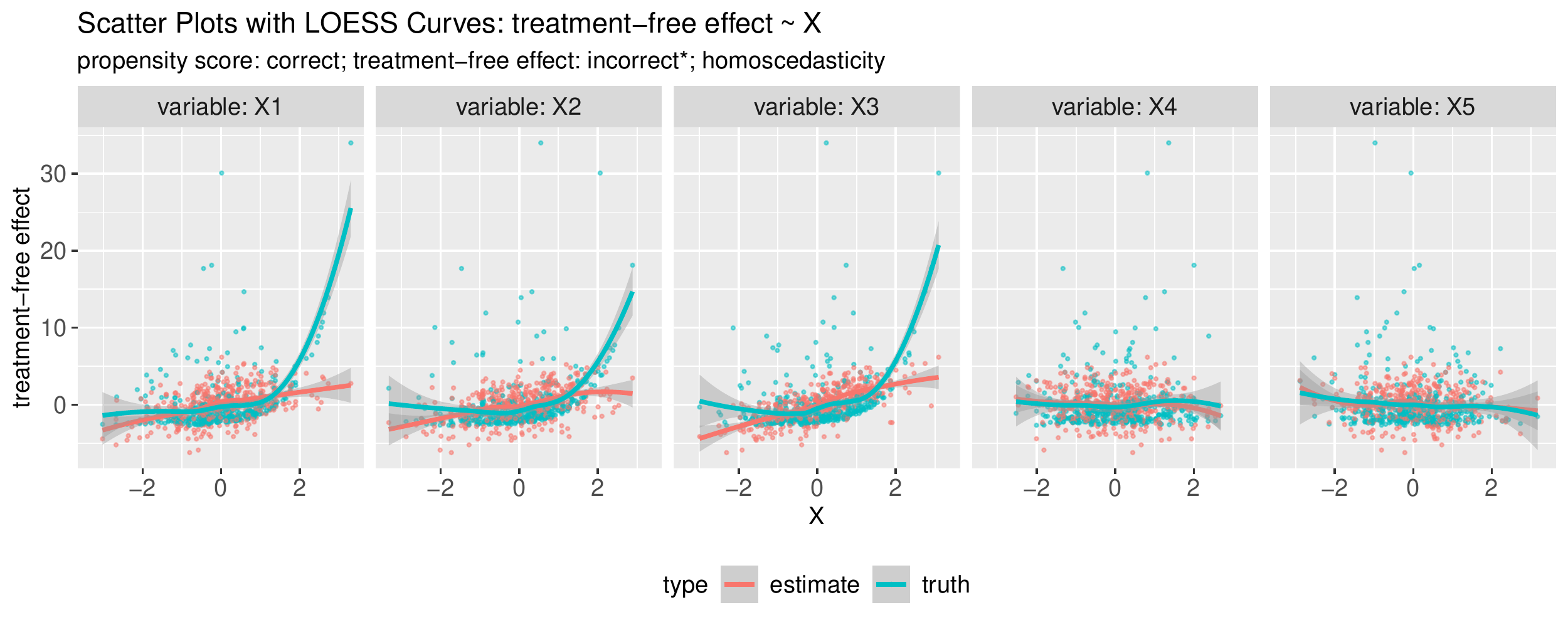}
	\caption[Simulation, Treatment-Free Effect Plots]{\footnotesize Fitted treatment-free effect plots with respect to $ X_{k} $ for $ 1 \le k \le 5 $ for the simulation studies (Section \ref{sec:simulation}) with $ n = 400 $, $ p = 10 $, $ K = 3 $. Curves are fitted by the \textit{LOcally wEighted Scatterplot Smoothing (LOESS)} of cubic spline. When the treatment-free effect model is correctly specified (Rows 1 and 2), it can be consistently estimated. Note that the treatment-free effect estimation utilizes the estimated propensity scores according to Section \ref{sec:tf}. The correctness of the treatment-free effect is not affected by the correctness of the propensity score model. When the treatment-free effect model is misspecified (Row 3), the estimated treatment-free effect deviates from the truth.}
	\label{fig:tf}
\end{figure}

\begin{figure}[p]
	\centering
	\includegraphics[width=0.95\linewidth]{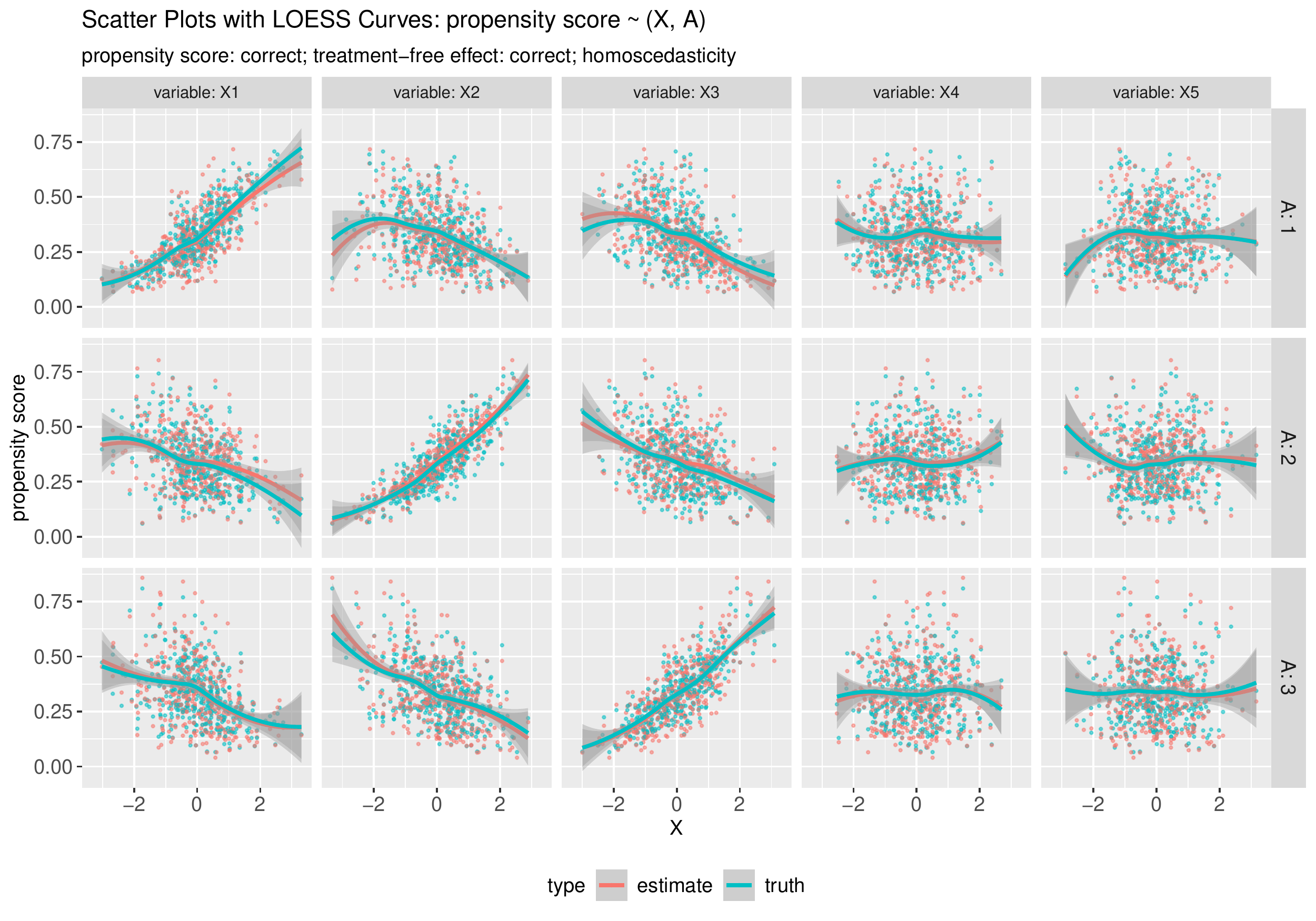}
	\includegraphics[width=0.95\linewidth]{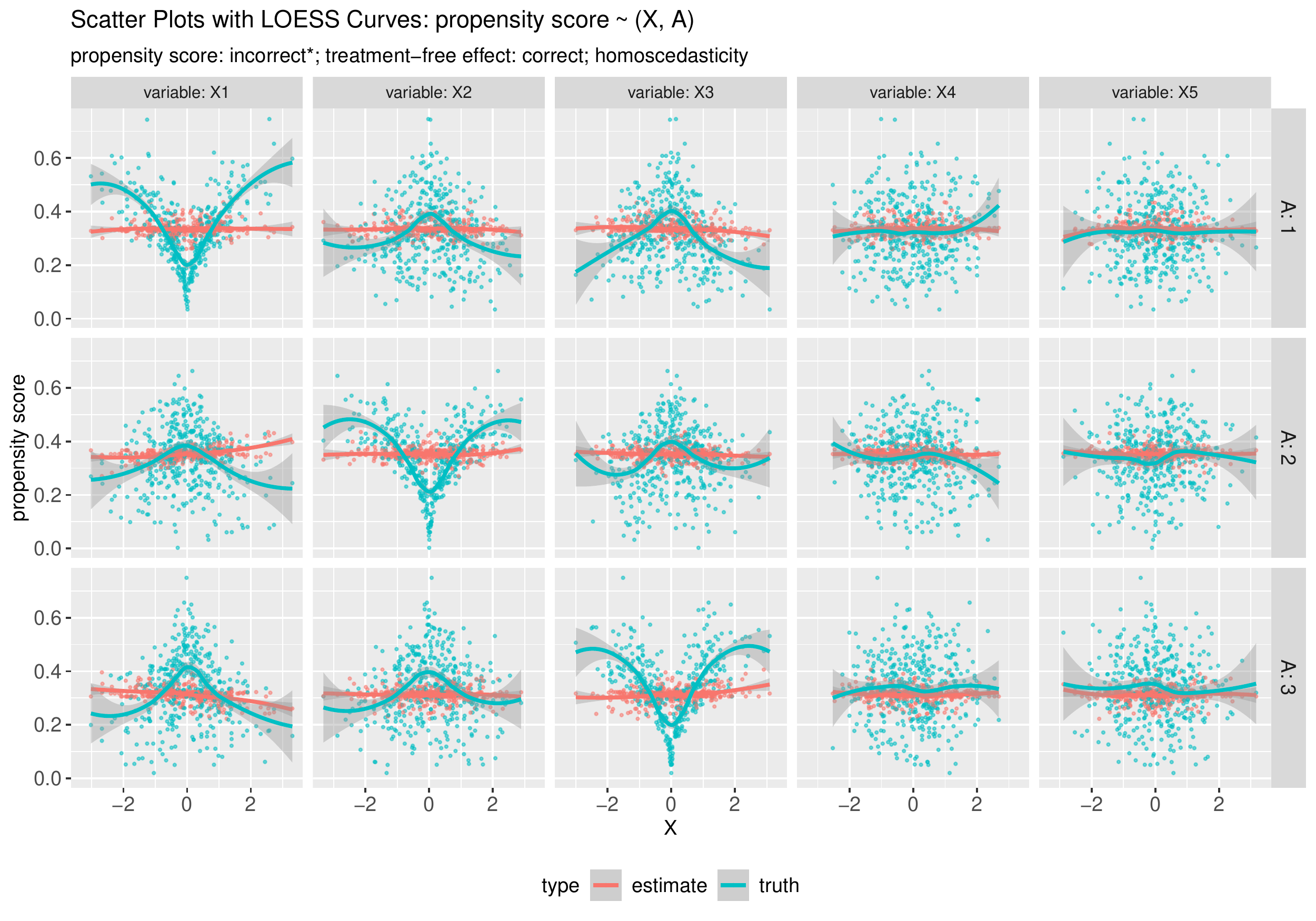}
	\caption[Simulation, Propensity Score Plots]{Fitted propensity score plots with respect to $ (X_{k},A) $ for $ 1 \le k \le 5 $ for the simulation studies (Section \ref{sec:simulation}) with $ n = 400 $, $ p = 10 $, $ K = 3 $. Curves are fitted by the LOESS of cubic spline. When the propensity score model is correctly specified (Panel 1), it can be consistently estimated. When the propensity score model is misspecified (Panel 2), the estimated propensity score deviates from the truth.}
	\label{fig:prop}
\end{figure}

\begin{figure}[p]
	\centering
	\includegraphics[width=\linewidth]{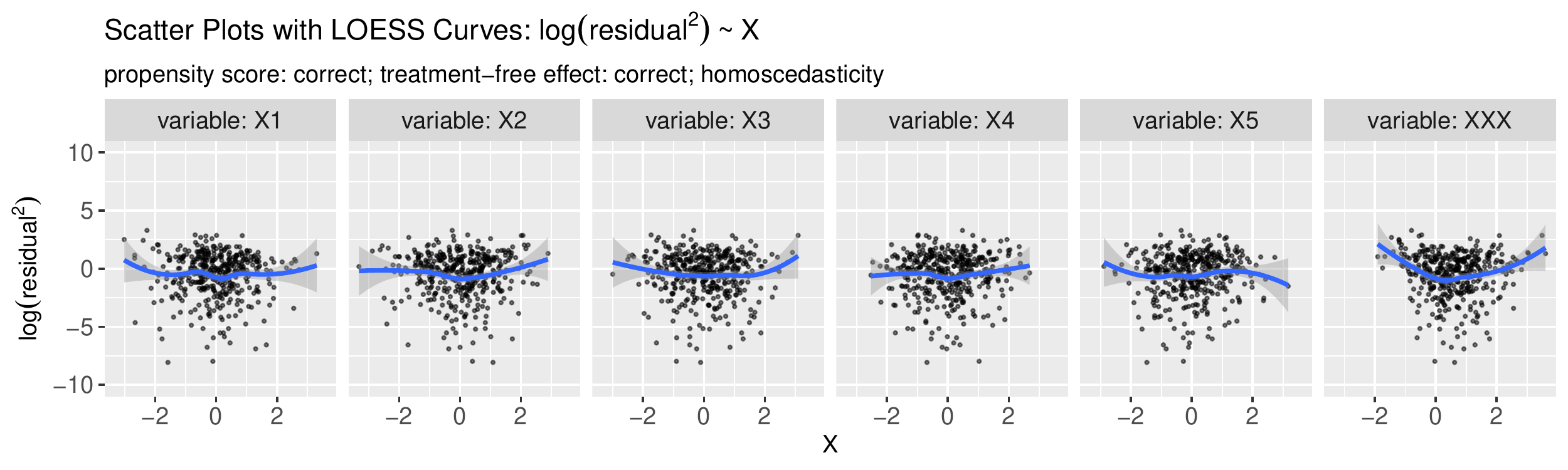}
	\includegraphics[width=\linewidth]{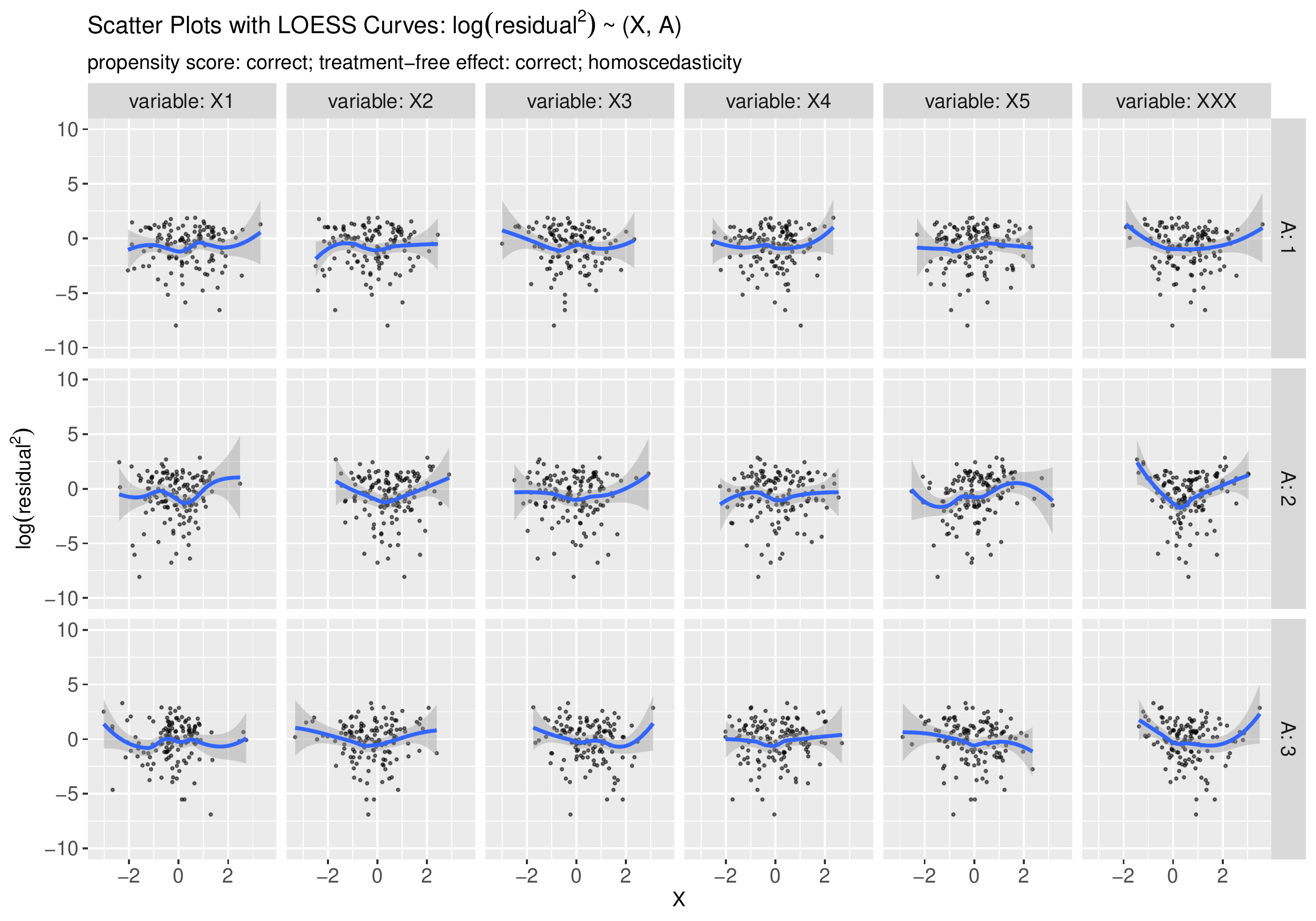}
	\caption[Simulation, Residual Plots, Correct Treatment-Free Effect, Homoscedasticity]{\footnotesize Residual plots with respect to $ X_{k} $ and $ (X_{k},A) $ for $ 1 \le k \le 5 $ for the simulation studies (Section \ref{sec:simulation}) with $ n = 400 $, $ p = 10 $, $ K = 3 $, correctly specified treatment-free effect and homoscedasticity. Define $ XXX := \log\left[ {1 \over 3}\left( e^{\sqrt{2}X_{1}} + e^{\sqrt{2}X_{2}} + e^{\sqrt{2}X_{3}} \right) \right] $. Residuals are computed from the fitted E-Learning. Curves are fitted by the LOESS of cubic spline. It shows no patterns of $ \log(\texttt{residual}^{2}) $ with respect to $ \bX $ or $ A $.}
	\label{fig:resid2_1}
\end{figure}

\begin{figure}[p]
	\centering
	\includegraphics[width=\linewidth]{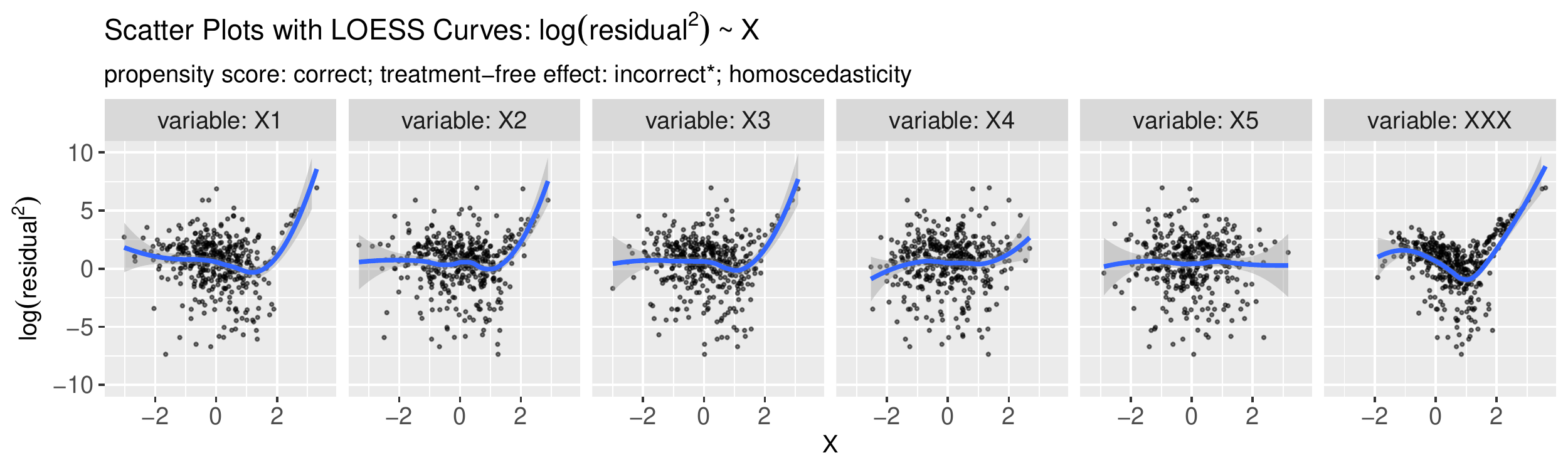}
	\includegraphics[width=\linewidth]{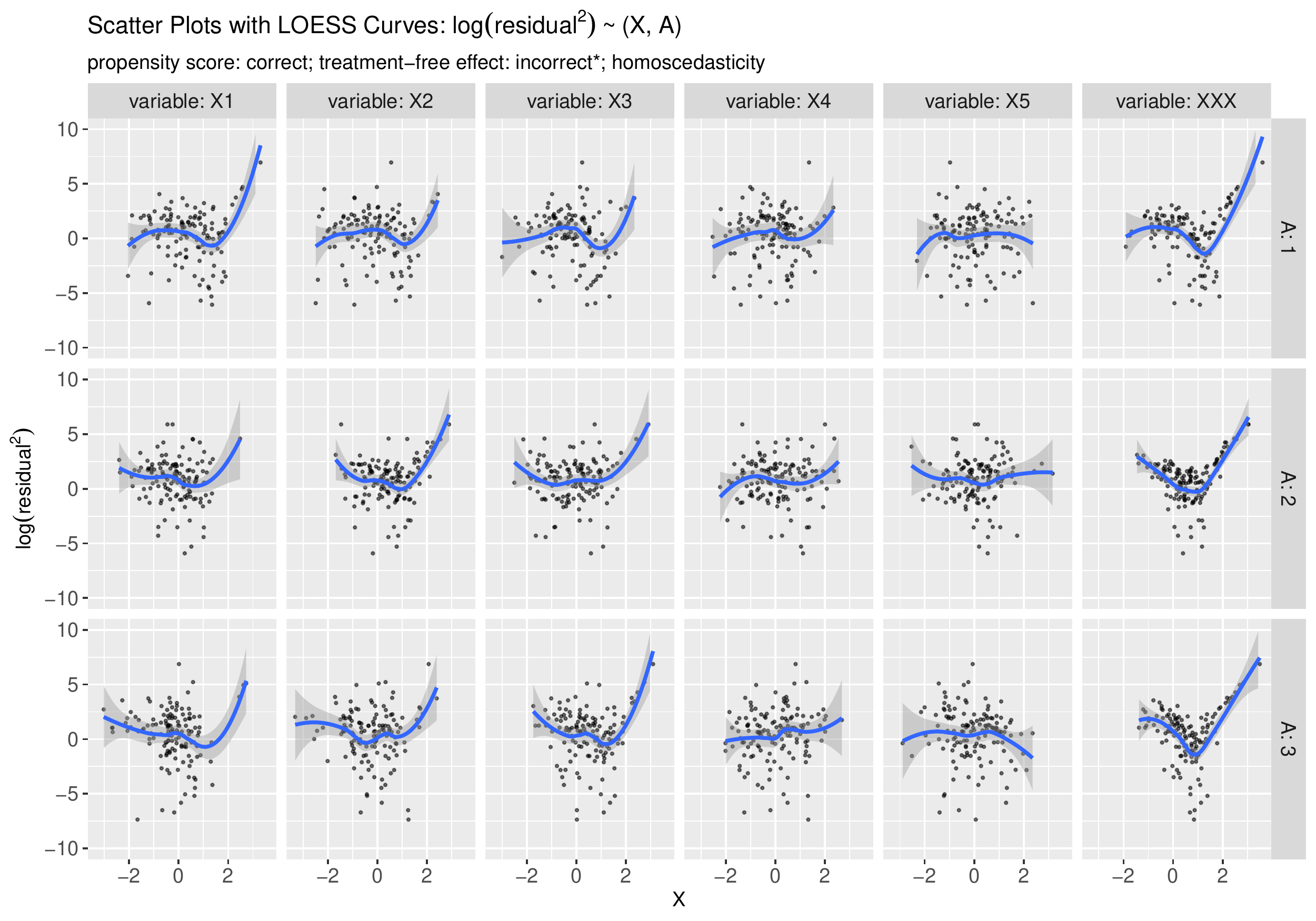}
	\caption[Simulation, Residual Plots, Incorrect Treatment-Free Effect, Homoscedasticity]{\footnotesize Residual plots with respect to $ X_{k} $ and $ (X_{k},A) $ for $ 1 \le k \le 5 $ for the simulation studies (Section \ref{sec:simulation}) with $ n = 400 $, $ p = 10 $, $ K = 3 $, misspecified treatment-free effect and homoscedasticity. Define $ XXX := \log\left[ {1 \over 3}\left( e^{\sqrt{2}X_{1}} + e^{\sqrt{2}X_{2}} + e^{\sqrt{2}X_{3}} \right) \right] $. Residuals are computed from the fitted E-Learning. Curves are fitted by the LOESS of cubic spline. It shows patterns of $ \log(\texttt{residual}^{2}) \sim X_{k} $ for $ k = 1,2,3 $ and $ \log(\texttt{residual}^{2}) \sim XXX $.}
	\label{fig:resid2_2}
\end{figure}

\begin{figure}[p]
	\centering
	\includegraphics[width=\linewidth]{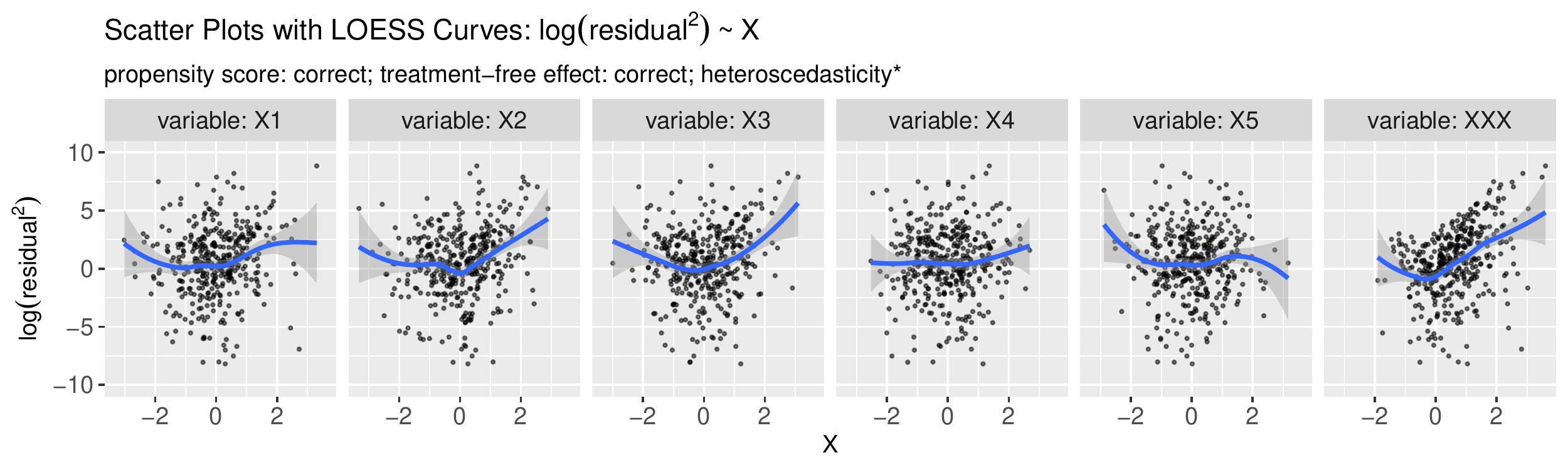}
	\includegraphics[width=\linewidth]{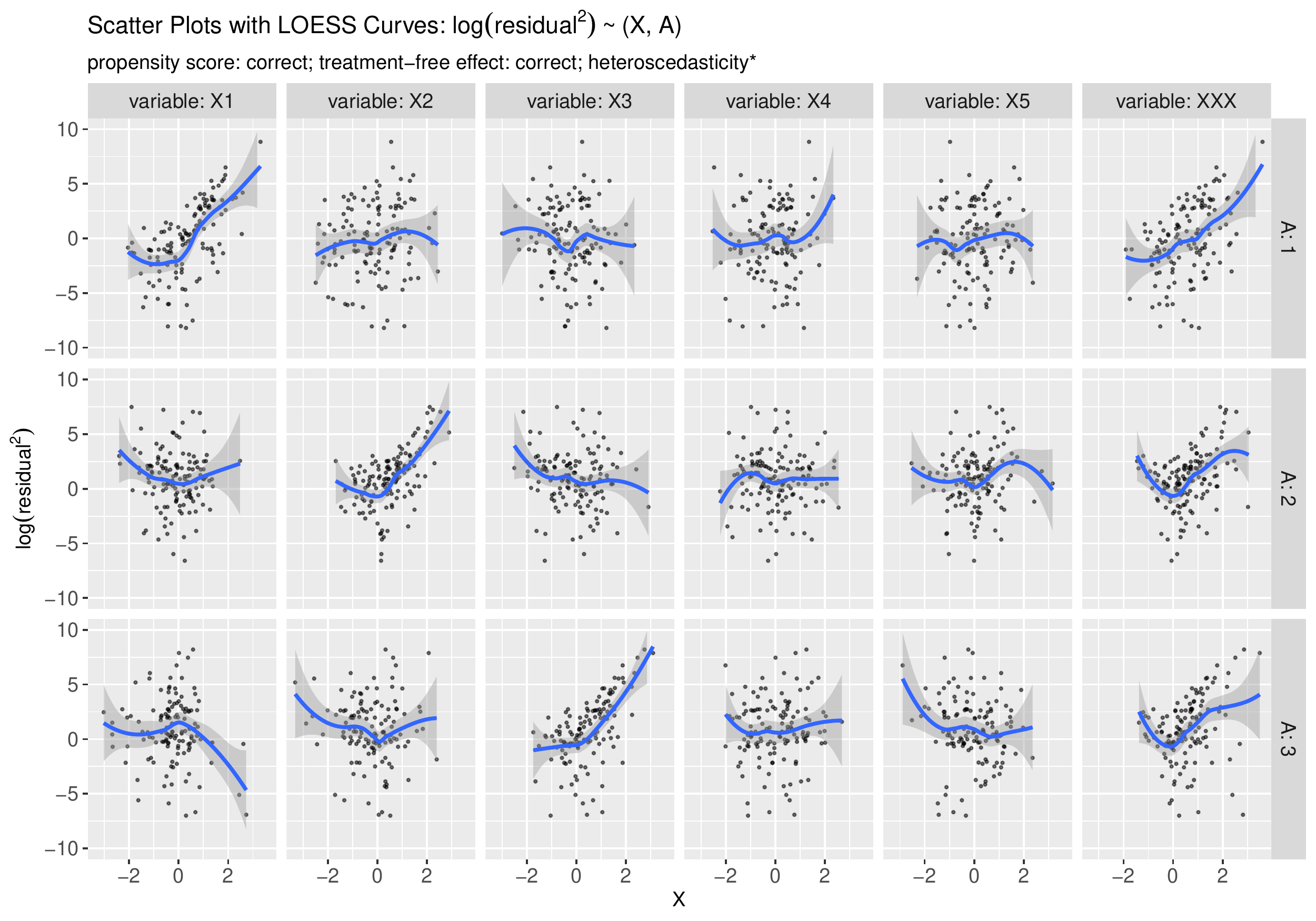}
	\caption[Simulation, Residual Plots, Correct Treatment-Free Effect, Heteroscedasticity]{\footnotesize Residual plots with respect to $ X_{k} $ and $ (X_{k},A) $ for $ 1 \le k \le 5 $ for the simulation studies (Section \ref{sec:simulation}) with $ n = 400 $, $ p = 10 $, $ K = 3 $, correctly specified treatment-free effect and heteroscedasticity. Define $ XXX := \log\left[ {1 \over 3}\left( e^{\sqrt{2}X_{1}} + e^{\sqrt{2}X_{2}} + e^{\sqrt{2}X_{3}} \right) \right] $. Residuals are computed from the fitted E-Learning. Curves are fitted by the LOESS of cubic spline. It shows patterns of $ \log(\texttt{residual}^{2}) \sim X_{k} $ on $ A = k $ for $ k = 1,2,3 $ and $ \log(\texttt{residual}^{2}) \sim XXX $.}
	\label{fig:resid2_3}
\end{figure}

\begin{figure}[p]
	\centering
	\includegraphics[width=\linewidth]{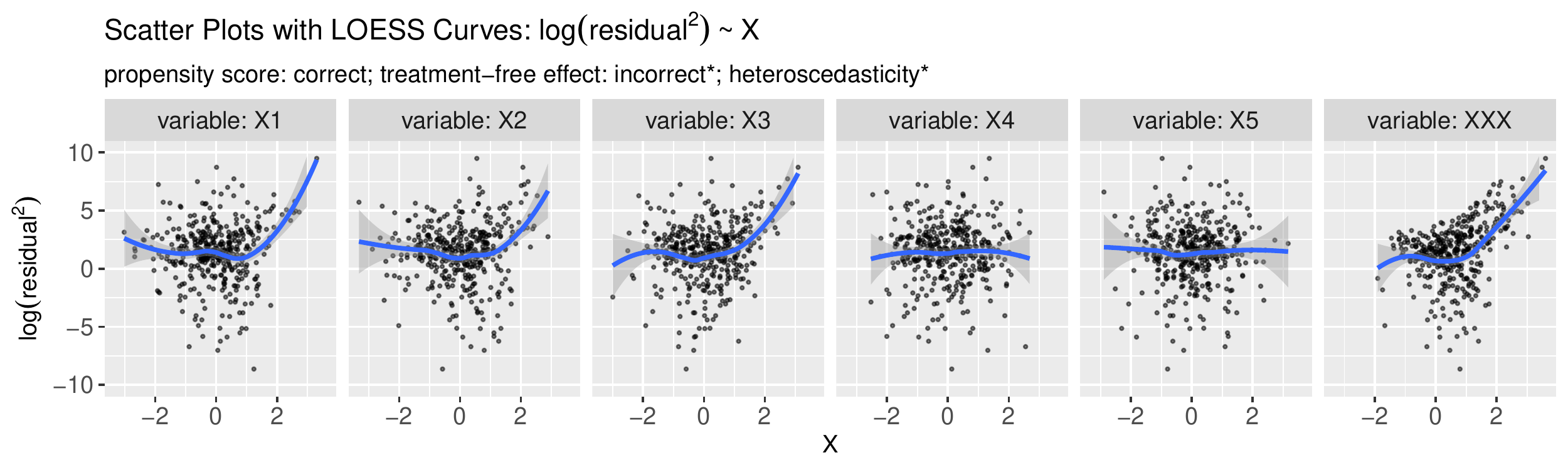}
	\includegraphics[width=\linewidth]{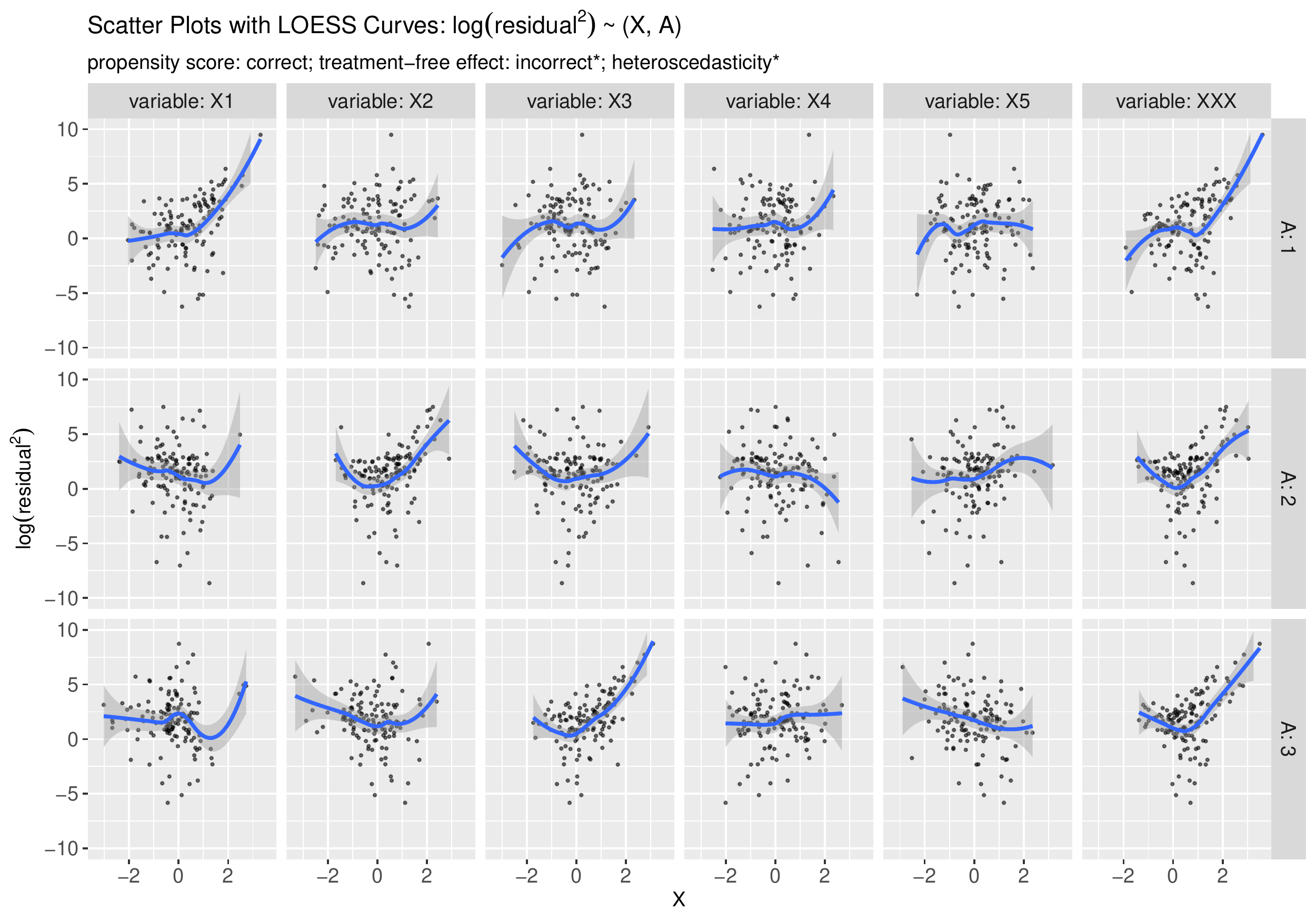}
	\caption[Simulation, Residual Plots, Incorrect Treatment-Free Effect, Heteroscedasticity]{\footnotesize Residual plots with respect to $ X_{k} $ and $ (X_{k},A) $ for $ 1 \le k \le 5 $ for the simulation studies (Section \ref{sec:simulation}) with $ n = 400 $, $ p = 10 $, $ K = 3 $, misspecified treatment-free effect and heteroscedasticity. Define $ XXX := \log\left[ {1 \over 3}\left( e^{\sqrt{2}X_{1}} + e^{\sqrt{2}X_{2}} + e^{\sqrt{2}X_{3}} \right) \right] $. Residuals are computed from the fitted E-Learning. Curves are fitted by the LOESS of cubic spline. It shows patterns of $ \log(\texttt{residual}^{2}) \sim X_{k} $ on $ A = k $ for $ k = 1,2,3 $ and $ \log(\texttt{residual}^{2}) \sim XXX $.}
	\label{fig:resid2_4}
\end{figure}


\begin{figure}[p]
	\centering
	\includegraphics[width=\linewidth]{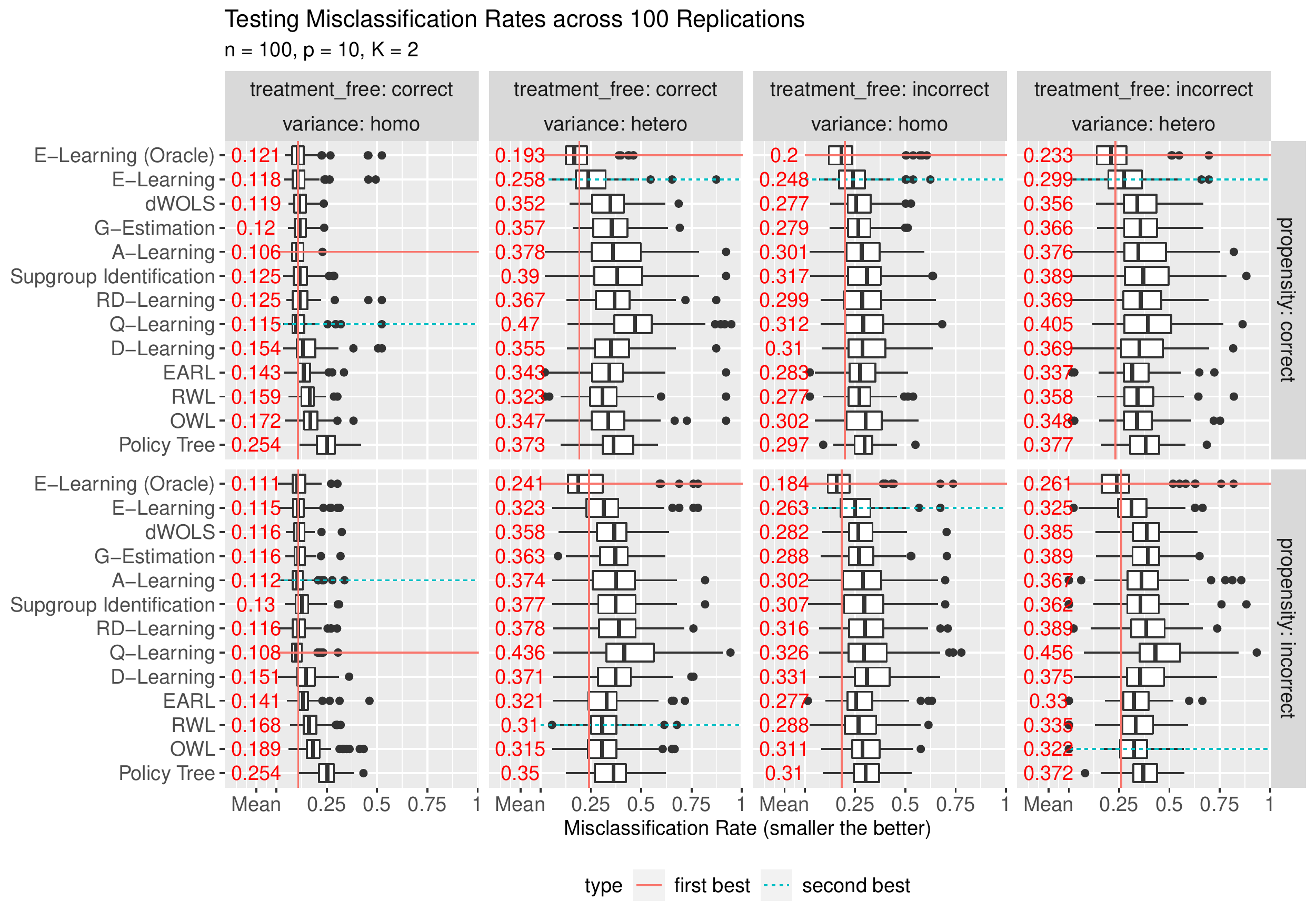}
	\caption[Simulation, Misclassification Rates, $ n = 100 $]{Testing misclassification rates (smaller the better) for $ n = 100 $, $ p = 10 $, $ K = 2 $ and each of the model specification scenarios in Table \ref{tab:model}. Methods in Table \ref{tab:compare} are compared, where \textit{E-Learning (Oracle)} corresponds to E-Learning with the oracle working variance function, and \textit{Policy Tree} corresponds to \textit{Policy Learning} with decision trees. First and second best methods in terms of the averaged misclassification rates are annotated in horizontal lines, while the minimal averaged misclassification rate is annotated in the vertical line.}
	\label{fig:all_n=100_misclass}
\end{figure}

\begin{figure}[p]
	\centering
	\includegraphics[width=\linewidth]{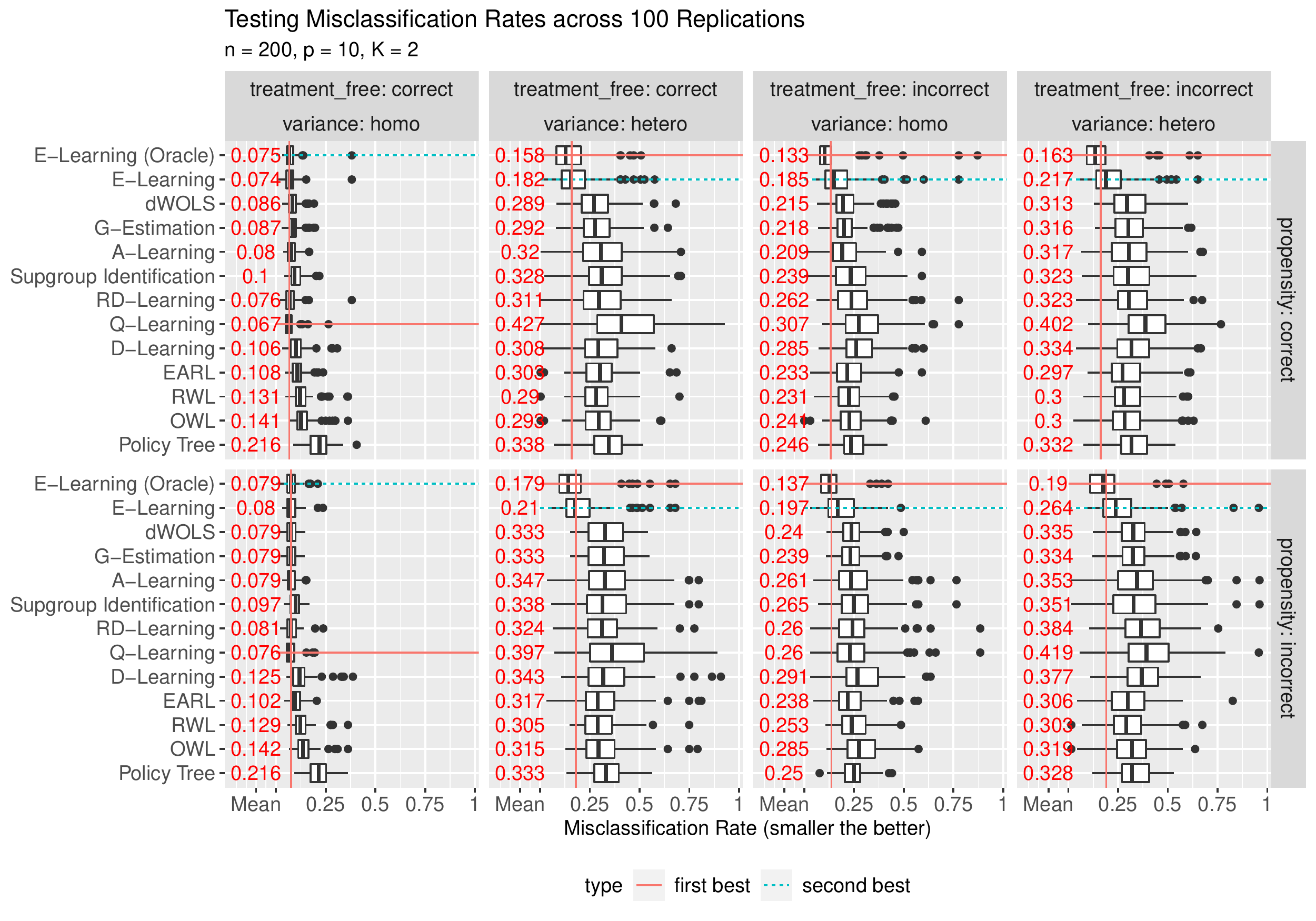}
	\caption[Simulation, Misclassification Rates, $ n = 200 $]{Testing misclassification rates (smaller the better) for $ n = 200 $, $ p = 10 $, $ K = 2 $ and each of the model specification scenarios in Table \ref{tab:model}. Methods in Table \ref{tab:compare} are compared, where \textit{E-Learning (Oracle)} corresponds to E-Learning with the oracle working variance function, and \textit{Policy Tree} corresponds to \textit{Policy Learning} with decision trees. First and second best methods in terms of the averaged misclassification rates are annotated in horizontal lines, while the minimal averaged misclassification rate is annotated in the vertical line.}
	\label{fig:all_n=200_misclass}
\end{figure}

\begin{figure}[p]
	\centering
	\includegraphics[width=\linewidth]{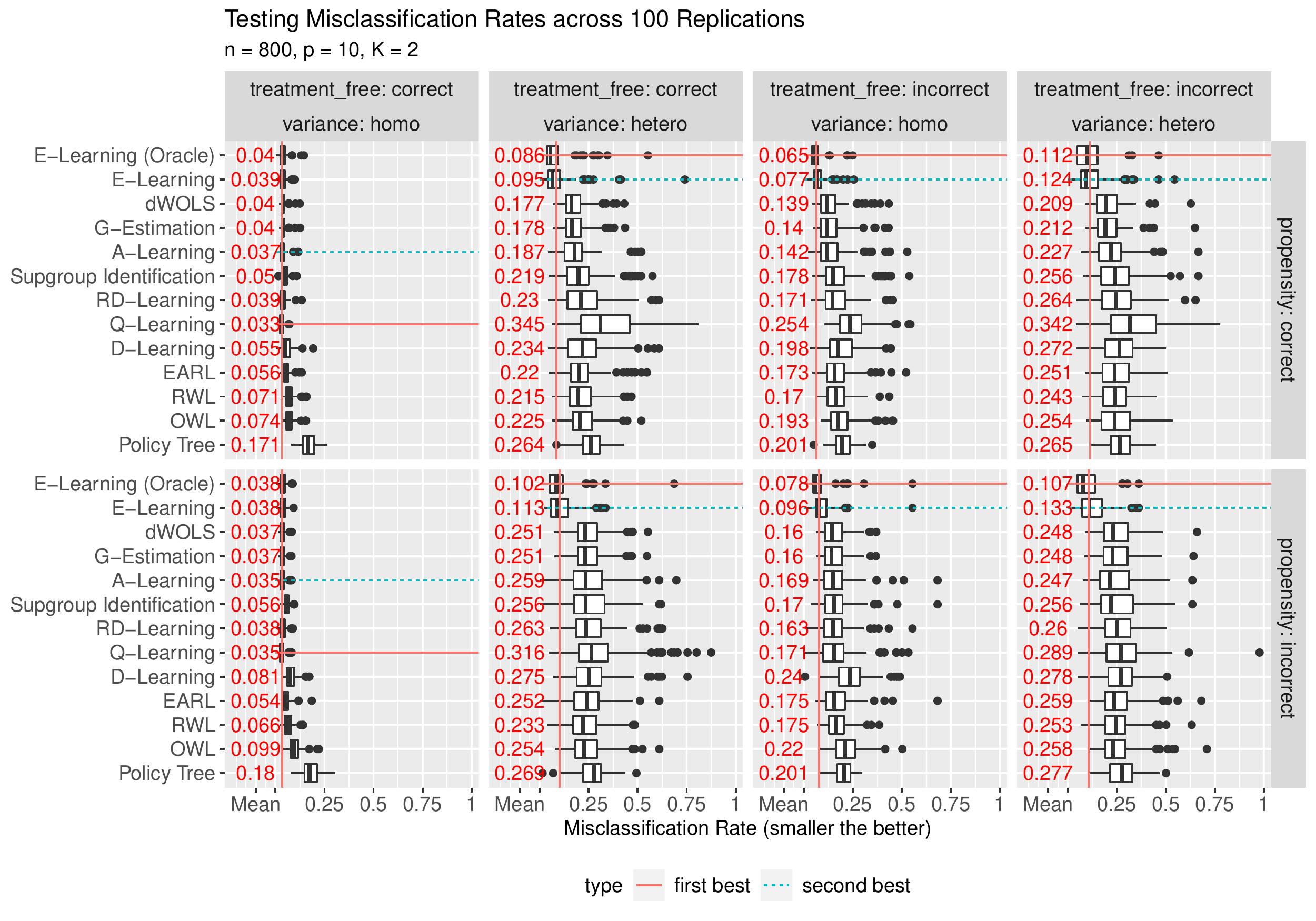}
	\caption[Simulation, Misclassification Rates, $ n = 800 $]{Testing misclassification rates (smaller the better) for $ n = 800 $, $ p = 10 $, $ K = 2 $ and each of the model specification scenarios in Table \ref{tab:model}. Methods in Table \ref{tab:compare} are compared, where \textit{E-Learning (Oracle)} corresponds to E-Learning with the oracle working variance function, and \textit{Policy Tree} corresponds to \textit{Policy Learning} with decision trees. First and second best methods in terms of the averaged misclassification rates are annotated in horizontal lines, while the minimal averaged misclassification rate is annotated in the vertical line.}
	\label{fig:all_n=800_misclass}
\end{figure}

\begin{figure}[p]
	\centering
	\includegraphics[width=\linewidth]{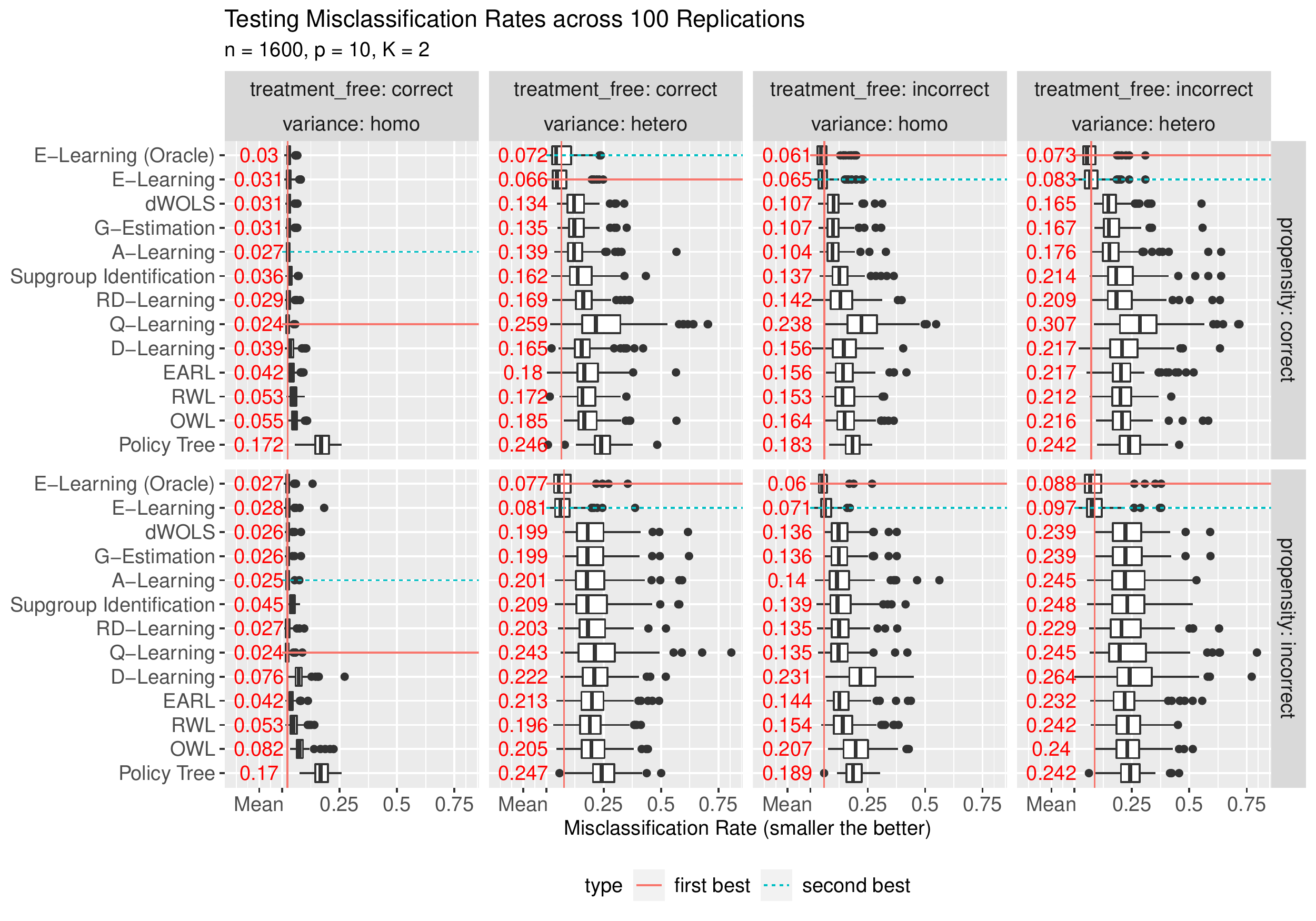}
	\caption[Simulation, Misclassification Rates, $ n = 1600 $]{Testing misclassification rates (smaller the better) for $ n = 1600 $, $ p = 10 $, $ K = 2 $ and each of the model specification scenarios in Table \ref{tab:model}. Methods in Table \ref{tab:compare} are compared, where \textit{E-Learning (Oracle)} corresponds to E-Learning with the oracle working variance function, and \textit{Policy Tree} corresponds to \textit{Policy Learning} with decision trees. First and second best methods in terms of the averaged misclassification rates are annotated in horizontal lines, while the minimal averaged misclassification rate is annotated in the vertical line.}
	\label{fig:all_n=1600_misclass}
\end{figure}

\begin{figure}[p]
	\centering
	\includegraphics[width=\linewidth]{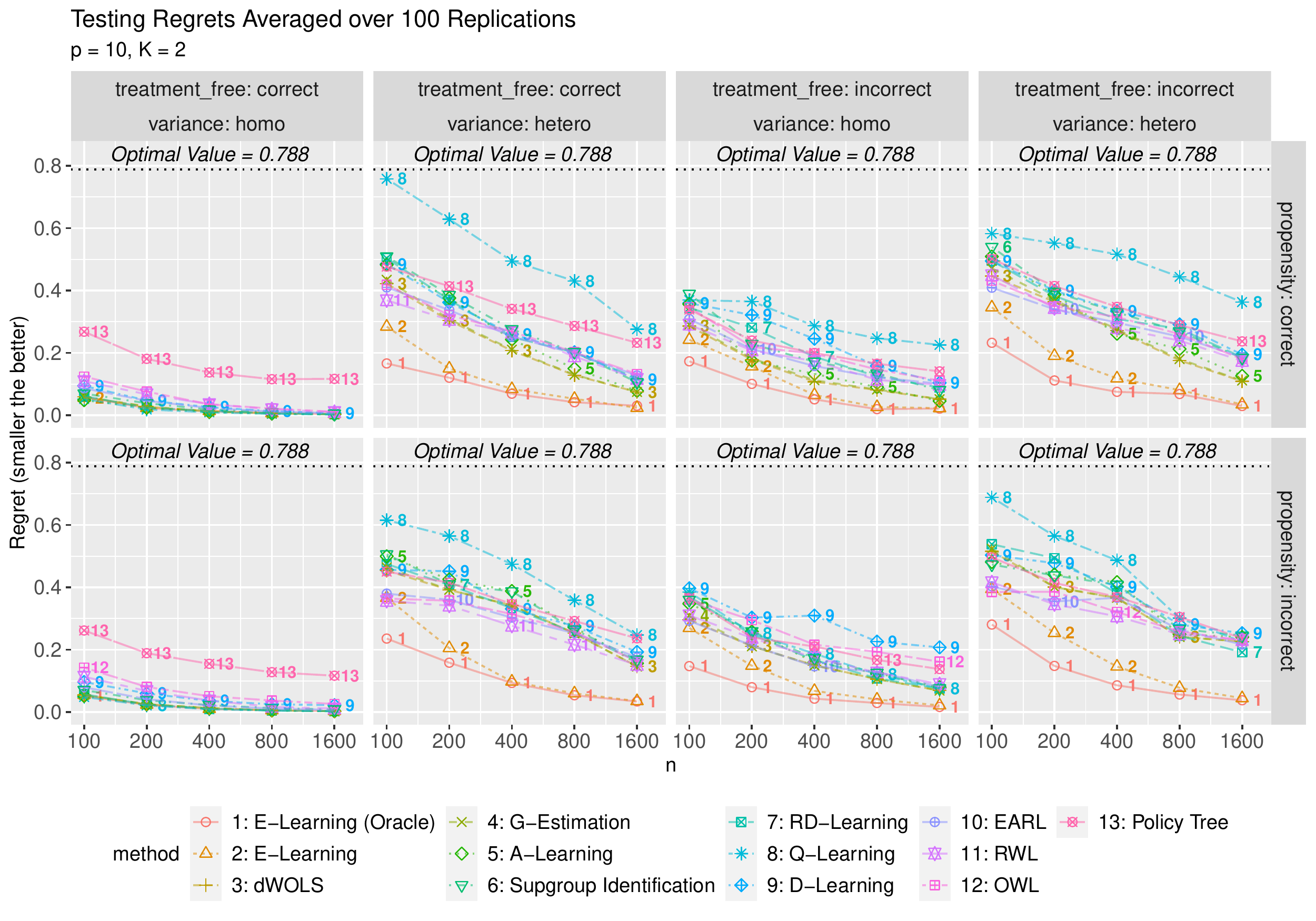}
	\caption[Simulation, Regrets]{Testing regrets (smaller the better) for $ n = \{100,200,400,800,1600\}  $, $ p = 10 $, $ K = 2 $ and each of the model specification scenarios in Table \ref{tab:model}. Methods in Table \ref{tab:compare} are compared, where \textit{E-Learning (Oracle)} corresponds to E-Learning with the oracle working variance function, and \textit{Policy Tree} corresponds to \textit{Policy Learning} with decision trees.}
	\label{fig:all_regret}
\end{figure}

\begin{figure}[p]
	\centering
	\includegraphics[width=\linewidth]{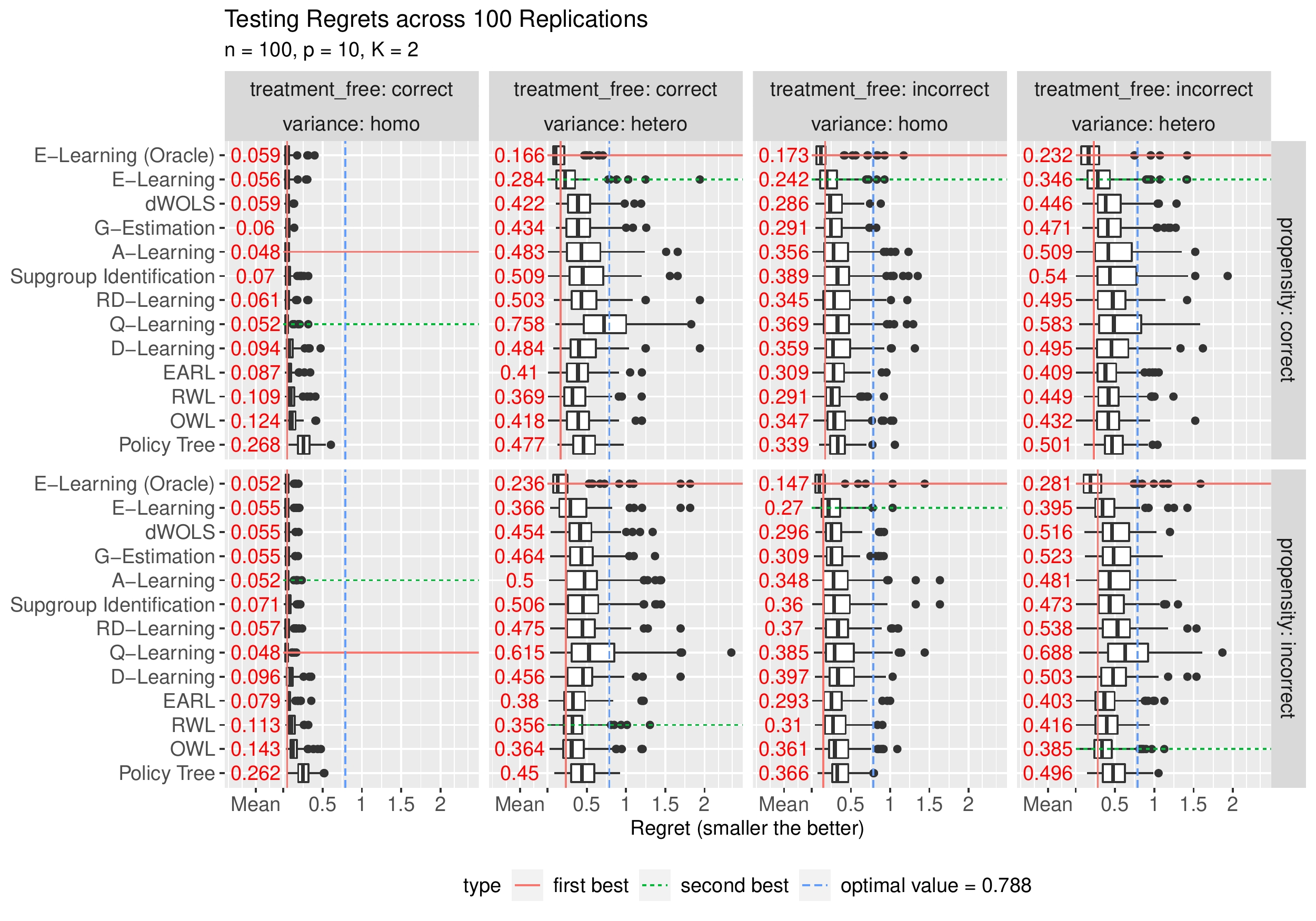}
	\caption[Simulation, Regrets, $ n = 100 $]{Testing regrets (smaller the better) for $ n = 100 $, $ p = 10 $, $ K = 2 $ and each of the model specification scenarios in Table \ref{tab:model}. Methods in Table \ref{tab:compare} are compared, where \textit{E-Learning (Oracle)} corresponds to E-Learning with the oracle working variance function, and \textit{Policy Tree} corresponds to \textit{Policy Learning} with decision trees. First and second best methods in terms of the averaged regrets are annotated in horizontal lines, while the minimal averaged regret is annotated in the vertical line. The optimal value is 0.788 and is annotated in the vertical long dashed line.}
	\label{fig:all_n=100_regret}
\end{figure}

\begin{figure}[p]
	\centering
	\includegraphics[width=\linewidth]{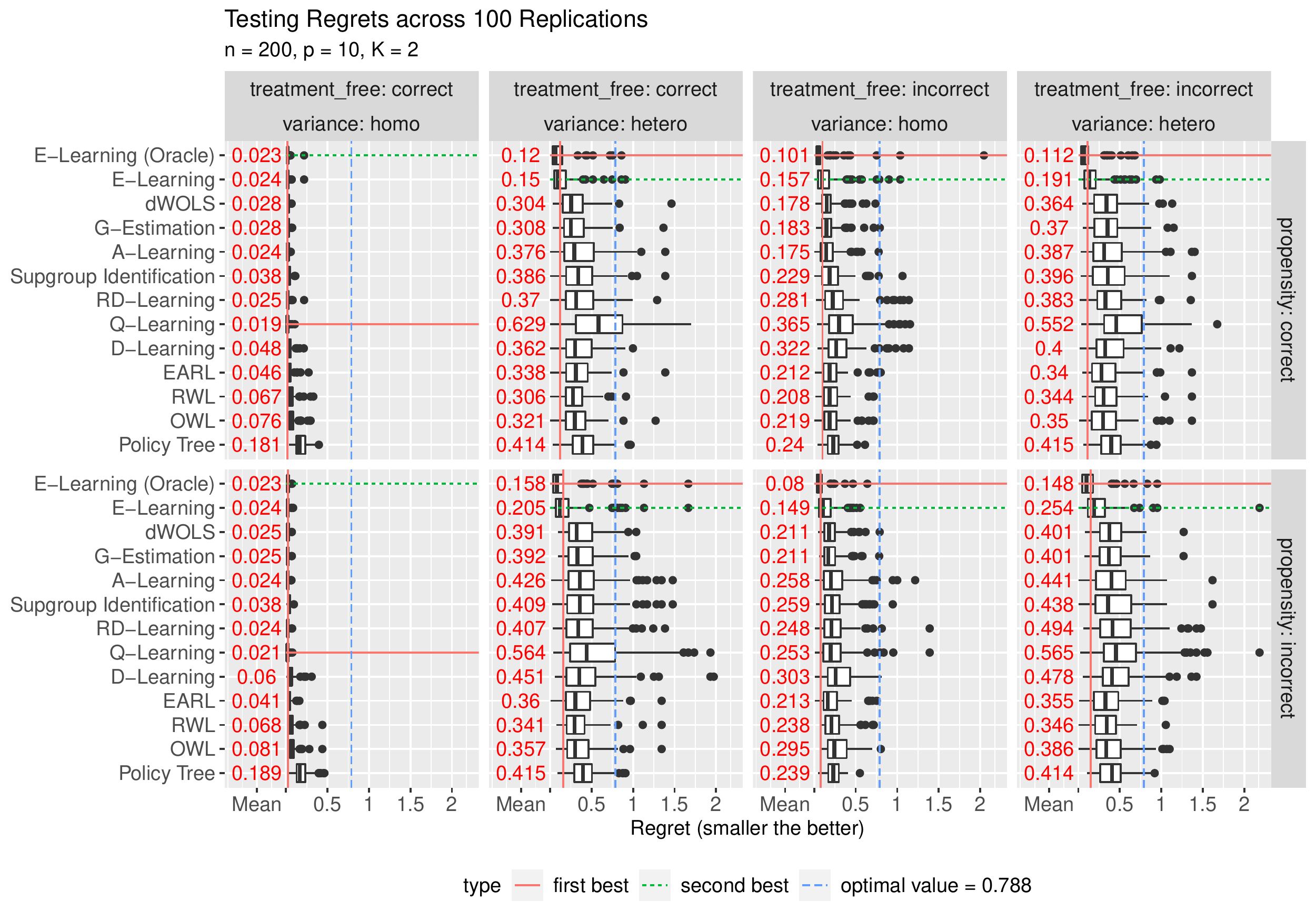}
	\caption[Simulation, Regrets, $ n = 200 $]{Testing regrets (smaller the better) for $ n = 200 $, $ p = 10 $, $ K = 2 $ and each of the model specification scenarios in Table \ref{tab:model}. Methods in Table \ref{tab:compare} are compared, where \textit{E-Learning (Oracle)} corresponds to E-Learning with the oracle working variance function, and \textit{Policy Tree} corresponds to \textit{Policy Learning} with decision trees. First and second best methods in terms of the averaged regrets are annotated in horizontal lines, while the minimal averaged regret is annotated in the vertical line.  The optimal value is 0.788 and is annotated in the vertical long dashed line.}
	\label{fig:all_n=200_regret}
\end{figure}

\begin{figure}[p]
	\centering
	\includegraphics[width=\linewidth]{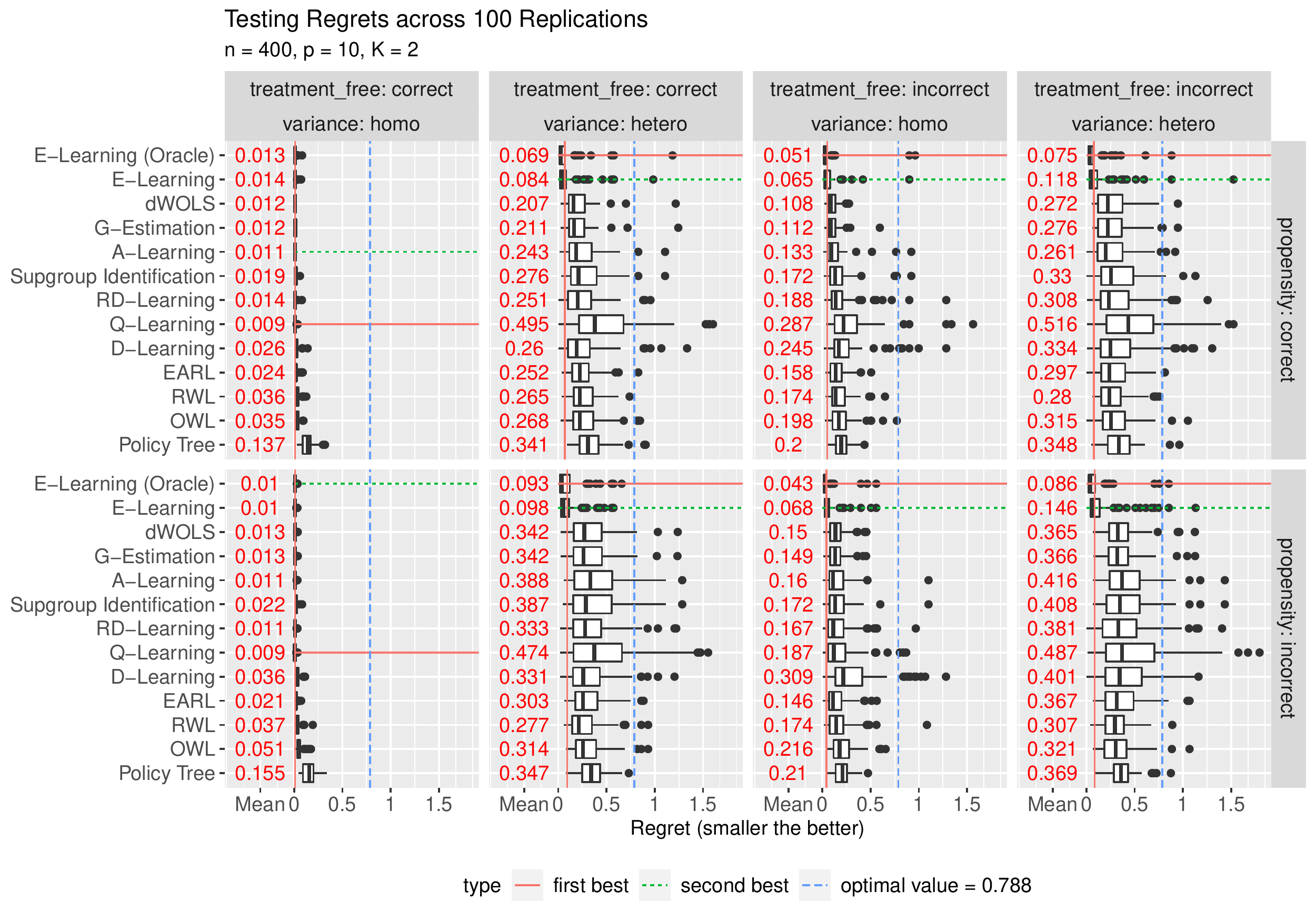}
	\caption[Simulation, Regrets, $ n = 400 $]{Testing regrets (smaller the better) for $ n = 400 $, $ p = 10 $, $ K = 2 $ and each of the model specification scenarios in Table \ref{tab:model}. Methods in Table \ref{tab:compare} are compared, where \textit{E-Learning (Oracle)} corresponds to E-Learning with the oracle working variance function, and \textit{Policy Tree} corresponds to \textit{Policy Learning} with decision trees. First and second best methods in terms of the averaged regrets are annotated in horizontal lines, while the minimal averaged regret is annotated in the vertical line. The optimal value is 0.788 and is annotated in the vertical long dashed line.}
	\label{fig:all_n=400_regret}
\end{figure}

\begin{figure}[p]
	\centering
	\includegraphics[width=\linewidth]{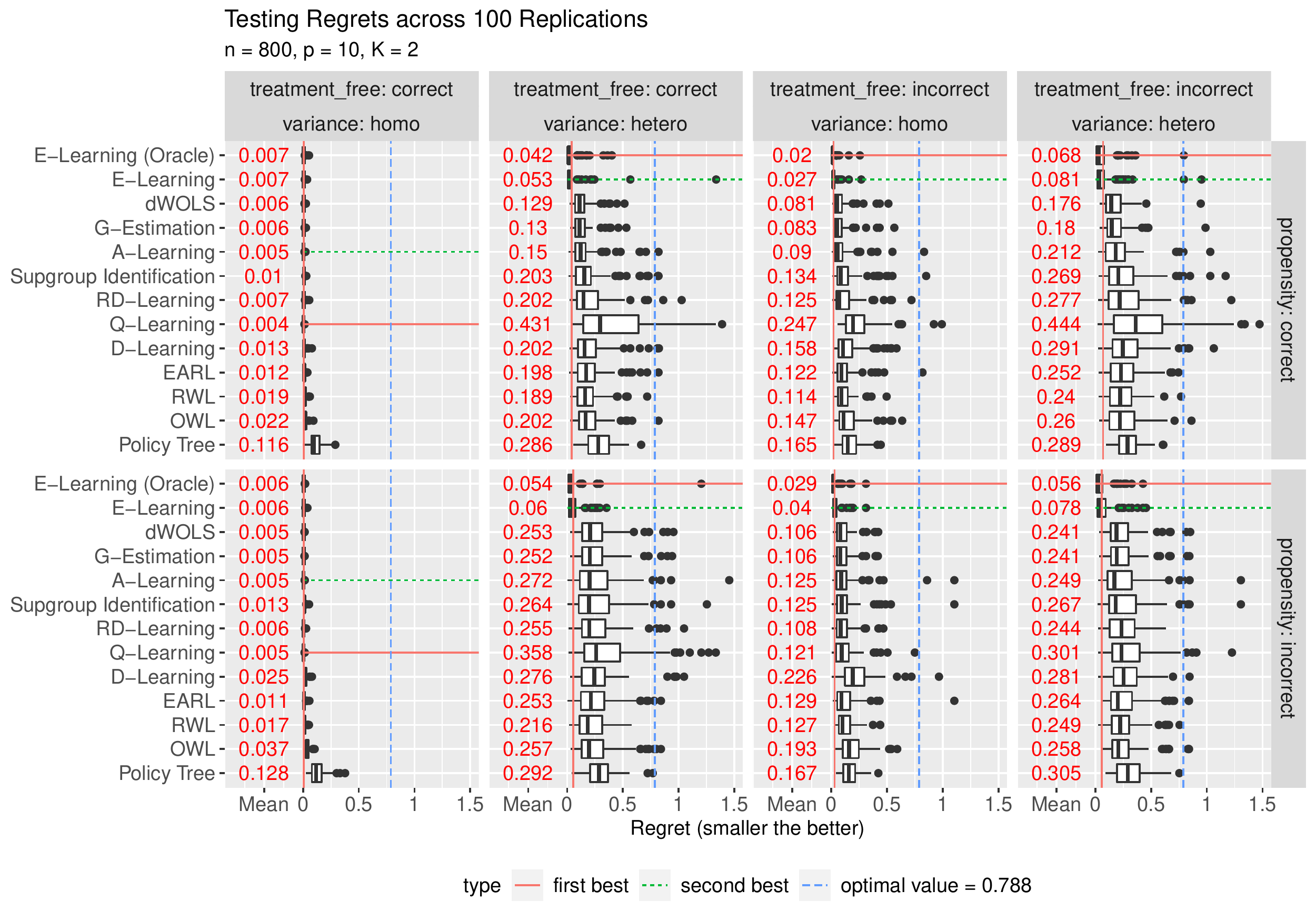}
	\caption[Simulation, Regrets, $ n = 800 $]{Testing regrets (smaller the better) for $ n = 800 $, $ p = 10 $, $ K = 2 $ and each of the model specification scenarios in Table \ref{tab:model}. Methods in Table \ref{tab:compare} are compared, where \textit{E-Learning (Oracle)} corresponds to E-Learning with the oracle working variance function, and \textit{Policy Tree} corresponds to \textit{Policy Learning} with decision trees. First and second best methods in terms of the averaged regrets are annotated in horizontal lines, while the minimal averaged regret is annotated in the vertical line. The optimal value is 0.788 and is annotated in the vertical long dashed line.}
	\label{fig:all_n=800_regret}
\end{figure}

\begin{figure}[p]
	\centering
	\includegraphics[width=\linewidth]{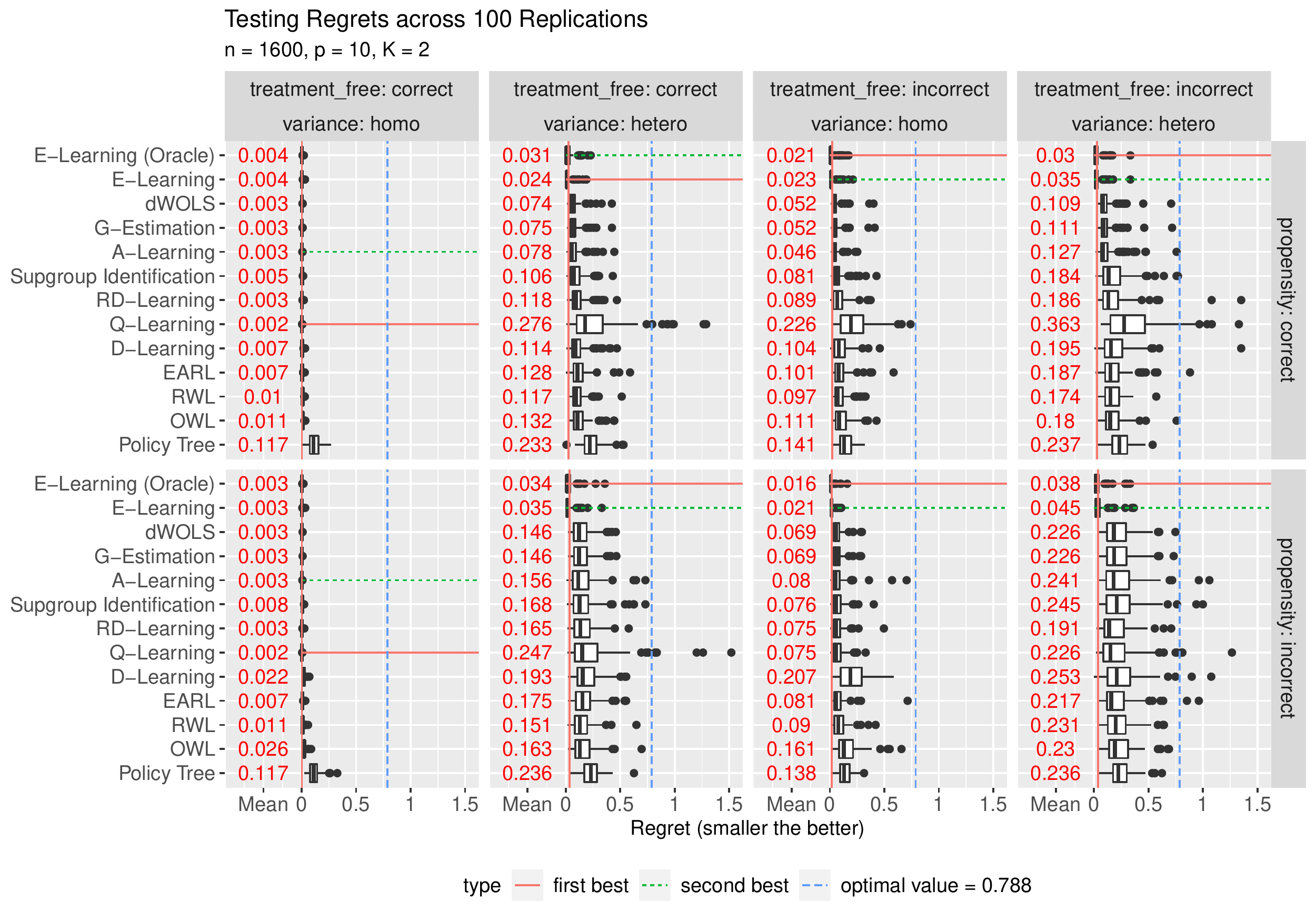}
	\caption[Simulation, Regrets, $ n = 1600 $]{Testing regrets (smaller the better) for $ n = 1600 $, $ p = 10 $, $ K = 2 $ and each of the model specification scenarios in Table \ref{tab:model}. Methods in Table \ref{tab:compare} are compared, where \textit{E-Learning (Oracle)} corresponds to E-Learning with the oracle working variance function, and \textit{Policy Tree} corresponds to \textit{Policy Learning} with decision trees. First and second best methods in terms of the averaged regrets are annotated in horizontal lines, while the minimal averaged regret is annotated in the vertical line. The optimal value is 0.788 and is annotated in the vertical long dashed line.}
	\label{fig:all_n=1600_regret}
\end{figure}

\begin{figure}[p]
	\centering
	\includegraphics[width=\linewidth]{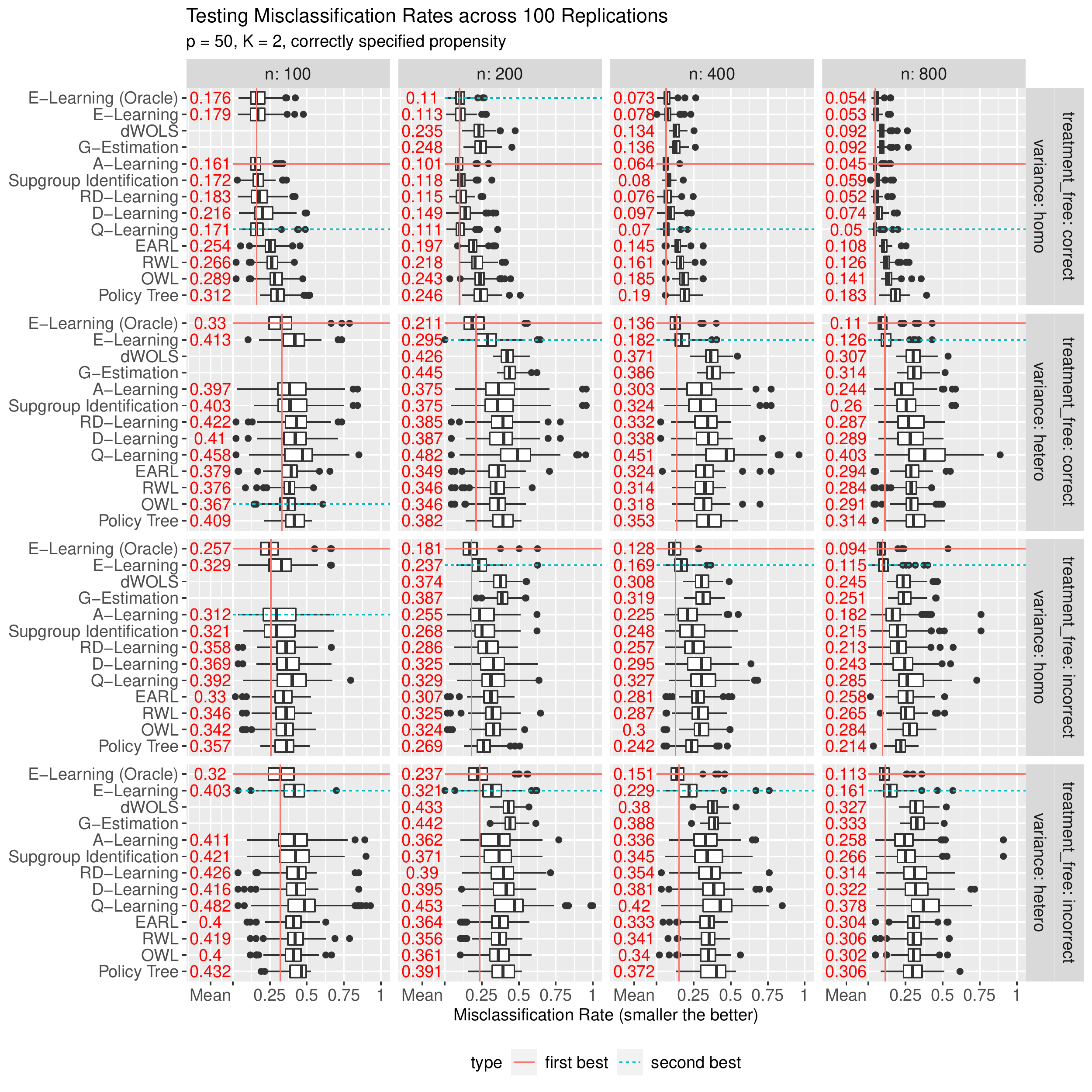}
	\caption[Simulation, Misclassification Rates, $ p = 50 $]{Testing misclassification rates (smaller the better) for $ n \in \{100,200,400,800\} $, $ p = 50 $, $ K = 2 $ and each of the model specification scenarios (correct propensity score) in Table \ref{tab:model}. The optimal value is 0.788. Methods in Table \ref{tab:compare} are compared, where \textit{E-Learning (Oracle)} corresponds to E-Learning with the oracle working variance function, and \textit{Policy Tree} corresponds to \textit{Policy Learning} with decision trees. \textit{dWOLS} and \textit{G-Estimation} for $ n = 100 $ cannot be implemented due to more number of parameters $ 2(p+1) $ than the training sample size $ n $. First and second best methods in terms of the averaged misclassification rates are annotated in horizontal lines, while the minimal averaged misclassification rates is annotated in the vertical line. }
	\label{fig:all_p_misclass}
\end{figure}

\begin{figure}[p]
	\centering
	\includegraphics[width=\linewidth]{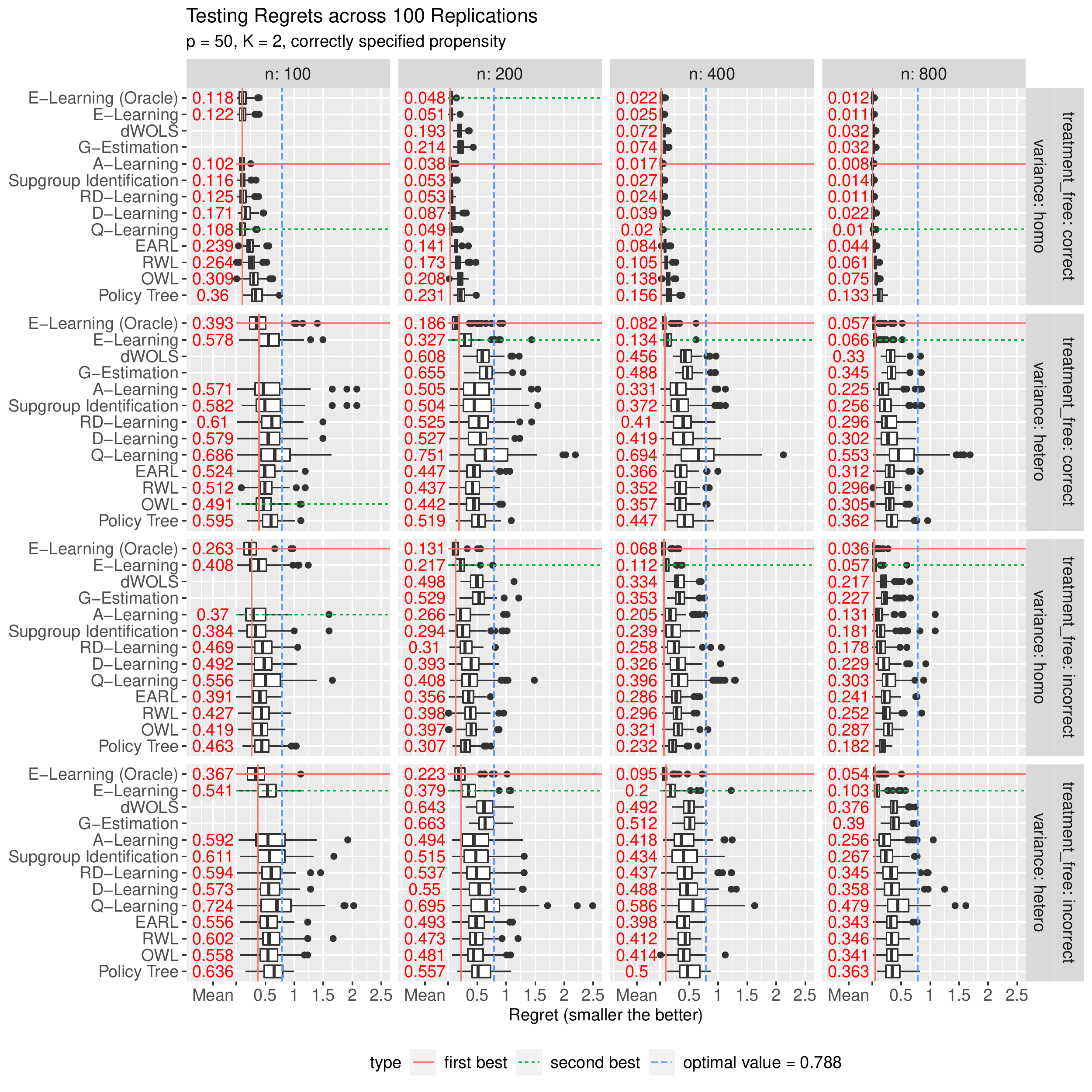}
	\caption[Simulation, Regrets, $ p = 50 $]{Testing regrets (smaller the better) for $ n \in \{100,200,400,800\} $, $ p = 50 $, $ K = 2 $ and each of the model specification scenarios (correct propensity score) in Table \ref{tab:model}. The optimal value is 0.788. Methods in Table \ref{tab:compare} are compared, where \textit{E-Learning (Oracle)} corresponds to E-Learning with the oracle working variance function, and \textit{Policy Tree} corresponds to \textit{Policy Learning} with decision trees. \textit{dWOLS} and \textit{G-Estimation} for $ n = 100 $ cannot be implemented due to more number of parameters $ 2(p+1) $ than the training sample size $ n $. First and second best methods in terms of the averaged regrets are annotated in horizontal lines, while the minimal averaged regret is annotated in the vertical line.}
	\label{fig:all_p_regret}
\end{figure}

\begin{figure}[p]
	\centering
	\includegraphics[width=0.95\linewidth]{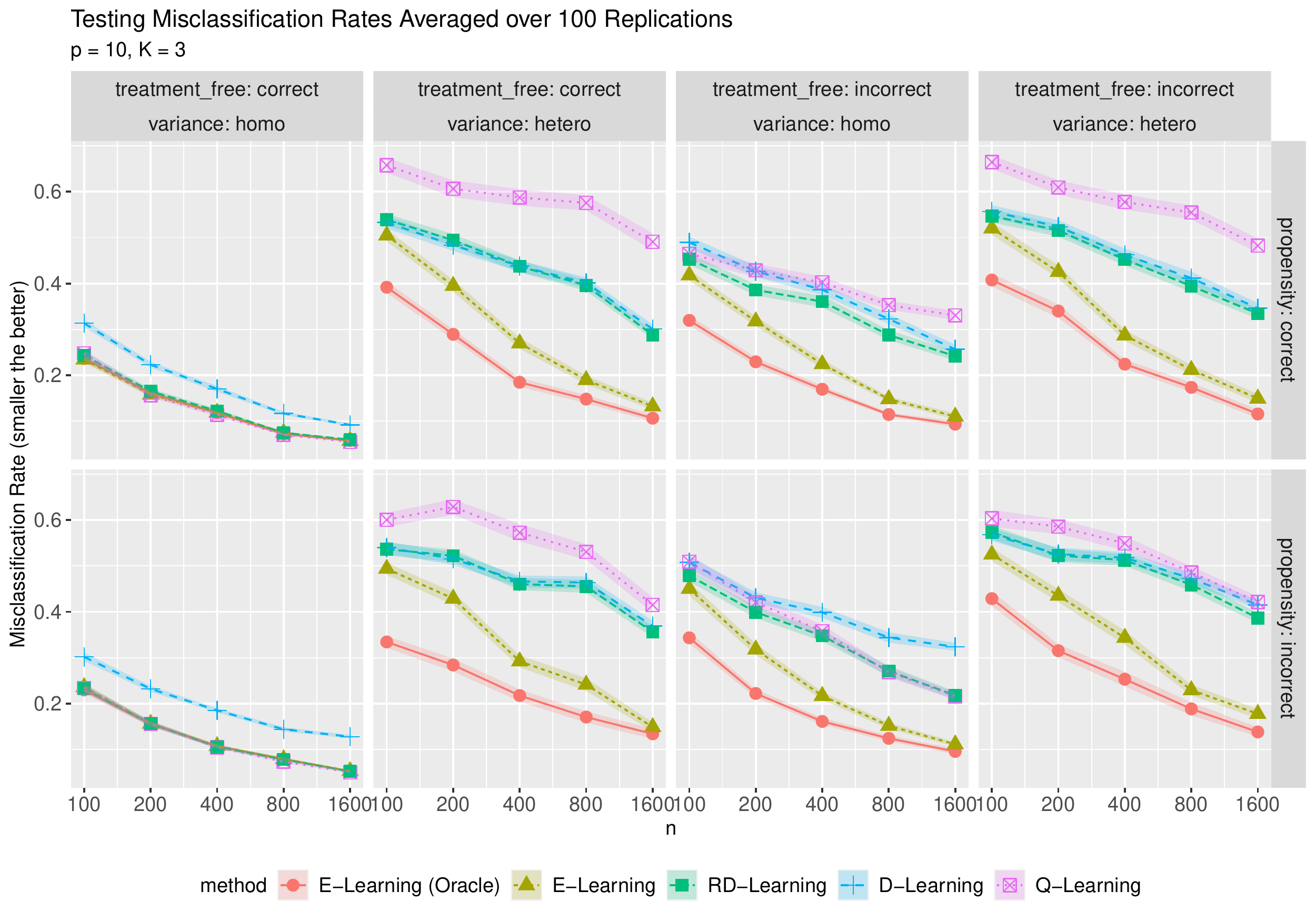}
	\includegraphics[width=0.95\linewidth]{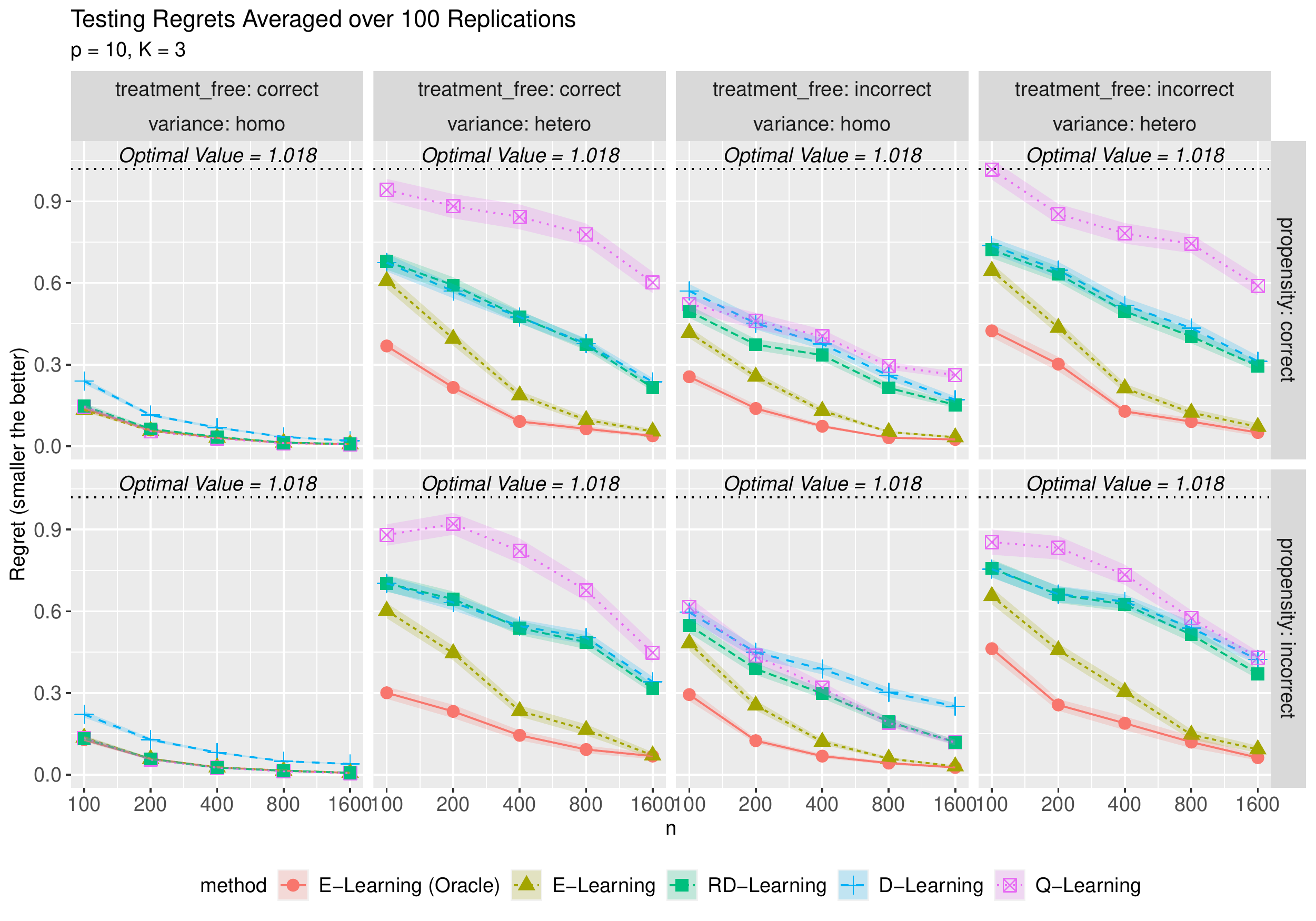}
	\caption[Simulation, $ K = 3 $]{\footnotesize Testing misclassification rates and regrets (smaller the better) for $ n = \{100,200,400,800,1600\}  $, $ p = 10 $, $ K = 3 $ and each of the model specification scenarios in Table \ref{tab:model}. \textit{E-Learning (Oracle)} corresponds to E-Learning with the oracle working variance function, and \textit{E-Learning} corresponds to E-Learning with the working variance function estimated by regression forest.}
	\label{fig:n}
\end{figure}

\begin{figure}[p]
	\centering
	\includegraphics[width=0.95\linewidth]{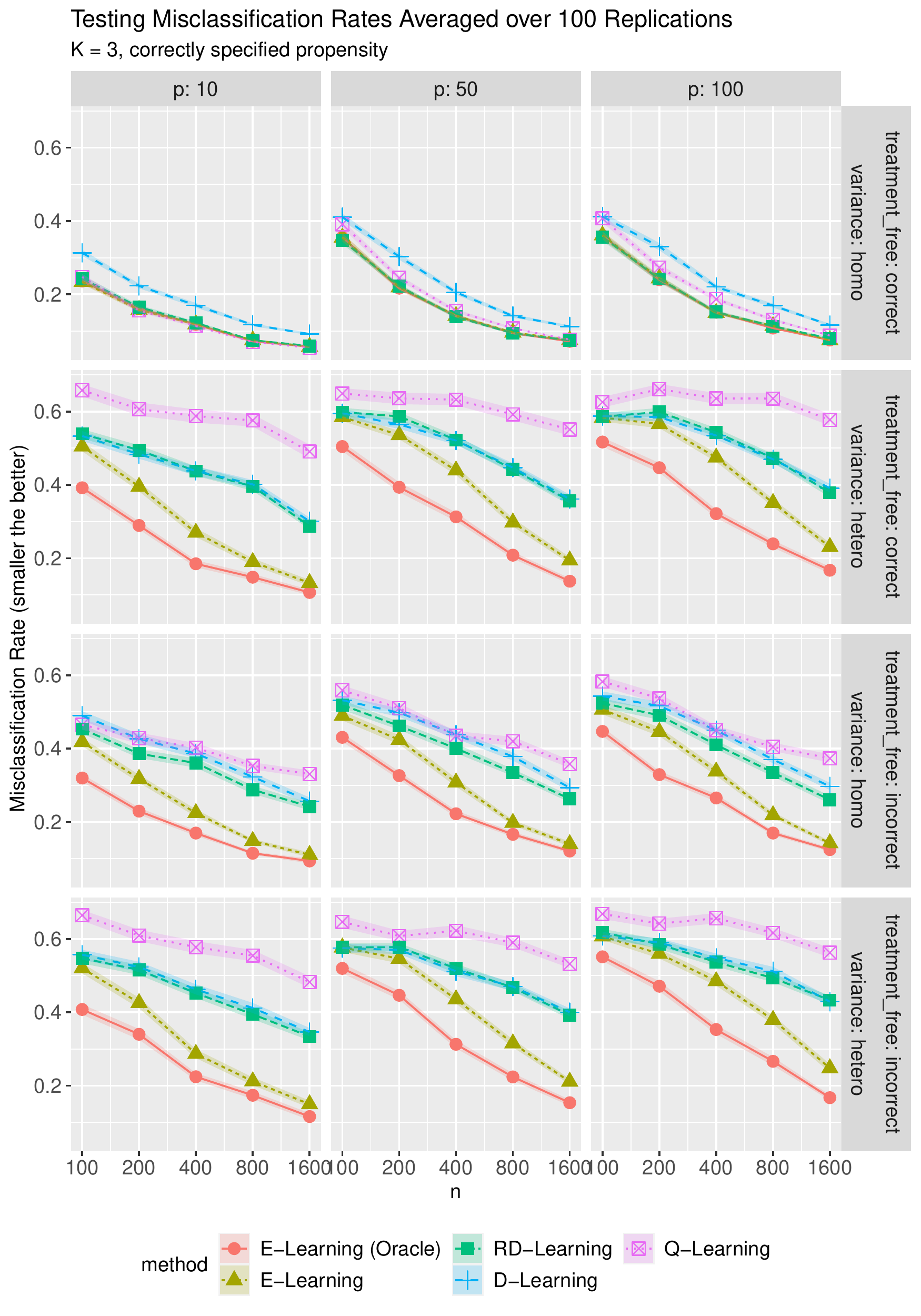}
	\caption[Simulation, Misclassification Rates, $ p = 50, 100 $, $ K = 3 $]{\footnotesize Testing misclassification rates (smaller the better) for $ n = \{100,200,400,800,1600\}  $, $ p \in \{ 10,50,100 \} $, $ K = 3 $ and each of the model specification scenarios with correctly specified propensity score in Table \ref{tab:model}. \textit{E-Learning (Oracle)} corresponds to E-Learning with the oracle working variance function, and \textit{E-Learning} corresponds to E-Learning with the working variance function estimated by regression forest.}
	\label{fig:p_misclass}
\end{figure}

\begin{figure}[p]
	\centering
	\includegraphics[width=0.95\linewidth]{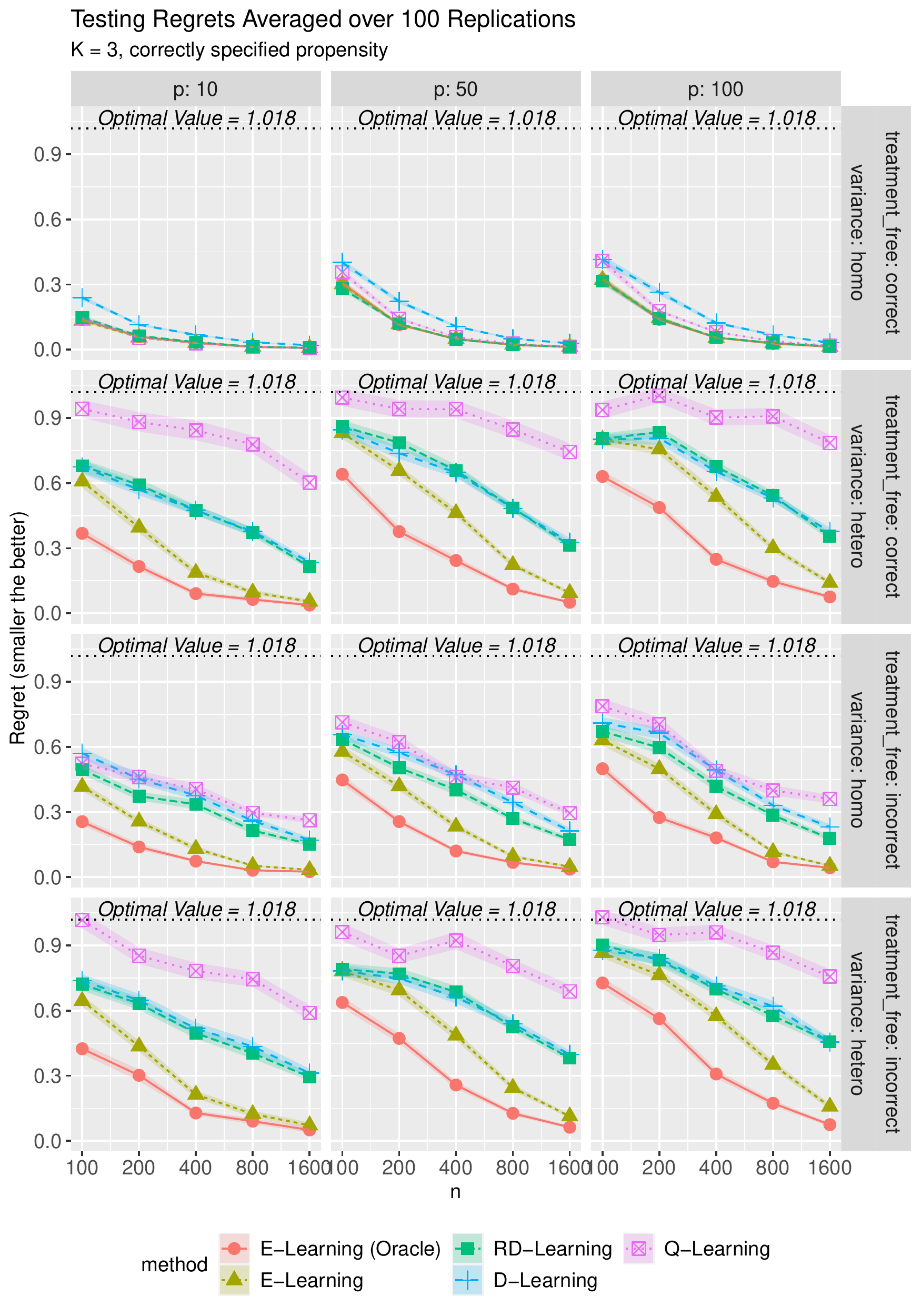}
	\caption[Simulation, Regrets, $ p = 50, 100 $, $ K = 3 $]{\footnotesize Testing regrets (smaller the better) for $ n = \{100,200,400,800,1600\}  $, $ p \in \{ 10,50,100 \} $, $ K = 3 $ and each of the model specification scenarios with correctly specified propensity score in Table \ref{tab:model}. \textit{E-Learning (Oracle)} corresponds to E-Learning with the oracle working variance function, and \textit{E-Learning} corresponds to E-Learning with the working variance function estimated by regression forest.}
	\label{fig:p_regret}
\end{figure}

\begin{figure}[p]
	\centering
	\includegraphics[width=\linewidth]{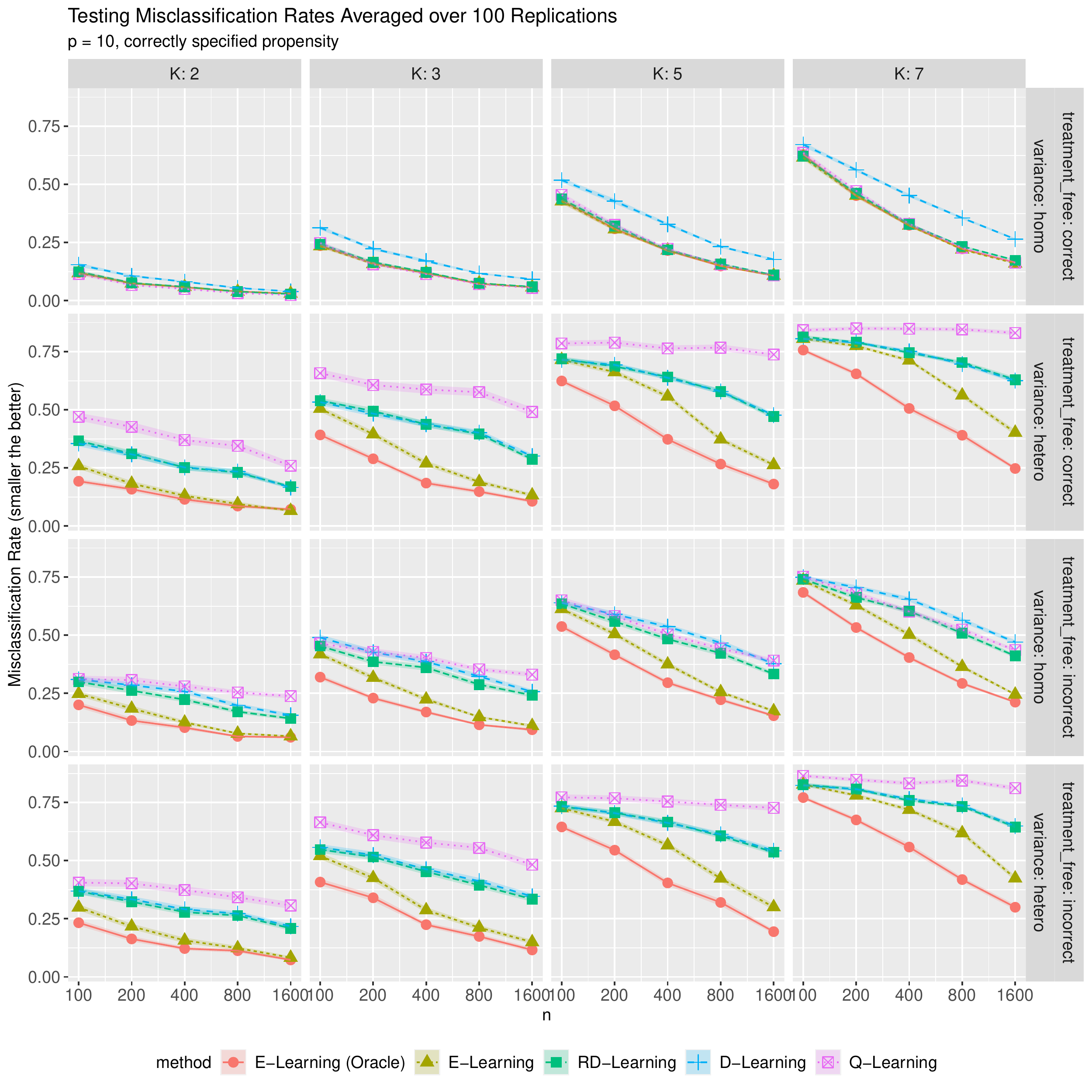}
	\caption[Simulation, Misclassification Rates, $ K = 2,3,5,7 $]{Testing misclassification rates (smaller the better) for $ n = \{100,200,400,800,1600\}  $, $ p = 10 $, $ K \in \{ 2,3,5,7 \} $ and each of the model specification scenarios with correctly specified propensity score in Table \ref{tab:model}. \textit{E-Learning (Oracle)} corresponds to E-Learning with the oracle working variance function, and \textit{E-Learning} corresponds to E-Learning with the working variance function estimated by regression forest.}
	\label{fig:K_misclass}
\end{figure}

\begin{figure}[p]
	\centering
	\includegraphics[width=\linewidth]{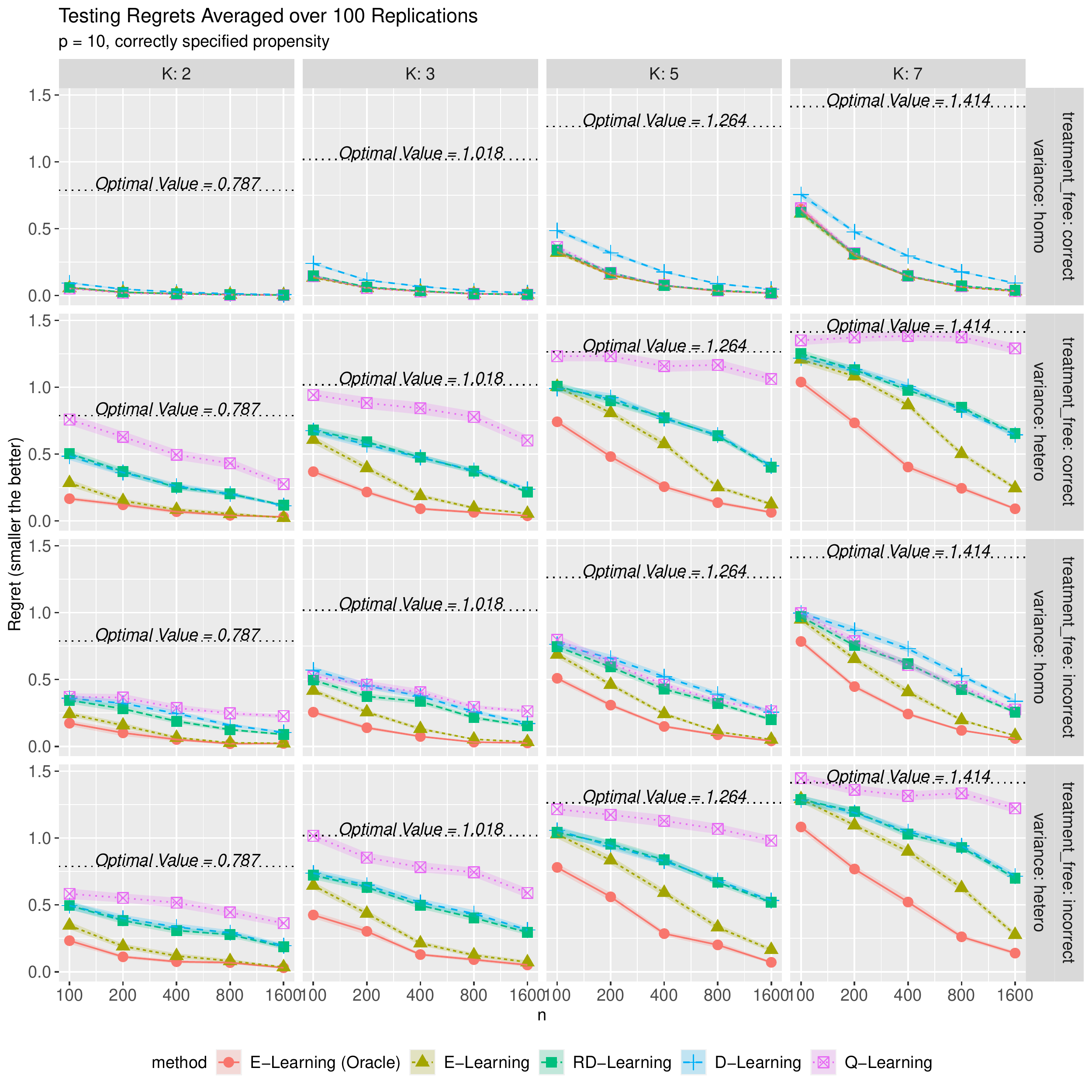}
	\caption[Simulation, Regrets, $ K = 2,3,5,7 $]{Testing regrets (smaller the better) for $ n = \{100,200,400,800,1600\}  $, $ p = 10 $, $ K \in \{ 2,3,5,7 \} $ and each of the model specification scenarios with correctly specified propensity score in Table \ref{tab:model}. \textit{E-Learning (Oracle)} corresponds to E-Learning with the oracle working variance function, and \textit{E-Learning} corresponds to E-Learning with the working variance function estimated by regression forest.}
	\label{fig:K_regret}
\end{figure}

\begin{figure}[p]
	\centering
	\includegraphics[width=0.95\linewidth]{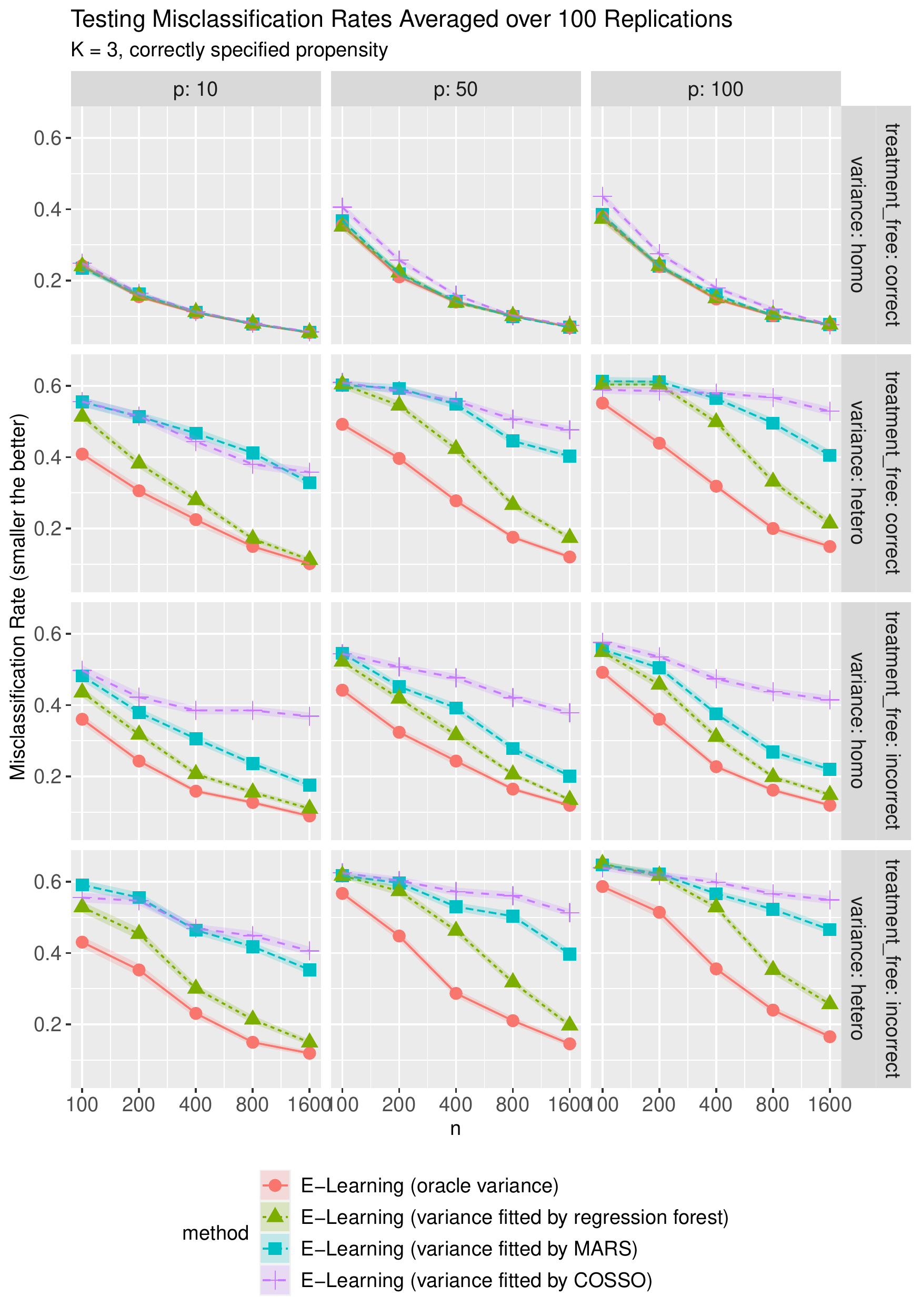}
	\caption[Simulation, Misclassification Rates, Variance Function Estimates]{\footnotesize Testing misclassification rates (smaller the better) for $ n = \{100,200,400,800,1600\}  $, $ p \in \{ 10,50,100 \} $, $ K = 3 $ and each of the model specification scenarios with correctly specified propensity score in Table \ref{tab:model}. The E-Learning procedures with different nonparametric estimation methods for variance function are compared.}
	\label{fig:var_misclass}
\end{figure}

\begin{figure}[p]
	\centering
	\includegraphics[width=0.95\linewidth]{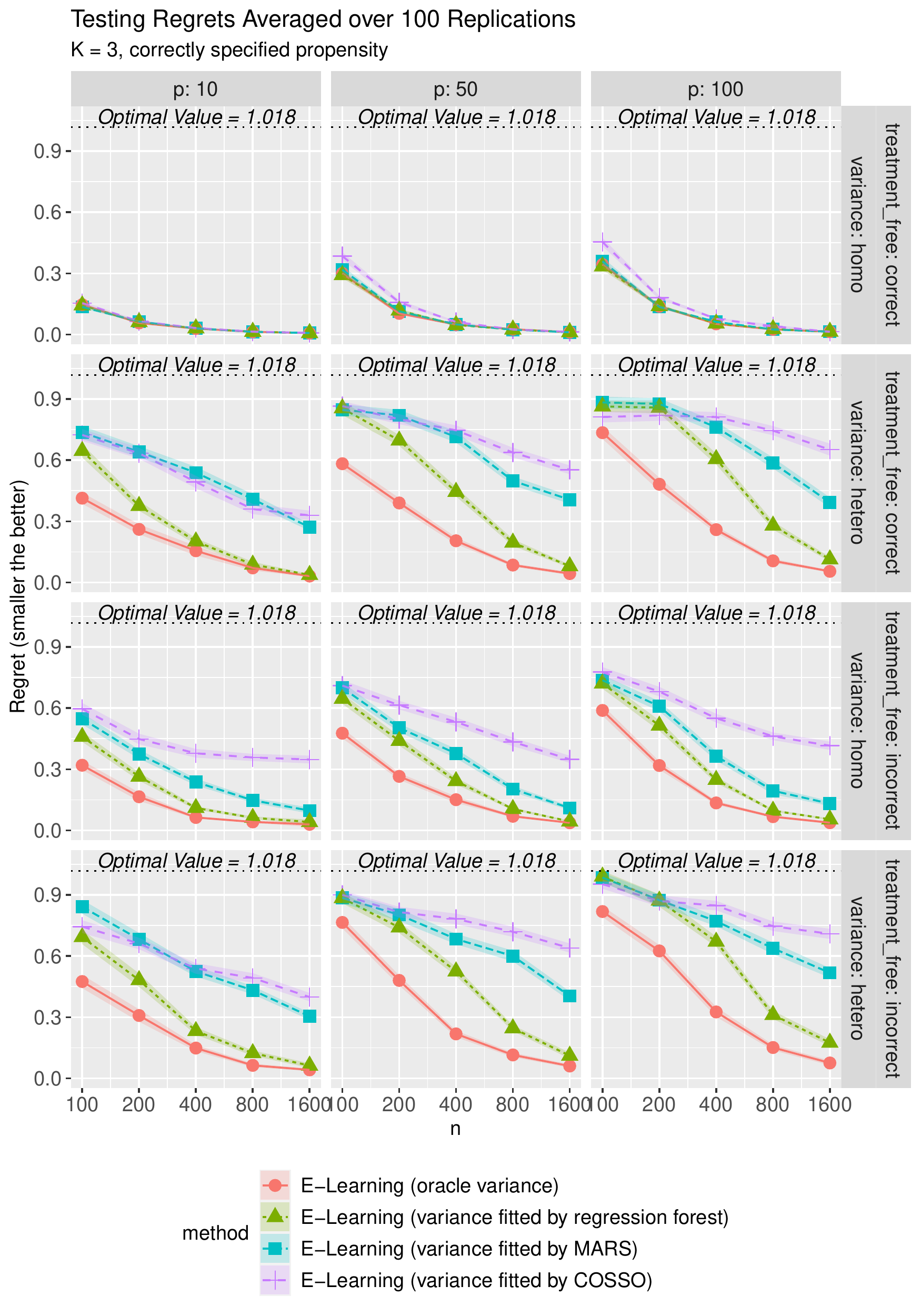}
	\caption[Simulation, Regrets, Variance Function Estimates]{\footnotesize Testing regrets (smaller the better) for $ n = \{100,200,400,800,1600\}  $, $ p \in \{ 10,50,100 \} $, $ K = 3 $ and each of the model specification scenarios with correctly specified propensity score in Table \ref{tab:model}. The E-Learning procedures with different nonparametric estimation methods for variance function are compared.}
	\label{fig:var_regret}
\end{figure}

\clearpage

\bibliography{bibfile.bib}
\bibliographystyle{asa}